\documentclass[11pt]{article}
\pdfoutput=1

\newlength{\abstractwidth}
\usepackage{amsmath}
\usepackage{amsfonts}
\usepackage{amssymb}
\usepackage{latexsym}
\usepackage{appendix}
\usepackage{verbatim}

\numberwithin{equation}{section}

\usepackage{bbold}

\usepackage{epsf}
\usepackage{color}
\usepackage{graphicx}
\usepackage{dsfont}
\usepackage{subfigure}

\usepackage[export]{adjustbox}
\usepackage{hyperref}
\definecolor{darkred}{rgb}{0.8,0.1,0.1}
\hypersetup{colorlinks=true, linkcolor=darkred, citecolor=blue, linktoc=page}

\usepackage{caption}
\usepackage{subcaption}

\graphicspath{{Figures/}}

\newcommand{\be}{\begin{equation}}
\newcommand{\ee}{\end{equation}}

\renewcommand{\>}{\rangle}

\def\cA{{\cal A}}

\def\cM{{\cal M}}
\def\cN{{\cal N}}

\def\zbar{{\bar z}}

\def\bsb{{\boldsymbol{b}}}
\def\bsx{{\boldsymbol{x}}}
\def\bsy{{\boldsymbol{y}}}
\def\bsz{{\boldsymbol{z}}}
\def\bsw{{\boldsymbol{w}}}

\def\bsk{{\boldsymbol{k}}}
\def\bsq{{\boldsymbol{q}}}
\def\bsp{{\boldsymbol{p}}}
\def\bsd{{\boldsymbol{\delta}}}
\def\bel{{\boldsymbol{\ell}}}
\def\bszero{{\boldsymbol{0}}}

\def\z{{\zeta}}

\def\Re{{\rm Re \,}}
\def\Im{{\rm Im \,}}

\def\Tr{{\rm Tr}}

\def\p{\partial}

\def\a{\alpha}
\def\b{\beta}

\def\eps{\epsilon}

\def\ep{\varepsilon}

\def\l{\lambda}
\def\d{\delta}
\def\s{\sigma}
\def\m{\mu}
\def\n{\nu}
\def\){\right)}
\def\({\left( }
\def\]{\right] }
\def\[{\left[ }

\def\no{\nonumber}

\makeatother

\begin{document}

\vskip 0.3in

\begin{center}
{\Large \bf  QCD-Gravity double copy in  Regge asymptotics:\\[10pt] from $2\rightarrow n$ amplitudes to radiation in shockwave collisions}
\vskip 0.4in
{\large   Himanshu Raj$^a$ and  Raju Venugopalan$^{a,b}$} 
\vskip .2in

$^a$ {\it Center for Frontiers in Nuclear Science, Department of Physics and Astronomy,}\\
{\it Stony Brook University, Stony Brook, NY 11794, USA}\\[0.5cm]

$^b${\it  Department of Physics, Brookhaven National Laboratory,}\\
{\it  Upton, NY 11973, USA} \\ 
\vskip 1.2in
 
\begin{abstract}
These lectures discuss multi-particle production in QCD and in gravity at ultrarelativistic energies, their double copy relations, and strong parallels in emergent shockwave dynamics.  Dispersive techniques are applied to derive the BFKL equation for multi-gluon production in Regge asymptotics. Identical methods apply in gravity and are captured by a gravitational Lipatov equation. The building blocks in both cases are Lipatov vertices and reggeized propagators satisfying double copy relations; in gravity, Weinberg's soft  theorem is recovered as a limit of the Lipatov framework. BFKL evolution in QCD generates wee parton states of maximal occupancy characterized by an emergent semi-hard saturation scale. Renormalization group equations in the Color Glass Condensate (CGC)  EFT describe wee parton correlations and their rapidity evolution. A  shockwave picture of deeply inelastic scattering and hadron-hadron collisions follows, with multi-particle production described by Cutkosky's rules in strong time-dependent fields. Gluon radiation in the CGC EFT has a double copy in gravitational shockwave collisions, with a similar correspondence applicable between gluon and graviton shockwave propagators. Possible extensions of this semi-classical double copy are outlined for computing multi-particle production in gravitational shockwave collisions, self-force and tidal contributions, and  classical and quantum noise in the focusing of geodesics.

\end{abstract}
\end{center}

\baselineskip=16pt
\setcounter{equation}{0}
\setcounter{footnote}{0}

\newpage
\tableofcontents

\section{Introduction}
\label{sec:1}

An outstanding problem in QCD is to arrive at a first principles understanding of $2\to n$ multi-particle production at collider energies. The Relativistic Heavy Ion Collider (RHIC) at Brookhaven National Laboratory and the Large Hadron Collider (LHC) at CERN are multi-particle factories, with a single heavy-ion collision  generating thousands of sub-atomic particles. In this deeply Lorentzian regime of the theory, nonperturbative Euclidean methods are invalid. The bulk of the multi-particle spectrum at colliders is ultrasoft, with transverse momenta $p_\perp < \Lambda_{\rm QCD}$ where perturbative QCD is inapplicable. This is an embarrassment in praxis\footnote{A Greek word, whose meaning in this context can be translated as the ``desire for active engagement and purposeful endeavor" of physicists with experiment. } even for this ``nearly perfect" quantum field theory since we have limited reliable access to a vast phase space of rich many-body quark-gluon phenomena\footnote{This comment also relevant for the phase diagram of QCD at finite temperatures and baryon chemical potentials.}. 
However with increasing ultrarelativistic center of mass energies ($\sqrt{s}$) the hadron spectrum broadens with average $\langle p_\perp\rangle > \Lambda_{\rm QCD}$. Since the underlying gluon and quark degrees of freedom generating this spectrum are significantly harder, the parton picture is increasingly viable, and due to asymptotic freedom, fully robust for $p_\perp \rightarrow \sqrt{s}$.

These transitions are cleaner  in deeply inelastic scattering (DIS) experiments where one has independent control over the resolution\footnote{Its equivalent in the language of $2\to 2+n$ amplitudes is the Mandelstam variable $t \equiv (p_1-\ell_0)^2$, where $\ell_0$ is the momenta of one of the incoming particles after the collision--see Fig. \ref{2-to-n-amplitude}.}($Q^2$) and the center-of mass-energy. Precision DIS experiments at HERA show that at fixed large $Q^2$ (where 
the QCD coupling $\alpha_S(Q^2)\ll 1$), the inclusive cross section grows rapidly with decreasing Bj\"{o}rken $x_{\rm Bj}\approx Q^2/s$, which, like $Q^2$ is a Lorentz scalar. In Feynman's parton model, $x_{\rm Bj}\sim x\ll 1$, the longitudinal momentum fraction of a ``wee" struck parton. 
Thus with decreasing $x$, one has access to a $2\to n$ regime (albeit spacelike) where weak coupling methods may be employed. 

The paradigmatic approach in this ``small $x$ Regge asymptotics" is the Balitsky-Fadin-Kuraev-Lipatov (BFKL) framework that provides a systematic approach to multi-particle production when $n\gg 1$. Since the number of Feynman diagrams explode factorially with $n$, it is truly remarkable that any sort of quantitative approach is reliable, a feat achieved by BFKL through powerful application of dispersive methods, and systematic resummation of leading contributions in $x$ at each order in perturbation theory.  There are many subtleties and caveats to be attached to this approach, but they too can be quantified. We will attempt in these lectures to elucidate the BFKL approach to $2\rightarrow n$ scattering, outline its regime of validity, and describe the rich many-body physics that emerges when the framework breaks down as $x\rightarrow 0$.

Equally strikingly, the BFKL paradigm has a quantitative counterpart in $2\rightarrow n$ gravitational scattering in the trans-Planckian regime. This was worked out in a pair of remarkable papers by Lipatov more than 40 years ago, who showed that the dispersive methods developed for QCD apply identically to gravity in the ultrarelativistic regime. Further, Lipatov noticed a double copy between the effective vertices in QCD and in gravity that are the fundamental objects in the construction of $2\to n$ amplitudes. This work predates the observation by  Kawai, Lewellen and Tye (KLT)~\cite{Kawai:1985xq} of  a double copy between Yang-Mills theory and Einstein gravity in the low energy limit of string theory, and the subsequent explosion of interest in double copies, and their relevance for gravitational radiation, following the seminal work of Bern, Carrasco and Johansson (BCJ) that we will briefly discuss later. 

Implicit in Lipatov's work is the understanding that the ultrarelativistic limit of Weinberg's soft graviton theorem arises as limit of his framework, providing a smooth extension of this theorem to the regime where Mandelstam $t$ is not ultrasoft. This realization connects his work to the burgeoning literature on ``celestial amplitudes", spelling out the ``triangle" of connections between soft theorems, asymptotic symmetries, and memory effects, as first shown by Strominger to be generic both to gauge theories and gravity. Lipatov understood further that his $2\to n$ scattering amplitude results for both QCD and gravity could be recast as an effective field theory (EFT), whose dynamics is captured by the interaction of emergent reggeon degrees of freedom with gluons and gravitons, respectively. Identical conclusions for gravity from the perspective of trans-Planckian superstring amplitudes, were reached by Amati, Ciafaloni and Veneziano (ACV) who employed this EFT description to great effect in quantifying classical and quantum contributions to scattering amplitudes. We will discuss some aspects of their work in these lectures. 

All of the aforementioned work provides essential insight into a first principles $S$-matrix approach, as articulated by `t Hooft, to the problem of black hole formation in quantum field theory. Important ingredients, as we will discuss, are $t$-channel fractionation and $s$-channel classicalization, leading to a dominantly classical picture, with sub-leading quantum corrections that are relevant for black hole formation and subsequent evaporation. In particular, some elements of the Lipatov construction have a much smaller window of applicability in gravity than in QCD, which others remain important in the emergent classical framework; this subtlety is sometimes a source of confusion. 

The emergent paradigm, both in QCD (in Regge asymptotics) and in gravity, is of shockwave scattering and multi-particle production in these collisions. In QCD, this dynamics is captured in a Color Glass Condensate (CGC) EFT which incorporates essential features of the BFKL framework. This is a theory of stochastic static  (large $x$) color sources coupled to dynamical (small $x$) gauge fields; small $x$ operators are computed for each 
static configuration of color charges, and averaged over with a nonperturbative gauge invariant weight functional (density matrix) containing nontrivial information on large $x$ modes. Shockwave propagators are computed from semi-classical fluctuations in the CGC background. With this, leading $\alpha_S\ln(1/x)\sim O(1)$ contributions can be extracted and summed to all orders. This summation is described by renormalization group (RG) equations for the rapidity evolution of lightlike Wilson line correlators, which capture the many-body dynamics of wee parton correlations. The BFKL equation results from taking a low parton density limit of the RG equation for two-point correlators. 

In gravity, we will show similarly to QCD that the Lipatov vertex emerges in shockwave scattering. Likewise, shockwave propagators satisfy a double copy relation. Thus emboldened, we can extend our double copy insights to multi-particle production. In doing so, we make use of a systematic dilute-dilute, dilute-dense and dense-dense power counting scheme, where one expands in ratios of densities of the shockwave sources relative to 
the impact parameter corresponding to inelastic emissions. This ordering, combined with the ultrarelativistic limit of shockwave scattering, allows us some measure of analytic insight into radiation that occurs in the strong field regime. In the dilute-dense approximation, one should be able to compute rescattering contributions, in particular self-force and tidal effects that formally appear at high order in the standard post-Minkowskian expansion. In the  dense-dense shockwave scattering regime, as in QCD, analytical computations may not be feasible; nevertheless, numerical simulations in this EFT framework may be more computationally efficient than full-blown numerical relativity simulations. 

From a conceptual perspective, the holy grail would be to follow the $2\to n$ framework all the way through black hole formation and evaporation. A key element, which we touch on briefly is congruence of null geodesics resulting from copious inelastic production  and rescattering of gravitons in the strong field regime. Classically, this is described by the Raychaudhuri equation, and a useful goal would be a RG based understanding this equation and how quantum effects influence it. This line of inquiry may be valuable in the search for quantum imprints on gravitational radiation. Ironically, the search at colliders in the context of $2\rightarrow n$ scattering is the identification of a robust many-body semi-classical QCD regime. 

Finally, we should emphasize the limited scope of these lectures, with our focus being on the weakly coupled strong field regimes of QCD and gravity. These regimes of course do not exist in isolation from the genuinely strongly coupled regimes of both theories. A powerful approach towards making progress in strongly coupled gauge theories is the well-known AdS/CFT correspondence~\cite{Maldacena:1997re}, and its many descendants. In the context of $2\to n$ scattering, particularly noteworthy  is the BPST framework of Brower, Polchinski, Strassler and Tan~\cite{Polchinski:2002jw,Polchinski:2001tt,Brower:2006ea}, which has spawned a significant literature. There is much common ground to explore, in particular in the context of holography, and in  general features of scattering amplitudes, that may be universal to strong field dynamics irrespective of coupling strength and matter content. Similarly,  in gravity, there is a vast literature on Planck scale physics - for a sense of the various strands in the literature, we refer the reader to the compilation of viewpoints in \cite{Buoninfante:2024yth}. Another loose strand in our coverage is the connection of our work to quantum information science. We have addressed this briefly in the concluding section but clearly this topic deserves a more comprehensive treatment.

\section{\texorpdfstring{$2\to 2+n$}{} processes in the Regge limit of QCD}
\label{sec:2}

We will discuss in this section the  radiation of gluons within the high energy scattering framework of Fadin, Kuraev and Lipatov \cite{Kuraev:1976ge}, and of Balitsky and Lipatov~\cite{Balitsky:1978ic}, together known by the acronym BFKL. Their work made use of dispersive techniques; we will largely follow this approach in the present section. Our treatment is strongly influenced by and complements excellent other reviews on BFKL~\cite{DelDuca:1995hf,Forshaw:1997dc,Ioffe:2010zz,Kovchegov:2012mbw}. We will focus on the leading-logarithm-approximation to $2\rightarrow n$ scattering amplitudes in multi-Regge kinematics. There is a considerable body of work on BFKL at next-to-leading logarithmic accuracy that we will only discuss briefly. For a recent state-of-the art review, we refer the reader to \cite{DelDuca:2022skz}. Subsequent to our discussion of the derivation of the BFKL equation, and a discussion of its solution, we will discuss the breakdown of the BFKL framework and the emergent phenomenon of gluon saturation. The latter is described in the Color Glass Condensate (CGC) effective field theory, which will  be discussed at length in Section \ref{sec:CGC}.  

\subsection{Diagrammatic approach to BFKL}
\label{sec:2.1}

\begin{figure}[ht]
\centering
\includegraphics[scale=1]{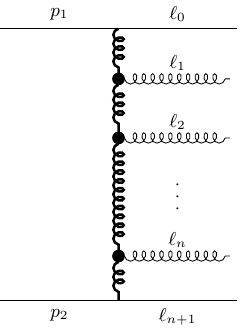}
\caption{Multi-gluon production amplitude in multi-Regge kinematics depicting the two key components in BFKL renormalization group evolution of the cross-section with rapidity: (1) The dark blobs represent the nonlocal Lipatov effective vertices, and (2) the thick vertical gluon lines represent t-channel reggeized gluon propagators incorporating all-order (leading logarithmic in $x$) virtual corrections. The external lines can be any source of glue.}
\label{2-to-n-amplitude}
\end{figure}

The discussion here, and in the following sub-sections, will demonstrate that the imaginary part of the elastic scattering amplitude of two gluons in the QCD Regge limit is dominated in the leading $\log s$ approximation by uncrossed ``effective" ladder diagrams, where the vertices of the ladder are nonlocal ``Lipatov" vertices instead of bare 3-point vertices, with the internal dominant t-channel propagators corresponding to the exchange of reggeized gluons. The color singlet projection of the exchange of two t-channel reggeized gluons is called the BFKL pomeron. The structure of the corresponding $2
\rightarrow 2+n$ inelastic amplitude (half-ladder) is shown schematically in Fig. \ref{2-to-n-amplitude}.

We begin by evaluating the elastic gluon-gluon scattering amplitude in the Regge regime to leading logarithmic accuracy. The latter is a kinematic approximation which is defined as
\begin{equation}
\label{ll-accuracy}
\alpha_S \ln(1/x_i) \sim 1~,\qquad \alpha_S \ll 1~,
\end{equation}
where $\alpha_S=g^2/4\pi$, with $g$ the QCD coupling. Referring to Fig.~\ref{2-to-n-amplitude}, $x_i$ is the longitudinal momentum fraction $x_{i}\sim |\bel_i|/\sqrt{s}$ of the $i$-th  emitted gluon, with transverse momenta $\bel_i$, and the squared center-of-mass energy $s= 2p_1\cdot p_2 $. In the Regge asymptotics of BFKL, one assumes that $\sqrt{s}\gg |{\bel_n}| \gg \Lambda_{\rm QCD}$, where $\Lambda_{\rm QCD}$ is the intrinsic QCD scale; in these kinematics, one can evaluate scattering amplitudes via Feynman diagrams where only leading logarithmic contributions in $\alpha_S \ln(1/x_i)$ are retained.
 
With this in mind, the computation of such terms can be reduced to the computation of the $s$-channel discontinuities of scattering amplitudes\footnote{For a recent modern introduction to the general properties of the analytic S-matrix, we refer readers to \cite{Mizera:2023tfe}.}. For the $2\to 2$ amplitude, the discontinuity is given in terms of its imaginary part by 
\begin{align}
\label{unitarity-condition-0}
    -i\text{Disc}_s \mathcal{A}_{2\to 2} = 2\, \Im \mathcal{A}_{2\to 2} = \sum_{n=0}^\infty \int d({\rm P.S.}^{n+2}) \sum_{\rm{color, polarizations}}\mathcal{A}_{2\to n+2}(\{k\})\mathcal{A}_{2\to n+2}^\dagger(\{k-q\})~.
\end{align}
where $q$ is the total momentum transfer in the amplitude $\mathcal{A}_{2\to 2}$ and $\{k\}$ denotes the momentum labels of the exchanged gluons in the ladder.  Here the sum extends over all possible intermediate states $n$ and $\mathcal{A}_{2\to n+2}$ is the amplitude for the transition $2\to n+2$. The factor $d({\rm P.S.}^{n+2})$ corresponds to the phase space density of the $n$ intermediate particles; we will define it explicitly and simplify it in multi-Regge kinematics (MRK) shortly in Sec.~\ref{sec:n-particle-intermediate}. 
From the discontinuity on the l.h.s, one can reconstruct the full amplitude to the required leading log accuracy  by the replacement
\begin{align}
\label{log-discontinuities}
     s\log^{n}(s/t) \to -\frac{1}{\pi i (n+1) } s \log^{n+1}(-s/t) ~.
\end{align}

For $n=0$ this gives the usual identity (we take the branch cut of the logarithm along the positive real axis)
$$
\Im \log(-s/t) = \frac 12 \text{Disc}_s \log(-s/t) = \frac 12\lim_{\ep\to 0} (\log(-s/t+i\ep)- \log(-s/t-i\ep)) = -i \pi~.
$$
For higher $n$, the exact result for the discontinuity is
\begin{align}
\frac 12 \text{Disc}_s \log^{n+1}(-s/t) &= \frac 12\lim_{\ep\to 0} (\log^{n+1}(-s/t+i\ep)- \log^{n+1}(-s/t-i\ep))\no\\
&= \frac12\big(\log^{n+1}(s/t)-(\log (s/t)+2\pi i)^{n+1} \big) ~.
\end{align}
Since in the Regge limit we have $\log(s/t)\gg 1$, the last equality can be expanded to leading order as $-\pi i(n+1) \log^n(s/t)$ ~\cite{Kuraev:1976ge,Forshaw:1997dc}.

\begin{figure}[ht]
\centering
\subfigure[]{\raisebox{1.45ex}{\includegraphics[scale=1]{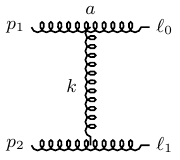}}}
\qquad\qquad
\centering
\subfigure[]{\includegraphics[scale=1]{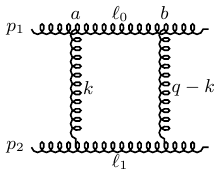}}
\qquad\qquad
\subfigure[]{\includegraphics[scale=1]{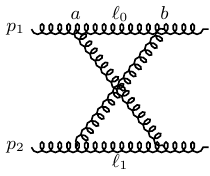}}
\caption{$2\to 2$ gluon scattering amplitude at order $g^2$ and $g^4$.  Fig.~(a) represents the leading order one-gluon exchange Born diagram. Figs.~(b) and (c) are the O($g^4$) two-gluon exchanges that give the lowest order contribution to color-singlet pomeron exchange.}
\label{one-loop-gluon-exchange}
\end{figure}

We will first study the $2\to 2$ scattering of gluons. The incoming gluons carrying momentum $p_1$ and $p_2$, along with respective helicities $\lambda_1$ and $\lambda_2$, scatter into final state gluons with momenta $p_3 = p_1-q$ and $p_4 = p_2+q$, with respective helicities $\lambda_1'$ and $\lambda_2'$. As we will discuss shortly, elastic scattering in the Regge regime (defined as $s\gg -t$, where $s=(p_1+p_2)^2$ and $t=q^2$) proceeds via the exchange of a color singlet object called the {\it pomeron} that carries the quantum number of the vacuum. Since the elementary gluon constituents of QCD carry color charge in the adjoint (octet) representation of $SU(3)$ color, the lowest order contribution that gives rise to pomeron exchange is the one-loop contribution shown in Fig.~\ref{one-loop-gluon-exchange} (b) and (c).

These diagrams have an imaginary part that can be computed using Cutkosky rules. First, we need the tree-level Born amplitude in the Regge limit, shown in Fig.~\ref{one-loop-gluon-exchange} (a). This can be evaluated as follows. Recall that the three-point gluon vertex can be expressed as 
\begin{align}
\label{eq:vertex-Gamma-0}
    \Gamma_{\m\m'\rho}^{\a\a'a} &= -ig f^{\a\a'a}\(\eta_{\mu\mu'}\left(-p_1-\ell_0\right)_{\rho}+\eta_{\m\rho}\left(k+p_1\right)_{\m'}+\eta_{\m'\rho}\left(\ell_0-k\right)_{\mu}\)\no\\[5pt]
    &=-ig f^{\a\a'a}\(\eta_{\mu\mu'}\left(-p_1-\ell_0\right)_{\rho}+\eta_{\m\rho}\left(2k+\ell_0\right)_{\m'}+\eta_{\m'\rho}\left(p_1-2k\right)_{\mu}\)~.
\end{align}
where $\eta_{\m\n}$ is the metric tensor of  Minkowski spacetime. The convention used here is that the momenta $p_1$ is incoming and $\ell_0$ and $k=p_1-\ell_0$ outgoing. Since the external particles have high energies, their momenta undergo only small changes in the scattering. Taking the momenta of the external particles at the vertex to be $p_1$ and $\ell_0$, we can get rid of the $k$ dependence in the above formula. Further, since the momenta $p_1$ and $\ell_0$ are on-shell in the Born amplitude, the contributions proportional to $p_1^{\m}$ and $\ell_0^{\m'}$ can be omitted since they ultimately vanish upon contracting with polarization tensors $\eps_\m(p_1)$ and $\eps_{\m'}(\ell_0)$. We then have
\begin{align}
\label{eq:vertex-Gamma-eik}
    \Gamma_{\m\m'\rho}^{\a\a' a} \approx 2ig f^{\a\a'c}\eta_{\mu\mu'}p_{1,\rho}~.
\end{align}
Similarly, the eikonal vertex associated with the second particle with momentum $p_2$ is $2ig f^{\b\b'a'}\eta_{\nu\nu'}p_{2,\rho'}$. Sewing together the two vertices via the gluon propagator (in  Feynman gauge) $\delta^{aa'}g_{\rho\rho'}/(k^2+i\eps)$, gives the following result for the Born amplitude in the Regge limit:
\be
\label{gluon-born-amplitude}
\mathcal{A}_{0,{\a\a'\b\b'}}^{\m\m'\n\n'}(s,t) = 2g^2\(\frac{s}{-t}\)\eta^{\m\m'}\eta^{\n\n'} (G_0)_{\a\a'\b\b'}~.
\ee
Here $s=2p_1\cdot p_2,~ t=-k^2=-(p_1-\ell_0)^2$, $T^a_{\a\a'}=if_{a\a\a'}$,  and $(G_0)_{\a\a'\b\b'}\equiv (T^a \otimes T^a)_{\a\a'\b\b'} = \sum_a T^a_{\a\a'} T^a_{\b\b'}$ is the associated color factor with $\a,\b,\a',\b'$ being the color indices of the incoming and outgoing gluons.  To avoid clutter, we will suppress the Lorentz and color indices on $\mathcal{A}_{0,{\a\a'\b\b'}}^{\m\m'\n\n'}$ when these labels are self-explanatory. 

In order to compute the imaginary part of the diagrams in Fig. \ref{one-loop-gluon-exchange} (b) and (c), we use Eq.~\eqref{unitarity-condition-0} in a systematic perturbative expansion of the r.h.s. Towards this end,  we begin with the ($n=0$) explicit expression for the 2-body phase space integral,
\begin{align}
\int d({\rm P.S.}^{2}) = \int \frac{d^4\ell_0}{\(2\pi\)^3} \delta(\ell_0^2) \frac{d^4\ell_1}{\(2\pi\)^3} \delta(\ell_1^2) \(2\pi\)^4 \delta^{(4)}\(p_1+p_2-\ell_0-\ell_1\)~.
\end{align}
Next, performing the change of integration variables,
\begin{align}
\begin{split}
    &\ell_0 = p_1 - k~,\qquad \ell_{1} = p_2 + k~,
\end{split}
\end{align}
the phase space measure  becomes
\begin{align}
    \int d({\rm P.S.}^{2}) = \frac{1}{\(2\pi\)^{2}}\int d^4k ~ \delta\[\(p_1-k\)^2\] \delta\[\(p_2+k\)^2\]~.
\end{align}
We introduce further a ``Sudakov decomposition" for the $k$ momenta,
\be
k = \rho p_1 + \lambda p_2 + k_{\perp}~,
\ee
where the Sudakov parameters $\rho$ and $\lambda$ are bounded between $(0, 1)$ and $(-1,0)$ respectively. 
From this decomposition, we see that
\be
d^4k = \frac s2 d\rho d\lambda d^2\bsk~.
\ee
Note that here $p_1 = (p_1^+,0,0,0)$ and $p_2=(0,p_2^-,0,0)$, with $s= 2\, p_1^+ p_2^-$. Using the on-shell condition for $\ell_0$ and $\ell_1$, the phase space measure in Sudakov variables can be expressed as 
\begin{align}
\begin{split}
    \int d({\rm P.S.}^{2}) &= \frac{s}{2\(2\pi\)^{2}}\int d\rho d\l d^2\bsk ~  \delta\[-s\l(1-\rho)-\bsk^2\] 
     \delta\[s\rho(\l+1)-\bsk^2\] \,.
  \end{split}
\end{align}
In the MRK regime, the leading logarithmic contributions in $x$ come from the regions of integration corresponding to 
\begin{align}
\label{MRK-regime}
\begin{split}
1\gg \rho \sim\frac{\bsk^2}{s} ~,\qquad 1\gg |\l| \sim\frac{\bsk^2}{s}~.
\end{split}
\end{align}
Therefore the $2$-body phase space measure in these kinematics can be approximated by
\begin{align}
    \int d({\rm P.S.}^{2}) &= \frac{s}{2\(2\pi\)^{2}}\int d\rho d\l d^2\bsk ~ \delta\(-s\l-\bsk^2\)\delta\(s\rho -\bsk^2\)\,.
\end{align}
Further, using 
\begin{align}
    \mathcal{A}_{2\to2,\a\a''\b\b''}^{\l_1\l_1''\l_2\l_2''}(s,t) = \mathcal{A}_{0,{\a\a''\b\b''}}^{\m\m''\n\n''}(s,t)\epsilon_{\m}^{\l_1} \epsilon_{\m''}^{\l_{1}''}  \epsilon_{\n}^{\l_2} \epsilon_{\n''}^{\l_{2}''} \,,
\end{align}
and the polarization sum $\sum_{\zeta}\epsilon_{\m}^{\zeta} \epsilon_{\n}^{\zeta}  = -\eta_{\m\n}$, we find that the imaginary part of Fig.~\ref{one-loop-gluon-exchange}(b), as given by Eq.~\eqref{unitarity-condition-0}, is 
\begin{align}
\Im \mathcal{A}_I^{\m\m'\n\n'} = \frac{g^4}{\(2\pi\)^2}s^3 \eta^{\m\m'}\eta^{\n\n'} G_I \int d\rho d\l d^2\bsk ~ & \delta\(-s\l-\bsk^2\) \delta\(s\rho-\bsk^2\) \frac{1}{k^2(q-k)^2}~.
\end{align}
Here $G_{I}\equiv (T^aT^b)\otimes(T^aT^b)$ (in terms of color indices: $G_{I,\a\a'\b\b'}= \sum_{ab\a''\b''} T^a_{\a\a''}T^b_{\a''\a'}T^a_{\b\b''}T^b_{\b''\b'}$). From this expression, a consequence of the delta function constraints in Regge asymptotics is  $-\rho\l s\ll\bsk^2$ which, in turn, implies $k^2\approx -\bsk^2$, $(k-q)^2\approx-(\bsk-\bsq)^2$ and $t= -\bsq^2$. Hence,
\be
\Im \mathcal{A}_I^{\m\m'\n\n'}= g^4\frac st \eta^{\m\m'}\eta^{\n\n'} G_I \int \frac{d^2\bsk}{\(2\pi\)^2} \frac{-\bsq^2}{\bsk^2(\bsq-\bsk)^2}~.
\ee
From this, one can reconstruct the complete amplitude in Fig.~\ref{one-loop-gluon-exchange}(b) using Eq.~\eqref{log-discontinuities}: 
\be
\label{gluon-one-loop-I}
\mathcal{A}_I^{\m\m'\n\n'} = \frac{g^4}{\pi} \frac st\log\(\frac {s}{-t}\) \eta^{\m\m'}\eta^{\n\n'} G_I \int \frac{d^2\bsk}{\(2\pi\)^2} \frac{-\bsq^2}{\bsk^2(\bsq-\bsk)^2} 
\ee
Similarly, the amplitude in Fig.~\ref{one-loop-gluon-exchange}(c) is
\be
\label{gluon-one-loop-II}
\mathcal{A}_{II}^{\m\m'\n\n'} = \frac{g^4}{\pi} \frac ut\log\(\frac {u}{-t}\) \eta^{\m\m'}\eta^{\n\n'} G_{II} \int \frac{d^2\bsk}{\(2\pi\)^2} \frac{-\bsq^2}{\bsk^2(\bsq-\bsk)^2} ~,
\ee
where $G_{II} = (T^aT^b)\otimes(T^bT^a)$. The total amplitude at this order is the sum of $\mathcal{A}_I$ and $\mathcal{A}_{II}$. 

To compute the pomeron exchange contribution to the elastic amplitude, we need to project these amplitudes onto the singlet representation for which we take the traces of $G_I$ and $G_{II}$ as $G_I\to \Tr(T^aT^b)\Tr(T^aT^b)$ and $G_{II}\to \Tr(T^aT^b)\Tr(T^bT^a)$. After this projection, the two color factors are clearly identical. Therefore the real parts of the amplitudes I and II cancel between the $s$- and $u$- channel contributions (after using $u \approx -s$ in the Regge limit), and we are just left with the imaginary part. Hence the lowest order contribution to the elastic (pomeron exchange)  amplitude is given by
\begin{align}
\label{lowest-order-gluon}
    \mathcal{A}^{(0),\m\m'\n\n'} = iN_c^2(N_c^2-1) g^4\frac st \eta^{\m\m'}\eta^{\n\n'} \int \frac{d^2\bsk}{\(2\pi\)^2} \frac{-\bsq^2}{\bsk^2(\bsq-\bsk)^2}~,
\end{align}
where we used $\Tr(T^aT^b) = N_c \delta_{ab}$. Note that at this order in the coupling, the imaginary part of the amplitude grows as $s$. At higher orders in the coupling that we turn to next, this behavior get corrected by $\log(s)$ terms. 

\begin{figure}[ht]
\centering
\includegraphics[scale=0.75]{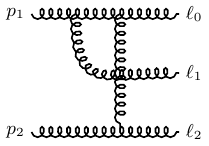}
\qquad
\includegraphics[scale=0.75]{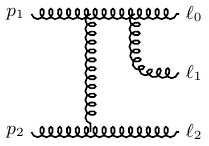}
\qquad
\includegraphics[scale=0.75]{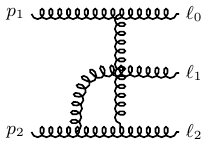}
\qquad
\includegraphics[scale=0.75]{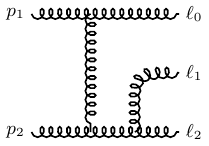}
\qquad
\includegraphics[scale=0.75]{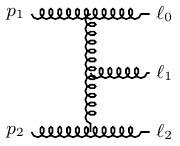}
\caption{Feynman graphs for the process $gg\to ggg$ that contribute to the imaginary part of the 2-to-2 scattering at order $g^6$.}
\label{LV-constituents}
\end{figure}

At next order in perturbation theory, the imaginary part of the elastic amplitude for color singlet exchange receives (a) purely real, and  (b) the interference of real and virtual contributions. 
To extract the real contribution to the imaginary part of the amplitude at this $O(g^6)$, we need to compute the  amplitude for the incoming gluons with momenta $p_1, p_2$ to scatter into a gluon of momentum $\ell_0 = p_1-k_1$, a second  gluon of momentum $\ell_2 = p_2+k_2$, and the third with momentum $\ell_1 = k_1-k_2$. The Feynman graphs that contribute to this 
$gg\rightarrow ggg$ process are shown in Fig~\ref{LV-constituents}. Notice that in principle there can be diagrams that involve 4-gluon interaction terms. The only such diagrams at this order are the ones where the 4-point vertex are associated to the external lines. However, because these vertices do not carry any energy factors, their contribution is suppressed in the Regge limit\footnote{We will discuss sub-leading contributions later in this section. We will see in the next section on gravity in the Regge limit that such diagrams do contribute (see  Fig.~\ref{grav-LV-constituents}) to the construction of the gravitational Lipatov vertex because  3- and 4-point vertices in gravity contribute with the same power of the energy.}.

As previously for the $2\rightarrow 2$ case, we now need the 3-body phase space measure in order to compute the imaginary part at order $g^6$. We shall compute below this measure in a way that is naturally generalizable to the n-body phase space measure which will be useful later. 
The three-body phase space measure is
\begin{align}
\int d({\rm P.S.}^{3}) = \int \prod_{i=0}^{2} \[\frac{d^4\ell_i}{\(2\pi\)^3} \delta(\ell_i^2)\] \(2\pi\)^4 \delta^{(4)}\(p_1+p_2-\sum_{i=0}^{2}\ell_i\)~.
\end{align}
The momenta $\ell_i$ correspond to the momenta of the produced on-shell gluons, as shown in Fig. \ref{LV-constituents}. We first integrate over $\ell_{2}$, which results in 
\begin{align}
    \int d({\rm P.S.}^{3}) = \int \prod_{i=0}^{1} \[\frac{d^4\ell_i}{\(2\pi\)^3} \delta(\ell_i^2)\] 2\pi \delta\(\[p_1+p_2-\sum_{i=0}^{1}\ell_i\]^2\)~.
\end{align}
We next perform the change of integration variables
\begin{align}
\begin{split}
    &\ell_0 = p_1 - k_1~,\qquad \ell_1 = k_1 - k_2~,\qquad \ell_2 = p_2 + k_2~.
\end{split}
\label{eq:int-variables}
\end{align}
The phase space measure in $k_i$ variables then becomes
\begin{align}
    \int d({\rm P.S.}^{3}) = \frac{1}{\(2\pi\)^{5}}\int d^4k_1 d^4k_2~ \delta\[(k_1-k_2)^2)\] \delta\[\(p_1-k_{1}\)^2\] \delta\[\(p_2+k_2\)^2\]~.
\end{align}
Introducing, as previously, the Sudakov decomposition for the $k_i$ momenta $k_i = \rho_i p_1 + \lambda_i p_2 + k_{i\perp}$, with $d^4k_i = \frac s2 d\rho_i d\lambda_i d^2\bsk_i$, and the on-shell conditions $\ell_0^2=\ell_1^2=\ell_2^2=0$, the phase space measure in terms of Sudakov variables can be expressed as 
\begin{align}
\begin{split}
    \int d({\rm P.S.}^{3}) = \frac{s^{2}}{4\(2\pi\)^{5}}\int \prod_{i=1}^{2} d\rho_i d\l_i d^2\bsk_i ~  &\delta\[-s\l_1(1-\rho_1)-\bsk_1^2\] \delta\[s\rho_{2}(\l_{2}+1)-\bsk_{2}^2\] \\
  &\times \delta\[s(\rho_1-\rho_{2})(\l_1-\l_{2})-(\bsk_{1}-\bsk_{2})^2\] ~.
  \end{split}
\end{align}
In the MRK regime, the leading logarithmic contributions in $x$ come from the regions of integration,
\begin{align}
\label{MRK-regime-1}
\begin{split}
1\gg \rho_1 \gg \rho_2  \sim \bsk^2/s~,\qquad  1\gg |\l_{2}| \gg |\l_1| \sim \bsk^2/s~.
\end{split}
\end{align}
where $\bsk$ is a generic transverse momentum whose magnitude is much smaller than $\sqrt{s}$. We we combine the above  with the delta-function constraints, we get  $\bsk_1^2/s = |\lambda_1|\approx \bsk^2/s$ which implies $\bsk_1^2\approx \bsk^2$. Similarly, we find that $\bsk_2^2\approx \bsk^2$ which follows from $\bsk_2^2/s = \rho_2 \approx \bsk^2/s$. Therefore we can approximate all transverse momenta in the delta-function as $\bsk_1^2\approx \bsk_2^2\approx (\bsk_1-\bsk_2)^2 \approx \bsk^2$. The $3$-body phase space measure, in turn can be  approximated as
\begin{align}
\begin{split}
    \int d({\rm P.S.}^{3}) = \frac{s^{2}}{4\(2\pi\)^{5}}\int \prod_{i=1}^{2} d\rho_i d\l_i d^2\bsk_i ~  &\delta\[-s\l_1-\bsk^2\] \delta\[s\rho_{2} -\bsk^2\] \delta\[-\rho_1\l_2s-\bsk^2\] ~.
  \end{split}
\end{align}

\begin{figure}[ht]
\centering
\includegraphics[scale=1]{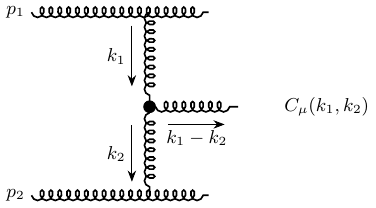}
\caption{The nonlocal central gluon emission Lipatov vertex, represented by the black blob.}
\label{LV-QCD}
\end{figure}

Having computed the 3-body phase space measure, we now need the expression for the amplitude $\mathcal{A}_{2\to 3}$. This is achieved through the explicit evaluation of the five graphs in  Fig.~\ref{LV-constituents}. Using the commutation relation $[T^a, T^b] = if^{abc}T_c$, the contribution from the first two figures in Fig.~\ref{LV-constituents} in the kinematic regime corresponding to Eq.~\eqref{MRK-regime-1} is
\be
-2i g^3 s \frac{2 p_1^\sigma}{\bsk_{\mathbf{2}}^2 \lambda_2 s} f_{a b c} T^a \otimes T^b g_{\m\m'}g_{\n\n'}~.
\ee
Likewise, the contribution from the third and fourth figures in Fig. \ref{LV-constituents} is
\be
-2i g^3 s \frac{2 p_2^\sigma}{\bsk_1^2 \rho_1 s} f_{a b c} T^a \otimes T^b g_{\m\m'}g_{\n\n'}~.
\ee
Finally, the last Figure in \ref{LV-constituents} gives,
\be
-\frac{2i g^3 s}{\bsk_1^2 \bsk_2^2}\left[\rho_1 p_1^\sigma+\lambda_2 p_2^\sigma-\left(k_1+k_2\right)_{\perp}^\sigma\right] g_{\m\m'}g_{\n\n'} f_{a b c} T^a \otimes T^b~.
\ee
Adding up all the contributions, we arrive at the following expression for the $\mathcal{A}_{2\to3}$ 
amplitude:
\begin{align}
\label{QCD-Lipatov-vertex}
    \mathcal{A}_{2\to3,b}^{\m\m'\n\n'\sigma}(k_1,k_2) = -\frac{2ig^3s}{\bsk_1^2\bsk_2^2} \eta^{\m\m'}\eta^{\n\n'}f_{acb} T^a \otimes T^c C^\sigma(k_1,k_2)~.
\end{align}
Here, $C^\sigma(k_1,k_2)$ is the effective ``central gluon emission vertex" (also known as the Lipatov vertex) illustrated in Fig.~\ref{LV-QCD},  and is given by
\begin{align}
\label{QCD-Lipatov-vertex-1}
    C^\mu(k_1,k_2) = -(\bsk_1+\bsk_2)^\mu+  p_1^\mu \left(\rho_1+\frac{2 \bsk_1^2}{\lambda_2 s}\right)+ p_2^\mu \left(\lambda_2+\frac{2 \bsk_2^2}{\rho_1 s}\right)~.
\end{align}

The contribution of these real graphs to  $\Im \mathcal{A}_1^{(1)}$, the imaginary part of the $2\rightarrow 2$ amplitude at order $g^6$, is then given by
\be
\Im \mathcal{A}_1^{(1)\mu\mu'\n\n'} = \frac{1}{2} \int d(P.S.^{3}) \sum_{\rm{color, polarizations}}\mathcal{A}^{\m\n}_{2\to 3}(k_1,k_2)\mathcal{A}^{\m'\n'\dagger}_{2\to 3}(k_1-q, k_2-q)~,
\ee
where
\be
\mathcal{A}^{\m\n, \l_0\l_1\l_2}_{2\to 3}(k_1,k_2) = \mathcal{A}_{2\to3,b}^{\m\m''\n\n''\sigma} \epsilon_{\m''}^{\l_0}\epsilon_{\sigma}^{\l_1}\epsilon_{\n''}^{\l_2}(k_1,k_2)~.
\ee
The sum over the color factors give (with the trace is taken to obtain the color singlet projection):
\be
\label{n=1-color-sum}
\operatorname{Tr}\left(T_a T_b\right) \operatorname{Tr}\left(T_c T_d\right) f_{a c e} f_{b d e} \equiv G_0^{(1)} = N_c^3(N_c^2-1)\,.
\ee
Performing the integration over $\l_i$ and $\rho_i$ (the lower limit of $\rho_2$ integration being $\bsk^2/s$), one finds
\begin{align}
   & \Im \mathcal{A}_1^{(1)\mu\mu'\n\n'} 
=  - \frac{ N_c g^6}{2\pi} \eta^{\m\m'} \eta^{\n\n'}G_0^{(1)} s \ln \left(s / \bsk^2\right)\no \\
\times & \int \frac{d^2\bsk_1}{\(2\pi\)^2} \frac{d^2 \bsk_2}{\(2\pi\)^2}   {\left[\frac{\bsq^2}{\bsk_{1}^2 \bsk_{2}^2\left(\bsk_1-\bsq\right)^2\left(\bsk_{2}-\bsq\right)^2}-\frac{1}{\bsk_{1}^2\left(\bsk_1-\bsk_{2}\right)^2\left(\bsk_{2}-\bsq\right)^2}\right.} 
 \left.-\frac{1}{\bsk_{2}^2\left(\bsk_1-\bsq\right)^2\left(\bsk_{1}-\bsk_{2}\right)^2}\right] .
\end{align}

\begin{figure}[ht]
\centering
\includegraphics[scale=1]{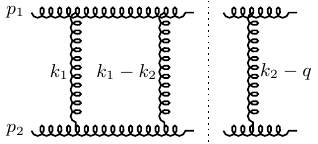}
\qquad
\qquad
\includegraphics[scale=1]{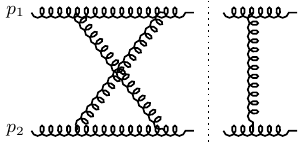}
\caption{Cut Feynman graphs contributing to $\Im \mathcal{A}$ at order $g^6$.}
\label{leading-virtual}
\end{figure}

We will now compute the contribution of the interference terms (shown in Fig.~\ref{leading-virtual}) to the imaginary part of the $2\rightarrow 2$ amplitude. 
Using the results from the computation of previous order diagrams, the contribution to the imaginary part from these diagrams is
\begin{align}
\Im \mathcal{A}_{2}^{(1)\mu\mu'\n\n'}= & - \frac{N_c g^6 }{4\pi} \eta^{\m\m'} \eta^{\n\n'}G_0^{(1)} s \ln \left(s / \bsk^2\right) \no\\
&\times \int \frac{ d^2 \bsk_1}{(2\pi)^2} \frac{d^2 \bsk_2}{(2\pi)^2} \[\frac{1}{\bsk_1^2\left(\bsk_1-\bsk_{2}\right)^2\left(\bsk_{2}-\bsq\right)^2}+\frac{1}{\bsk_2^2\left(\bsk_2-\bsk_1\right)^2\left(\bsk_1-\bsq\right)^2}\]\,.
\end{align}
The second term in this expression comes from the diagrams where we cut the amplitude with the one gluon exchange contribution to the left of the cut. 

Combining the two contributions $\mathcal{A}_1^{(1)}$ and $\mathcal{A}_2^{(1)}$, the full expression at order $g^6$, to  leading logarithmic accuracy, is 
\begin{align}
    &\Im \mathcal{A}^{(1)\mu\mu'\n\n'} 
=  - \frac{ N g^6}{2\pi} \eta^{\m\m'} \eta^{\n\n'}G_0^{(1)} s \ln \left(s / \bsk^2\right) \int \frac{d^2\bsk_1}{\(2\pi\)^2} \frac{d^2 \bsk_2}{\(2\pi\)^2}\no \\
\times &    {\left[\frac{\bsq^2}{\bsk_{1}^2 \bsk_{2}^2\left(\bsk_1-\bsq\right)^2\left(\bsk_{2}-\bsq\right)^2}-\frac12 \frac{1}{\bsk_{1}^2\left(\bsk_1-\bsk_{2}\right)^2\left(\bsk_{2}-\bsq\right)^2}\right.} 
 \left.-\frac12 \frac{1}{\bsk_{2}^2\left(\bsk_1-\bsq\right)^2\left(\bsk_{1}-\bsk_{2}\right)^2}\right] .
\end{align}
As noted earlier, we see that the imaginary part of the $2\to 2$ amplitude at order $g^6$ has $s \log(s)$ behavior. The BFKL equation we will discuss below resums these logarithms to all orders in the coupling. 

Before discussing higher order contributions, it will be convenient to introduce the Mellin transform of the imaginary part $2\rightarrow 2$ amplitude at order $r$,
\begin{align}
\int_1^{\infty} d\left(\frac{s}{\bsk^2}\right)\left(\frac{s}{\bsk^2}\right)^{-\ell-1}  \frac{\Im \mathcal{A}^{(r)\mu\mu'\n\n'}(s, t)}{s} \equiv 4 i \alpha_S^2 \eta^{\m\m'} \eta^{\n\n'}G_0^{(1)} \int \frac{d^2 \bsk_{1} d^2 \bsk_{2}}{\bsk_{2}^2\left(\bsk_{1}-\bsq\right)^2} f_{r+1}\left(\ell, \bsk_{1}, \bsk_{2}, \bsq\right)\,.
\end{align}
Here the  Mellin amplitude $f_{r+1}(\ell, \bsk_1, \bsk_2, \bsq)$ at order $r$ is a function of the variable $\ell$ which is conjugate to the Mandelstam variable $s$. From our previous computations, we find that the Mellin amplitude for the first two nontrivial orders is given by
\begin{align}
\label{Mellin-amplitude-leading-orders}
\begin{split}
    f_1\left(\ell, \bsk_{1}, \bsk_{2}, \bsq\right) &= \frac{1}{\ell} \delta(\bsk_1-\bsk_2)~,\\
    f_2\left(\ell, \bsk_{1}, \bsk_{2}, \bsq\right) &= -\frac{\bar{\alpha}_s}{2\pi} \frac{1}{\ell^2} \left[\frac{\bsq^2}{\bsk_1^2\left(\bsk_2-\bsq\right)^2}-\frac{1}{2} \frac{1}{\left(\bsk_1-\bsk_2\right)^2}\left(1+\frac{\bsk_2^2\left(\bsk_1-\bsq\right)^2}{\bsk_1^2\left(\bsk_2-\bsq\right)^2}\right)\right]\,,
\end{split}
\end{align}
where $\bar{\alpha}_S = N\alpha_S/\pi$.

In Sec.~\ref{BFKL:derivation}, we will write down the recursive BFKL integral equation which computes the Mellin amplitude to all orders including the leading logarithmic contributions from all the intermediate particle cuts that contribute to the imaginary part of the $2\to 2$ gluon scattering amplitude. Towards this end, we will first assemble the machinery to evaluate higher order contributions. This requires the $n$-body phase space associated with multi-particle cuts in Regge asymptotics, and the generalization of the leading contributions of $n$-particle amplitudes, to all orders at leading logarithmic accuracy, to obtain the structure of the effective ladder diagram of the type shown in Fig.~\ref{2-to-n-amplitude}. 

\subsection{Higher order contributions and the BFKL ladder}

\subsubsection{The \texorpdfstring{$n$}{}-particle phase space in multi-Regge asymptotics}
\label{sec:n-particle-intermediate}

Recall that the contribution to the imaginary part of the $2\to 2$ amplitude due to $n+2$ intermediate particle exchange is 
\begin{align}
\label{unitarity-condition}
    \Im \mathcal{A}_{2\to 2} = \sum_{n=0}^\infty \frac12 \int d({\rm P.S.}^{n+2}) \sum_{\rm{color, polarizations}}\mathcal{A}_{2\to n+2}(\{k\})\mathcal{A}_{2\to n+2}^\dagger(\{k\}-q)~,
\end{align}
where $d({\rm P.S.}^{n+2})$ is the measure for the $(n+2)$-body phase space given by
\begin{align}
\int d({\rm P.S.}^{n+2}) = \int \prod_{i=0}^{n+1} \[\frac{d^4\ell_i}{\(2\pi\)^3} \delta(\ell_i^2)\] \(2\pi\)^4 \delta^{(4)}\(p_1+p_2-\sum_{i=0}^{n+1}\ell_i\)~.
\end{align}
The momenta $\ell_i$ correspond to the momenta of the produced on-shell gluons; they are related to the $k_i$'s (represented by $\{k\}$) by the change in integration variables generalizing Eq.~\eqref{eq:int-variables}. Following the method outlined in the previous analysis, we need to approximate this measure in the MRK regime in order to obtain the leading log contribution. The MRK regime is defined as
\begin{align}
\label{MRK-regime-2}
\begin{split}
&1\gg \rho_1 \gg \rho_2 \gg \cdots \gg \rho_{n+1} \sim \bsk^2/s~,\\
&1\gg |\l_{n+1}| \gg |\l_n| \gg \cdots \gg |\l_{1}| \sim \bsk^2/s~.
\end{split}
\end{align}
As mentioned previously, $\bsk$ is a generic transverse momentum whose magnitude is much smaller than $\sqrt{s}$.

The $n$-body phase space measure, approximated thus, reads as 
\begin{align}
    \int d({\rm P.S.}^{n+2}) \approx& \frac{s^{n+1}}{2^{n+1}\(2\pi\)^{3n+2}}\int \prod_{i=1}^{n+1} d\rho_i d\l_i d^2\bsk_i ~  \delta\[-s\l_1-\bsk_1^2\] \delta\[s\rho_{n+1}-\bsk_{n+1}^2\] \no\\
  &\times \prod_{i=1}^n \delta\[-s\rho_i\l_{i+1}-(\bsk_{i}-\bsk_{i+1})^2\] ~.
\end{align}
Performing the integrals in the $\l_i$'s, we get\footnote{In the expressions that we will encounter henceforth, the integrand will not be a function of $\l_i$; it is sufficient to order the longitudinal momenta as fractions $x_i$ of $p_1^+$.}
\begin{align}
\label{n-body-phase-space-measure}
    \int d({\rm P.S.}^{n+2}) \approx \frac{1}{2^{n+1}\(2\pi\)^{3n+2}}\int \prod_{i=1}^{n} \(\frac{d\rho_i}{\rho_i} d^2\bsk_i \)d\rho_{n+1}d^2\bsk_{n+1}  \delta\[s\rho_{n+1}-\bsk^2\]~.
\end{align}
Equipped with this formula for the multi-particle phase space,  we will now turn to compute the contribution of the $n+2$ intermediate particles to the imaginary part of the $2\to2$ amplitude. 

\subsubsection{Generalization of the \texorpdfstring{$2\to 3$}{} amplitude to \texorpdfstring{$2\to 2+n$}{} in multi-Regge kinematics}
\label{sec:2-to-n-amplitude}

In the previous subsection, we computed the imaginary part of the $2\rightarrow 2$ amplitude up to to $O(g^6)$. We shall now generalize this analysis to all orders, restricting the discussion to real contributions alone; diagrammatically, this corresponds to generalizing Fig.~\ref{LV-QCD} to $n$-gluon emissions. The generalization involving the virtual contributions will be dealt with in Section \ref{BFKL:derivation}.

We recall that the structure of Fig.~\ref{LV-QCD} with the central gluon emission Lipatov vertex follows from summing over the five Feynman graphs in Fig.~\ref{LV-constituents}. The generalization that we seek is one where there are $n$ additional gluons emitted in multi-Regge kinematics.  As noted, formally they appear to be suppressed in fixed-order perturbation theory. However the presence of large logs in $x_n$ provide $O(1)$ contributions at each order. 
Performing this generalization using Feynman diagrams will be extremely cumbersome. An efficient alternative approach is that of dispersive techniques that rely on results at prior orders. Before considering the general $n$-emissions case, we will first present this alternative derivation for the Lipatov vertex ($n=1$ case).

We begin with the observation that the result for the 2-to-2 gluon amplitude with a single gluon exchange in the t-channel \cite{Lipatov:1976zz,Lipatov:1982vv,Lipatov:1982it} can be expressed as
\begin{align}
\label{general-born-amplitude}
    \mathcal{A}_{2\to 2, p_1+p_2\to \ell_0+\ell_1}^{\a\a'\b\b'} = \Gamma_{p_1\ell_0}^{\alpha\alpha'c} \frac{g^2s}{t} \Gamma_{p_2\ell_1}^{\beta\beta'c}~,
\end{align}
where the vertex $\Gamma$ is given by \cite{Ioffe:2010zz}:
\begin{align}
\label{eq:vertex-Gamma}
    \Gamma_{p_1\ell_0}^{\alpha\alpha'c} = -\sqrt{2}if^{\a\a'c}\(-g_{\m\m'}+\frac{p_{2,\m}p_{1,\m'}+p_{2,\m'}\ell_{0,\m}}{p_2\cdot p_1}+ (p_1-\ell_0)^2\frac{p_{2,\m}p_{2,\m'}}{2(p_2\cdot p_1)^2}\)\epsilon ^\m(p_1)\epsilon^{\m'}(\ell_0)~.
\end{align}
Since this is not a standard expression, it will be useful to understand how it is derived. First, we need the expression of the three-point gluon vertex in the eikonal approximation. This was derived earlier in Eq.\eqref{eq:vertex-Gamma-eik} to be 
\begin{align}
\label{eq:vertex-Gamma-3}
    \Gamma_{\m\m'\rho}^{abc} \sim 2ig f^{abc}\eta_{\m \m'}p_{1,\rho}~.
\end{align}
We next contract this vertex with polarization vectors. For the polarizations associated with incoming and outgoing gluons, we will impose the gauge where $\eps(p_1)\cdot p_2 = \eps'(\ell_0)\cdot p_2 = 0$ as follows:
\begin{align}
\label{gauge-pol-condition}
    \eps(p_1) \to \tilde \eps(p_1) = \eps - \frac{p_2\cdot \eps}{p_2\cdot p_1} p_1~,\qquad \eps'(\ell_0) \to \tilde \eps'(\ell_0) = \eps' - \frac{p_2\cdot \eps'}{p_2\cdot \ell_0} \ell_0~.
\end{align}
Finally, the magnitude of the polarization vector of the exchanged gluon with momenta $p_1-\ell_0$ is given by $\sqrt{2}p_2/s$. (We will discuss why this is the case further below.) Contracting these polarization vectors with the vertex in Eq.~\eqref{eq:vertex-Gamma-3} gives
\begin{align}
\label{gamma-construction}
\begin{split}
    \Gamma_{p_1\ell_0}^{abc}&\equiv\frac{1}{g}\Gamma_{\m\m'\rho}^{abc} \tilde \eps^{\m} \tilde \eps'^{\m'}\frac{\sqrt{2}p_2^{\rho}}{s} =\\[5pt]
    &2if^{abc}\(\eps\cdot \eps' -\frac{(p_2\cdot \eps') (\ell_0\cdot \eps)}{p_2\cdot\ell_0}-\frac{(p_2\cdot \eps) (p_1\cdot \eps')}{p_1\cdot p_2}+\frac{(p_2\cdot \eps) (p_2\cdot \eps')}{(p_1\cdot p_2) (p_2\cdot \ell_0)} p_1\cdot \ell_0\)\frac{\sqrt{2}p_1\cdot p_2}{s}~,\\[5pt]
    =&\sqrt{2}if^{abc} \(g_{\m\n} -\frac{p_{2\nu}\ell_{0\mu}}{p_2\cdot\ell_0}-\frac{p_{2\mu} p_{1\nu}}{p_1\cdot p_2}+\frac{p_{2\mu}p_{2\nu}}{(p_1\cdot p_2) (p_2\cdot \ell_0)} p_1\cdot \ell_0 \)\eps^{\mu}\eps'^{\nu}~,\\[5pt]
    =&-\sqrt{2}if^{abc}\(-g_{\m\m'}+\frac{p_{2\m}p_{1\m'}+p_{2\m'}\ell_{0\m}}{p_2\cdot p_1}+ (p_1-\ell_0)^2\frac{p_{2\m}p_{2\m'}}{2(p_2\cdot p_1)^2}\)\epsilon ^\m(p_1)\epsilon^{\m'}(\ell_0)~.
\end{split}
\end{align}
To arrive at the last line, we performed the following steps: (1) we relabeled the Lorentz indices, (2) used the approximation $p_2\cdot\ell_0 \approx p_2\cdot p_1$, (3) used the fact that $-p_1\cdot \ell_0 = \frac{1}{2}(p_1-\ell_0)^2$, and finally, (4) wrote $\eps'^{\mu'}$ as $\eps^{\mu'}(\ell_0)$ to avoid clutter in the formulas. This completes the derivation of Eq.~\eqref{eq:vertex-Gamma}.

In the Regge limit, where the center-of-mass energy $s=2p_1\cdot p_2$ is much larger than any other scale, the formula in Eq.~\eqref{general-born-amplitude} reduces to the Born amplitude (using $T^a_{\a\a'}=if^a_{\a\a'}$) given in Eq.~\eqref{gluon-born-amplitude}. We will repeatedly using the expression in Eq.~\eqref{general-born-amplitude}. 

\begin{figure}[ht]
\centering
\subfigure[]{\includegraphics[scale=1]{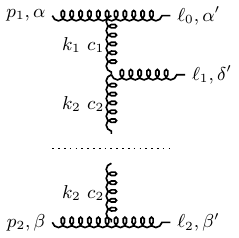}}
\qquad\qquad
\subfigure[]{\includegraphics[scale=1]{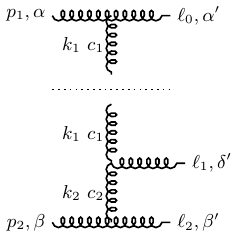}}
\caption{Cut diagrams that contribute to pole reconstruction of the Lipatov vertex in the $2\to 3$ amplitude.}
\label{1g-disp}
\end{figure}

Consider now the $2\to 3$ amplitude in Fig. \ref{LV-QCD} with incoming particles labeled by $p_1,p_2$ and the outgoing particles by their momenta $\ell_i$, $i=0,1,2$, with $\ell_{1}=k_1-k_2$ the momenta of the additional emitted particle. The residue of the $1/k_2^2$ pole of the amplitude $\mathcal{A}$, denoted by $P_{k_2^2}\mathcal{A}$, is
\be
P_{k_2^2}\mathcal{A}_{2\to2+1}^{\a\a'\b\b'\d'} =  \mathcal{A}_{p_1+(-k_2)\to \ell_0+\ell_1}^{\a\a'c_2\d'} sg^2 \Gamma^{\b\b'c_2}_{p_2\ell_2}~.
\ee
Here $\mathcal{A}_{p_1+(-k_2)\to \ell_0+\ell_1}^{\a\a'c_2\d'}$ is the $2\to 2$ amplitude in which $k_2$ is the cut (on-shell) gluon with momentum $k_2$, as depicted in Fig. \ref{1g-disp} (left), whose polarization vector is replaced by $\sqrt{2}p_2/s$. (As noted, the reason for this replacement  will be clarified at the end of the subsection.) The minus sign in front of $k_2$ is because it an outgoing momentum from the central emission vertex. Then using Eq.~\eqref{general-born-amplitude}, we obtain
\begin{align}
    &\mathcal{A}_{p_1+(-k_2)\to \ell_0+\ell_1}^{\a\a'c_2\d'} = g\Gamma_{p_1\ell_0}^{\alpha\alpha'c_1} \frac{(p_1-k_2)^2}{k_1^2} \no\\[5pt]
    & \times \[-\sqrt{2}igf^{c_2\d'c_1}\(-g_{\m\n}+\frac{-p_{1,\m}k_{2,\n}+p_{1,\n}\ell_{1,\m}}{-p_1\cdot k_2}+ (-k_2-\ell_1)^2\frac{p_{1,\m}p_{1,\n}}{2(p_1\cdot k_2)^2}\)\epsilon^\m(-k_2)\epsilon^{\n}(\ell_1) \]~,
\end{align}
using in addition Eq.~\eqref{eq:vertex-Gamma}. Since $(p_1-k_2)^2 \approx -\lambda_2 s$ (a valid approximation in the MRK regime), collecting all terms together, we get
\begin{align}
    P_{k_2^2}\mathcal{A}_{2\to2+1}^{\a\a' \b\b' \d'} &= -\sqrt{2}ig^3 \frac{-\lambda_2 s}{k_1^2}  s \Gamma_{p_1\ell_0}^{\alpha\alpha'c_1}  \Gamma^{\b\b'c_2}_{p_2\ell_2}f^{c_2\d'c_1}\no\\[5pt]
    &\times \(-g_{\m\n}+\frac{-p_{1,\m}k_{2,\n}+p_{1,\n}\ell_{1,\m}}{-p_1\cdot k_2}+ (-k_2-\ell_1)^2\frac{p_{1,\m}p_{1,\n}}{2(p_1\cdot k_2)^2}\)\epsilon^\m(-k_2)\epsilon^{\n}(\ell_1)~.
\end{align}
Next, substituting $\epsilon^\m(-k_2)\to \sqrt{2}p_2^\m /s$,
\begin{align}
    P_{k_2^2}\mathcal{A}_{2\to2+1}^{\a\a' \b\b' \d'} &= -2ig^3 \frac{-\lambda_2}{k_1^2}  s \Gamma_{p_1\ell_0}^{\alpha\alpha'c_1}  \Gamma^{\b\b'c_2}_{p_2\ell_2}f^{c_2\d'c_1}\no\\[5pt]
    &\times \(-p_{2,\nu}+\frac{(p_{1}\cdot p_2) k_{2,\n}-(\ell_{1}\cdot p_2)p_{1,\n}}{ p_1\cdot k_2}+ k_1^2\frac{(p_{1}\cdot p_2)}{2(p_1\cdot k_2)^2}p_{1,\n}\) \epsilon^{\n}(\ell_1)\,.
\end{align}
Pulling in $-2\l_2$ through the parenthesis, and identifying $\lambda_2 = \frac{p_1\cdot k_2}{p_1\cdot p_2}$, gives
\begin{align}
    P_{k_2^2}\mathcal{A}_{2\to2+1}^{\a\a' \b\b' \d'} &= -ig^3\frac{s}{k_1^2} \Gamma_{p_1\ell_0}^{\alpha\alpha'c_1}  \Gamma^{\b\b'c_2}_{p_2\ell_2}f^{c_2\d'c_1} \nonumber \\
    &\times \(2p_{2,\nu} \frac{p_1\cdot k_2}{p_1\cdot p_2} - 2k_{2,\n}+2\frac{\ell_{1}\cdot p_2}{p_1\cdot p_2}p_{1,\n}- \frac{k_1^2}{p_1\cdot k_2}p_{1,\n}\)
     \epsilon^{\n}(\ell_1)\,.
\end{align}
Finally, using $(k_1-k_2)\cdot \epsilon_{\lambda'} (\ell_1)=0$ and $p_1\cdot \ell_1 \approx -p_1 \cdot k_2$ (recall $\ell_1 = k_1-k_2$), we get
\begin{align}
\label{k2-pole}
    P_{k_2^2}\mathcal{A}_{2\to2+1}^{\a\a' \b\b' \d'} &= -ig^3\frac{s}{k_1^2} \Gamma_{p_1\ell_0}^{\alpha\alpha'c_1}  \Gamma^{\b\b'c_2}_{p_2\ell_2}f^{c_2\d'c_1}\nonumber \\
    &\times\(-2p_{2,\nu} \frac{p_1\cdot \ell_1}{p_1\cdot p_2} - (k_1+k_2)_\n+2\frac{\ell_{1}\cdot p_2}{p_1\cdot p_2}p_{1,\n}+\frac{k_1^2}{p_1\cdot \ell_1}p_{1,\n}\) \epsilon^{\n}(\ell_1)\,.
\end{align}
In the same manner, one can perform the computation of the residue of the $1/k_1^2$ pole of the $\mathcal{A}_{2\to2+1}$ amplitude shown in Fig.~\ref{1g-disp} (right), with the result
\begin{align}
\label{k1-pole}
    P_{k_1^2}\mathcal{A}_{2\to2+1}^{\a\a' \b\b' \d'} &= ig^3\frac{s}{k_2^2} \Gamma_{p_1\ell_0}^{\alpha\alpha'c_1}  \Gamma^{\b\b'c_2}_{p_2\ell_2}f^{c_1\d'c_2} \nonumber \\
    &\times\(2p_{1,\nu} \frac{p_2\cdot \ell_1}{p_1\cdot p_2} - (k_1+k_2)_\n-2\frac{\ell_{1}\cdot p_1}{p_1\cdot p_2}p_{2,\n}-\frac{k_2^2}{p_2\cdot \ell_1}p_{2,\n}\) \epsilon^{\n}(\ell_1)\,.
\end{align}
From the results in Eq.~\eqref{k2-pole} and Eq.~\eqref{k1-pole}, one can read off the simultaneous residue of the $1/(k_1^2k_2^2)$ pole. The reconstructed amplitude then reads
\begin{align}
\label{2-3-final}
    &\mathcal{A}_{2\to2+1}^{\a\a' \b\b' \d'} = ig^3\frac{s}{k_1^2 k_2^2} \Gamma_{p_1\ell_0}^{\alpha\alpha'c_1}  \Gamma^{\b\b'c_2}_{p_2\ell_2}f^{c_1\d'c_2} C_\n(k_1, k_2) \epsilon^{\n}(\ell_1)~,
\end{align}
where $C_\nu(k_1, k_2)$ is
\begin{align}
\label{QCD-LV-cov}
    C_\nu(k_1, k_2) = -(k_1+k_2)_\nu + p_{1,\nu}\(\frac{2p_2\cdot \ell_{1}}{p_1\cdot p_2}+\frac{k_1^2}{p_1\cdot \ell_{1}}\) - p_{2,\nu}\(\frac{2p_1\cdot \ell_{1}}{p_1\cdot p_2}+\frac{k_2^2}{p_2\cdot \ell_{1}}\)\,,
\end{align}
which is precisely the expression for the covariant form of the Lipatov vertex. Indeed, performing the Sudakov decomposition of $k_{1,2}$ reveals that this expression is identical to the result we obtained in Eq.~\eqref{QCD-Lipatov-vertex-1} from explicit computation of the relevant Feynman diagrams: 
\begin{align}
    C_\nu(k_1, k_2) &\approx -(\bsk_1+\bsk_2)_\nu - \rho_1 p_1 - \lambda_2 p_2 + 2p_{1,\nu}\(\rho_1+\frac{\bsk_1^2}{\lambda_2 s}\) + 2p_{2,\nu}\(\lambda_2+\frac{\bsk_2^2}{\rho_1 s}\)\no\\[5pt]
    & = -(\bsk_1+\bsk_2)_\nu + p_{1,\nu}\(\rho_1+\frac{2\bsk_1^2}{\lambda_2 s}\) + p_{2,\nu}\(\lambda_2+\frac{2\bsk_2^2}{\rho_1 s}\)~.
\end{align}
We note that this method of reconstructing the $2\to 3$ amplitude from its pole structure leaves some ambiguity in the amplitude since there is the freedom to add a term proportional to $k_1^2k_2^2$. Analyticity and dimensional analysis fixes the largest possible such contribution to be either proportional to $k_1^2k_2^2 p_1/(\ell_0\cdot \ell_1)^2$  or $k_1^2k_2^2 p_2/(\ell_2\cdot \ell_1)^2$. Since these contributions are small in the MRK regime, their addition to the $2\to 3$ amplitude do not contribute to leading order processes. 

\begin{figure}[ht]
\centering
\includegraphics[scale=1]{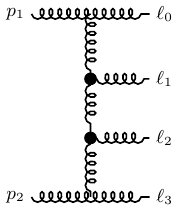}
\caption{The effective (half) ladder for two gluon production in the MRK regime.}
\label{fig:LVn2}
\end{figure}

\begin{figure}[ht]
\centering
\includegraphics[scale=0.75]{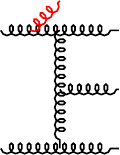}
\hspace{0.1in}
\includegraphics[scale=0.75]{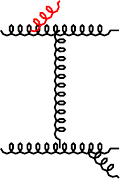}
\hspace{0.1in}
\includegraphics[scale=0.75]{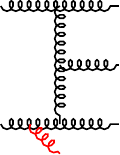}
\hspace{0.1in}
\includegraphics[scale=0.75]{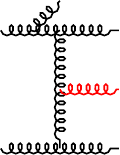}
\hspace{0.1in}
\includegraphics[scale=0.75]{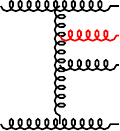}
\hspace{0.1in}
\includegraphics[scale=0.75]{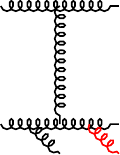}
\hspace{0.1in}
\includegraphics[scale=0.75]{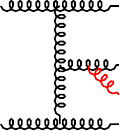}
\caption{A subset of bare Feynman graphs that contribute to the effective diagram for two gluon production in Fig.~\ref{fig:LVn2}. Here MRK kinematics is to be understood as the red gluon being of a smaller rapidity than the emitted gluon in black. The last diagram (along with any crossed diagram) is subleading in the MRK regime and does not contribute in the leading log approximation. See the discussion in Sec. \ref{sec:2.2.3} below.}
\label{LVn2-constituents}
\end{figure}

The aforementioned method of computing the effective Lipatov vertex can now be generalized to the case of two gluon emissions that we discuss next~\cite{Kuraev:1976ge}. The effective (half) ladder in the $n=2$ case is shown in Fig.~\ref{fig:LVn2}.
As in the $n=1$ case, this effective diagram comes from summing over several bare Feynman graphs. A {\it small subset} is shown in Fig. \ref{LVn2-constituents}. 
As is clear, the sum over all such $2\rightarrow 4$ Feynman diagrams (even in the Regge limit) is a laborious task. Instead, as we will now discuss, the use of dispersive techniques is far more efficient and can be extended to higher order in $n$. 

\begin{figure}[ht]
\centering
\subfigure[]{\includegraphics[scale=1]{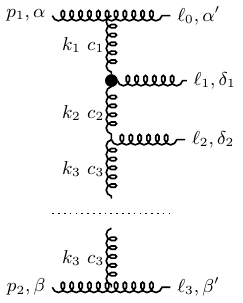}}
\qquad\qquad
\subfigure[]{\includegraphics[scale=1]{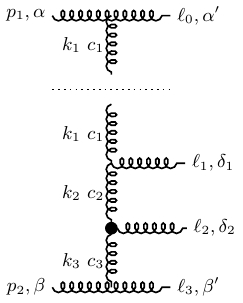}}
\qquad\qquad
\subfigure[]{\includegraphics[scale=1]{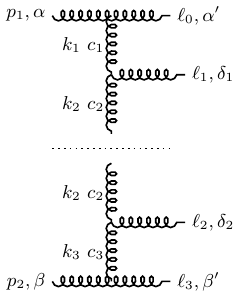}}
\caption{Cut diagrams that contribute to pole reconstruction of two Lipatov vertices in the $2\to 4$ amplitude.}
\label{2g-disp}
\end{figure}

As before, we label the incoming gluons by their momenta $p_1, p_2$ and the outgoing gluons by $\ell_i$, $i=0,1,2,3$. Here $\ell_{1,2}$ are the additional produced gluons. First, we seek to determine the residue of the $k_3^2$ pole of the amplitude associated with the bottom vertical gluon in Fig. \ref{fig:LVn2}. This pole is given by (for example, the leftmost figure in Fig.~\ref{2g-disp})
\be
P_{k_3^2}\mathcal{A}_{2\to2+2}^{\a\a' \b\b' \d_1\d_2} = \mathcal{A}_{p_1+(-k_3)\to \ell_0+\ell_1+\ell_2}^{\a\a' c_3 \d_1\d_2} gs \Gamma_{p_2\ell_3}^{\b\b'c_3}. 
\ee
Here $\mathcal{A}_{p_1+(-k_3)\to \ell_0+\ell_1+\ell_2}^{\a\a' c_3 \d_1\d_2}$ is the $2\to 3$ amplitude for which we computed the answer above -- we just need to replace $p_2\to -k_3$ appropriately. As a result, one obtains (the sum over $c_1$ and $c_2$ is implied below)
\begin{align}
    &\mathcal{A}_{p_1+(-k_3)\to \ell_0+\ell_1+\ell_2}^{\a\a' c_3 \d_1\d_2} = - \frac{s\lambda_3 g^2}{k_1^2k_2^2} \Gamma^{\a\a'c_1}_{p_1\ell_0}\gamma^{c_2\delta_1 c_1}(k_1, k_2) \(-\sqrt{2}igf^{c_3\d_2 c_2}\)  \no\\[5pt]
    &\times \(-g_{\m\m'}+\frac{-p_{1,\m}k_{3,\m'}+p_{1,\m'}\ell_{2,\m}}{-p_1\cdot k_3} + (-k_3-\ell_2)^2\frac{p_{1,\m}p_{1,\m'}}{2(p_1\cdot k_3)^2}\)\epsilon ^\m(-k_3)\epsilon^{\m'}(\ell_2)\,,
\end{align}
where $\lambda_3$ is the Sudakov decomposition parameter associated with $k_3$ and $\gamma^{c_2\delta_1 c_1}(k_1, k_2)$ is defined as
\begin{align}
\label{gamma-lipatov}
    \gamma^{c_2\delta_1 c_1}(k_1, k_2) \equiv i f^{c_2\delta_1c_1}C_\nu(k_1, k_2)\epsilon^\nu (\ell_1)\,.
\end{align}
Following the same steps as previously, we can simplify this expression further:
\begin{align}
    \mathcal{A}_{p_1+(-k_3)\to \ell_0+\ell_1+\ell_2}^{\a\a' c_3 \d_1\d_2} &= - \frac{ig^3}{k_1^2k_2^2} \Gamma^{\a\a'c_1}_{p_1\ell_0}\gamma^{c_2\delta_1 c_1}(k_1, k_2) f^{c_3\d_2 c_2} \nonumber \\
    &\times \[ -(\bsk_2+\bsk_3)_{\m'} + p_{1\mu'}\(\rho_2-\frac{2k_2^2}{\lambda_3 s}\) +\lambda_3 p_{2\mu'}\] \epsilon^{\m'}(\ell_2)~.
\end{align}
Therefore the net result for the residue of $1/k_3^2$ pole of the $\mathcal{A}_{2\to2+2}$ amplitude is 
\begin{align}
\label{k3-pole-2-4}
    P_{k_3^2}\mathcal{A}_{2\to2+2}^{\a\a' \b\b' \d_1\d_2} = & -\frac{ig^4s}{k_1^2k_2^2} \Gamma^{\a\a'c_1}_{p_1\ell_0}\gamma^{c_2\delta_1 c_1}(k_1, k_2) f^{c_3\d_2 c_2}  \Gamma_{p_2\ell_3}^{\b\b'c_3} \no\\[5pt] 
    & \times\[ -(\bsk_2+\bsk_3)_{\m'} + p_{1\mu'}\(\rho_2-\frac{2k_2^2}{\lambda_3 s}\) +\lambda_3 p_{2\mu'}\] \epsilon^{\m'}(\ell_2)\,.
\end{align}
In a similar manner, we can determine the residue of the $1/k_1^2$ pole to be
\begin{align}
\label{k1-pole-2-4}
    P_{k_1^2}\mathcal{A}_{2\to2+2}^{\a\a' \b\b' \d_1\d_2} =& \frac{ig^4s}{k_2^2k_3^2} \Gamma^{\a\a'c_1}_{p_1\ell_0}\gamma^{c_3\delta_2 c_2}(k_2, k_3)  f^{c_1\d_1 c_2}  \Gamma_{p_2\ell_3}^{\b\b'c_3} \no\\[5pt] 
    &\times \[ -(\bsk_1+\bsk_2)_{\m'} + p_{2\mu'}\(\lambda_2-\frac{2k_2^2}{\rho_1 s}\) +\rho_1 p_{1\mu'}\] \epsilon^{\m'}(\ell_1)\,.
\end{align}
Finally, we need to compute the residue of $1/k_2^2$ pole (rightmost diagram in Fig. \ref{2g-disp})
\begin{align}
\label{eq:k2-pole}
    P_{k_2^2}\mathcal{A}_{2\to2+2}^{\a\a' \b\b' \d_1\d_2} = \mathcal{A}_{p_1+(-k_2)\to \ell_0+\ell_1}^{\a\a' c_2 \d_1} s \mathcal{A}_{k_2+p_2\to \ell_2+\ell_3}^{\b\b' c_2 \d_2}\,.
\end{align}
The two 2-to-2 sub-amplitudes $\mathcal{A}_{p_1+(-k_2)\to \ell_0+\ell_1}^{\a\a' c_2 \d_1}$ and $\mathcal{A}_{k_2+p_2\to \ell_2+\ell_3}^{\b\b' c_2 \d_2}$ can be written in a manner similar to Eq.~\eqref{general-born-amplitude},
\begin{align}
    \mathcal{A}_{p_1+(-k_2)\to \ell_0+\ell_1}^{\a\a' c_2 \d_1} = g^2\Gamma^{\a\a'c_1}_{p_1\ell_0} \frac{(p_1-k_2)^2}{k_1^2} \Gamma^{c_2\d_1 c_1}_{-k_2\ell_1}~,\qquad \mathcal{A}_{k_2+p_2\to \ell_2+\ell_3}^{\b\b' c_2 \d_2} = g^2\Gamma^{c_2\delta_2c_3}_{k_2\ell_2} \frac{(p_2+k_2)^2}{k_3^2} \Gamma^{\b\b'c_3}_{p_2\ell_3}~,
\end{align}
where (after some simplification) the sub-amplitudes read 
\begin{align}
    &\mathcal{A}_{p_1+(-k_2)\to \ell_0+\ell_1}^{\a\a' c_2 \d_1} = \Gamma^{\a\a'c_1}_{p_1\ell_0} \frac{-g}{k_1^2} \[igf^{c_2\delta_1c_1}\(\l_2 p_{2,\m'}-(\bsk_{1,\m'}+\bsk_{2,\m'})+\rho_1 p_{1,\m'} - \frac{2k_1^2}{s\lambda_2}p_{1,\m'}\)\epsilon^{\m'}(\ell_1)\]~,\no\\
    &\mathcal{A}_{k_2+p_2\to \ell_2+\ell_3}^{\b\b' c_2 \d_2} = \Gamma^{\b\b'c_3}_{p_2\ell_3} \frac{g}{k_3^2} \[igf^{c_2\delta_2c_3}\(\rho_2 p_{1,\m'}-(\bsk_{2,\m'}+\bsk_{3,\m'})+\lambda_3p_{2,\m'}-\frac{2k_3^2}{s\rho_2}p_{2,\m'}\) \epsilon^{\m'}(\ell_2)\]~.
\end{align}
Collecting all terms together, and substituting these in Eq.~\eqref{eq:k2-pole}, we obtain the residue of the $1/k_2^2$ pole of the $\mathcal{A}_{2\to2+2}^{\a\a' \b\b' \d_1\d_2}$ amplitude to be
\begin{align}
\label{k2-pole-2-4}
    P_{k_2^2}\mathcal{A}_{2\to2+2}^{\a\a' \b\b' \d_1\d_2} &= -\frac{sg^2}{k_1^2 k_3^2}\Gamma^{\a\a'c_1}_{p_1\ell_0}\Gamma^{\b\b'c_3}_{p_2\ell_3}\[igf^{c_2\delta_1c_1}\(\l_2 p_{2,\m'}-(\bsk_{1,\m'}+\bsk_{2,\m'})+\rho_1 p_{1,\m'} - \frac{2k_1^2}{s\lambda_2}p_{1,\m'}\)\epsilon^{\m'}(\ell_1)\]\no\\[5pt]
    &\times\[igf^{c_2\delta_2c_3}\(\rho_2 p_{1,\m'}-(\bsk_{2,\m'}+\bsk_{3,\m'})+\lambda_3p_{2,\m'}-\frac{2k_3^2}{s\rho_2}p_{2,\m'}\) \epsilon^{\m'}(\ell_2)\]~.
\end{align}
Now examining the results in Eqs.~\eqref{k3-pole-2-4}, \eqref{k1-pole-2-4} and \eqref{k2-pole-2-4}, we can read off the simultaneous pole in $1/(k_1^2k_2^2k_3^2)$. The final result for the $2\to 2+2$ MRK amplitude is
\begin{align}
\label{2-4-final}
    \mathcal{A}_{2\to2+2}^{\a\a' \b\b' \d_1\d_2} = &-\frac{sg^2}{k_1^2k_2^2k_3^2}\Gamma^{\a\a'c_1}_{p_1\ell_0}\Gamma^{\b\b'c_3}_{p_2\ell_3}\no\\
    &\times \[igf^{c_2\delta_1c_1}\(-(\bsk_{1,\m'}+\bsk_{2,\m'})+p_{1,\m'}\[\rho_1 - \frac{2k_1^2}{s\lambda_2}\] + \[\l_2 -\frac{2k_2^2}{\rho_1 s}\]p_{2,\m'} \)\epsilon^{\m'}(\ell_1)\]\no\\[5pt]
    &\times \[igf^{c_2\delta_2c_3}\(-(\bsk_{2,\m'}+\bsk_{3,\m'})+p_{2,\m'}\[\lambda_3-\frac{2k_3^2}{s\rho_2}\]+p_{1,\m'}\[\rho_2 -\frac{2k_2^2}{\lambda_3 s}\]\) \epsilon^{\m'}(\ell_2)\]\no\\[10pt]
    = &~ \frac{s g^4}{k_1^2k_2^2k_3^2}\Gamma^{\a\a'c_1}_{p_1\ell_0}\gamma^{c_1\delta_1c_2}(k_1,k_2)\gamma^{c_2\delta_2c_3}(k_2,k_3)\Gamma^{\b\b'c_3}_{p_2\ell_3} \,.
\end{align}
This neat (and compact) result is what is shown diagrammatically in Fig. \ref{fig:LVn2}. 

\begin{figure}[ht]
\centering
\includegraphics[scale=1]{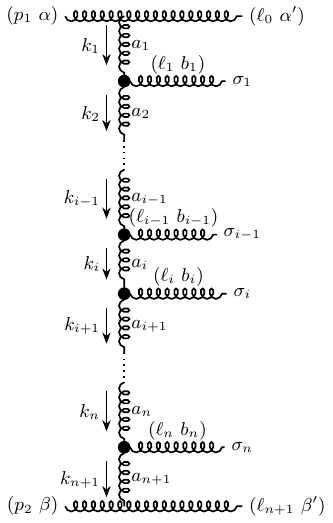}
\caption{Tree level multi-gluon production amplitude in multi-Regge kinematics. Black blobs represent the nonlocal Lipatov effective vertices.}
\label{QCD-half-ladder}
\end{figure}

The above method can be straightforwardly generalized to the $\mathcal{A}_{2\to 2+n}$ MRK amplitude and we find that the result is of the same form as those in Eqs.~\eqref{2-3-final} and \eqref{2-4-final}. The $n$ Lipatov vertices are sandwiched between the external vertices $\Gamma$ along with the associated propagators
\begin{align}
\label{eq:2plusn-QCD-Born}
    \mathcal{A}_{2\to n+2}^{\a\a' \b\b' \d_1\cdots\d_n} = & \frac{s g^{n+2}}{\prod_{i=1}^{n+1}k_i^2}\Gamma^{\a\a'c_1}_{p_1\ell_0}\left(\prod_{i=1}^{n}\gamma^{c_i\d_ic_{i+1}}(k_i,k_{i+1})\right) \Gamma^{\b\b'c_{n+1}}_{p_2\ell_{n+1}}~,
\end{align}
where the relation of $\gamma^{c_i\d_ic_{i+1}}$ to the Lipatov vertex was specified in Eq.~\eqref{gamma-lipatov}.
The amplitude is represented by the Feynman ladder graph shown in Fig. \ref{QCD-half-ladder} with bare gluon propagators being exchanged in the t-channel and effective Lipatov vertices for each emitted gluon.  
This ladder compactly sums $n!$ bare Feynman graphs (with bare three-point gluon vertices), albeit only those corresponding to the leading contributions in MRK kinematics\footnote{Eq.~\eqref{eq:2plusn-QCD-Born} is an expression for a $2 \to 2+n$ tree-level gluon amplitude with arbitrary helicity configurations for the external gluons. When considering maximally helicity-violating (MHV) configurations, the tree-level gluon amplitude for arbitrary kinematics is described by the Parke-Taylor formula \cite{Parke:1986gb}. The MRK limit of the Parke-Taylor amplitude was obtained in \cite{DelDuca:1993pp} and it was further demonstrated in \cite{DelDuca:1995zy} that this limit coincides with the MHV projection of the tree-level amplitude in Eq.~\eqref{eq:2plusn-QCD-Born}.}. In the remainder of this section, we will discuss yet another elegant proof\footnote{A general method for constructing $2\to 2+n$ tree-level gluon amplitudes is the off-shell current method of Berends and Giele~\cite{Berends:1987me}, resulting in recursion relations for $n$-point  tree-level gluon amplitudes in terms of lower point amplitudes; a proof of the Parke-Taylor formula~\cite{Parke:1986gb} for MHV amplitudes that we noted earlier is obtained in this approach. An alternative on-shell framework is that of Britto-Cachazo-Feng-Witten (BCFW)~\cite{Britto:2004ap, Britto:2005fq} where the recursion relations of the n-point tree-level amplitudes are given in terms of sum over products of lower point on-shell tree-level amplitudes convoluted with off-shell propagators sandwiched in the product. The construction we discuss here~\cite{Gribov:1983ivg} appears to be MRK limit of this framework. For further reviews on these topics, see \cite{Elvang:2013cua,Dixon:1996wi, Dixon:2013uaa}.}. 

\begin{figure}[ht]
\centering
\includegraphics[scale=1]{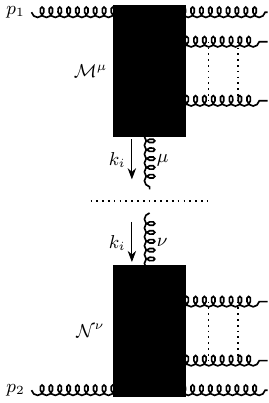}
\caption{Illustration of the $2\rightarrow n+2$ amplitude $\mathcal{A}_{2\to 2+n}(p_1, k_1,k_2,\cdots k_{n+1},p_2)$ factorized into two sub-ampitudes $\mathcal{M}^\mu(p_1, k_1, k_2, \cdots k_i)$ and $\mathcal{N}^\nu(k_i, k_{i+1}, \cdots k_{n+1}, p_2)$ separated by a cut propagator represented by the horizontal dashed line.}
\label{QCD-effective-vertices}
\end{figure}

In Fig.~\ref{QCD-half-ladder}, there are $n+1$ intermediate propagators $(ig_{\m\n}/\bsk_i^2), ~ (i = 1,\cdots, n+1)$. Consider cutting the $i$'th propagator, as we discussed previously. This separates the amplitude\\ 
$\mathcal{A}_{2\to 2+n}(p_1, k_1,k_2,\cdots k_{n+1},p_2)$ into two parts: $\mathcal{M}^\mu(p_1, k_1, k_2, \cdots k_i)$ and $\mathcal{N}^\nu(k_i, k_{i+1}, \cdots k_{n+1}, p_2)$ as depicted in Fig. \ref{QCD-effective-vertices}.
Since the $i$'th gluon was cut, and therefore on-shell, the sub-amplitudes $\cM^\m$ and $\cN^\mu$ satisfy the Ward identities
\be
k_{i\mu}\mathcal{M}^\mu(p_1, k_1, k_2, \cdots k_i) = 0~,\qquad k_{i\mu}\mathcal{N}^\mu(k_i, k_{i+1}, \cdots k_{n+1}, p_2) = 0~.
\ee
In addition, these sub-amplitudes satisfy the identities
\be
p_{1\mu}\mathcal{M}^\mu(p_1, k_1, k_2, \cdots k_i) = 0~,\qquad p_{2\mu}\mathcal{N}^\mu(k_i, k_{i+1}, \cdots k_{n+1}, p_2) = 0~.
\ee
{\it These identities are satisfied in the eikonal approximation where $\mathcal{M}^\mu$ ($\mathcal{N}^\mu$) is proportional to the largest momenta $p_1^\mu$ ($p_2^\mu$), with the sub-leading corrections suppressed by the magnitudes of these momenta.}

Further, taking advantage of the Sudakov decomposition of $k_i^
\mu$, we obtain 
\begin{align}
\label{WIs-1}
\begin{split}
    &k_{i\perp\mu}\mathcal{M}^\mu(p_1, k_1, k_2, \cdots k_i)  = -\lambda_i\, p_{2\mu}\, \mathcal{M}^\mu(p_1, k_1, k_2, \cdots k_i)~,\\[5pt]
    &k_{i\perp\nu}\mathcal{N}^\nu(k_i, k_{i+1}, \cdots k_{n+1}, p_2)  = -\rho_i\, p_{1\nu}\, \mathcal{N}^\nu(k_i, k_{i+1}, \cdots k_{n+1}, p_2)~.
\end{split}
\end{align}

As a next step in the derivation, we observe that inside the full $2\to n+2$ amplitude $\mathcal{A}_{2\to n+2}$, the sub-amplitudes $\cM^\m$ and $\cN^\n$ were glued together via the numerator $g^{\m\n}$ of the uncut propagator.
In the eikonal approximation, where we retain only the piece proportional to the external momenta, $\cM^\m$  is contracted with the largest piece of $\cN^\m$ ($p_{2}^\m$); likewise, $\cN^\n$ is contracted with the largest piece of $\cM^\n$ ($p_{1}^\n$). This implies that the numerator of the uncut propagator can be replaced by $g^{\m\n}\to 2p_1^\m p_2^\n/s$, with the factor  $2/s$ to ensure unit normalization of the metric. (Note that  since $\sum_{\text{pol}}\epsilon^\mu\epsilon^\nu\sim g^{\m\n}$, the aforementioned eikonal counting is what was responsible for the replacements of the polarization vectors by $p_{1,2}/\sqrt{s}$ earlier in this subsection.) Now, using Eq.~\eqref{WIs-1}, this replacement can be reexpressed as  
\be
\label{prop-replacement}
g^{\m\n} \to \frac{2k_{i\perp\mu}k_{i\perp\nu}}{\l_i\rho_i s}~.
\ee

Stated more transparently, we assign a factor of $\sqrt{2/s}\,k_{i\perp\mu}/\l_i$ with the vertex at the top end of an intermediate gluon line and a factor of $\sqrt{2/s}\,k_{i\perp\nu}/\rho_i$ with the vertex at the bottom end of the intermediate gluon line. Since the index $i$ corresponding to the cut is arbitrary in this discussion, the argument can be repeated for any of the {\it intermediate} gluon lines. As a result, the $\cA_{2\to 2+n}$ amplitude can be expressed as 
\begin{align}
\label{2-to-n-born}
    \cA_{2\to 2+n}^{\m\n\m'\sigma_1\cdots\sigma_n \n'} =& 2ig^{2+n}s \eta^{\m\m'}\eta^{\n\n'} G_n(b_1, \cdots, b_n) \frac{i}{\bsk_1^2}\\
    \times& \prod_{i=1}^n \frac{i}{\bsk_{i+1}^2}\frac{2k_{i\perp}^{\m_i}k_{i+1\perp}^{\n_i}}{2\l_{i+1}\rho_i s} \[g_{\m_i\n_i}(-k_i-k_{i+1})^{\s_i}+g^{\s_i}_{\m_i}(2k_i-k_{i+1})_{\n_i}+g^{\s_i}_{\n_i}(2k_{i+1}-k_i)_{\m_i}\]~,\no
\end{align}
where we used the bare three-point gluon vertex and  $G_n(b_1,\cdots b_n)$ is the color factor
\be
\label{gluon-color-factor}
G_n(b_1,\cdots b_n) = T^{a_1}\otimes T^{a_{n+1}} \prod_{i=1}^n f_{a_i a_{i+1} b_i}~,
\ee
with all the $a$-type indices contracted. The factor $\frac{2k_{i\perp}^{\m_i}k_{i+1\perp}^{\n_i}}{2\l_{i+1}\rho_i s}$ associated with the vertex $i$ comes from the replacement in Eq.~\eqref{prop-replacement} of the two vertical gluon propagators that are adjacent to the vertex. The factor of $\sqrt{2/s}\,k_{i\perp\mu}/\l_i$ comes from the $i$'th propagator and the factor $\sqrt{2/s}\,k_{i+1\perp\nu}/\rho_{i+1}$ comes from the $i+1$'th propagator. Next, inserting the Sudakov decomposition for the $k_i$ it is straightforward to show that (up to additive terms proportional to $k_i-k_{i+1}$) 
\be
\frac{2k_{i\perp}^{\m_i}k_{i+1\perp}^{\n_i}}{\l_{i+1}\rho_i s} \[g_{\m_i\n_i}(-k_i-k_{i+1})^{\s_i}+g^{\s_i}_{\m_i}(2k_i-k_{i+1})_{\n_i}+g^{\s_i}_{\n_i}(2k_{i+1}-k_i)_{\m_i}\] = C^{\s_i}(k_i, k_{i+1})\,,
\ee
where $C^{\sigma_i}$ is the effective Lipatov vertex in QCD given in Eq.~\eqref{QCD-Lipatov-vertex-1}. Plugging this expression back into Eq.~\eqref{2-to-n-born},  we recover the Born-level MRK amplitude for $2\to 2+n$ scattering in Eq.~\eqref{eq:2plusn-QCD-Born}. 
This is striking because it implies that if one takes the standard three-gluon vertex and projects on it the above mentioned specific form of the eikonal polarization vectors, one recovers the Lipatov vertex!

As we will see later in Sec. \ref{sec:CGC}, the filled black rectangles in  Fig.~\ref{QCD-effective-vertices} will be mapped to the classical color charge densities in the CGC framework, and the iterative procedure described above can likewise  mapped on to the Wilsonian renormalization group evolution of these classical sources, described by the BFKL equation. The latter framework also includes virtual corrections to the Born amplitude, which we will discuss shortly in Sec. \ref{BFKL:derivation}. 
Further, as we will discuss in Sec. \ref{sec:CGC}, the CGC framework allows one to go beyond BFKL when the source densities become large and nonperturbative.

\subsubsection{Other possible diagrams which are subleading in the Regge limit}
\label{sec:2.2.3}
\begin{figure}[ht]
\centering
\includegraphics[scale=1.25]{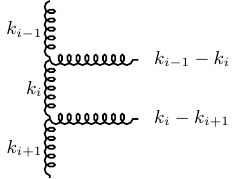}
\qquad
\qquad
\includegraphics[scale=1.25]{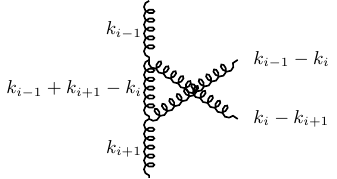}
\caption{Feynman diagrams representing the  $i$'th segment of uncrossed and crossed ladders.}
\label{cross-ladder}
\end{figure}

In the preceding discussion on the construction of the BFKL ladder, we did not consider all the tree-level diagrams that could possibly contribute to the n-gluon final state. Here we will see that the contribution from diagrams that we ignored are subleading in multi-Regge kinematics. One such diagram is the cross-ladder diagram shown in Fig. \ref{cross-ladder}
\begin{figure}[ht]
\centering
\subfigure[]{\includegraphics[scale=1]{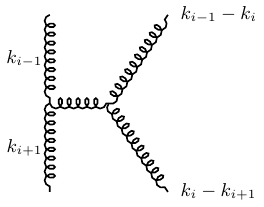}}
\qquad\qquad
\subfigure[]{\includegraphics[scale=1]{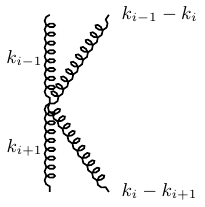}}
\caption{$i$'th segment of the ladder with a three-point gluon fusion and a quartic vertex.}
\label{k-diagrams}
\end{figure}
representing the $i$'th segment of the uncrossed and crossed ladder diagrams. In the uncrossed case, this segment of the ladder is proportional to the product of two Lipatov vertices $C^{\sigma_{i-1}}C^{\sigma_{i}}$. This product, when expanded out, yields several tensor factors. For instance, the term proportional to $p_1^{\sigma_{i-1}}p_2^{\sigma_{i}}$ in the product scales as $\rho_{i-1} \lambda_{i+1} p_1^{\sigma_{i-1}} p_2^{\sigma_i}$. We can compare this piece in the product with the corresponding factor in the crossed-ladder diagram, proportional to 
$(\rho_i^2/\rho_{i-1}^2)\,\rho_{i-1} \lambda_{i+1}\, p_1^{\sigma_{i-1}} p_2^{\sigma_i}$, which is suppressed in the MRK regime as indicated by Eq.~\eqref{MRK-regime-2}. Furthermore, the propagator in the uncrossed case is just $1/\bsk_i^2$, whereas in the crossed case, it is is proportional to $1/(\rho_{i-1}\lambda_{i+1}s)$. Since the on-shell constraint on the produced gluons implies that $\bsk_i^2 \sim s\rho_{i-1}\lambda_{i}$, this further suppresses the crossed contribution in the MRK regime. More crossed rungs will give rise to additional suppression factors.

Let us now look at the left diagram in Fig.~\ref{k-diagrams}. This was the rightmost diagram in Fig.~\ref{LVn2-constituents} that we will argue to be sub-leading in the MRK regime. Here we compare to the coefficient of $k_{i-1 \perp}^{\sigma_{i-1}} k_{i+1 \perp}^{\sigma_i}$ in the product of Lipatov vertices in the uncrossed diagram of Fig. \ref{cross-ladder}, which is unity. In Fig.~\ref{k-diagrams} (a), the coefficient is $\rho_i/\rho_{i+1}$. Furthermore, there is an extra suppression due to the internal gluon propagator, proportional to $1/(\rho_{i-1}\lambda_{i+1}s)$. Likewise, for the diagram with the 4-point vertex in Fig.~\ref{k-diagrams} (b), the coefficient of  $k_{i-1 \perp}^{\sigma_{i-1}} k_{i+1 \perp}^{\sigma_i}$ is $1/(\rho_{i-1}\lambda_{i+1}s) \approx \rho_i/(\rho_{i-1}\bsk^2)$. When we take the effect of the propagator of the vertical gluon line in the uncrossed diagram, we again see a suppression by a factor of $\rho_i/\rho_{i-1}$. 
Thus the power counting in the MRK regime establishes the claim that the dominant contribution to the scattering of two gluons into $n+2$ gluons in the multi-Regge regime is given by the uncrossed ladder in Fig.~\ref{QCD-half-ladder}. 

So far, we only discussed diagrams involving purely gluons. In QCD, one must consider contributions from fermion (quark) fields as well, in particular diagrams analogous to those in Fig.~\ref{cross-ladder} and \ref{k-diagrams} (a) where, in a section of the ladder, two of the gluons are replaced by a fermion-antifermion pair. This is depicted in Fig.~\ref{diagrams-fermion} below. These diagrams are again suppressed with respect to typical terms that arise from the purely gluonic ladder by a factor of $\rho_i/\rho_{i-1}$. In more detail, the contribution from Fig. \ref{diagrams-fermion} (a) is 
$$\frac{1}{\rho_{i-1} \lambda_{i+1} s} \bar{u}\left(k_{i-1}-k_i\right) \gamma \cdot k_{i-1 \perp} \frac{\gamma \cdot k_i}{k_i^2} \gamma \cdot k_{i+1 \perp} u\left(k_{i+1}-k_i\right).$$ 
Here $u$ ($\bar{u}$) is the fermion (anti-fermion) wavefunction and $\gamma^\mu$ are the Dirac gamma matrices. 
We further use the earlier argument that the lower part of $i-1$'th gluon propagator gives a factor of $\sqrt{2/s}\,k_{i-1\perp}/\lambda_{i-1}$ and the upper part of the $i+1$'th gluon propagator gives a factor of $\sqrt{2/s}\,k_{i+1\perp}/\rho_{i+1}$.  The contribution from Fig.~\ref{diagrams-fermion} (a) is therefore of the order $\frac{\bsk^2}{\rho_{i-1} \lambda_{i+1} s}$ which, as we have seen previously, is of the order $\rho_i/\rho_{i+1}$. Thus we obtain a contribution that is $\rho_i/\rho_{i+1}$ suppressed relative to a typical term from the purely gluonic ladder. Likewise, the contributions from the graphs (b) and (c) are similarly suppressed.  

Therefore, in the leading logarithmic approximation, we can neglect fermion-antifermion pair production in the final state for the computation of the imaginary part of the $2\to 2$ elastic amplitude. Beyond leading logarithmic accuracy, such fermion contributions, along with those shown in Figs.~\ref{cross-ladder} and \ref{k-diagrams}, can  contribute to the BFKL ladder.

\begin{figure}[ht]
\centering
\subfigure[]{\includegraphics[scale=1]{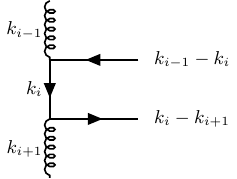}}
\qquad\qquad
\subfigure[]{\includegraphics[scale=1]{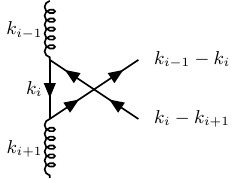}}
\qquad\qquad
\subfigure[]{\includegraphics[scale=1]{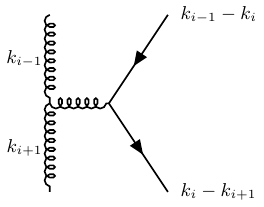}}
\caption{$i$'th segment of the ladder with fermionic lines.}
\label{diagrams-fermion}
\end{figure}
\subsection{Leading virtual graphs: Reggeization}
\label{BFKL:virtual}

In addition to multiple tree level contributions to the $2\rightarrow n$ amplitude that we discussed thus far (culminating in Eq.~\eqref{2-to-n-born}), there are also  virtual corrections that contribute to the same leading logarithmic accuracy in MRK kinematics. These must therefore be fully taken into account. 

\begin{figure}[ht]
  \centering
  \raisebox{-30pt}{\includegraphics[scale=1]{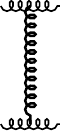}}
  =
  \raisebox{-30pt}{\includegraphics[scale=1]{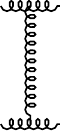}}
  +
  \raisebox{-30pt}{\includegraphics[scale=1]{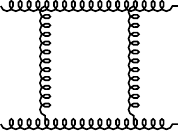}}
  +
  \raisebox{-30pt}{\includegraphics[scale=1]{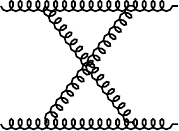}}
  +
  $\cdots$
  \caption{The reggeized gluon, depicted as a thick gluon line, resums the leading double logarithmic corrections to the bare one-gluon exchange amplitude.}
  \label{gluon-reggeization}
\end{figure}

The leading virtual terms originate from multiple t-channel gluon exchange (``horizontal ladder") diagrams shown in Fig. \ref{gluon-reggeization}. The contribution from the uncrossed and crossed diagrams exponentiates in the MRK regime, with each additional virtual gluon exchange introducing a leading logarithmic factor of  $\alpha_S \log(s/|t|) \alpha(t)$, where $\alpha(t)$ is the Regge trajectory that will be defined shortly. The result is nontrivial. It depends on the following: a) employing the Sudakov decomposition of the four internal propagators, b) performing a contour integral over one of the Sudakov variables, and explicitly performing the integral over the second one, c) using crossing symmetry, and d) taking the octet projection of the product of the color structure factors\footnote{This derivation is discussed at length (specifically, pages 34-36) in \cite{DelDuca:1995hf} so we not repeat the argument here but refer the interested reader to that discussion.}. 

We obtained these one-loop results in Eqs.~\eqref{gluon-one-loop-I} and \eqref{gluon-one-loop-II}, and the result for the one-loop gluon Regge trajectory $\alpha(t)$ reads 
\begin{align}
\label{one-loop-gluon-trajectory}
    \alpha(t) =  \alpha_S N_c\,t \int \frac{d^2\bsq}{(2\pi)^2} \frac{1}{\bsq^2(\bsk-\bsq)^2} = \frac{\alpha_S N_c}{2\pi} \log\(\frac{-t}{\lambda^2}\)~,  \quad\quad\quad \left(t=-\bsk^2\right)~,
\end{align}
where $\lambda$ is an infrared cutoff.
This ``dressing" of the bare t-channel gluon propagator is equivalent to replacing the propagators for the vertical gluon lines in the ladder with the reggeized propagators:
\be
\label{reggeized-gluon-prop}
\frac{i g_{\mu \nu}}{\bsk_{i}^2} \to \frac{i g_{\mu \nu}}{\bsk_{i}^2}\left(\frac{\hat{s}_i}{\bsk^2}\right)^{\alpha\left(k_i^2\right)}~.
\ee
In this expression, $\hat{s}_i$ is the invariant constructed out of the momenta of the $i-1$ and $i$'th emitted gluon from the half-ladder: $\hat{s}_i = (\ell_{i-1}+\ell_{i})^2$. Here we reiterate that the momentum $\bsk^2$ which appears in the denominator in the parenthesis is a transverse momentum whose magnitude is much smaller than $\sqrt{s}$--see Eq.~\eqref{MRK-regime-1} and the discussion below it.

Note that the Regge trajectory has a logarithmic dependence on the IR cutoff $\lambda$. As we will see in the following sub-section, this divergence will cancel against the contribution from the real terms. Before we get there, we will first present an alternative way of incorporating the virtual corrections to the tree-level $2\to 2+n$ amplitude that also explains the origin of the trajectory in Eq.~\eqref{one-loop-gluon-trajectory} in an elegant way. 

\begin{figure}[ht]
\centering
\includegraphics[scale=1]{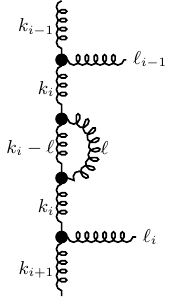}
\caption{Contribution of a soft virtual gluon to the tree-level MRK amplitude. One of the Lipatov vertices corresponds to the summation over all Feynman diagrams corresponding to leading gluon emission (absorption) as shown in Fig.~\ref{LV-constituents}, while the other  corresponds to the summation over all leading absorption (emission) Feynman diagram contributions.  }
\label{virtual-gluon}
\end{figure}

To obtain the radiative corrections to the tree-level $2\to 2+n$ MRK amplitude, we first start with the $2\to 2+n+2$ amplitude, with the two additional gluons of momentum $\ell$ and $-\ell$ with rapidities in between those of the gluons of momenta $\ell_{i-1}$ and $\ell_{i}$. Specifically, we assume that with $\ell = \rho p_1+\lambda p_2+\bel$, we have the ordering $\cdots \rho_{i-1} \gg \rho \gg \rho_{i} \cdots$ and $\cdots \lambda_{i-1} \ll \lambda \ll \lambda_{i} \cdots$. However, instead of putting the two additional momenta on-shell, we make them virtual by replacing the $\delta(\ell^2)$ function with the propagator $1/\ell^2$ and integrating over $\ell$. This construction is depicted in 
Fig.~\ref{virtual-gluon}. 

A point to emphasize is that the additional two gluons are connected to the ladder via the Lipatov vertex as opposed to bare 3-point gluon vertices. This therefore includes the automatic sum over all the relevant (bare) Feynman graphs with a soft internal gluon line. In contrast, one would have needed to sum over all the graphs manually had we used bare three-point vertices. 
The effect of adding this virtual line is to insert in the tree-level amplitude the factor $\sigma(\bsk_i)$ given by
\begin{align}
\sigma(\bsk_i) &= \frac{g^2N_c}{2}\frac{i}{k_i^2}\int \frac{d^4 \ell}{(2\pi)^4}  \frac{i}{\ell^2}\frac{i}{(k-\ell)^2} C^\mu(k_i, k_i-\ell)C_\mu(k_i-\ell,k_i)~.
\end{align}
The factor of $N_c$ arises from contraction of the $SU(N_c)$ structure constants $f_{abc}f_{abd} = N_c\delta_{cd}$ associated with each pair of adjacent Lipatov vertex that are contracted with each other. The factor of $1/2$ is included to avoid double counting the Feynman diagrams corresponding to the different ways in which a soft line can be attached. (This can be visualized as the combinatorics of the contributions that generate the crossed and uncrossed exchanges we discussed previously.)  

In the product of these Lipatov vertices, we will keep only terms that have a pole in $s$, since the contributions from the remaining terms do not give a large $\log s$ factor. The relevant term in the product is 
\begin{align}
C^\mu(k_i, k_i-\ell)C_\mu(k_i-\ell,k_i) \approx \frac{2s\bsk^4_i}{((\lambda_i-\lambda) s+i\epsilon)((\rho_i-\rho) s - i\epsilon)}~.
\end{align}
Here we reinstated the $i\epsilon$'s which are important for carrying out the integration over the Sudakov variables. Using the Sudakov decomposition of $\ell$, and writing the measure as $d^4\ell = \frac{s}{2}d\rho d\lambda d^2\bel$, we get\footnote{Recall that $k_i^2 = s\rho_i\lambda_i-\bsk_i^2\approx -\bsk_i^2$ because $s\lambda_i=\bsk_i^2$ is enforced by the phase space $\delta$-function.}
\begin{align}
\sigma(\bsk_i) = -\frac{g^2N_cs}{4}\frac{i}{\bsk_i^2}\int \frac{d^2 \bel}{(2\pi)^2}&\int \frac{d\rho d\lambda}{(2\pi)^2}  \frac{i}{s\lambda\rho-\bel^2- i\epsilon}\frac{i}{-s\rho(\lambda_i-\lambda)-(\bsk_i-\bel)^2-i\epsilon}\nonumber\\
&\times \frac{2s\bsk^4_i}{((\lambda_i-\lambda) s+i\epsilon)((\rho_i-\rho) s - i\epsilon)}~.
\end{align}
Integrating over $\lambda$ using Cauchy's theorem gives
\begin{align}
\sigma(\bsk_i) &= -\frac{g^2N_c}{2}\int \frac{d^2 \bel}{(2\pi)^2} \frac{\bsk_i^2}{\bel^2(\bsk_i-\bel)^2} \int \frac{d\rho}{2\pi}\frac{1}{\rho} ~.
\end{align}
In obtaining this result, we used $\rho\gg \rho_i$ and $\bel^2 \sim \bsk^2_i$. Integrating over $\rho$ from $\rho_{i}$ to $\rho_{i-1}$, we obtain finally 
\begin{align}
\sigma(\bsk_i) = -\frac{g^2N_c}{4\pi}\log\(\frac{\rho_{i-1}}{\rho_i
}\) \int \frac{d^2 \bel}{(2\pi)^2} \frac{\bsk_i^2}{\bel^2(\bsk-\bel)^2} 
=  \log\(\frac{\rho_{i-1}}{\rho_i}\) \alpha(\bsk_i^2)
\approx  \log\(\frac{\hat{s}_i}{\bsk^2}\) \alpha(\bsk_i^2)~.
\label{g-traj-alt}
\end{align}
Here the last approximation is obtained as follows: The invariant $\hat{s}_i = (k_{i-1}-k_{i+1})^2 \approx \frac{\rho_{i-1}}{\rho_i} (\bsk_i-\bsk_{i+1})^2$ where we used the on-shell condition for the $i$'th outgoing gluon. In MRK kinematics, since all the transverse momenta are of the same order we can replace the factor $(\bsk_i-\bsk_{i+1})^2$ above by a typical transverse momentum squared $\bsk^2$ (without the subscript $i$). Consequently, the ratio $\hat{s}_i/\bsk^2$ becomes $\rho_{i-1}/\rho_i$. 

In the above expression for a single virtual gluon insertion, we recovered the expression for the 1-loop gluon trajectory along with the associated large log. When taking all virtual gluon insertions into account, it was conjectured in \cite{Kuraev:1976ge} that the one-loop result in Eq.~\eqref{g-traj-alt} can be exponentiated when leading log corrections are included to all-loop order\footnote{Apart from the nested integrals, the justification for the exponentiation also involves nontrivial rearrangements for the color factors associated with multiple Lipatov vertices that can appear in all possible permutations;  one needs to make use of identities such as $\Tr(T^aT^bT^aT^c)=\frac12N^2\delta^{bc}$.}; this is the phenomenon of gluon reggeization. 
Namely, every internal vertical propagator $1/\bsk_i^2$ is multiplied by the factor $(\frac{\hat{s}_i}{\bsk^2})^{\alpha(\bsk_i^2)}$. This generalizes the replacement in Eq.~\eqref{reggeized-gluon-prop} to all orders in MRK kinematics.
\subsection{Derivation of the BFKL equation}
\label{BFKL:derivation}
Following this replacement, the $2\to n+2$ multi-Regge gluon scattering amplitude, including both real and virtual contributions, reads

\begin{align}
\label{2twonQCD}
    \mathcal{A}_{2\to n+2}^{\m\m'\n\n'\sigma_1\cdots\sigma_n} = 2is g^{n+2} \eta^{\m\m'}\eta^{\n\n'} G_n(b_1,\cdots b_n) \frac{i}{\bsk_1^2} \(\frac{1}{\rho_1}\)^{\alpha(\bsk_1^2)} \prod_{i=1}^n C^{\sigma_i}(k_i, k_{i+1})\frac{i}{\bsk_{i+1}^2}\(\frac{\rho_{i-1}}{\rho_{i}}\)^{\alpha(\bsk_{i}^2)},
\end{align}
where $G_n$ is the color factor in Eq.~\eqref{gluon-color-factor}.
\begin{figure}[ht]
\centering
\includegraphics[scale=1]{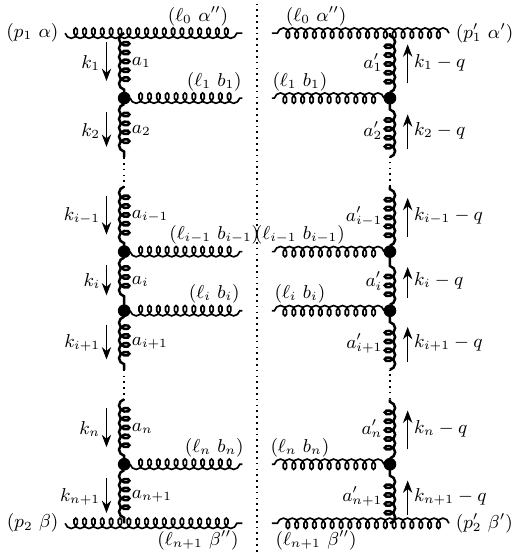}
\caption{The n-rung ladder contribution to the imaginary part of the amplitude, where the bold vertical gluon lines indicate reggeized gluons and the black dots indicate Lipatov vertices.}
\label{fig:n-ladder-cut}
\end{figure}
To compute the imaginary part of the $2\to 2$ scattering amplitude, we insert this MRK amplitude into the unitarity condition (optical theorem) in Eq.~\eqref{unitarity-condition}. After taking the color sum projection onto the singlet representation, we obtain 
\begin{align}
\label{2-2-gluon-im-part}
    \Im \mathcal{A}_{2\to 2}^{\m\m'\n\n'} = & \eta^{\m\m'} \eta^{\n\n'} \sum_{n=0}^\infty\frac{4s^2 g^{2n+4}(-1)^n}{2^{n+2}\(2\pi\)^{3n+2}}  \int \prod_{i=1}^{n} \(\frac{d\rho_i}{\rho_i} d^2\bsk_i \)d\rho_{n+1}d^2\bsk_{n+1}  \delta\[s\rho_{n+1}-\bsk_{n+1}^2\] \no \\
    &\times P_S\[ \sum_{b_i} G_{n}(b_1,\cdots, b_i)G_{n}(b_1,\cdots, b_i) \]\frac{1}{\bsk_1^2(\bsq-\bsk_1)^2}\(\frac{1}{\rho_1}\)^{\a(\bsk_1^2)+\a((\bsq-\bsk_1)^2)}\\
    &\times\prod_{i=1}^n C^{\mu_i}(k_i, k_{i+1})C_{\mu_i}(q-k_i, q-k_{i+1}) \frac{1}{\bsk_{i+1}^2(\bsq-\bsk_{i+1})^2}  \(\frac{\rho_{i}}{\rho_{i+1}}\)^{\a(\bsk_{i+1}^2)+\a((\bsq-\bsk_{i+1})^2)}\,. \no
\end{align}
The factor of $(-1)^n$ is from the sum over gluon polarizations. From the sum over the color factors in Eq.~\eqref{unitarity-condition-0}, one gets 
\begin{align}
\label{color-1}
     &\sum_{\alpha''\beta'' b_1,\cdots b_n}f_{\a a_1 \a''}f_{\b'' a_{n+1}\b}f_{\a'' a'_1 \a'}f_{\b' a'_{n+1}\b''}\prod_{i=1}^n f_{a_ia_{i+1}b_i}f_{a'_ia'_{i+1}b_i}\no\\[5pt]
     =&\sum_{\alpha''\beta'' b_1,\cdots b_n}\(T^{a_1}T^{a'_1}\)_{\a \a'} \(T^{a'_{n+1}}T^{a_{n+1}}\)_{\b' \b} \prod_{i=1}^n f_{a_ia_{i+1}b_i}f_{a'_ia'_{i+1}b_i}\,,
\end{align}
where in the last line we used $f_{abc} = i(T^b)_{ac}$ with $T^b$ being the generator of the gauge group in the adjoint representation. 

This expression is manifestly in a tensor product representation of two adjoint representations of the gauge group. For $SU(N)$, this product decomposes into a singlet, an octet plus other higher dimensional representations\footnote{For $N=3$ the decomposition is $\boldsymbol{8}\otimes\boldsymbol{8} = (\boldsymbol{1}\oplus \boldsymbol{8}\oplus\boldsymbol{27})_S \oplus (\boldsymbol{8}\oplus \boldsymbol{10}\oplus\boldsymbol{\bar{10}})_A$ where the subscript S (A) stands for symmetric (anti-symmetric) representation.}. To obtain the single representation, we contract $\a$ with $\a'$ and  $\b$ with $\b'$. This projection onto the singlet representation is what we denoted by $P_S[\cdots]$ in Eq.~\eqref{2-2-gluon-im-part}. It evaluates to
\begin{align}
P_S\[\sum_{b_i} G_{n}(b_1,\cdots, b_i)G_{n}(b_1,\cdots, b_i)\] & = \Tr(T^{a_1}T^{a_1'})\Tr(T^{a_{n+1}}T^{a'_{n+1}})\prod_{i=1}^n f_{a_ia_{i+1}b_i}f_{a'_ia'_{i+1}b_i}~,\no\\
&=N_c^2 \delta^{a_1, a_1'}\delta^{a_{n+1}a'_{n+1}} \prod_{i=1}^n f_{a_ia_{i+1}b_i}f_{a'_ia'_{i+1}b_i}~,\no\\
&=N_c^{n+2}(N_c^2-1)~.
\end{align}
One can also obtain the octet $\boldsymbol{8}_A$ representation from Eq.~\eqref{color-1} by contracting the hanging color indices by $f_{\a c \a'}f_{\b c \b'}$. This projection is relevant for the demonstration of gluon reggeization \cite{DelDuca:1995hf}.

For $n=0,1$ this gives the result in Eq.~\eqref{lowest-order-gluon} and Eq.~\eqref{n=1-color-sum} respectively. Furthermore, the contraction over the Lipatov vertices give
\be
C^{\mu_i}(k_i, k_{i+1})C_{\mu_i}(q-k_i, q-k_{i+1}) = 2\(\bsq^2 -\frac{\bsk_i^2 (\bsq-\bsk_{i+1})^2+    \bsk_{i+1}^2 (\bsq-\bsk_{i})^2}{(\bsk_i-\bsk_{i+1})^2}\)~.
\ee
Putting everything together, and expressing the imaginary part in terms of the Born amplitude $\mathcal{A}_0^{\m\m'\n\n'}(s,t) = 2g^2 \frac{s}{-t} \eta^{\m\m'}\eta^{\n\n'} $,
we get
\begin{align}
\label{cr0}
\begin{split}
    &\frac{\Im \mathcal{A}^{\m\m'\n\n'}_{2\to 2}}{\mathcal{A}^{\m\m'\n\n'}_0(s,t)} =  \frac{s\,t}{2}\sum_{n=0}^\infty\frac{ g^{2n+2}N_c^{n+2}(N_c^2-1)(-1)^{n+1}}{\(2\pi\)^{3n+2}} \int \prod_{i=1}^{n} \(\frac{d\rho_i}{\rho_i} d^2\bsk_i \)d\rho_{n+1}d^2\bsk_{n+1}  \delta\[s\rho_{n+1}-\bsk_{n+1}^2\] \\
    & \times \frac{1}{\bsk_1^2(\bsq-\bsk_1)^2}\(\frac{1}{\rho_1}\)^{\a(\bsk_1^2)+\a((\bsq-\bsk_1)^2)} 
    \prod_{i=1}^n \(\bsq^2 -\frac{\bsk_i^2 (\bsq-\bsk_{i+1})^2+    \bsk_{i+1}^2 (\bsq-\bsk_{i})^2}{(\bsk_i-\bsk_{i+1})^2}\) \\
    &\times\frac{1}{\bsk_{i+1}^2(\bsq-\bsk_{i+1})^2}  \(\frac{\rho_{i}}{\rho_{i+1}}\)^{\a(\bsk_{i+1}^2)+\a((\bsq-\bsk_{i+1})^2)} \,.
\end{split}
\end{align}

This expression can be simplified greatly as follows. We first take its  Mellin transform, which can be expressed as 
\begin{align}
    \mathcal{M}_\ell(\bsq^2) \equiv \int_1^\infty d\(\frac{s}{\bsk^2}\)\frac{\Im \mathcal{A}^{\m\m'\n\n'}_{2\to 2}(s,t)}{\mathcal{A}^{\m\m'\n\n'}_0(s,t)}\(\frac{s}{\bsk^2}\)^{-\ell-1}~.
\end{align}
A straightforward integration over $s$ (recall that $\bsq^2 = -t$) gives
\begin{align}
\begin{split}
    &\mathcal{M}_\ell(\bsq^2) = \frac{-t}{2}\sum_{n=0}^\infty \frac{ g^{2n+2}N_c^{n+2}(N_c^2-1)}{\(2\pi\)^{3n+2}} \int \prod_{i=1}^{n} \(\frac{d\rho_i}{\rho_i} d^2\bsk_i \)d\rho_{n+1}d^2\bsk_{n+1} \rho_{n+1}^{\ell-1} \frac{1}{\bsk_1^2(\bsq-\bsk_1)^2}\(\frac{1}{\rho_1}\)^{\a(\bsk_1^2)+\a((\bsq-\bsk_1)^2)} \\[5pt]
    &\times (-1)^n \prod_{i=1}^n \(\bsq^2 -\frac{\bsk_i^2 (\bsq-\bsk_{i+1})^2+    \bsk_{i+1}^2 (\bsq-\bsk_{i})^2}{(\bsk_i-\bsk_{i+1})^2}\) \frac{1}{\bsk_{i+1}^2(\bsq-\bsk_{i+1})^2}  \(\frac{\rho_{i}}{\rho_{i+1}}\)^{\a(\bsk_{i+1}^2)+\a((\bsq-\bsk_{i+1})^2)} \,.
\end{split}
\end{align}
Next, we successively integrate over the $\rho$ variables keeping in mind that the domain of the integration for $\rho_i$ is ($\rho_{i+1}, \rho_{i-1}$) and the MRK condition forces us to drop contributions from the lower limit of the integral. The result of this step is
\begin{align}
    \mathcal{M}_\ell(\bsq^2) &= \frac{-t}{2} \sum_{n=0}^\infty \frac{g^{2n+2}N_c^{n+2}(N_c^2-1)}{\(2\pi\)^{3n+2}} \int \prod_{i=1}^{n+1} d^2\bsk_i \frac{1}{\bsk_i^2(\bsq-\bsk_i)^2}\frac{1}{\ell-\a(\bsk_{i}^2)-\a((\bsq-\bsk_{i})^2)} \no \\[10pt]
    &\times (-1)^n \prod_{i=1}^n \(\bsq^2 -\frac{\bsk_i^2 (\bsq-\bsk_{i+1})^2+\bsk_{i+1}^2 (\bsq-\bsk_{i})^2}{(\bsk_i-\bsk_{i+1})^2}\)~, \\[10pt]
    &=   2\pi\bsq^2 \a_s N_c^2(N_c^2-1)  \sum_{n=0}^\infty (-2\a_s N)^{n} \int \prod_{i=1}^{n+1} \frac{d^2\bsk_i}{\(2\pi\)^2} \frac{1}{\bsk_i^2(\bsq-\bsk_i)^2}\frac{1}{\ell-\a(\bsk_{i}^2)-\a((\bsq-\bsk_{i})^2)}\no \\[10pt]
    &\times \prod_{i=1}^n \(\bsq^2 -\frac{\bsk_i^2 (\bsq-\bsk_{i+1})^2+    \bsk_{i+1}^2 (\bsq-\bsk_{i})^2}{(\bsk_i-\bsk_{i+1})^2}\)~.
    \label{mellin-amplitude-1}
\end{align}
Remarkably, this equation can be rexpressed simply as 
\begin{align}
\label{eq:Mellin-amplitude-2}
    \mathcal{M}_\ell(\bsq^2) = 2 \pi\bsq^2 \a_s N_c^2(N_c^2-1) \int \frac{d^2\bsk}{\(2\pi\)^2} \frac{1}{\bsk^2(\bsq-\bsk)^2} f_\ell(\bsk, \bsq)~,
\end{align}
where $f_\ell(\bsk, \bsq)$ satisfies the integral equation 
\begin{align}
\label{integral-eq-qcd}
    (\ell-\alpha(\bsk^2)-\alpha((\bsq-\bsk)^2)) f_\ell(\bsk, \bsq) = 1 - 2\a_s N_c \int \frac{d^2\bsk'}{\(2\pi\)^2} \frac{f_\ell(\bsk', \bsq)}{\bsk'^2(\bsq-\bsk')^2} \(\bsq^2 -\frac{\bsk^2 (\bsq-\bsk')^2+    \bsk'^2 (\bsq-\bsk)^2}{(\bsk-\bsk')^2}\)\,.
\end{align}
It can be checked that this recursive relation generates the $n$-sum in the expression for the Mellin transformed amplitude $\mathcal{M}_\ell$ in Eq.~\eqref{mellin-amplitude-1} by expanding out in $\alpha_S$. Eq.~\eqref{integral-eq-qcd} is the celebrated BFKL equation, whose kernel is the BFKL Hamiltonian\footnote{The BFKL Hamiltonian describes the evolution in rapidity of color singlet compound states of two reggeized gluons, with remarkably properties such as holomorphic separability and conformal $SL(2,C)$ invariance. A particular generalization to multi-reggeon compound states is the Bartels-Kwiecinski-Praszalowicz (BKP) equation~\cite{Bartels:1980pe,Kwiecinski:1980wb}. In the large $N_c$ limit, this Hamiltonian is that of an $SL(2,C)$ Heisenberg magnet~\cite{Lipatov:1993yb,Faddeev:1994zg} For a review, see \cite{Belitsky:2004cz}.}. As a consistency check, one can see from a perturbative expansion of $f_\ell$ in $\alpha_S$ that one one obtains Eq.~\eqref{Mellin-amplitude-leading-orders} for the first two orders in the perturbative expansion. 

We end this sub-section with few remarks about Eq.~\eqref{integral-eq-qcd}. Firstly, the solutions of the BFKL equation are free of any IR divergences. The IR divergence coming from the phase space integral of the kernel (the term in the parenthesis in the right hand side) is canceled by IR divergence of the one-loop gluon Regge trajectory. Secondly, for $\bsq=0$, this equation computes the total cross-section for two-gluon scattering. A significant consequence is that the solution of the BFKL equation (which we will discuss shortly) predicts a rapid growth of deeply inelastic scattering (DIS)  electron-hadron scattering cross-section at high energies such as those accessed at the HERA collider and the future Electron-Ion Collider (EIC)~\cite{Aschenauer:2017jsk}.  This growth in the cross-section is far more rapid than observed at HERA. It is tamed both by including next-to-leading logarithmic contributions and by gluon saturation effects~\cite{Morreale:2021pnn} as we will discuss further in Sec.~\ref{sec:NLL-gluon-sat} and at length in Sec. \ref{sec:CGC}.

\subsection{Solution of the BFKL Equation}

We will now outline the solution of Eq.~\eqref{integral-eq-qcd}. We first reexpress the equation as 
\begin{align}
    \ell f_\ell(\bsk,\bsq)-1 &= -\alpha_SN_c  \int \frac{d^2\bsk'}{(2\pi)^2} \bigg[ \frac{2}{\bsk'^2(\bsq-\bsk')^2} \(\bsq^2 -\frac{\bsk^2 (\bsq-\bsk')^2+\bsk'^2 (\bsq-\bsk)^2}{(\bsk-\bsk')^2}\)f_\ell(\bsk', \bsq)\no\\
    & +\(\frac{\bsk^2}{\bsk'^2(\bsk'-\bsk)^2}+ \frac{(\bsk-\bsq)^2}{\bsk'^2(\bsk'-\bsq+\bsk)^2}\)f_\ell(\bsk,\bsq)\bigg]\,,
\end{align}
where we used the expression for the one-loop gluon trajectory in Eq.~\eqref{one-loop-gluon-trajectory}. Defining $\tilde\bsk = \bsq-\bsk$ and $\tilde\bsk' = \bsq-\bsk'$, this becomes 
\begin{align}
    &\ell f_\ell(\bsk,\bsq)-1 = \no\\
    &-\alpha_SN_c  \int \frac{d^2\bsk'}{(2\pi)^2} \bigg[ \frac{2}{\bsk'^2\tilde\bsk'^2} \(\bsq^2 -\frac{\bsk^2 \tilde\bsk'^2+\bsk'^2 \tilde\bsk^2}{(\bsk-\bsk')^2}\)f_\ell(\bsk', \bsq) +\(\frac{\bsk^2}{\bsk'^2(\bsk'-\bsk)^2}+ \frac{\tilde\bsk^2}{\bsk'^2(\bsk'-\tilde\bsk)^2}\)f_\ell(\bsk,\bsq)\bigg]\,.
\end{align}
In order to handle the divergences in the integral when $\bsk' \to \bsk$ and $\bsk' \to \tilde\bsk$, we replace~\cite{Forshaw:1997dc} 
\be
\label{fraction-substitution}
\frac{1}{\bsk'^2(\bsk-\bsk')^2} \to \frac{2}{(\bsk-\bsk')^2\(\bsk'^2+(\bsk-\bsk')^2\)} \,,
\ee
and after a change of integration variables, one finds
\begin{align}
    \ell f_\ell(\bsk,\bsq)-1 &= -2\alpha_SN_c  \int \frac{d^2\bsk'}{(2\pi)^2} \frac{1}{(\bsk-\bsk')^2}\bigg[ \(\bsq^2\frac{(\bsk-\bsk')^2}{\bsk'^2\tilde\bsk'^2} -\frac{\bsk^2 }{\bsk'^2}-\frac{\tilde\bsk^2}{\tilde\bsk'^2} \)f_\ell(\bsk', \bsq) \no\\
    &+\( \frac{\bsk^2}{\(\bsk'^2+(\bsk-\bsk')^2\)} + \frac{\tilde\bsk^2}{\(\tilde\bsk'^2+(\bsk-\bsk')^2\)}\)f_\ell(\bsk,\bsq)\bigg]\,.
\end{align}
A possible IR divergence arises when $\bsk'\to \bsk$. However we see that the terms in the square brackets cancel completely in this limit for any value of $\bsq$. This result shows that the imaginary part of  $2\to 2$ gluon scattering is free of IR divergences. 

As previously mentioned, the solution of this equation at $\bsq=0$ gives the total cross-section. 
The equation at $\bsq=0$ reads (dropping the argument $\bsq$ in the partial amplitudes $f_\ell$)
\begin{align}
\label{BFKL-eq-q0}
    \ell f_\ell(\bsk)-1 = 4\alpha_SN_c \int \frac{d^2\bsk'}{(2\pi)^2}\frac{1}{(\bsk-\bsk')^2}\[\frac{\bsk^2}{\bsk'^2}f_\ell(\bsk')-\frac{\bsk^2}{\bsk'^2+(\bsk-\bsk')^2}f_\ell(\bsk)\]~.
\end{align}
This is an integral equation with a self-adjoint kernel and can be solved by recasting it into an eigenvalue problem. We first write
\begin{align}
    f_\ell(\bsk) \equiv \int \frac{d^2\bsp}{(2\pi)^2}~ \frac{\bsk^2}{\bsp^2}g_\ell(\bsk,\bsp)~\qquad,\qquad 1= \int \frac{d^2\bsp}{(2\pi)^2} ~(2\pi)^2\delta^{(2)}(\bsk-\bsp)~.
\end{align}
Eq.~\eqref{BFKL-eq-q0} in these variables becomes 
\begin{align}
\label{BFKL-eq-q0-1}
    \ell g_\ell(\bsk, \bsp) - ~(2\pi)^2\delta^{(2)}(\bsk-\bsp) = 4\alpha_S N_c \int \frac{d^2\bsk'}{(2\pi)^2}\frac{1}{(\bsk-\bsk')^2}\[ g_\ell(\bsk', \bsp)-\frac{\bsk^2}{\bsk'^2+(\bsk-\bsk')^2}g_\ell(\bsk, \bsp)\]~.
\end{align}
One seeks a solution of the form
\begin{align}
\label{solution-ansatz}
    g_\ell(\bsk,\bsp) = \frac{1}{\sqrt{\bsp^2\bsk^2}}\sum_{n=-\infty}^\infty \int_{-\infty}^\infty d\n ~a_\ell(\n,n) e^{i\n(\lambda_k-\lambda_p) }e^{in(\phi_k-\phi_p)}~,
\end{align}
motivated by the $\delta$-function representation,
\begin{align}
\label{delta-function}
    \delta^{(2)}(\bsk-\bsp)=\frac{1}{2\pi^2\sqrt{\bsp^2\bsk^2}}\sum_{n=-\infty}^\infty\int_{-\infty}^\infty d\nu~ e^{i\nu (\lambda_k-\lambda_p)} e^{in(\phi_k-\phi_p)}~,
\end{align}
where $\lambda_k=\log(\bsk^2/\m^2), ~\lambda_p=\log(\bsp^2/\m^2)$, with $\m$ being an arbitrary IR scale, and $\phi_{k,p}$ are the azimuthal angles of the vectors $\bsk,\bsp$, respectively. To deal with the convolution, we write the r.h.s of Eq.~\eqref{BFKL-eq-q0-1} as 
\begin{align}
    \label{BFKL-ev-eq-q0}
    &4\alpha_S N_c \int \frac{d^2\bsk'}{(2\pi)^2}\frac{1}{(\bsk-\bsk')^2}\[ g_\ell(\bsk', \bsp)-\frac{\bsk^2}{\bsk'^2+(\bsk-\bsk')^2}g_\ell(\bsk, \bsp)\]\no\\[5pt]
    =& \frac{1}{\sqrt{\bsp^2\bsk^2}}\sum_{n=-\infty}^\infty \int_{-\infty}^\infty d\n ~a_\ell(\n,n)\omega(\n,n) e^{i\n(\lambda_k-\lambda_p) }e^{in(\phi_k-\phi_p)}\,,
\end{align}
allowing us to express the solution of Eq.~\eqref{BFKL-eq-q0-1} for the unknown coefficients $a_\ell(\n , n)$ in terms of the BFKL eigenvalue $\omega(\n,n)$ as 
\begin{align}
    \ell a_\ell(\n, n) - 2 = a(\n, n) \omega(\n, n) \implies a_\ell(\n,n) = \frac{2}{\ell - \omega(\n, n)}~.
\end{align}

All that remains now is to compute the BFKL eigenvalue $\omega(\n,n)$. Simplifying Eq.~\eqref{BFKL-ev-eq-q0}, we find 
\begin{align}
\label{BFKL-ev-eq1-q0}
    \omega(\n,n) = 4\alpha_S N_c \int \frac{d^2\bsk'}{(2\pi)^2}\frac{1}{(\bsk-\bsk')^2}\[\(\frac{\bsk^2}{\bsk'^2}\)^{1/2}e^{i\nu (\lambda_{k'}-\lambda_k)}e^{in(\phi_{k'}-\phi_k)}-\frac{\bsk^2}{\bsk'^2+(\bsk-\bsk')^2}\]\,.
\end{align}
In order to handle the singularities near $\bsk' = \bsk$, we can rewrite this equation as 
\begin{align}
\label{BFKL-ev-eq2-q0}
    \omega(\n,n) = 4\alpha_S N_c \int \frac{d^2\bsk'}{(2\pi)^2}\[\frac{e^{in(\phi_{k'}-\phi_k)}}{(\bsk-\bsk')^2}\(\frac{\bsk^2}{\bsk'^2}\)^{i\nu +\frac12}-\frac{\bsk^2}{\bsk'^2}\(\frac{1}{(\bsk-\bsk')^2} - \frac{1}{\bsk'^2+(\bsk-\bsk')^2}\)\] \,.
\end{align}
The next step is to carry out the angular integrals. These need to be performed in the regions $\bsk'^2<\bsk^2$ and $\bsk'^2>\bsk^2$ separately. First define the variable $x$ as 
\begin{align}
\label{x-def}
    x= \begin{cases}\bsk'^2 / \bsk^2 & \text { for } \bsk'^2<\bsk^2 \\ \bsk^2 / \bsk'^2 & \text { for } \bsk'^2>\bsk^2\end{cases}~.
\end{align}
Carrying out the (somewhat involved) angular integrals in Eq.~\eqref{BFKL-ev-eq2-q0} and adding the result from the two regions on gets
\begin{align}
\label{BFKL-ev-eq3-q0}
    \omega(\n,n) = \frac{\alpha_S N_c}{\pi} \int_0^1 dx \[ 2 \,\Re\(\frac{x^{\frac{|n|-1}{2}+i\n}}{1-x}\)-\(\frac{1}{x (1-x)}+\frac{1}{1-x}-\frac{1}{x \sqrt{4 x^2+1}}-\frac{1}{\sqrt{x^2+4}}\)\]\,,
\end{align}
In Eq.~\eqref{BFKL-ev-eq3-q0}, inside the square brackets, the term in the first parenthesis comes from the angular integral of the first term in square brackets in Eq.~\eqref{BFKL-ev-eq2-q0}. The reason we get the real part is because the angular integrals in the $\bsk'^2<\bsk^2$ and $\bsk'^2>\bsk^2$ regions gives complex conjugate terms with the same sign.  The four terms in the second parenthesis of Eq.~\eqref{BFKL-ev-eq3-q0} comes from the second set of terms in the square brackets of Eq.~\eqref{BFKL-ev-eq2-q0}. The first and the third of these four terms come from the $\bsk'^2<\bsk^2$ region, while the others come from the $\bsk'^2>\bsk^2$ region. Finally, we need to perform the integration in $x$.

\begin{figure}[ht]
\centering
\includegraphics[scale=0.5]{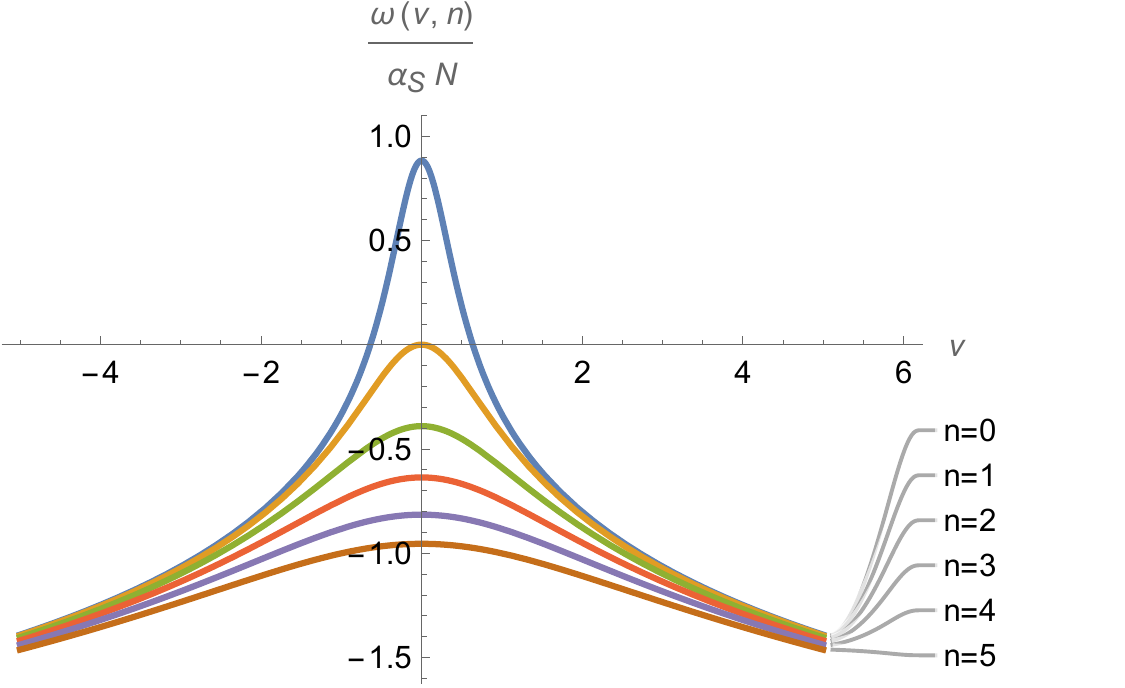}
\caption{Plot of the BFKL eigenvalues in Eq.~\eqref{BFKL-ev-eq5-q0} for $n=0,1,2,3,4,5$. The maximum is attained for $n=0$ at $\nu=0$, with the maximum value $\omega^* = 4\alpha_S N_c \ln(2)/\pi$.}
\label{fig:eigenvalues}
\end{figure}

Rearranging Eq.~\eqref{BFKL-ev-eq3-q0} a bit further, one gets via a partial fraction decomposition,
\begin{align}
\label{BFKL-ev-eq4-q0}
    \omega(\n,n) = \frac{\alpha_S N_c}{\pi} \int_0^1 dx \[ 2\,\Re\(\frac{x^{\frac{|n|-1}{2}+i\n}-1}{1-x}\)-\(\frac{1}{x}-\frac{1}{x \sqrt{4 x^2+1}}-\frac{1}{\sqrt{x^2+4}}\)\]\,.
\end{align}
One can check the integral of the second parenthesis vanishes! This is a consequence of the UV and IR finiteness of the $\bsk'$ integral of the BFKL equation. One finally gets for the BFKL 
eigenvalue, the result
\begin{align}
\label{BFKL-ev-eq5-q0}
    \omega(\n,n) = -\frac{2\alpha_SN_c}{\pi} \Re\(\psi\left(\frac{|n|+1}{2} +i\nu\right)+\gamma_E\)~,
\end{align}
where $\psi$ is the digamma function and $\gamma_E\sim 0.577$ is the Euler constant. A plot of the eigenvalues for different values of $n$ is shown in Fig.~\ref{fig:eigenvalues}. The maximum value of $\omega^*(\n,n)$ is attained for $\n=0, n=0$, given by
\be
\label{eq:max-BFKL-eigenvalue}
\omega^*(\n=0,n=0) = \frac{4}{\pi}\alpha_S N_c \ln 2~.
\ee
Plugging this result back into the expression for the partial waves, we can determine the $2\to 2$ cross-section,  
\begin{align}
    \sigma = \frac{1}{2s} \Im \mathcal{A}_{2\to 2} (s, t=0)~,
\end{align}
with the leading contribution to the imaginary part of the forward amplitude given by  $\Im \mathcal{A}_{2\to 2} (s, t=0) \sim s^{1+\omega^*}$. One therefore obtains the LLx BFKL contribution to the total cross-section to be 
\begin{equation}
\sigma \sim s^{\omega^*} = s^{\frac{4}{\pi}\alpha_S N_c \ln 2}\,\approx s^{0.5} \,\, {\rm for}\,\, \alpha_S=0.2 \,\, {\rm and}\,\, N_c=3\,.
\end{equation}
Here $\alpha_S=0.2$ provides a rough estimate of the running coupling in the small-$x$ kinematics of the HERA DIS data, where the rapid growth in cross-sections was first observed, providing the segue for the discussion below of small $x$
evolution in DIS. 

\subsection{Leading and next-to-leading log \texorpdfstring{$x_{\rm Bj}$}{} DIS cross-sections and gluon saturation}
\label{sec:NLL-gluon-sat}

 It is useful to connect our prior discussion of $2\rightarrow N$ gluon scattering to a specific underlying process in QCD for which it is relevant. The simplest example is deeply inelastic scattering (DIS) which we alluded to previously at the end of Sec.~\ref{BFKL:derivation} and shall discuss further here as well as in Section~\ref{sec:CGC} in the context of the CGC EFT. Much of the DIS discussion is worked out in textbooks~\cite{Kovchegov:2012mbw}, so we will provide here the briefest of primers sufficient for non-experts to follow the material of relevance to us. 
 
 The DIS process corresponds to a high energy electron scattering off a proton or heavier nuclei, exchanging a spacelike virtual photon with high resolution $q^2=-Q^2>0$, that probes the structure of matter inside the hadron. In the DIS ``dipole" frame, $q^\mu= (-Q^2/2q^-,q^-,{\bf 0}_\perp)$, the proton four momentum $P^\mu = (P^+,0,{\bf 0}_\perp)$ and the DIS center-of-mass energy is $s= 2 P\cdot q$. A key kinematic variable is Bjorken $x_{\rm Bj} = Q^2/(2P\cdot q)$, which in the quark-parton model is $x_{\rm Bj}\equiv x$, the fraction of the lightcone fraction of the momentum $P^+$ of the hadron carried by the struck quark or antiquark. The cross-section for the scattering (where only the incoming plus outgoing electron energies, and their relative scattering angle are measured) can be expressed as the convolution $L_{\mu\nu} W^{\mu\nu}$, where the former is the lepton tensor corresponding to the product of the lepton current in the amplitude with its complex conjugate amplitude, while the latter is the product of the amplitude of the electromagnetic current $J^\mu$ in the hadron (due to the struck quark within), and its complex conjugate. It can be expressed as 
\begin{eqnarray}
\label{eq:DIS-hadron-tensor}
    W^{\mu\nu} &=& 2 \,{\rm Disc.}\, T^{\mu\nu} \equiv \frac{1}{2\pi}{\rm Im}\int d^4 y\, e^{iq\cdot y}\, \langle P|T(J^\mu(y)J^\nu(0))|P\rangle\no\\
    &=& \frac{1}{2\pi}{\rm Im}\int d^4 y\, e^{iq\cdot y}\, {\rm Tr} (\gamma^\mu G_A(y,0)\gamma^\nu G_A(0,y))\,,
\end{eqnarray}
where $T^{\mu\nu}$ is the DIS forward Compton scattering amplitude and $G_A(y,0) = -i\langle \psi(y){\bar \psi}(0)\rangle_A$ is the Green function for the struck quark that is weakly coupled to the QCD background field $A$ in the proton. The last expression is obtained using Wick's theorem\footnote{Note that a product of tadpole terms ${\rm Tr}(\gamma^\mu G_A(y)){\rm Tr}(\gamma^\nu G_A(0))$ present in the decomposition of the product of currents does not appear in the above equation because it lacks an imaginary part.}. One can further decompose  $W^{\mu\nu}$ 
as
\begin{eqnarray}
    M_N W^{\mu\nu}= -\left(g^{\mu\nu}-\frac{q^\mu q^\nu}{q^2}\right) F_1 +\left(P^\mu-\frac{q^\mu (P\cdot q)}{q^2}\right)\left(P^\nu -\frac{q^\nu (P\cdot q)}{q^2}\right)F_2\,,
\end{eqnarray}
where $M_N$ is the nucleon mass, and $F_1$ and $F_2$ are the DIS form factors of the hadron measured in experiment. (In the QCD parton model, $F_2$ is the sum of the quark and anti-quark distributions in the proton and $F_L=F_2 -2 x_{\rm Bj} F_1$ is proportional to the gluon distribution inside the proton.) Following this decomposition, one can write the DIS cross-section as kinematic factors times the total cross-section for the virtual photon to scatter off the proton/nucleus 
\begin{eqnarray}
    \sigma_{\gamma^* A}(x_{\rm Bj},Q^2) = \frac{4\pi^2 \alpha_{\rm em}}{Q^2} F_2(x_{\rm Bj},Q^2)\,. 
\end{eqnarray}
Regge asymptotics in this DIS context corresponds to $P^+\rightarrow\infty$, with $x_{\rm Bj}\approx Q^2/s\rightarrow 0$, for fixed $Q^2$. In this kinematics, and for $Q^2\gg \Lambda_{\rm QCD}^2$ where parton degrees of freedom are manifest, the cross-section above can be factorized as~\cite{Bjorken:1970ah}
\begin{eqnarray}
\label{eq:DIS-cross-section}
    \sigma_{\gamma^* A}(x_{\rm Bj},Q^2)= \int_0^1 dz \int d^2 r_\perp |\Psi(z,r_\perp,Q^2)|^2 \sigma_{\rm dipole}(x_{\rm Bj},r_\perp) \,,
\end{eqnarray}
where $|\Psi(z,r_\perp,Q^2)|^2$ is the probability for the virtual photon to split into a quark-antiquark ``dipole" at relative separation $r_\perp$, with the quark (antiquark) carrying a longitudinal momentum fraction $z$ ($1-z$) of the photon's momentum, and $\sigma_{\rm dipole}(x_{\rm Bj},r_\perp) $ is the cross-section for the $q\bar q$-dipole of size $r_\perp\ll 1/\Lambda_{\rm QCD}$ to scatter off the hadron/nucleus. 

This factorized expression is valid in the eikonal approximation, where the dynamics of the QCD gauge fields in the hadron are localized static configurations in light cone on the time scales of the interaction with the dipole probe. (We have also assumed that the gauge field configurations probed are homogeneous and isotropic.) Indeed, an explicit realization of this expression is obtained by computing the quark propagators in Eq.~\ref{eq:DIS-hadron-tensor} in such semi-classical backgrounds~\cite{McLerran:1998nk,Venugopalan:1999wu}.  
\begin{figure}[ht]
\centering
\includegraphics[scale=1.3]{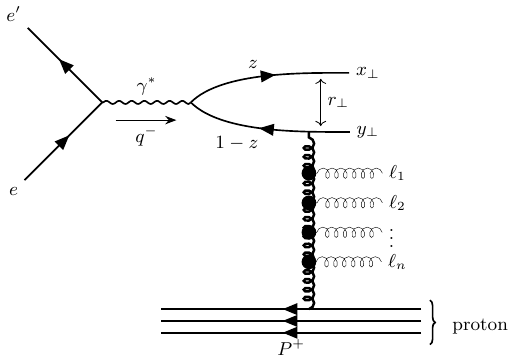}
\caption{Illustration of the underlying $2\rightarrow N$ structure of deeply inelastic scattering (DIS) at small x (high energies).}
\label{fig:DIS-fig}
\end{figure}

The DIS process is illustrated in Fig.~\ref{fig:DIS-fig}. The inclusive cross-section contains within, the $2\rightarrow N$ gluon amplitude we have been discussing. In particular, the dipole cross-section can be expressed as
\begin{eqnarray}
    \sigma_{\rm dipole}(x_{\rm Bj},r_\perp)= -\int  \frac{d^2 k_\perp}{2\pi}\, e^{ik_\perp\cdot r_\perp}\,\nabla_{k_\perp}^2{\phi}(x_{\rm Bj},k_\perp^2)\,,
\end{eqnarray}
where the BFKL amplitude\footnote{\label{footnote:two-dists}In the literature, $\phi(x_{\rm Bj},k_\perp^2)$ is called the Weizs\"{a}ker-Williams gluon distribution, which can be distinguished in general~\cite{Kharzeev:2003wz} from a dipole gluon distribution we will introduce in Sec.~\ref{sec:CGC}. At large transverse momenta $k_\perp\gg Q_S$, the two distributions agree.} ${\phi}(x_{\rm Bj},k_\perp^2)$ is the inverse Mellin transform of ${\cal M}_l(k_\perp^2)$, defined simply in  Eq.~\ref{eq:Mellin-amplitude-2} in terms of the solution to the BFKL equation given in Eq.~\ref{integral-eq-qcd}. The LLx energy dependence of ${\phi}$ is given by the maximal  eigenvalue in Eq.~\eqref{eq:max-BFKL-eigenvalue}:
${\phi}(x_{\rm Bj},Q^2)\sim (\frac{s}{Q^2})^{\omega^*} \equiv x_{\rm Bj}^{-\omega*}$ with $\omega^*=\frac{4}{\pi}\alpha_S N_c \ln 2$. The full solution of the LLx BFKL equation~\cite{Forshaw:1997dc}, reexpressed in terms of the dipole cross-section, gives
\begin{eqnarray}
\label{eq:BFKL-DIS-solution}
    \sigma_{\rm dipole}(x_{\rm Bj},r_\perp)\approx \sqrt{r_\perp^2 Q_0^2}\,\frac{e^{\omega^* Y}}{\sqrt{2\pi\beta {\bar \alpha_S Y}}}\exp\left(-\frac{\ln^2(r_\perp^2 Q_0^2)}{2\beta\bar\alpha_S Y}\right)\,,
\end{eqnarray}
where $Y= \ln(1/x_{\rm Bj})$, $\beta=28\,\zeta(3)\sim 33.67$, and $\bar \alpha_S = \alpha_S N_c/\pi$.
Further, $Q_0^2\sim \Lambda_{\rm QCD}^2$ is the initial nonperturbative scale in the proton. This BFKL solution leads to a very rapid rise ($\omega^*\sim 0.5$ for $\alpha_S=0.2$ and $N_c=3$) in the DIS cross-section. It mirrored the rapid rise in the DIS data observed at HERA, and revitalized interest\footnote{The papers \cite{Kuraev:1977fs} and \cite{Balitsky:1978ic}, cited on the order of 4000 citations each in the INSPIRE data base, had received only approximately 100 citations  prior to 1991.} in the BFKL framework beginning in the mid-1990s. 

The next-to-leading logarithmic in $x$ (NLLx) BFKL equation, that resums contributions of the form 
$(\alpha_S^{n+1} \ln^n(1/x_{\rm Bj}))$ beyond the LLx BFKL resummation of $(\alpha_S^n \ln^n(1/x_{\rm Bj}))$ contributions, was derived in \cite{Fadin:1998py,Ciafaloni:1998gs}. The principal elements of the NLLx BFKL formalism are the same as the LLx BFKL. The new ingredients are the next order $\alpha_S$ correction to the gluon Regge trajectory and to the Lipatov vertex. The Regge trajectory is extended to two-loop, while the central emission vertex gets corrections from two gluons which are {\it not} strongly ordered in rapidity.  The NLLx BFKL equation still has the structure of an integral equation, with the kernel now including enhancements such as gluon-and-quark pair emissions and radiative corrections to the central emission vertex, enriching the iterative structure of the equation. The resulting NLL BFKL is again free of IR divergences and the solution of this equation is  characterized by eigenvalues 
$(\omega_{\nu n})$ that receive higher order corrections in $\alpha_S$. 

The results in strict MRK kinematics for the NLLx BFKL eigenfunctions and eigenvalues have unphysical solutions, leading to the realization that finite energy constraints are important even in multi-Regge kinematics. In other words, for physical processes, one needs to impose lifetime ordering in addition to rapidity ordering to obtain sensible results. This requirement leads to large double logs in transverse momenta ($\alpha_S \ln^2(Q^2/Q_0^2)$) at NLO that when resummed give sensible results for the NLLx BKFL cross-section~\cite{Ciafaloni:2001db,Ciafaloni:1999yw,Ciafaloni:2003rd} and phenomenological results that are closer to the values seen in data. Beyond NLLx, the gluon reggeization framework we discussed breaks down; despite a significant body of work, a complete picture of the NNLx extension of BFKL is lacking. A review of the status of this work can be found in 
\cite{DelDuca:2022skz}.

The great simplifications described here to $2\rightarrow n$ scattering follow, as we showed, from implementing MRK kinematics, corresponding to strict $x$ (or rapidity) ordering in the amplitude. This simplicity however comes at a price. Already at LLx, the full solution of the BFKL equation for the DIS case in Eq.~\eqref{eq:BFKL-DIS-solution} reveals a diffusive property in $r_\perp$ with increasing $Y$, to the UV and the IR, where the theory is strongly coupled, leading to a breakdown of the framework at large rapidities. The diffusion in rapidity is not surprising since the BFKL equation has the structure of a reaction-diffusion equation where the spatial coordinate is the log of the transverse momenta and time is the rapidity~\cite{Gelis:2010nm}.  This diffusive property is not cured at NLLx~\cite{Ciafaloni:2002xk} and is only cured by the phenomenon of gluon saturation~\cite{Mueller:2002zm} which we now discuss.

The BFKL ladder construction, as is transparent in the DIS case, is obtained within a leading twist framework, where the rapidity evolution is of leading twist transverse momentum dependent operators, with contributions of more nontrivial higher twist operators (in operator product expansion (OPE) language) suppressed by powers of $1/Q^2$. 
\begin{figure}[ht]
\centering
\includegraphics[scale=1]{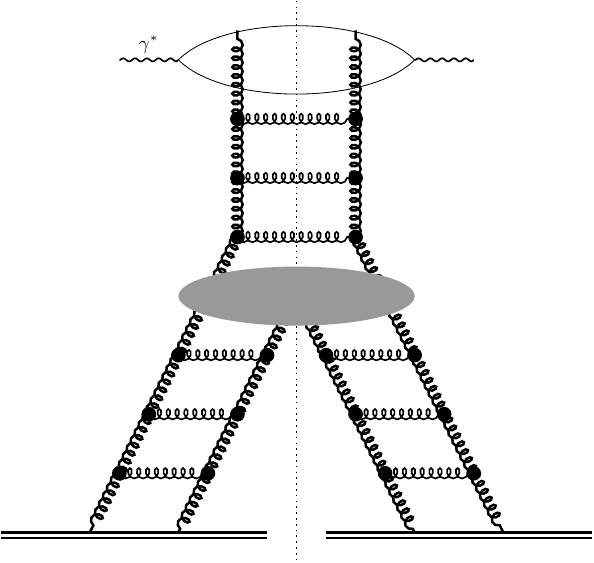}
\caption{A twist-four $1\rightarrow 2$ pomeron ``fan" diagram contributing to the DIS dipole cross-section.}
\label{fig:fan-diagram}
\end{figure}
However with increasing rapidity, as $x_{\rm Bj}\rightarrow 0$, the rapid growth in the dipole cross-section can lead to increasingly large contributions from higher-twist operators. An example of such a contribution is illustrated by the fan diagram in Fig.~\ref{fig:fan-diagram}, corresponding to twist-four operators in the hadron, whose evolution with rapidity is still determined by multi-Regge kinematics. 

As first discussed in \cite{Gribov:1983ivg,Mueller:1985wy}, there are several cuts of the fan diagram contributing to a given final state multiplicity. The cut shown by the dashed line is a so-called ``diffractive cut" where particles are produced for rapidities above that of the gray blob in Fig.~\ref{fig:fan-diagram}, but the two ladders below are uncut color singlet ``pomerons", corresponding to a rapidity gap in the phase space of the DIS scattering. One one side of the cut, this diagram allows for two gluons at lower rapidities fuse to produce a gluon at a higher rapidity. There are other contributions where both ladders are cut, or an interference contribution where only one ladder below the blob is cut while the other is not. As noted in \cite{Mueller:1985wy}, such contributions involve a two-gluon distribution ${\cal F}(x_1,x_2, k_{\perp,1}^2, k_{\perp,2}^2$); for large nuclei, and large $N_c$, this two-body distribution factorizes into the square of the momentum dependent gluon distribution ${\phi}(Y,k_\perp^2)$. As we will discuss further in Section~\ref{sec:CGC}, one obtains the Balitsky-Kovchegov (BK) equation~\cite{Balitsky:1995ub,Kovchegov:1999yj,Kovchegov:1999ua}, which in momentum space\footnote{The equivalent equation for the dipole distribution (see footnote~\ref{footnote:two-dists}) was obtained previously in \cite{Kutak:2003bd,Bartels:2004ef,Bartels:2007dm}. We thank Krzysztof Kutak for reminding us of this work.} takes the form~\cite{Marquet:2005zf}
\begin{eqnarray}
\label{eq:JIMWLK-dipole-momentum-space}
    \frac{\partial {\phi}(Y,k_\perp^2)}{\partial Y} = {\bar \alpha}_S (K_{\rm BFKL}\otimes {\phi}(Y,k_\perp^2))
    -{\bar \alpha}_S\,{\phi}^2(Y,k_\perp^2)\,,
\end{eqnarray}
where $K_{\rm BFKL}\otimes {\phi}$ is the integral equation we 
obtained in Eq.~\eqref{BFKL-eq-q0}, with its kernel represented by $K_{\rm BFKL}$. The linear version of this equation generates the BFKL ladder and the solution to the BFKL equation we discussed earlier. However, for any $k_\perp^2 \approx Q^2$, there is always a value of the rapidity (or $x_{\rm Bj}$) for which the r.h.s of the equation is zero, leading to a saturation of the growth of ${\phi}$ with rapidity. The line in the $x_{\rm Bj}$-$Q^2$  corresponding to this saturation of the gluon distribution is characterized by a {\it saturation scale}
\begin{equation}
    Q_S^2(x_{\rm Bj}) = Q_0^2 \,e^{\lambda_s {\bar Y}}\,,
\end{equation}
where ${\bar Y}= \ln(1/x_{\rm Bj})$, and $\lambda_s$ is a constant that can be computed analytically. Both the BK equation and the saturation scale will be discussed at length  in Sec.~\ref{sec:JIMWLK}. 

A few comments are in order here. Firstly, the most general observation is that the OPE breaks down in the sense that higher-twist operators become important~\cite{Mueller:1996hm} and cannot be ignored even if $Q^2$ is very large, when one takes $Y\rightarrow \infty$. Secondly, the factorization of multi-point gluon distributions is not guaranteed -- however, as argued previously~\cite{Mueller:1989st,McLerran:1993ka,McLerran:1993ni}, it is justified for large nuclei, where the dominant higher twist contributions come from fans emanating from separate nucleons coherently scattering off the DIS probe. In this case, $Q_S^2 \sim A^{1/3} \Lambda_{\rm QCD}^2$. Not least, the nonlinearities manifest in the BK equation, prevent the diffusion of the solution to the infrared, leading to a self-consistent weak coupling solution. 

We end this section with a discussion of the spacetime picture of gluon saturation. For a dipole of fixed size $r_\perp\sim 1/Q$, the BFKL driven exponential growth in the gluon distribution in the hadron, with increasing rapidity, leads to a large phase space occupancy at a critical rapidity $\bar Y$. This  corresponds to the largest possible squared field strengths in QCD, of $O(1/\alpha_S)$. The physical picture is one of close packing of gluons on the distance scale $1/Q_S(x)$ much smaller than the size of the hadron. Because the large field strength is driven by large occupancy, the corresponding gauge fields that couple to the quark dipole are classical configurations: $A\rightarrow A_{\rm classical}\sim 1/g$. In the rest frame of the dipole, the phase space inside the boosted hadron is that of a collection of ``hot spots" with squared field strengths $1/\alpha_S$, and size $1/Q_S(x_{\rm Bj})$. If a dipole at fixed impact parameter has a size $r_\perp \ll 1/Q_S$, it interacts very weakly with the hadron, as dictated by perturbative QCD. On the other hand, dipoles of size $r_\perp \geq 1/Q_S$ interact with unit probability. Thus classicalization and unitarization of cross-sections are both features of gluon saturation. This picture is realized quantitatively when the hadron is a large nucleus in a Color Glass Condensate effective field theory that we will discuss at length in Section~\ref{sec:CGC}. 

We note finally that the presence of a large color charge density in a nucleus at high energies provides an additional scale that suppresses fluctuations that can potentially drive the system away from unitarization at fixed impact parameter. These are so-called ``pomeron loop" configurations. Instead of the fusion of two ladders from the target shown in Fig.~\ref{fig:fan-diagram}, they correspond to a pomeron from the target that splits into two or more pomerons which can then merge with the fan diagrams to form a closed loop - beginning at the projectile with one ladder and ending at the target with another. For a nice introductory discussion of this physics, see \cite{Triantafyllopoulos:2005cn}. It has been argued in a toy model computation that such loops are suppressed by running coupling effects; the full story remains to be understood~\cite{Dumitru:2007ew}. 

\section{\texorpdfstring{$2\to 2+n$}{} scattering in Einstein gravity}
\label{sec:3}

In the previous section, we discussed the building blocks of the $2\to 2+n$ scattering of gluons in the regime in Eq.~\eqref{ll-accuracy} which is dominated by the multi-Regge kinematics. The main elements are the nonlocal Lipatov vertex, the effective vertex of the $2\to 3$ amplitude, and the reggeized gluon propagator, obtained by  resumming  the IR divergent pieces of the virtual contributions to the $2\to 2$ amplitude. In \cite{Lipatov:1982vv, Lipatov:1982it}, Lipatov demonstrated a similar construction of the $2\to 2+n$ amplitude in Einstein gravity (GR). After providing some relevant background material, and a brief overview of relevant works in the literature, we will elaborate upon Lipatov's work in GR in some detail,  highlighting various subtle points in his derivation. 

In gravity, unlike QCD, the coupling constant $G$ is dimensionful. Its relation to the Planck mass and Planck length (in $\hbar=c=1$ units) is given by
\be
\frac{\kappa^2}{8\pi} = G = \frac{1}{M_{pl}^2} = \ell_{pl}^2~.
\ee 
The latter can be combined with $M_{pl}$ to give the dimensionless gravitational coupling
\be
\label{effectiveCoupling}
\lambda_{\rm GR}(Q) = \frac{Q^2}{M_{pl}^2}~,
\ee
which is more analogous to the QCD case, where the strength of the coupling also depends (via dimensional transmutation) on the resolution scale of a probe. An additional important scale in gravity is $R_S$, the characteristic Schwarzschild radius, which is set by the center-of-mass energy $\sqrt{s}$:
\be
R_S \equiv G\sqrt{s} =\frac{\sqrt{s}}{M_{pl}^2}~.
\ee
The so-called trans-Planckian scattering regime is specified by taking the center-of-mass energy $\sqrt{s}\gg M_{pl}$,  corresponding to $ \ell_p\ll R_S$. This separation of scales ensures that quantum gravity effects (characterized by corrections sensitive to $\ell_{pl}$) are small at the Schwarzschild scale $R_S$. For incoming particles with impact parameter smaller than $O(R_S)$, by classical arguments we expect that the final state will be dominated by a black hole \cite{Thorne:1972ji}, plus radiation in the form of gravitational waves \cite{Pretorius:2007jn}. Since the region $b\sim R_S$ is dominated by strong gravity that captures all the nonlinearities of GR, we will start our discussion in the weak coupling regime where the impact parameter $b$ is much larger than the Schwarzschild radius $b\gg R_S$.

\begin{figure}[ht]
\centering
\raisebox{-27pt}{\includegraphics[scale=1]{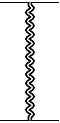} }
+
\raisebox{-27pt}{\includegraphics[scale=1]{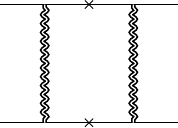} }
+
\raisebox{-27pt}{\includegraphics[scale=1]{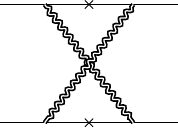} }
+
\raisebox{-27pt}{\includegraphics[scale=1]{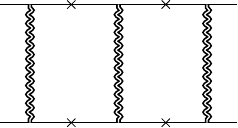} }
+~$\cdots$
\caption{The eikonal scattering series comprised of horizontal ladder and crossed ladder diagrams. The crosses  denote that the propagators of the high energy lines are approximated as $1/(p-k)^2\sim -1/(2p\cdot k)$.} 
\label{fig:221}
\end{figure}

It is well known that at large impact parameters $b\gg R_S$, the $2\to 2$ gravitational amplitude eikonalizes in the Regge limit \cite{Amati:1987uf, Muzinich:1987in, Kabat:1992tb}. The diagrams that contribute to eikonalization comes from resumming the horizontal ladder and cross-ladder terms in the series shown in Fig. \ref{fig:221}. In these diagrams, the eikonal approximation requires that the momenta of the exchanged gravitons are  small with respect to those of the external lines, namely, $(p_1-k)^2 \approx -2p_1\cdot k$ where $p_1$ is the momentum of one of the external lines and $k$ is the momentum of the exchanged graviton. This replacement is illustrated by the crosses in Fig.~\ref{fig:221}. The resummation of this series generates the eikonal amplitude:
\be
\label{EikAmp}
i\mathcal{M}_{\rm Eik}=2 s \int d^{2} \bsb~ e^{-i \mathbf{q} \cdot \mathbf{b}}\left(e^{i \chi(\bsb, s)}-1\right)~,
\ee
where the IR divergent eikonal phase $\chi(\bsb,s)$ is given by
\be
\label{leading-eikonal}
\chi(\bsb,s)=\frac{\kappa^2 s}{2\hbar} \int \frac{d^2 \bsk}{(2 \pi)^2} \frac{1}{\bsk^2} e^{i \bsb \cdot \bsk}  = -\frac{2Gs}{\hbar}\log(|\bsb|/L)~,
\ee
where $L$ is a long distance cutoff scale. 

The eikonal approach to gravitational amplitudes has become one of the central tools to compute classical observables via the scattering amplitude formalism where the classical terms undergo an exponentiation. This was also the main theme in the works of Amati, Ciafaloni and Veneziano (ACV) \cite{Amati:1987uf, Amati:1987wq, Amati:1990xe, Amati:1992zb, Amati:1993tb, Amati:2007ak} that addressed the problem of trans-Planckian gravitational scattering in Einstein gravity as well as string theory. It is evident however from Eqs.~\eqref{EikAmp} and  \eqref{leading-eikonal} that an amplitude based approach for computing the eikonal is problematic since perturbation theory breaks down already at leading order, since the expansion is effectively in $Gs\gg1$. One therefore needs to find a way of resumming certain infinite sets of diagrams, as for instance represented in Fig.~\ref{fig:221}. 

This resummation is in the form of an exponentiation where the coupling constant sits in the exponent. To obtain this form, it is then crucial that contributions from entire classes of loop-level diagrams appear with the appropriate power of the coupling and with the right combinatoric factors to {\it all loop} orders. The validity of the eikonal approximation and exponentiation thus needs to be checked on a case-by-case basis. For the leading eikonal in Eq.~\eqref{leading-eikonal}, this resummation was performed in \cite{Kabat:1992tb} where exponentiation was explicitly demonstrated in impact parameter space. From the resummed amplitude corresponding to the eikonal in Eq.~\eqref{EikAmp}, one can then compute leading-order classical observables such as the deflection angle \cite{Bjerrum-Bohr:2016hpa, Bjerrum-Bohr:2017dxw}, and the Shapiro time delay \cite{Ciafaloni:2014esa}.

To obtain higher order correction to the classical observables one needs to compute higher order corrections to the leading eikonal. In the ACV framework\footnote{In a more modern presentation, the elastic $2\to 2$ S-matrix in the classical limit can be expressed as 
\be
\label{elastic-S-matrix-1}
1+i\mathcal{M}(s, b) = \(1+i\Delta(s,b)\) e^{2i\tilde\delta(s,b)}~.
\ee
This way of writing the S-matrix distinguishes purely classical terms that should exponentiate (and therefore contribute to the phase $\tilde\delta\propto 1/\hbar$) from purely quantum terms (contributing to $\Delta$) that appear with non-negative powers of $\hbar$ and may not exponentiate. Though one can write $(1+i\Delta(s,b)) = e^{2i\delta_{\text{quantum}}}$, the form in Eq.~\eqref{elastic-S-matrix-1} is preferred since it isolates the purely classical terms that are completely captured by $\tilde \delta$ that can then be used to compute classical observables. See \cite{DiVecchia:2019kta} for a thorough discussion.
} the elastic $2\to 2$ S-matrix is written in the exponentiated form as $1+i\mathcal{M}(s, b) = e^{2i\delta(s,b)}$ with the phase $\delta(s, b)$ admitting an expansion in the gravitational coupling $G$: $\delta(s, b) = \delta_0 + \delta_1 + \delta_2 + \cdots$ where $\delta_j$ is an order $O(G^{j+1})$ term in the post-Minkowski expansion. The leading eikonal is $\delta_0=\chi$ that was given in \eqref{leading-eikonal} is a universal term in the ultra-relativistic limit in that it is not sensitive to the details of the theory. On the other hand, the subleading terms in $\delta$ are generically not universal. In Einstein gravity the term $\delta_1$ was computed by ACV \cite{Amati:1990xe, Amati:1992zb} 
\begin{align}
\label{delta-1}
\delta_1 = \frac{6}{\pi}\frac{G^2s}{b^2}\log(s b^2)~,
\end{align}
This term is of $O(\hbar^0)$. In the ultrarelativistic limit, it does not receive any classical contributions (terms proportional to $\hbar^{-1}$). However for massive colliding particles at finite velocities, there are nonvanishing classical contributions at this order. (See for instance \cite{KoemansCollado:2019ggb}.) Note that the above expression of $\delta_1$ is of the order $\ell_p^2/b^2$ with respect to the leading eikonal $\delta_0$ since $G\sim \ell_p^2$. Therefore in the regions $b\gg R_S$ the contribution from such a term is not important for computing physical observables. 

The next term $\delta_2$ is \cite{Amati:1990xe}: 
\begin{align}
\label{delta-2}
    \delta_2 =\frac{2G^3s^2}{b^2\hbar}\left[1+\frac{i}{\pi} \log (s b^2)\left(\log \frac{L^2}{b^2}+2\right)\right] ,
\end{align}
which is a classical contribution and is of the order $G^2s/b^2\propto R_S^2/b^2$  w.r.t. the leading eikonal. The real part of this term is universal in the ultrarelativistic limit \cite{DiVecchia:2020ymx}. Its imaginary part encodes the leading inelastic contribution that starts to contribute from two-loop order. Such contributions are obtained by gluing together two effective $2\to 3$ graviton emission diagrams, computed in the multi-Regge kinematics we discussed previously for QCD in Sec. \ref{sec:2}. As we will discuss in Sec.~\ref{sec:3.1}, there are several such diagrams that sum up into the effective graviton emission diagram, with an effective vertex which is the gravitational analog of the QCD Lipatov vertex. This gravitational Lipatov vertex takes the form 
\begin{align}
\label{gravitational-lipatov-vertex-0}
    C^{\mu\nu}(k_1, k_2) = \frac12 C^\mu(k_1, k_2)C^\nu(k_1, k_2) - \frac12 N^\mu(k_1, k_2)N^\nu(k_1, k_2)~,
\end{align}
where $C^\mu(k_1,k_2)$ is the QCD Lipatov vertex we discussed at length in Sec.~\ref{sec:2}. Further, 
\begin{align}
\label{QED-factor-0}
    N^\mu(k_1, k_2) &= ~\sqrt{k_1^2 k_2^2}\(\frac{p_1^\mu}{p_1\cdot\ell}-\frac{p_2^\mu}{p_2\cdot\ell}\)~,
\end{align}
which contains the QED bremsstrahlung factor within the parenthesis. Thus the effective gravitation emission amplitude remarkably is a bilinear comprised of their counterparts in QCD and QED. 

\begin{figure}[ht]
\centering
\includegraphics[scale=1]{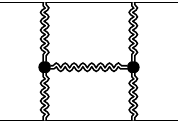}
\caption{The H-diagram, which gives the leading inelastic correction to the eikonal scattering series. The black blobs represent the gravitational Lipatov effective vertices.} \label{H-diagram}
\end{figure}

Gluing two such $2\to 3$ diagrams results in the so-called {\it H-diagram} \cite{Amati:1990xe}, depicted in Fig. \ref{H-diagram}. This diagram contributes to both the real and imaginary parts of $\delta_2$ in Eq.~\eqref{delta-2}.
The ACV Regge EFT results for $\delta_1$ and $\delta_2$ have been confirmed independently in an explicit computation of two-loop four-point massless amplitudes in Einstein gravity using powerful numerical unitarity methods~\cite{Bern:2020gjj,Abreu:2020lyk}.

In the previous section, for gluon scattering in the LLx approximation, both the Lipatov vertex and the reggeized gluon trajectory were crucial in the construction, and are the key elements in Lipatov's 2-D reggeon EFT. As discussed in \cite{Lipatov:1991nf}, this is also the case of gravity. Indeed, both contributions, as in the QCD case, are crucial for the cancellation of the IR divergence in the computation of the $2\to 2$ cross section, as we shall see below. Reggeization at leading-logarithmic order in gravity also arises from the iteration of Sudakov double-logs that appears in the calculation of the one-loop gluon scattering amplitude. Since in gravity classical contributions from loop-level diagrams dominate (genuine quantum contributions are hugely suppressed), one should carefully analyze whether the terms in the one-loop diagram contributing to the graviton trajectory are classical or quantum. For an example of such a discussion, see \cite{Holstein:2004dn}. As we will discuss further in the following sections both in the context of QCD, and gravity, one can alternatively formulate this problem cleanly within the Schwinger-Keldysh formalism.  

Putting aside this issue for now, it is important to understand graviton reggeization in gravity relative to QCD. Recall that in the latter  one resums logarithmic IR divergent contributions that arises in the horizontal ladder and cross-ladder series to all-loop order. In this resummation, the propagator of the gluon exchange is dressed by a factor of $(-s/t)^{\alpha(t)}$ with $\alpha(t)$ being the gluon Regge trajectory. However in the analogous computation in gravity (see Fig.~\ref{fig:221}),  graviton reggeization is not manifest. 

Why is this the case? In order to address this disparity, let us look at the full one-loop four-point amplitude in gravity. Specifically, we examine terms in the Regge limit that contain an IR divergence and are not sensitive to details of the theory\footnote{For the complete expression, see for instance Eqs.~(14) and (18) in \cite{Bartels:2012ra}.}. There are two contributions, 
\begin{align}
\label{one-loop-graviton}
    \mathcal{M}^{(1)} \sim \frac{\kappa^2}{8\pi^2}\( - i\pi  s \log\(\frac{-t}{\Lambda^2_{\text{IR}}}\) + t \log\(\frac{s}{-t}\)\log\(\frac{-t}{\Lambda_{\text{IR}}^2}\)\)~\,,
\end{align}
where $\Lambda_{\text{IR}}$ is an IR cutoff. Since both terms have an identical IR divergence in the Regge limit, the first term dominates over the second term by a factor of $(-s/t)$. The latter term (which is the Sudakov double log) is subleading as $-t/s \sim R_S^2/b^2$ and therefore not relevant for the large impact parameter regime $b\gg R_S$ \cite{Giddings:2011xs}. This is the origin of the difference from the perturbative QCD case where the double logs dominate over the eikonal phase contribution because the (color) prefactor of the former  is not suppressed relative to the latter. 

A color-kinematic double copy perspective of this difference between QCD and Gravity is discussed in \cite{Melville:2013qca}. If instead of the leading MRK contribution in Eq.~\eqref{one-loop-gluon-trajectory} one keeps the full result, one observes that it is of the same structure as Eq.~\eqref{one-loop-graviton}, with the $s$ and $t$ prefactors of the logs replaced by $T_s^2 = (T_a+T_b)^2$ and $T_t^2 = (T_a+T_{a'})^2$, respectively, with $T_a$ and $T_b$ being the color factors associated with the two incoming gluons and $T_{a'}$ is the color factor associated to one of the outgoing gluons.  Clearly then, the term responsible for gluon reggeization ($\ln(\frac{s}{-t})\ln(\frac{-t}{\Lambda_{\text{IR}}^2})$) dominates over the eikonal term ($-i\pi s$) in QCD. For gravity in the Regge limit, the kinematical replacements $T_s^2\to s~,~T_t^2\to t$ are what result in the graviton reggeization term being $t/s$ suppressed relative to the eikonal term. 

Nevertheless, the double logs are important at impact parameters approaching $R_S$, indeed when inelastic radiation {\it a la Lipatov} is becoming important. They play a similarly crucial role, as in QCD, in the construction of the $2\rightarrow 2+n$ inelastic amplitude. This is because the IR divergences from the loop terms are what cancel the contributions from the real emission amplitude in the scattering cross-section. This cancellation is identical to the QCD case and therefore important for the same reason as in multi-particle production in perturbative QCD (and in QED). 
Further, reggeization goes through in the same manner as in QCD, and along with the gravitational Lipatov vertex, provide the building blocks for the construction of the $2\rightarrow 2+n$ amplitude to all orders to leading logarithmic accuracy. 

In \cite{Lipatov:1982vv,Lipatov:1982it}, Lipatov computed the one-loop graviton Regge trajectory\footnote{See \cite{Grisaru:1975tb} for earlier work on graviton reggeization in pure Einstein gravity.} and the gravitational Lipatov vertex. In the following subsections, we will give a streamlined derivation of these quantities and we will then use these to derive the gravitational Lipatov equation. Our presentation will mirror the derivation of the BFKL equation in QCD discussion in the previous section. In particular, we will discuss throughout the double copy relations to the QCD case. We will also discuss the infrared structure of the GR Lipatov equation and outline its solution, and comment on the similarities and differences to the QCD discussion of the same. We will next demonstrate explicitly the little appreciated fact that the soft limit of the Lipatov result gives the ultrarelativistic limit of the soft graviton theorem derived by Weinberg~\cite{Weinberg:1965nx}. This connection is especially relevant to our discussion above since the Weinberg derivation too relies on the cancellation of real and virtual soft divergences in the cross-section. 

In Sec.~~\ref{sec:classical-double-copy}, we will show how the gravitational emission amplitude containing the Lipatov vertex can be obtained as a classical double copy of the gluon emission amplitude obtained in the scattering of classical scalar color charges employing so-called Yang-Mills+Wong equations~\cite{Wong:1970fu}. One similarly obtains Weinberg's result from this classical double copy in the soft limit. When $b\rightarrow R_S$, copious gravitational radiation driven by the Lipatov RG, and coherent multiple scattering, can in principle, as in QCD, lead to the formation of high occupancy states. We will discuss this issue further in the QCD context in Sec.~\ref{sec:CGC} and in GR in Sec.~\ref{sec:GR-shockwave-formalism}. A general discussion of black hole formation in the S-matrix approach was given in \cite{tHooft:1984kcu,tHooft:1987vrq}. In Sec.~\ref{sec:classicalization}, we will outline an alternative derivation~\cite{Addazi:2016ksu,Dvali:2014ila} of $2\rightarrow n$ scattering that does not employ MRK kinematics, and discuss the connection of this framework to that of ACV, as spelt out in \cite{Addazi:2016ksu}.

\subsection{Gravitational Lipatov vertex}
\label{sec:3.1}

We will work in the Regge limit of gravity, where the appropriate modification of Eq.~\eqref{ll-accuracy} in the QCD case is 
\be
\label{grav-coupling-regime}
 \lambda_{\rm GR}(t)\log\(\frac{s}{-t}\) \sim O(1)~,\qquad \kappa^2 t \equiv \lambda_{\rm GR}(t) \ll 1~,
\ee
where $\lambda_{\rm GR}(t)$ is the dimensionless gravitational coupling defined in Eq.~\eqref{effectiveCoupling}. This is the limit in which we will construct the $2\to 2+n$ amplitude in gravity\footnote{This limit corresponds to 
$s\gg |t|\gg M_{pl}^2$. Unlike QCD, this corresponds to a much narrower window of applicability in gravity. We will address this point further in Sec.~\ref{sec:classicalization} and Sec.~\ref{sec:GR-shockwave-formalism}.}. 

\begin{figure}[ht]
  \centering
  \raisebox{-25pt}{\includegraphics[scale=1]{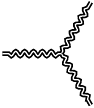}}
  \qquad
  \raisebox{-25pt}{\includegraphics[scale=1]{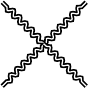}}
  \qquad
  \raisebox{-25pt}{\includegraphics[scale=1]{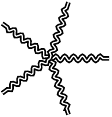}}
  \qquad
  \raisebox{-25pt}{\includegraphics[scale=1]{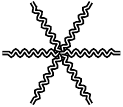}}
  \qquad
  $\cdots$
  \caption{In general relativity, there are infinitely many higher point interactions that are suppressed by higher powers of the gravitational coupling.}
  \label{grav-higher-pt-diagrams}
\end{figure}

\begin{figure}[ht]
  \centering
  \raisebox{-30pt}{\includegraphics[scale=1]{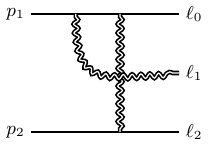}}
  +
  \raisebox{-30pt}{\includegraphics[scale=1]{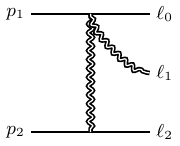}}
  +
  \raisebox{-30pt}{\includegraphics[scale=1]{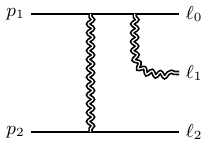}}
  +
  \raisebox{-30pt}{\includegraphics[scale=1]{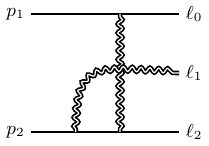}}
  +
  \raisebox{-30pt}{\includegraphics[scale=1]{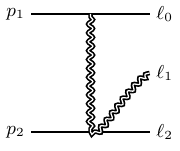}}
  +
  \raisebox{-30pt}{\includegraphics[scale=1]{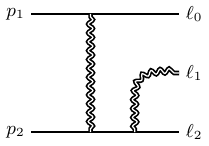}}
  +
  \raisebox{-30pt}{\includegraphics[scale=1]{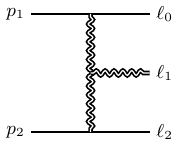}}
  \\[5pt]
  = \raisebox{-47pt}{\includegraphics[scale=1]{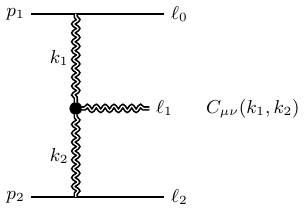}}
\caption{Feynman graphs for the $2\to 3$ tree-level graviton scattering that sum up to the gravitational Lipatov vertex in MRK kinematics. There are two additional diagrams relative to the QCD case since the four-point vertex in gravity is not suppressed in energy.}
\label{grav-LV-constituents}
\end{figure}

As opposed to QCD, gravity contains infinitely many higher order contact interaction terms involving graviton lines\footnote{While every vertex in Fig.~\ref{grav-higher-pt-diagrams} carries two powers of the momenta of the participating legs, they are suppressed by a power of coupling $\kappa$ for every additional leg. The explicit Feynman rules for the three-point and four-point interaction vertices were first computed in \cite{DeWitt:1967yk, DeWitt:1967ub, DeWitt:1967uc}. See also \cite{Sannan:1986tz}.}, as shown in Fig. \ref{grav-higher-pt-diagrams}. As a consequence, more diagrams contribute to the construction of the effective $2\to 2+n$ ladder relative to the QCD case. As a relevant example, as shown in Fig.~\ref{grav-LV-constituents}, there are two more diagrams contributing to the gravitational Lipatov vertex than in QCD. These additional diagrams are of the contact type, and one can expect many more such terms with increasing $n$.

In \cite{SabioVera:2011wy} (see also \cite{Rothstein:2024nlq}), an explicit derivation shows that the sum over the graphs in Fig.~\ref{grav-LV-constituents} in multi-Regge kinematics sums up to the effective gravitational emission vertex introduced by Lipatov in \cite{Lipatov:1982vv,Lipatov:1982it}. However such a diagrammatic technique, with increasing $n$, is even more cumbersome to work with in gravity than in QCD  because of the appearance of higher point vertices. The dispersive techniques we discussed in the QCD context, which avoid explicit computation of Feynman graphs, are therefore especially useful here.

In order to use these techniques, we need to first compute the $2\to 2$ process in the Born approximation with a graviton exchange in the $t$-channel. In particular, we need the expression for the (off-shell) three-point graviton vertex. This expression in the de Donder gauge (specified by $\partial_\mu h_\nu^\mu-\frac{1}{2} \partial_\nu h_\lambda^\lambda=0$, where $h_{\m\n}$ is the graviton field) is \cite{Donoghue:2017pgk}

\begin{figure}[ht]
\centering
\includegraphics[scale=1]{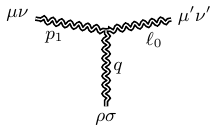}
\caption{Three point graviton vertex.}
\label{grav-3pt-feyn}
\end{figure}

\begin{align}
\label{graviton-3pt-vertex}
\begin{split}
V_{\m\n\m'\n'}^{\rho\sigma}(p_1,q)=\frac{i \kappa}{2}\bigg\{&P_{\mu \nu\mu' \nu'}\left[p_1^\rho p_1^\sigma+(p_1-q)^\rho(p_1-q)^\sigma+q^\rho q^\sigma-\frac{3}{2} \eta^{\rho \sigma} q^2\right] \\
& +2 q_\lambda q_\sigma\left[I^{\lambda \sigma}{ }_{\mu \nu} I^{\rho \sigma}{ }_{\mu' \nu'}+I^{\lambda \sigma}{ }_{\mu' \nu'} I^{\rho \sigma}{ }_{\mu \nu}-I^{\lambda \rho}{ }_{\mu \nu} I^{\sigma \sigma}{ }_{\mu' \nu'}-I^{\sigma \sigma}{ }_{\mu \nu} I^{\lambda \sigma}{ }_{\mu' \nu'}\right] \\
& +\left[q_\lambda q^\rho\left(\eta_{\mu \nu} I^{\lambda \sigma}{ }_{\mu' \nu'}+\eta_{\mu' \nu'} I^{\lambda \sigma}{ }_{\mu \nu}\right)+q_\lambda q^\sigma\left(\eta_{\mu \nu} I^{\lambda \rho}{ }_{\mu' \nu'}+\eta_{\mu' \nu'} I^{\lambda \rho}{ }_{\mu \nu}\right)\right. \\
& \left.-q^2\left(\eta_{\mu \nu} I^{\rho \sigma}{ }_{\mu' \nu'}+\eta_{\mu' \nu'} I^{\rho \sigma}{ }_{\mu \nu}\right)-\eta^{\rho \sigma} q^\lambda q^\sigma\left(\eta_{\mu \nu} I_{\mu' \nu', \lambda \sigma}+\eta_{\mu' \nu'} I_{\mu \nu, \lambda \sigma}\right)\right] \\
& +\left[2 q ^ { \lambda } \left(I^{\sigma \sigma}{ }_{\mu \nu} I_{\mu' \nu', \lambda \sigma}(p_1-q)^\rho+I^{\sigma \rho}{ }_{\mu \nu} I_{\mu' \nu', \lambda \sigma}(p_1-q)^\sigma\right.\right. \\
& \left.-I^{\sigma \sigma}{ }_{\mu' \nu'} I_{\mu \nu, \lambda \sigma} p_1^\rho-I^{\sigma \rho}{ }_{\mu' \nu'} I_{\mu \nu, \lambda \sigma} p_1^\sigma\right) \\
& \left.+q^2\left(I^{\sigma \rho}{ }_{\mu \nu} I_{\mu' \nu', \sigma}{ }^\sigma+I_{\mu \nu, \sigma}{ }^\sigma I^{\sigma \rho}{ }_{\mu \nu'}\right)+\eta^{\rho \sigma} q^\lambda q_\sigma\left(I^{\rho \sigma}{ }_{\mu' \nu'} I_{\mu \nu, \lambda \rho}+I^{\rho \sigma}{ }_{\mu \nu} I_{\mu' \nu', \lambda \rho}\right)\right] \\
& +\bigg[\left(p_1^2+(p_1-q)^2\right)\left(I^{\sigma \rho}{ }_{\mu \nu} I_{\mu' \nu', \sigma}{ }^\sigma+I^{\sigma \sigma}{ }_{\mu \nu} I_{\mu' \nu', \sigma}{ }^\rho-\frac{1}{2} \eta^{\rho \sigma} P_{\mu \nu, \mu' \nu'}\right) \\
& -p_1^2 \eta_{\mu' \nu'} I^{\rho \sigma}{ }_{\mu \nu}-(p_1-q)^2 \eta_{\mu \nu} I^{\rho \sigma}{ }_{\mu' \nu'}\bigg]\bigg\}.
\end{split}
\end{align}
Here the tensors $P_{\m\n\m'\n'}$ and $I_{\m\n\m'\n'}$ are identity operators in the space of symmetric traceless and symmetric matrices, respectively: 
\begin{align}
    P_{\m\n\m'\n'} = \frac12 \(\eta_{\m\m'}\eta_{\n\n'}+\eta_{\m\n'}\eta_{\n\m'}-\eta_{\m\n}\eta_{\m'\n'}\)~,\qquad I_{\m\n\m'\n'} = \frac12 \(\eta_{\m\m'}\eta_{\n\n'}+\eta_{\m\n'}\eta_{\n\m'} \)
\end{align}

While the expression in Eq.~\eqref{graviton-3pt-vertex} is complicated to work with, it simplifies considerably in the eikonal approximation, and as a result of $p_1$ and $\ell_0$ being on-shell-- as in the discussion in the QCD case in Sec.~\eqref{sec:2.1}. Under these approximations, the graviton three-point vertex is (deleting all occurrence of $q$ and setting $p_1^2=0$ in Eq.~\eqref{graviton-3pt-vertex}),
\begin{align}
    V_{\m\n\m'\n'}^{\rho\sigma}(p_1,q) \approx i \kappa P_{\mu \nu, \mu' \nu'}p_1^\rho p_1^\sigma~,
\end{align}
which is a remarkable simplification of Eq.~\eqref{graviton-3pt-vertex}. 
With this result, and the expression for the graviton propagator in the de Donder gauge
\begin{align}
    G^{\rho \sigma \rho' \sigma'}(q) =\frac{P^{\rho \sigma \rho' \sigma'}}{q^2}~,
\end{align}
the four-point graviton scattering amplitude is\footnote{Graviton scattering amplitudes will be denoted by $\mathcal{M}$ to differentiate them from the QCD case where they were denoted by $\mathcal{A}$. The four-point amplitude of massive identical scalars interacting via gravity in the Born approximation is 
\be
   \mathcal{M}_{p_1+p_2\to \ell_0+\ell_1}^{\text{Born}}= \frac{\kappa^2}{t}\(\frac{1}{2}(s^2+u^2)-6m^4\)~,
\ee
which is identical to the four-point graviton scattering amplitude in Eq.~\eqref{graviton-born-amplitude} in the high energy limit $-u\approx s\gg m^2$. In the nonrelativistic limit $s\rightarrow 4m^2,~ u\to0$, corresponding to the Born scattering amplitude of a massive scalar particle of mass $m$, one obtains $\mathcal{M}_{p_1+p_2\to \ell_0+\ell_1}^{\text{Born}}\rightarrow 4 \int V(r) e^{iq\cdot r}d^3 r$ with $q^2=-t$. For the gravitational potential $V(r)=-Gm^2/r$, equating the two expressions gives $\kappa^2 = 8\pi G$.
}
\begin{align}
\label{graviton-born-amplitude}
    \mathcal{M}_{\m\n\m'\n'\a\b\a'\b'}^{\text{Born}}(s, t) &= \(i\kappa P_{\mu \nu, \mu' \nu'}p_1^\rho p_1^\sigma\) \frac{P^{\rho \sigma \rho' \sigma'}}{q^2} \(i\kappa P_{\a \b \a' \b'} p_2^{\rho'} p_2^{\sigma'}\)~,\no\\[5pt]
    & = \frac{1}{4}\frac{\kappa^2 s^2}{t}P_{\mu \nu\mu'\nu'}P_{\a \b \a' \b'}
\end{align}
Here $s=(p_1+p_2)^2 = 2p_1\cdot p_2$ and $t=-q^2 =-(p_1-l_0)^2$. Contracting $P_{\m\n\m'\n'}$ with the graviton polarization tensors $\eps^{\m\n}_\z$ (in the helicity eigenstate), one gets
\begin{align}
    P_{\m\n\m'\n'}\eps^{\m\n}_\zeta(p_1)\eps^{\m'\n'}_{\zeta'}(\ell_0) \equiv P_{\m\n\m'\n'}\eps^{\m\n}_\zeta(p_1)\eps_{\m'\n', \zeta'}(p_1-q) = \delta_{\zeta\zeta'} + O(q)~,
\end{align}
where we made use of their orthonormality property, $\eps^{\m\n}(p)_\zeta\eps_{\m\n,\zeta'}(p) = \delta_{\zeta\zeta'}$. Upon contracting the Born amplitude with the polarization tensor of the external gravitons  one finds
\begin{align}
    \mathcal{M}^{\text{Born}}_{\zeta_1\zeta'_1\zeta_2\zeta'_2}(s,t) = \frac14 \frac{\kappa^2 s^2}{t} \delta_{\zeta_1\zeta'_1}\delta_{\zeta_2\zeta'_2} + O(q)~.
\end{align}
From this expression, we see that the helicity of the scattered gravitons is preserved in the small $t$ limit of the Born amplitude. 

In the next subsection, we will need this expression for the Lorentz covariant Born amplitude to construct the $2\to 2+n$ MRK gravitational amplitude. We will follow the same line of reasoning as was done for QCD in Sec. \ref{sec:2-to-n-amplitude}. The first thing to note is that one can construct gravitational polarization tensors from the gluon polarization vectors as follows:
\begin{align}
\label{graviton-pol}
    \epsilon^{\mu \nu}_{\l \l'} = \frac{1}{2}\left(\epsilon^\mu_{\l} \epsilon^\nu_{\l'}+\epsilon^\mu_{\l'} \epsilon^\nu_{\l}-\epsilon^\mu_{\omega} \epsilon^{\nu,\omega}\delta_{\l\l'} \right)~.
\end{align}
The helicity eigenstate polarization tensor $\eps^{\m\n}_\z$ is a linear combination of $\epsilon^{\mu \nu}_{\l \l'}$ (for different values of $\l$ and $\l'$) constructed  as a bilinear product of gluon polarization vectors. Next, we project the polarization tensor of the incoming and outgoing external graviton to the gauge where $\eps^{\m\n}(p_1) p_{2\m} = \eps^{'\m\n}(\ell_0) p_{2\m} = 0$. These conditions can be imposed on Eq.~\eqref{graviton-pol} using the gauge constraint on the gluon polarization vector we introduced in Eq.~\eqref{gauge-pol-condition}. Further, for the graviton carrying the momentum transfer $p_1-\ell_0$, the associated polarization tensor is $2p_2^\rho p_2^\sigma/s^2$. The three-point graviton vertex contracted with the respective polarization tensors is then 
\begin{align}
    V_{\m\n\m'\n'}^{\rho\sigma} \tilde{\eps}^{\m\n}(p_1)\tilde{\eps}^{\m'\n'}(\ell_0) \frac{2p_{2\rho} p_{2,\sigma}}{s^2}~,
\end{align}
where $\tilde{\eps}^{\m\n}$ is obtained from $\eps^{\m\n}$ by replacing $\eps^\m$ with $\tilde \eps^\m$ in Eq.~\eqref{graviton-pol} with $\tilde \eps^\m$ as defined in Eq.~\eqref{gauge-pol-condition}.  Straightforward algebra gives 
\begin{align}
\label{grav-gamma-construction}
    &V_{\m\n\m'\n'}^{\rho\sigma} \tilde{\eps}^{\m\n}_{\l_1 \l'_1}(p_1)\tilde{\eps}^{\m'\n'}_{\l_2 \l'_2}(\ell_0) \frac{2p_{2\rho} p_{2,\sigma}}{s^2} =\frac{i\kappa}{2} \tilde{\eps}^{\m\n}_{\l_1 \l'_1}(p_1)\tilde{\eps}_{\m\n, \l_2 \l'_2}(\ell_0)=\\[5pt]
    =&2i\kappa \[\Gamma^{\m\m'}_{p_1,\ell_0}\Gamma^{\n\n'}_{p_1,\ell_0}+ \Gamma^{\m\n'}_{p_1,\ell_0}\Gamma^{\n\m'}_{p_1,\ell_0}- \(\delta^{\m\n} -\frac{p_1^\m p_2^\n+p_1^\n p_2^\m}{p_1\cdot p_2}\)\(\delta^{\m'\n'} -\frac{\ell_0^{\m'} p_2^{\n'}+\ell_0^{\n'} p_2^{\m'}}{\ell_0\cdot p_2}\)\] \epsilon^{\mu \nu}_{\l_1 \l'_1}\epsilon^{\mu' \nu'}_{\l_2 \l'_2}\,.\no
\end{align}
The tensor $\Gamma^{\m\m'}_{p_1,\ell_0}$ in this expression was constructed in Eq.~\eqref{gamma-construction}, with the form 
\begin{align}
    \Gamma^{\m\m'}_{p_1,\ell_0} =  -\eta^{\m\m'}+\frac{p_{2}^\m p_{1}^{\m'}+p_{2}^{\m'}\ell_{0}^\m}{p_2\cdot p_1}+ (p_1-\ell_0)^2\frac{p_{2}^\m p_{2}^{\m'}}{2(p_2\cdot p_1)^2} ~.\no
\end{align}
In deriving the second line of Eq.~\eqref{grav-gamma-construction}, we used $\tilde{\eps}^\m_{\lambda_1}\tilde{\eps}_{\m,\lambda_1}=\Gamma^{\m\m'}_{p_1,\ell_0}{\eps}^\m_{\lambda_1}{\eps}_{\m',\lambda_1} $ obtained from Eq.~\eqref{gamma-construction}, the completeness relation for the polarization vector
\begin{align}
    \sum_\omega \eps^\m_\omega(p_1)\eps^{\n}_\omega(p_1) = \eta^{\m\n}-\frac{p_1^\m p_2^\n+p_1^\n p_2^\m}{p_1\cdot p_2} = \delta_\perp^{\m\n}~,
\end{align}
and the transversality properties $p_{1,\m'}\Gamma^{\m\m'}_{p_1\ell_0}=\ell_{0,\m'}\Gamma^{\m\m'}_{p_1\ell_0}=0$, $\Gamma^{\m\m'}_{p_1\ell_0}\Gamma^{\n\n'}_{p_1\ell_0}\eta_{\m'\n'}=\delta^{\m\n}_\perp$. We further made the approximaton that $O(q)$ terms are small. 

We will now denote the rank-four tensor\footnote{This tensor has the following properties:
\begin{equation}
    p_{1,\m}{\tilde {\mathcal G}}^{\m\n\m'\n'}_{p_1, \ell_0}= \ell_{0,\m'}{\tilde {\mathcal G}}^{\m\n\m'\n'}_{p_1, \ell_0}=0\,,
\end{equation}
\begin{equation}
    \eta_{\m\n}{\mathcal{\tilde G}}^{\m\n\m'\n'}_{p_1, \ell_0} = \eta_{\m'\n'}{\mathcal{\tilde G}}^{\m\n\m'\n'}_{p_1, \ell_0} =0\,,
\end{equation}
and 
\begin{align}
\label{graviton-polarization-prop}
    {\tilde{\mathcal G}}^{\m\n\m'\n'}_{\ell, \ell'} {\tilde {\mathcal G}}^{\m''\n''}_{\m'\n'; \ell, \ell'} = \frac12\(\delta_\perp^{\m\m''}\delta_\perp^{\n\n''}+\delta_\perp^{\m\n''}\delta_\perp^{\n\m''}-\delta_\perp^{\m\n}\delta_\perp^{\m''\n''}\)~,
\end{align}
where $\delta_\perp$ is the two-dimensional Kronecker delta.} corresponding to the argument of the square bracket in the second line of Eq.~\eqref{grav-gamma-construction} as ${\tilde{\mathcal G}}^{\m\n\m'\n'}_{p_1, \ell_0}$. 
With this, the Born amplitude can equivalently be expressed as 
\begin{align}
\label{2-to-2-graviton-scattering}
    \mathcal{M}_{p_1+p_2\to \ell_0+\ell_1}^{\text{Born}} =  V_{p_1,\ell_0} \frac{\kappa^2s^2}{t} V_{p_2,\ell_1}~,
\end{align}
where the vertex $V_{p_1,\ell_0}$ is given by
\begin{align}
\label{2-2-GR-building-block}
    V_{p_1,\ell_0} =~ & 2\,{\mathcal{\tilde G}}^{\m\n\m'\n'}_{p_1, \ell_0}  \epsilon_{\m\n}(p_1)\epsilon_{\m'\n'}(\ell_0) \equiv 4\,\epsilon_{\m\n}(p_1)\epsilon_{\m'\n'}(\ell_0) \\[5pt]
    &\times\frac12\[ \Gamma^{\m\m'}_{p_1,\ell_0}\Gamma^{\n\n'}_{p_1,\ell_0}+ \Gamma^{\m\n'}_{p_1,\ell_0}\Gamma^{\n\m'}_{p_1,\ell_0}- \(\delta^{\m\n} -\frac{p_1^\m p_2^\n+p_1^\n p_2^\m}{p_1\cdot p_2}\)\(\delta^{\m'\n'} -\frac{\ell_0^{\m'} p_2^{\n'}+\ell_0^{\n'} p_2^{\m'}}{\ell_0\cdot p_2}\)\]\,.\no
\end{align}
The Born amplitude in Eq.~\eqref{2-to-2-graviton-scattering}, with Eq.~\eqref{2-2-GR-building-block} as its principal feature, results from a  projection of the $2\rightarrow 2$ graviton scattering amplitude on to the physical two-dimensional subspace spanned by the gravitational polarization vectors, in complete analogy to the QCD case. As we will now show explicitly, this structure will enable us to compute the $2\rightarrow 3$ amplitude. This computation will then be generalized to determine the $2\rightarrow n$ inelastic gravitational amplitude in MRK kinematics.

\subsubsection{Reconstructing the \texorpdfstring{$2\rightarrow 3$}{} amplitude from the Born amplitude}
\label{sec: Gravitational-Lipatov}

The construction of the $2\rightarrow 3$ amplitude is identical to that used in the previous section, a result being its expression in terms of the Lipatov vertex in gauge theory. We will similarly compute the $2\to 3$ amplitude by reconstructing the simultaneous residue of the $1/(k_1^2k_2^2)$ pole. As briefly mentioned earlier, this procedure leaves some ambiguity in the amplitude due to the freedom of adding terms proportional to $k_1^2k_2^2$. However, as discussed below, this ambiguity is fixed by the unitarity of the amplitude.

Denoting the residue of the pole in $k_i^2$ ($i=1,2$) of the $2\to 3$ amplitude by $P_{k_i^2} \mathcal{M}_{2\to 2+1}$ gives
\begin{align}
    P_{k_1^2} \mathcal{M}_{2\to 2+1} =  V_{p_1,\ell_0} \kappa s^2\mathcal{M}_{k_1+p_2\to \ell_1+\ell_2}~,\qquad P_{k_2^2} \mathcal{M}_{2\to 2+1} = \mathcal{M}_{p_1+(-k_2)\to \ell_0+\ell_1}\kappa s^2 V_{p_2,\ell_2}~.
\end{align}
As we showed in  Eq.~\eqref{2-to-2-graviton-scattering}, the $2\to 2$ amplitudes in these formulas can be expressed as 
\begin{align}
    \mathcal{M}_{k_1+p_2\to \ell_1+\ell_2} =  V_{k_1,\ell_1} \frac{\kappa^2(p_2+k_1)^4}{k_2^2} V_{p_2,\ell_2}~,\qquad \mathcal{M}_{p_1+(-k_2)\to \ell_0+\ell_1} =  V_{p_1,\ell_0} \frac{\kappa^2(p_1-k_2)^4}{k_1^2} V_{-k_2,\ell_1}~.
\end{align}
As a next step, we will simplify the vertices $V_{k_1,\ell_1}$ and $V_{-k_2,\ell_1}$ by replacing the  polarization tensors as $\epsilon^{\m\n}(k_1)\to 2p_1^\m p_1^\n/s^2$ and $\epsilon^{\m\n}(-k_2)\to 2p_2^\m p_2^\n/s^2$. This follows from the replacement rule for gluon polarizations  we discussed previously towards the end of Sec.~\ref{sec:2-to-n-amplitude}, coupled with the fact that the polarization tensors for gravitons and gluons are related by $\epsilon^{\m\n}(k) \sim \epsilon^{\m}(k)\epsilon^{\n}(k)$. 
We see already at this step a glimpse of the double-copy relation between QCD and gravity.

Performing some straightforward algebra, one finds that the residue of the $1/k_i^2$ poles can be reexpressed as 
\begin{align}
    &P_{k_1^2} \mathcal{M}_{2\to 2+1} = \kappa^3\frac{s^2}{k_2^2} V_{p_1,\ell_0}  C_{1}^{\m\n}(k_1, k_2)\epsilon_{\m\n}(\ell_1) V_{p_2,\ell_2}~,\no\\[5pt]
    &P_{k_2^2} \mathcal{M}_{2\to 2+1} = \kappa^3\frac{s^2}{k_1^2} V_{p_1,\ell_0}  C_{2}^{\m\n}(k_1, k_2)\epsilon_{\m\n}(\ell_1) V_{p_2,\ell_2}~,
\end{align}
where 
\begin{align}
    C_{i}^{\m\n}(k_1, k_2) = \frac12 C_{i}^{\m}(k_1, k_2)C_{i}^{\n}(k_1, k_2)~,
\end{align}
and $C_{i}^{\m}(k_1, k_2)$ is proportional to the residue of the $1/k_i^2$ pole that we encountered previously:
\begin{align}
\begin{split}
    C_{1}^{\m}(k_1, k_2) &=2 \frac{p_2\cdot \ell_1}{p_1\cdot p_2}p_{1}^\m - (k_1+k_2)^\m-2\frac{\ell_{1}\cdot p_1}{p_1\cdot p_2}p_{2}^\m-\frac{k_2^2}{p_2\cdot \ell_1}p_{2}^\m~,\\[10pt]
    C_{2}^{\m}(k_1, k_2) &=-2 \frac{p_1\cdot \ell_1}{p_1\cdot p_2}p_{2}^\m - (k_1+k_2)^\m+2\frac{\ell_{1}\cdot p_2}{p_1\cdot p_2}p_{1}^\m+\frac{k_1^2}{p_1\cdot \ell_1}p_{1}^\m~.
\end{split}
\end{align}
Since our goal is to construct the $2\to 3$ inelastic amplitude from the residues of the $1/k_i^2$ poles, inspecting the above formulas, one can read off the simultaneous residue of the $1/(k_1^2 k_2^2)$ pole to be 
\begin{align}
\label{2-to-3-grav-1}
    \mathcal{M}_{p_1+p_2\to \ell_0+\ell_1+\ell_2} = \kappa^3\frac{s^2}{k_1^2 k_2^2} V_{p_1,\ell_0} \tilde C^{\m\n}(k_1, k_2)\epsilon_{\m\n}(\ell_1) V_{p_2,\ell_2}~.
\end{align}
where $\tilde C^{\m\n}(k_1, k_2)$ is 
\begin{align}
    \tilde C^{\m\n}(k_1, k_2) = \frac12 C^{\m}(k_1, k_2)C^{\n}(k_1, k_2)~.
\end{align}
Here $C^{\m}(k_1, k_2)$ is the QCD Lipatov vertex whose covariant form was given in Eq.~\eqref{QCD-LV-cov}. 

However the result in Eq.~\eqref{2-to-3-grav-1} has unphysical terms corresponding to overlapping poles $\propto 1/((p_1\cdot \ell_1) (p_2\cdot \ell_1))$, which violate the unitarity of the amplitude~\cite{Steinmann1, Steinmann2, Lipatov:1982it, SabioVera:2012zky, Johansson:2013nsa,Bartels:2008ce}. Therefore to obtain an amplitude that respects unitarity, we need to get rid of such unphysical poles. Towards this end,  we are reminded that there is an ambiguity in the reconstruction of Eq.~\eqref{2-to-3-grav-1} having to do with the freedom to add terms proportional to $k_1^2k_2^2$.  Lipatov used this ambiguity to construct a unitarity preserving amplitude by subtracting from the amplitude a double copy of the vector
\begin{align}
\label{QED-factor}
    N^\mu(k_1, k_2) &= ~\sqrt{k_1^2 k_2^2}\(\frac{p_1^\mu}{p_1\cdot\ell}-\frac{p_2^\mu}{p_2\cdot\ell}\)~,
\end{align}
which straightforwardly gets rid of the unphysical contributions. As a result, one obtains the gravitational analog of the QCD Lipatov vertex to instead be, 
\begin{align}
\label{gravitational-lipatov-vertex}
    C^{\mu\nu}(k_1, k_2) = \frac12 C^\mu(k_1, k_2)C^\nu(k_1, k_2) - \frac12 N^\mu(k_1, k_2)N^\nu(k_1, k_2)~,
\end{align}
allowing us to express the physical $2\to 3$ gravitational Regge amplitude in terms of this vertex as  
\begin{align}
\label{2-to-3-grav-2}
    \mathcal{M}_{p_1+p_2\to \ell_0+\ell_1+\ell_2} = \kappa^3\frac{s^2}{k_1^2 k_2^2} V_{p_1,\ell_0} C^{\m\n}(k_1, k_2)\epsilon_{\m\n}(\ell_1) V_{p_2,\ell_2}~.
\end{align}

\subsection{Graviton reggeization and generalization to the \texorpdfstring{$2\to n$}{} amplitude}
\label{sec:3.2}

The construction of the Lipatov ladder in gravity proceeds analogously to the BFKL case in QCD with the fundamental ingredients being the gravitational Lipatov vertex that we discussed already at length and the reggeized graviton propagator, which will be the focus of the discussion here.
Before we do so, we should state that at tree-level our discussion of the reconstruction of the Lipatov vertex from the poles of $2\to 3$ graviton scattering amplitude in the MRK regime can similarly be generalized\footnote{The generalization to the $2\to 4$ amplitude in the next-to-multi-Regge-kinematics was done recently in \cite{Barcaro:2025ifi}.} to $2\to n+2$ graviton amplitude in the MRK regime using precisely the same methodology as that used in Sec. \ref{sec:2-to-n-amplitude}. The result of the (tree-level) $2\to n+2$ amplitude is 
\begin{align}
\label{2-to-n-grav-tree}
    \mathcal{M}_{2\to n+2}^{\text{tree}} = \kappa^{n+2}\frac{s^2}{k_1^2} V_{p_1,\ell_0} \prod_{i=1}^n C^{\m_i\n_i}(k_i, k_{i+1})\epsilon_{\m_i\n_i}(\ell_i)\frac{1}{k_{i+1}^2} V_{p_2,\ell_{n+1}}~,
\end{align}
which can be compared to Eq.~\eqref{eq:2plusn-QCD-Born}.

With this construction of the tree-level $2\to 2+n$ MRK graviton scattering amplitude in hand, we will now examine the virtual corrections that will generate the reggeized graviton propagator \cite{Lipatov:1982vv, Lipatov:1982it}. To calculate the contribution of a single virtual graviton to the tree-level $2\to n+2$ graviton amplitude, we start with a $2\to n+2+``2"$ tree-level graviton amplitude in the MRK regime, with the specification that the two additional gravitons be emitted between final state gravitons of momenta $\ell_{i-1}$ and $\ell_i$. We denote the momenta of these gravitons to be $\ell$ and $-\ell$. Next, instead of putting their momenta on-shell, we make them virtual by replacing the on-shell $\delta(\ell^2)$ function propagator between them  with the off-shell propagator $1/\ell^2$, as illustrated for the QCD case in Fig.~\ref{virtual-gluon}. The result of this procedure, similarly to the QCD case, leads to the modification of the $2\to n+2$ tree-level graviton scattering amplitude by the factor
\begin{align}
\label{rho-factor-1}
    \sigma(k_i^2) = \frac{\kappa^2}{4}\int \frac{d^4\ell}{(2\pi)^4} \frac{i}{(k_i-\ell)^2}\frac{i}{\ell^2} \frac{i}{k_i^2} C^{\m\n}(k_i, k_i-\ell)C_{\m\n}(k_i-\ell,k_i)~.
\end{align}

The next step is to decompose $\ell$ in terms of the Sudakov parameters: $\ell = \rho p_1+\lambda p_2+\bel$ where $\rho$ and $\lambda$ satisfy the multi-Regge kinematic condition: $\cdots \rho_{i-1} \gg \rho \gg \rho_{i}\cdots$ and $\cdots \lambda_{i-1} \ll \lambda \ll \lambda_{i}\cdots$. Multiplying the Lipatov vertices, and keeping terms that are singular in $\rho s$ and $\lambda s$ (the other terms give sub-leading contributions in $\log s$), we get\footnote{The notation $(\bf A,\bf B)$ we use here represents the scalar product of the vectors $\bf A$ and $\bf B$.}
\begin{align}
\label{rho-factor-2}
    \sigma(k_i^2) = \frac{\kappa^2s}{(2\pi)^4}\int d\rho d\lambda d^2\bel \frac{i}{(k_i-\ell)^2}\frac{i}{\ell^2}ik_i^2 \(\frac{s^2(\bsk_i-\bel,\bel)^2}{(-s\lambda+i\epsilon)^2(s\rho+i\epsilon)^2}-\frac{s k_i^2}{(-s\lambda+i\epsilon)(s\rho+i\epsilon)}\)~.
\end{align}
We next write $\ell^2 = s\rho \lambda -\bel^2$ and $(k_i-\ell)^2 \approx s\rho\lambda-(\bsk_i-\bel)^2$ (employing MRK to approximate $(\rho_i-\rho)(\lambda_i-\lambda)\approx \rho\lambda$), and performing the integral over $\lambda$,  its residues provide the structure  
\begin{align}
\label{rho-factor-3}
    \sigma(k_i^2) = -\frac{\kappa^2s}{(2\pi)^3}\int d\rho d^2\bel ~ k_i^2 \(\frac{(\bsk_i-\bel,\bel)^2}{\bel^2(\bsk_i-\bel)^2 s\rho}\[\frac{1}{\bel^2}+\frac{1}{(\bsk_i-\bel)^2}\]+\frac{k_i^2}{\bel^2(\bsk_i-\bel)^2 s\rho}\)~.
\end{align}
The terms in the square bracket (within the parenthesis above) are from the first term in the parenthesis of Eq.~\eqref{rho-factor-2}, and likewise, the remaining terms in Eq.~\eqref{rho-factor-3} from the second term in the parenthesis of Eq.~\eqref{rho-factor-2}. Finally, the integral in $\rho$ between $\rho_{i-1}$ and $\rho_i$ gives
\begin{align}
\label{rho-factor-4}
    \sigma(k_i^2) = -\frac{\kappa^2k_i^2}{(2\pi)^3}\log\(\frac{\rho_{i-1}}{\rho_i}\)\int d^2\bel \frac{1}{\bel^2(\bsk_i-\bel)^2}  \((\bsk_i-\bel,\bel)^2\[\frac{1}{\bel^2}+\frac{1}{(\bsk_i-\bel)^2}\]-\bsk_i^2\)~.
\end{align}
Recalling from Sec.~\ref{sec:2} (see the discussion below Eq.~\eqref{g-traj-alt}) that $\ln(\frac{\rho_{i-1}}{\rho_i}) \approx \ln(\frac{s_i}{\bsk^2})$, to leading logarithmic accuracy, the contribution of a single soft graviton insertion between the $\ell_{i-1}$ and $\ell_{i}$ emitted gravitons  takes the form
\begin{align}
\label{rho-factor-5}
    \sigma(k_i^2) = \log\(\frac{s_i}{\bsk^2}\)^{\alpha(k_i^2)}~,
\end{align}
where the exponent $\alpha(k_i^2)$ is the one-loop graviton Regge trajectory
\begin{align}
\label{graviton-trajectory}
    \alpha(k_i^2) = \frac{\kappa^2\bsk_i^2}{(2\pi)^3}\int d^2\bel \frac{1}{\bel^2(\bsk_i-\bel)^2}  \((\bsk_i-\bel,\bel)^2\[\frac{1}{\bel^2}+\frac{1}{(\bsk_i-\bel)^2}\]-\bsk_i^2\)~.
\end{align}

The contribution of an arbitrary number of virtual gravitons is obtained by exponentiating $\sigma(k_i^2)$; this corresponds to multiplying every (internal) graviton propagator going down the tree-level $2\to 2+n$ MRK graviton ladder by the factor
\begin{align}
\label{regge-growth}
    \(\frac{s_i}{\bsk^2}\)^{\alpha(k_i^2)}\,.
\end{align}
Thereby incorporating the virtual leading ``reggeized" corrections, we obtain for the $2\to 2+n$ MRK graviton amplitude, 
\begin{align}
\label{gravitational-2-to-n-amplitude}
    \mathcal{M}_{2\to n+2} = \kappa^{n+2} \frac{s^2}{k_1^2}\(\frac{s_1}{\bsk^2}\)^{\alpha(k_1^2)} V_{p_1,\ell_0} \prod_{i=1}^n C^{\m_i\n_i}(k_i, k_{i+1})\epsilon_{\m_i\n_i}(\ell_i)\frac{1}{k_{i+1}^2} \(\frac{s_{i+1}}{\bsk^2}\)^{\alpha(k_{i+1}^2)}V_{p_2,\ell_{n+1}}~.
\end{align}
This result for the $2\to n+2$ amplitude for graviton scattering in the MRK kinematics that encodes both the real and virtual contributions is structurally identical to the QCD case in Eq.~\eqref{2twonQCD}. 

Using the form in Eq.~\eqref{gravitational-2-to-n-amplitude} of the $2\to 2+n$ amplitude, we can now compute the s-channel discontinuity in the $2\to 2$ amplitude due to the exchange of $n$ intermediate gravitons employing the formula in Eq.~\eqref{unitarity-condition-0},
\begin{align}
    &\Im_s \mathcal{M}_{2\to 2} (s, q) =  \frac12 \frac{\kappa^4}{4} s^4 V_{p_1,p_1'}V_{p_2,p_2'} \sum_{n=0}^\infty  \frac{\kappa^{2n}}{2^{n+1}\(2\pi\)^{3n+2}}\int \prod_{i=1}^{n} \(\frac{d\rho_i}{\rho_i} d^2\bsk_i \)d\rho_{n+1}d^2\bsk_{n+1}  \delta\[s\rho_{n+1}-\bsk_{n+1}^2\] \no\\[10pt]
    &\times \frac{1}{\bsk_1^2(\bsq-\bsk_1)^2}\(\frac{1}{\rho_1}\)^{\a(\bsk_1^2)+\a((\bsq-\bsk_1)^2)}
    \prod_{i=1}^n \frac{1}{\bsk_{i+1}^2(\bsq-\bsk_{i+1})^2}  \(\frac{\rho_{i}}{\rho_{i+1}}\)^{\a(\bsk_{i+1}^2)+\a((\bsq-\bsk_{i+1})^2)}\no\\  
    &\times C^{\m_i\n_i}(k_i,k_{i+1}) C_{\m_i\n_i}(q-k_i,q-k_{i+1})\,.
\end{align}
Expressing this as a ratio relative to the $2\to 2$ Born amplitude $\mathcal{M}_{p_1+p_2\to p'_1+p'_2}^{\text{Born}}$ in Eq.~\eqref{2-to-2-graviton-scattering}, we get
\begin{align}
    &\frac{\Im_s \mathcal{M}_{2\to 2} (s,q)}{\mathcal{M}_{p_1+p_2\to p'_1+p'_2}^{\text{Born}}} =  \kappa^2 s^2 t \sum_{n=0}^\infty  \frac{\kappa^{2n}}{2^{n+4}\(2\pi\)^{3n+2}}\int \prod_{i=1}^{n} \(\frac{d\rho_i}{\rho_i} d^2\bsk_i \)d\rho_{n+1}d^2\bsk_{n+1}  \delta\[s\rho_{n+1}-\bsk_{n+1}^2\] \no\\
    & \times \frac{1}{\bsk_1^2(\bsq-\bsk_1)^2}\(\frac{1}{\rho_1}\)^{\a(\bsk_1^2)+\a((\bsq-\bsk_1)^2)}
    \prod_{i=1}^n \frac{1}{\bsk_{i+1}^2(\bsq-\bsk_{i+1})^2}  \(\frac{\rho_{i}}{\rho_{i+1}}\)^{\a(\bsk_{i+1}^2)+\a((\bsq-\bsk_{i+1})^2)} \no\\
    &\times C^{\m_i\n_i}(k_i,k_{i+1}) C_{\m_i\n_i}(q-k_i,q-k_{i+1})\,.
\end{align}
Note that this expression has an extra factor of the center-of-mass energy $s$ relative to Eq.~\eqref{cr0}, the analogous expression in the QCD case, reflecting the spin-2 nature of the graviton. 

We now perform the Mellin integral transform of this expression,
\begin{align}
\label{integral-transform}
    \tilde{\mathcal{M}}_\ell(t) \equiv \int_1^\infty d\(\frac{s}{\bsk^2}\)\frac{\Im \mathcal{M}_{2\to 2}(s,q)}{\kappa^2 s\mathcal{M}_{p_1+p_2\to p'_1+p'_2}^{\text{Born}}}\(\frac{s}{\bsk^2}\)^{-\ell-1}~.
\end{align}
In doing so, we first divided the expression by a factor of $\kappa^2 s$ to make the integral over $s$ manifestly identical to the QCD case. As in the QCD case, this will greatly simply our manipulations, and subsequently, we can recover the total cross-section by performing the inverse Mellin transform\footnote{This expression follows from the identities:
$$
\frac{1}{2\pi i} \int_{-i\infty}^{i \infty} d\nu \(\frac{s}{\tilde s}\)^\nu = \delta\(\log\frac{s}{\tilde s}\)~,\qquad\qquad\qquad
\int_1^\infty dx~ f(x)\delta\(\log\frac xy\) = y f(y) ~~~(y>1)~.
$$
}
\begin{align}
\label{inverse-integral-transform}
\frac{\Im \mathcal{M}_{2\to 2}(s,q)}{\kappa^2 s\mathcal{M}_{p_1+p_2\to p'_1+p'_2}^{\text{Born}}} = \frac{1}{2\pi i} \int_{-i\infty}^{i\infty}d\ell~ \tilde{\mathcal{M}}_\ell(t) \(\frac{s}{\bsk^2}\)^{\ell}~.
\end{align}
Performing the integral in Eq.~\eqref{integral-transform}, followed by integration over the $\rho$ variables with the domain of integration for $\rho_i$ being ($\rho_{i+1}, \rho_{i-1}$), gives 
\begin{align}
    \tilde{\mathcal{M}}_\ell(t) = & \frac {t}{16} \sum_{n=0}^\infty  \(\frac{\kappa^2}{4\pi}\)^n \int \prod_{i=1}^{n+1} \frac{d^2\bsk_i}{\(2\pi\)^2} \frac{1}{\bsk_i^2(\bsq-\bsk_i)^2}\frac{1}{\ell-\a(\bsk_{i}^2)-\a((\bsq-\bsk_{i})^2)}\prod_{i=1}^n\mathcal{K}_G(k_i,k_{i+1})~,
\end{align}
where the Regge pole structure of the amplitude is now manifest.  Further, $\mathcal{K}_G(k_i,k_{i+1})$ represents the contraction of the gravitational Lipatov vertices, 
\be
 \mathcal{K}_G(k_i,k_{i+1}) = C^{\m_i\n_i}(k_i,k_{i+1}) C_{\m_i\n_i}(q-k_i,q-k_{i+1})~.
\ee
As in the QCD case, this equation can be written as
\begin{align}
    \tilde{\mathcal{M}}_\ell(t) = \frac{t}{16} \int \frac{d^2\bsk}{\(2\pi\)^2} \frac{1}{\bsk^2(\bsq-\bsk)^2} f_\ell(\bsk, \bsq)~,
\end{align}
where the amplitude $f_\ell(\bsk, \bsq)$ satisfies the integral equation
\be
\label{grav-BFKL-eq}
(\ell-\alpha(\bsk^2)-\alpha((\bsq-\bsk)^2)) f_\ell(\bsk, \bsq) = 1 + \frac{\kappa^2}{4\pi}\int \frac{d^2\bsk'}{\(2\pi\)^2} \frac{f_\ell(\bsk', \bsq)}{\bsk'^2(\bsq-\bsk')^2} \mathcal{K}_G(\bsk,\bsk')\,.
\ee
Since the structure and derivation of this equation is exactly analogous to the BFKL equation in QCD, we will 
henceforth call it the {\it gravitational BFKL equation}, albeit with the understanding that it was derived solely by 
Lipatov more than 40 years ago~\cite{Lipatov:1982it,Lipatov:1982vv,Lipatov:1991nf}. For a recent related discussion within the framework of Soft Collinear Effective Theory (SCET) \cite{Rothstein:2016bsq}, see \cite{Rothstein:2024nlq}.

\subsection{Divergences in the gravitational BFKL equation}
\label{sec:3.3}

A quick inspection of the gravitation BFKL equation reveals that there are possible divergences that arise from both the $|\bsk'|\to\infty$ and $|\bsk'|\to 0$ limits. In this subsection, we will analyze these limits of the $\bsk'$ integral in the gravitational BFKL equation carefully and conclude that there are no actual divergences. This is similar to the situation in gauge theory although there are some differences that we will uncover below. 

We first rewrite Eq.~\eqref{grav-BFKL-eq} explicitly, where we collect the virtual terms (from the one-loop Regge trajectory), and the real terms (from the square of the Lipatov vertex), under a single integral
\begin{align}
\label{grav-BFKL-2}
    \ell f_\ell(\bsk, \bsq) &= 1 + \frac{\kappa^2}{4\pi}\int \frac{d^2\bsk'}{\(2\pi\)^2}\bigg[\frac{1}{(\bsk-\bsk')^4} \bigg\{ \bsk'^2\tilde\bsk'^2 \bigg( \frac{\tilde\bsk^2}{\tilde\bsk'^2}+\frac{\bsk^2}{\bsk'^2}-\bsq^2 \frac{(\bsk-\bsk')^2}{\bsk'^2\tilde\bsk'^2}\bigg)^2 \no\\[10pt]
    &+4\big(\bsk^2\tilde\bsk^2-\frac{\bsk^2}{\tilde\bsk'^2} (\tilde\bsk\cdot \tilde\bsk')^2 - \frac{\tilde\bsk^2}{\bsk'^2} (\bsk\cdot \bsk')^2\big) \bigg\}f_\ell(\bsk',\bsq)\no\\[10pt]
    &+\frac{4}{\(\bsk-\bsk'\)^2}\bigg\{\frac{\bsk^2}{\bsk'^2+(\bsk-\bsk')^2}\[(\bsk'\cdot(\bsk-\bsk'))^2\(\frac{1}{\bsk'^2}+\frac{1}{(\bsk-\bsk')^2}\)-\bsk^2\]\no\\[10pt]
    &+\frac{\tilde\bsk^2}{\tilde\bsk'^2+\(\bsk-\bsk'\)^2}\[(\tilde\bsk'\cdot(\bsk-\bsk'))^2\(\frac{1}{\tilde\bsk'^2}+\frac{1}{(\bsk-\bsk')^2}\)-\tilde\bsk^2\]\bigg\}f_\ell(\bsk, \bsq)\bigg]\,.
\end{align} 
Here $\tilde\bsk = \bsq-\bsk$ and $\tilde\bsk' = \bsq-\bsk'$. In writing this equation, we made use of the identity in Eq.~\eqref{fraction-substitution} under the integral sign, which facilitates the analysis of divergences that we shall discuss next.

In Eq.~\eqref{grav-BFKL-2}, possible divergences arise when $\bsk'\to\bsk$ or $|\bsk'|\to\infty$. First we analyze the $\bsk'\to\bsk$ integration region. Because of the quartic and quadratic denominators in ($\bsk-\bsk'$) we need to expand out the respective terms in the curly parenthesis to quadratic and zeroth order in $(\bsk-\bsk')^2$. Denoting $\bsk-\bsk'\equiv \bsd$, the contribution of the $\bsk'\to\bsk$ region to the integral, after Taylor expansion, is
\begin{align}
\label{ir-divergence}
\begin{split}
    &\frac{\kappa^2}{4\pi}\int \frac{d^2\bsd}{\(2\pi\)^2}\bigg[\(\frac{4\bsk^2}{\bsd^2}+\frac{4\tilde\bsk^2}{\bsd^2}-\frac{6\bsq^2}{\bsd^2}+8\frac{(\bsk\cdot \bsd)(\tilde\bsk\cdot \bsd)}{\bsd^4}\)-4\(\frac{\bsk^2}{\bsd^2}+\frac{\tilde\bsk^2}{\bsd^2}-\frac{(\bsd\cdot\bsk)^2}{\bsd^4}-\frac{(\bsd\cdot\tilde\bsk)^2}{\bsd^4}\) \bigg]f_\ell(\bsk, \bsq)\\[10pt]
    =&\frac{\kappa^2}{4\pi}\int \frac{d^2\bsd}{\(2\pi\)^2}\bigg[-6\frac{\bsq^2}{\bsd^2}+4\frac{(\bsq\cdot \bsd)^2}{\bsd^4} \bigg]f_\ell(\bsk, \bsq) \sim \frac{\kappa^2}{4\pi^2} \bsq^2  f_\ell(\bsk,\bsq)\log\Lambda^2_{IR}~.
\end{split}
\end{align}
In the first line of the above equation, the terms in the first parenthesis follow from the first two lines in Eq.~\eqref{grav-BFKL-2} whereas the terms in the second parenthesis follows from the last two lines in Eq.~\eqref{grav-BFKL-2}. To obtain the result in Eq.~\eqref{ir-divergence}, we further needed to open up the squared parenthesis in the first line in Eq.~\eqref{grav-BFKL-2} and replace under the $\bsk'$ integral\footnote{Alternatively, one can integrate Eq.~\eqref{grav-BFKL-2} explicitly term-by-term, using a cut-off prescription, to obtain Eq.~\eqref{ir-divergence}.}
\begin{align}
    \frac{f(\bsk')}{\(\bsk-\bsk'\)^4 \bsk'^2}\to \frac{f(\bsk')}{\(\bsk-\bsk'\)^4 \(\bsk'^2+\(\bsk-\bsk'\)^2\)}+\frac{f(\bsk')+f(\bsk-\bsk')}{\(\bsk-\bsk'\)^2 \(\bsk'^2+\(\bsk-\bsk'\)^2\)^2}~.
\end{align}

We observe that for $\bsq\neq \bszero$ this integral has a logarithmic divergence,  reflecting the fact that gravity in 4 dimensions is IR divergent\footnote{For general spacetime dimensions, the relevant integral has the form $\int d^{D-2}\bsd\, \frac{1}{\bsd^2}$, which is manifestly IR finite for $D>4$.}. However for $\bsq=0$, provided the amplitude $f_\ell(\bsk,\bsq)$ is well-behaved as $\bsq\to0$, the $\bsk'$ integral in Eq.~\eqref{grav-BFKL-2} is free of IR divergences. This should indeed be the case since for $\bsq=0$ the amplitude $f_\ell(\bsk,\bsq)$ computes the inclusive cross-section which, as shown by Weinberg in \cite{Weinberg:1965nx}, is free from any IR divergence. We note that Weinberg's analysis involved using the soft graviton emission vertex for computing real and virtual divergences. We have here a generalization of that analysis to hard graviton emissions where instead of using the soft graviton emission vertex one uses Lipatov's gravitational vertex. Later, in Sec. \ref{sec:Weinberg}, we will demonstrate that these vertices smoothly connect to each other in the appropriate kinematic regimes.

The gravitational BFKL equation also suffers from UV divergences when $|\bsk'|\to\infty$, originating from  virtual contributions. (Inspecting Eq.~\eqref{grav-BFKL-2}, we see that the real contribution does not contain this UV divergence.). Evaluating the integrals, one finds that the one-loop Regge trajectory in Eq.~\eqref{graviton-trajectory} depends on the UV and IR cutoffs as 
\begin{align}
\label{alpha_grav}
    \alpha(\bsk^2) = \frac{\kappa^2\bsk^2}{8\pi^2} \[2\log\(\frac{\Lambda^2_{UV}}{\bsk^2}\)-\log\(\frac{\bsk^2}{\Lambda^2_{IR}}\)\]~.
\end{align}
While the dependence on $\Lambda_{IR}$ drops out the case of $\bsq=0$ as shown above, the dependence on  $\Lambda_{UV}$ does not.  However, when evaluating the virtual contributions in the ladder, the domain of the transverse momentum integration (see Eq.~\eqref{rho-factor-1}) should not violate the assumption of multi-Regge kinematics. The upper limit on squared transverse momentum should therefore be the center-of-mass energy.  Then, from Eq.~\eqref{rho-factor-1} we see that the inclusion of a virtual graviton gives a term that is proportional to $\log^2(s^2/\bsk^2)$. One of the factors come from $\log(\rho_{i-1}/\rho_i)$ (see below Eq.~\eqref{rho-factor-4}) and the other one comes from the upper limit of the transverse momentum integral. Such doubly-logarithmic terms in $s$ were already discussed by Lipatov in \cite{Lipatov:1982vv} and later by Bartels, Lipatov and Sabio Vera in \cite{Bartels:2012ra} where universal infrared sources of these double-logs that come from the integration region $|\bsk'| < |\bsk|$ were identified and resummed via an evolution equation in Mellin space~\cite{Kirschner:1983di}. The gravitational BFKL equation does not captures the latter source of these double logs and a careful inclusion of these terms requires further work perhaps along the lines of \cite{Caucal:2022ulg}-see also \cite{Salam:1998tj, Ciafaloni:1998iv, Ciafaloni:1999yw, Ciafaloni:2003rd}.
In the next subsection, we will solve the gravitational BFKL equation for $\bsq=0$ analytically in terms of its eigen values, which are structurally quite similar to the gauge theory case. 

\subsection{Analytical solution of the gravitational BFKL equation for \texorpdfstring{$\bsq=0$}{}}
\label{sec:3.4}

Let us first simplify Eq.~\eqref{grav-BFKL-2}, denoting $f_\ell(\bsk,\bszero)\equiv f_\ell(\bsk) $:
\begin{align}
\label{grav-BFKL-3}
    \ell f_\ell(\bsk) = 1 + &\frac{2\kappa^2}{\pi}\int \frac{d^2\bsk'}{\(2\pi\)^2}\bigg[\frac{1}{(\bsk-\bsk')^4} \(\bsk^4 - \frac{\bsk^2}{\bsk'^2} (\bsk\cdot \bsk')^2  \)f_\ell(\bsk')\no\\[10pt]
    +&\frac{\bsk^2}{(\bsk-\bsk')^2\(\bsk'^2+(\bsk-\bsk')^2\)} \((\bsk'\cdot(\bsk-\bsk'))^2\(\frac{1}{\bsk'^2}+\frac{1}{(\bsk-\bsk')^2}\)-\bsk^2\)f_\ell(\bsk)\bigg]~.
\end{align}
Following the methodology from the previous section, defining 
\begin{align}
    f_\ell(\bsk) \equiv \int \frac{d^2\bsp}{(2\pi)^2}~ g_\ell(\bsk,\bsp)~,\,\,\,{\rm and\, using}\,\,\,
    1= \int \frac{d^2\bsp}{(2\pi)^2} ~(2\pi)^2\delta^{(2)}(\bsk-\bsp)~,
\end{align}
allows us to write Eq.~\eqref{grav-BFKL-3} as 
\begin{align}
\label{grav-BFKL-4}
    \ell g_\ell(\bsk,\bsp) =& (2\pi)^2\delta^{(2)}(\bsk-\bsp) + \frac{2\kappa^2}{\pi}\int \frac{d^2\bsk'}{\(2\pi\)^2}\bigg[\frac{1}{(\bsk-\bsk')^4} \(\bsk^4 - \frac{\bsk^2}{\bsk'^2} (\bsk\cdot \bsk')^2  \)g_\ell(\bsk',\bsp)\no\\[10pt]
    +&\frac{\bsk^2}{(\bsk-\bsk')^2\(\bsk'^2+(\bsk-\bsk')^2\)} \((\bsk'\cdot(\bsk-\bsk'))^2\(\frac{1}{\bsk'^2}+\frac{1}{(\bsk-\bsk')^2}\)-\bsk^2\)g_\ell(\bsk,\bsp)\bigg]~.
\end{align}
As previously, we seek a solution of the form
\begin{align}
\label{solution-ansatz-grav}
    g_\ell(\bsk,\bsp) = \frac{1}{\bsp^2}\sum_{n=-\infty}^\infty \int_{-\infty}^\infty d\n ~a_\ell(\n,n) e^{i\n(\lambda_k-\lambda_p) }e^{in(\phi_k-\phi_p)}~,
\end{align}
Here $\lambda_k=\log(\bsk^2/\m^2), ~\lambda_p=\log(\bsp^2/\m^2)$ ($\m$ being an arbitrary small scale) and $\phi_{k,p}$ are the azimuthal angles of the vectors $\bsk,\bsp$ respectively. The above ansatz is motivated by the delta function representation,
\begin{align}
\label{delta-function-grav}
    \delta^{(2)}(\bsk-\bsp)=\frac{1}{2\pi^2\bsp^2}\sum_{n=-\infty}^\infty\int_{-\infty}^\infty d\nu~ e^{i\nu (\lambda_k-\lambda_p)} e^{in(\phi_k-\phi_p)}~.
\end{align}
with which one can write the r.h.s of Eq.~\eqref{grav-BFKL-4} as 
\begin{align}
\label{eigen-value-equation-1}
     &\frac{2\kappa^2}{\pi}\int \frac{d^2\bsk'}{\(2\pi\)^2}\bigg[\frac{1}{(\bsk-\bsk')^4} \(\bsk^4 - \frac{\bsk^2}{\bsk'^2} (\bsk\cdot \bsk')^2  \)g_\ell(\bsk',\bsp)+\no\\[10pt]
    +&\frac{\bsk^2}{2} \frac{1}{\(\bsk-\bsk'\)^2\bsk'^2} \((\bsk'\cdot(\bsk-\bsk'))^2\(\frac{1}{\bsk'^2}+\frac{1}{(\bsk-\bsk')^2}\)-\bsk^2\)g_\ell(\bsk,\bsp)\bigg]\no\\[10pt]
    =&\frac{1}{\bsp^2}\sum_{n=-\infty}^\infty \int_{-\infty}^\infty d\n ~a_\ell(\n,n)\omega(\n,n) e^{i\n(\lambda_k-\lambda_p) }e^{in(\phi_k-\phi_p)}~,
\end{align}
where $\omega(\n,n)$ is the eigenvalue of the BFKL integral operator on the l.h.s. Using this eigenvalue equation, along with the ansatz Eq.~\eqref{solution-ansatz-grav}, in Eq.~\eqref{grav-BFKL-4}, we find that the coefficients $a_\ell(\nu,n)$ of the ansatz are given in terms of the BFKL eigenvalue by 
\begin{align}
    a_\ell(\nu,n) = \frac{2}{\ell-\omega(\n,n)}~.
\end{align}
Hence the solution to  Eq.~\eqref{grav-BFKL-4} is given by
\begin{align}
\label{solution-ansatz-2}
    g_\ell(\bsk,\bsp) = \frac{2}{\bsp^2}\sum_{n=-\infty}^\infty \int_{-\infty}^\infty d\n ~\frac{1}{\ell-\omega(\n,n)} e^{i\n(\lambda_k-\lambda_p) }e^{in(\phi_k-\phi_p)}~,
\end{align}

Inserting the ansatz Eq.~\eqref{solution-ansatz-grav} for $g_\ell(\bsk,\bsp)$ into Eq.~\eqref{eigen-value-equation-1}, the gravitational BFKL eigenvalues can be expressed as 
\begin{align}
\label{eigen-value-equation-2}
 \omega(\n,n)=    &\frac{2\kappa^2}{\pi}\int \frac{d^2\bsk'}{\(2\pi\)^2}\bigg[\frac{1}{(\bsk-\bsk')^4} \(\bsk^4 - \frac{\bsk^2}{\bsk'^2} (\bsk\cdot \bsk')^2  \)e^{i\n(\lambda_{k'}-\lambda_k) }e^{in(\phi_{k'}-\phi_k)}\no\\[10pt]
    +&\frac{\bsk^2}{(\bsk-\bsk')^2\(\bsk'^2+(\bsk-\bsk')^2\)} \((\bsk'\cdot(\bsk-\bsk'))^2\(\frac{1}{\bsk'^2}+\frac{1}{(\bsk-\bsk')^2}\)-\bsk^2\)\bigg]~.
\end{align}
To regulate the divergence near $\bsk'=\bsk$, we split the prefactor in the second line as follows
\begin{align}
\label{eigen-value-equation-3}
 \omega(\n,n)=&\frac{2\kappa^2}{\pi}\int \frac{d^2\bsk'}{\(2\pi\)^2}\bigg[\frac{1}{(\bsk-\bsk')^4} \(\bsk^4 - \frac{\bsk^2}{\bsk'^2} (\bsk\cdot \bsk')^2  \)e^{i\n(\lambda_{k'}-\lambda_k) }e^{in(\phi_{k'}-\phi_k)}\no\\[10pt]
    +&\frac{\bsk^2}{\bsk'^2}\(\frac{1}{(\bsk-\bsk')^2}-\frac{1}{\(\bsk'^2+(\bsk-\bsk')^2\)}\) \((\bsk'\cdot(\bsk-\bsk'))^2\(\frac{1}{\bsk'^2}+\frac{1}{(\bsk-\bsk')^2}\)-\bsk^2\)\bigg]~.
\end{align}
Carrying out the angular integrals, we find 
\begin{align}
\label{eigen-value-equation-4}
    &\omega(\n,n) = -\frac{\kappa^2 \bsk^2}{4\pi^2} \Re\int_0^1 dx~e^{i\nu\log x}\[x^{\frac{|n|}{2}-1}\(|n|-\frac{1+x}{1-x}\)+\frac{\delta_{n,0}}{x}\]\\[5pt]
    -&\frac{ \kappa ^2 k^2}{4 \pi ^2} \int_0^1 dx \left[\left(\frac{1}{x}+\frac{1}{1-x}\right)+\left(\frac{1}{x}-\frac{2}{x \sqrt{4 x^2+1}}\right)-\left(\frac{4}{x}+\frac{1}{x-1}-1\right)-\left(\frac{2}{\sqrt{x^2+4}}-\frac{2}{x}+1\right)\right]~,\no
\end{align}
where the variable $x$ is given by
\begin{align}
\label{x-def-grav}
    x= \begin{cases}\bsk'^2 / \bsk^2 & \text { for } \bsk'^2<\bsk^2 \\ \bsk^2 / \bsk'^2 & \text { for } \bsk'^2>\bsk^2\end{cases}~.
\end{align}

The origins of the various terms that appear in Eq.~\eqref{eigen-value-equation-4} are as follows. The first line in Eq.~\eqref{eigen-value-equation-4} comes from the first line in Eq.~\eqref{eigen-value-equation-3}. The fact that we only have the real part follows from evaluating the $k'$ integration in the domains $(0,k)$ and $(k,\infty)$ separately. In the second line of Eq.~\eqref{eigen-value-equation-4} there are four parenthesis within the square brackets: (a) the terms in the first parenthesis come from the $1/(\bsk-\bsk')^2$ term in Eq.~\eqref{eigen-value-equation-3} and the integration region $k'\in (0,k)$, (b) the terms in the second parenthesis are from the $1/(\bsk'^2 +(\bsk-\bsk')^2)$ term in Eq.~\eqref{eigen-value-equation-3} and the integration region $k'\in (0,k)$, (c) the terms in the third parenthesis are from the $1/(\bsk-\bsk')^2$ term in Eq.~\eqref{eigen-value-equation-3} and the integration region $k'\in (k,\infty)$, and finally (d) the terms in the last parenthesis come from the $1/(\bsk'^2 +(\bsk-\bsk')^2)$ term in Eq.~\eqref{eigen-value-equation-3} and the integration region $k'\in (k,\infty)$. We should note here the terms (particularly those from $\bsk'^2>\bsk^2$) that diverge in the $x\to 0$ limit. These are present in the last two parenthesis in the second line of Eq.~\eqref{eigen-value-equation-4}. They diverge in the argument as $4/x$ and $-2/x$, giving a net  $-2/x$ contribution, whose logarithmic divergence can  be regulated with the cut-off $\Lambda_{UV}$. 

Isolating the  divergent terms, we obtain  
\begin{align}
\label{eigen-value-equation-6}
    \omega(\n,n) = -&\frac{\kappa^2 \bsk^2}{4\pi^2} \Re\int_0^1 dx~e^{i\nu\log x}\[x^{\frac{|n|}{2}-1}\(|n|-\frac{1+x}{1-x}\)+\frac{\delta_{n,0}}{x}+\frac{2}{1-x}\]\no\\[5pt]
    -&\frac{ \kappa ^2 k^2}{4 \pi ^2}\int_{0}^1 dx \left[\frac2x -\frac{2}{x \sqrt{4 x^2+1}}-\frac{2}{\sqrt{x^2+4}}\right]+\frac{ \kappa ^2 k^2}{4 \pi ^2}\int_{\bsk^2/\Lambda_{UV}^2}^1 dx \frac2x~.
\end{align}
Thanks to the cancellation of IR divergences, the result of integral in first line is finite and the net contribution from the terms within the square bracket in the second line vanishes. As a result, 
\begin{align}
\label{grav-BFKL-eigenvalues}
    \omega(\n,n) = -\frac{\kappa^2 \bsk^2}{2\pi^2} \[\Re \psi\left(i \nu +\frac{| n| }{2}+1\right)+\gamma_E+\frac{n^2-| n| }{4 \nu ^2+n^2}+\log\(\frac{\bsk^2}{\Lambda_{UV}^2}\)\]\,,
\end{align}
where $\gamma_E$ is  Euler's constant and $\psi(x)$ is the digamma function. 

\begin{figure}[ht]
\centering
\includegraphics[scale=0.5]{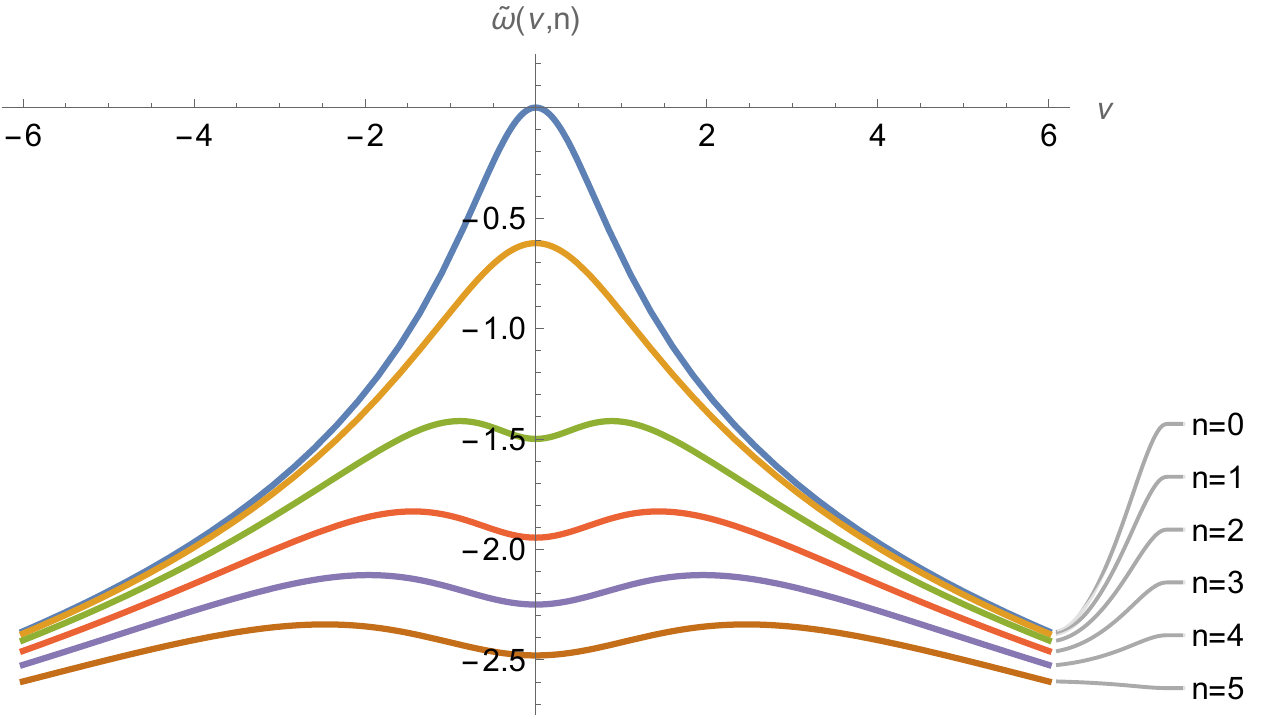}
\caption{Plot of $\tilde{\omega}(\n,n)=\frac{2\pi^2}{\kappa^2\bsk^2}\omega(\n,n)+\log(\frac{\bsk^2}{\Lambda_{UV}^2})$, where $\omega(\n,n)$ are the gravitational BFKL eigenvalues in Eq.~\eqref{grav-BFKL-eigenvalues} for $n=0,1,2,3,4,5$. $\tilde{\omega}(\n,n)$ is therefore independent of $\Lambda_{UV}^2$. As in the gauge theory case, the maximum is attained for $n=0$ at $\nu=0$, with maximal value $\tilde \omega^* = 0$.}
\label{fig:grav-eigenvalues}
\end{figure}

The GR BFKL eiegnvalues are plotted in Fig.~\ref{fig:grav-eigenvalues}. It is evident that the largest eigenvalue is attained for $\nu=0$ and $n=0$, just as in the gauge theory case. However, unlike the latter, the maximal value here is dependent on the UV cut-off. At $n=0$, the eigenvalue has the  small $\nu$ expansion 
\begin{align}
    \omega(\nu, n=0) &= -\frac{\kappa^2 \bsk^2}{2\pi^2}\[\log\(\frac{\bsk^2}{\Lambda_{UV}^2}\)+ \sum_{k=1}^\infty \frac{(-1)^k \psi ^{(2 k)}(1)}{(2 k)!}\nu^{2k}\]~\no\\[5pt]
    &=\frac{\kappa^2 \bsk^2}{2\pi^2}\[\log\(\frac{\Lambda_{UV}^2}{\bsk^2}\)- \nu ^2 \zeta (3) + O(\n^4)\]~.
\end{align}

Having computed the BFKL eigenvalues, one can now work backwards and compute the total $2\to 2$ cross-section of graviton-graviton scattering. However, there is a subtlety in our derivation of the eigenvalues. Notice that in deriving the eigenvalues we had to regulate the $\bsk'$ integral by $\Lambda_{UV}$. In the QCD case, there was no need of such a cut-off\footnote{This is true only for the BFKL equation at leading logarithmic accuracy. At NLLx order, the transverse momenta integral has $\sqrt{s}$ as the upper bound \cite{Salam:1998tj}- see the discussion in Sec.~\ref{sec:NLL-gluon-sat}.} since the BFKL equation there had no UV divergences. In gravity though, because the gravitational Regge trajectory has a UV divergence, the handling of the transverse momentum integrals are trickier. As noted, the natural cut-off is the center-of-mass energy $\sqrt{s}$ instead of $\Lambda_{UV}$. However, simply replacing $\Lambda_{UV}$ with $\sqrt{s}$ in the expression for the BFKL eigenvalue is an inconsistent procedure because one then takes the Mellin transform w.r.t. to $s$ before performing the transverse momentum integrals, whose upper bound is dependent on $s$. Therefore, one needs to go back and reanalyze the BFKL ladder diagram prior to  performing the Mellin transform. 

To appreciate this further, let us have a closer look at the expression for the gravitational one-loop Regge trajectory in Eq.~\eqref{graviton-trajectory}. This provides the source of the UV divergence in the gravitational BFKL equation. In light of the above discussion, and replacing $\Lambda_{UV}^2$ with $s$, from Eq.~\eqref{graviton-trajectory}, and the expression for the insertion of a soft graviton line  $\sigma(k)$ in Eq.~\eqref{rho-factor-4}, we see that in addition to the Sudakov log ($\log(-s/t)\log(-t/\Lambda_{IR})$) we also get a double log of the form $\log^2(-s/t)$ that needs to be resummed as well. There are two sources of this double log, one from virtual graviton lines along the ladder, and the other from virtual graviton lines that are emitted and absorbed by external legs. 

It was pointed out in \cite{Lipatov:1982vv} that the resummation of the ladder-type $\log^2(-s/t)$ contributions sum up to the modified Bessel function,

\begin{align}
\label{dl-resummed-1}
\begin{split}
    \mathcal{M}_{2\to 2, \text{ladder}}^{L^2} &\sim\sqrt{-\frac{8 \pi ^2}{3 \kappa^2 t \xi ^2}} I_1\left(2 \sqrt{-\frac{3 \kappa^2 t  \xi ^2}{8 \pi ^2}}\right)~,\\[5pt]
    & =  1 - \frac{3 \log ^2\left(-\frac{s}{t}\right)}{16 \pi ^2}\kappa^2 t+\frac{3 \log ^4\left(-\frac{s}{t}\right)}{256 \pi ^4}\kappa^4 t^2+\cdots~,
\end{split}
\end{align}
where $\xi=\log(-\frac st)$. On the other hand, the resummation of the non-ladder-type $\log^2(-s/t)$ terms (involving soft virtual graviton exchanges between external legs) was discussed in \cite{Bartels:2008ce} building upon a similar analysis in the context of quark-quark scattering in QCD \cite{Kirschner:1982qf, Kirschner:1983di}.  

The upshot of the analysis in \cite{Bartels:2008ce} is that the complete (ladder + non-ladder)  $\log^2(-s/t)$ contributions can be neatly expressed as a contour integral 
\be
\label{dl-contour}
\mathcal{M}_{2\to 2}^{L^2} \sim \int_{\delta-i \infty}^{\delta+i \infty} \frac{d \omega}{2 \pi i}\left(-\frac{s}{t}\right)^\omega \frac{f(\omega, t)}{\omega}\,,
\ee
where the contour is along the imaginary axis anchored at $\delta$ that is to the right of all the singularities of $f(\omega, t)$. The latter satisfies the Riccati equation
\begin{align}
\label{riccati}
    f(\omega, t) = 1-\frac{\kappa^2 t}{8\pi^2} \frac{d}{d \omega} \(\frac{f(\omega)}{\omega}\)-\frac{3\kappa^2 t}{8\pi^2} \frac{f(\omega)^2}{\omega^2}~.
\end{align}
On the r.h.s, the last term corresponds to the purely ladder contribution whereas the derivative term corresponds to the non-ladder contributions\footnote{The solution of Eq.~\eqref{riccati} in the absence of the derivative term 
recovers the resummed result of purely ladder-type $\ln^2(-s/t)$ contributions in Eq.~\eqref{dl-resummed-1}. We thank Anna Stasto for a discussion on this point.}. Such equations also appear in the double-log contributions in quark-quark scattering~\cite{Kirschner:1982qf, Kirschner:1983di}. Eq.~\eqref{riccati} admits an exact solution given by
\begin{align}
    f(\omega) = 1-\frac{2 \kappa^2 t }{3 \kappa^2 t + 8\pi^2\omega ^2}~.
\end{align}
We observe that the ratio $f(\omega)/\omega$ in Eq.~\eqref{dl-contour} has simple poles in $\omega$ at $\omega = \{0, \pm \sqrt{-3\kappa^2 t/8\pi^2}$\}. Computing the contour integral in Eq.~\eqref{dl-contour} therefore gives
\begin{align}
\begin{split}
\mathcal{M}_{2\to 2}^{L^2} \sim & ~\frac{1}{3} \left[1 + \(-\frac st\)^{\sqrt{-3\kappa^2 t/8\pi^2}} + \(-\frac st\)^{-\sqrt{-3\kappa^2 t/8\pi^2}} \right]~,\\[5pt]
=&~ 1 - \frac{\kappa^2 t}{8\pi^2}\log^2 \(-\frac{s}{t} \) + \frac{\kappa^4 t^2}{256 \pi ^4}\log ^4 \(-\frac{s}{t} \)+\cdots~.
\end{split}
\end{align}
The resummed result in the above equation shows that in the Regge limit the $\ln^2(-s/t)$ contribution to the growth of the amplitude behaves as $s^{\sqrt{-3\kappa^2t/8\pi^2}}$. We remark that this resummation decouples from the resummation of the Sudakov double logs ($\log(-s/t)\log(-t/\Lambda_{IR})$). The latter resums to  $(-s/t)^{\kappa^2 t\log(-t/\Lambda_{\rm IR})/8\pi^2}$, as previously discussed in Eq.~\eqref{regge-growth}.

In summary, taking into account both the resummation of Sudakov logs  and $\log^2(-s/t)$ contributions,  the complete  $2\to 2$ elastic graviton amplitude can be expressed as 
\begin{align}
\begin{split}
    \mathcal{M}_{2\to 2}^{\text{Sudakov} + L^2} & \sim \mathcal{M}^{\text{Born}}\(-\frac{s}{t}\)^{\kappa^2 t\log(-t/\Lambda_{\rm IR})/8\pi^2}\frac{1}{3} \left[1 + \(-\frac st\)^{\sqrt{-3\kappa^2 t/8\pi^2}} + \(-\frac st\)^{-\sqrt{-3\kappa^2 t/8\pi^2}} \right]~,\\[10pt]
    &= \mathcal{M}^{\text{Born}}\(1+\frac{\kappa ^2}{8 \pi ^2} t \log \left(-\frac{s}{t}\right) \left[\log \left(-\frac{t}{\Lambda _{\text{IR}}}\right)-\log \left(-\frac{s}{t}\right)\right]+\cdots\)~.
\end{split}
\end{align}
Here $\mathcal{M}^{\text{Born}}$ is the Born amplitude in Eq.~\eqref{graviton-born-amplitude}. We note that since $t<0$, the growth of the $2\to 2$ amplitude is slightly slower than $s^{2}$.

\subsection{The Weinberg regime}
\label{sec:Weinberg}

We will  discuss here the relation between Lipatov's (hard) graviton emission vertex and the Weinberg (soft) graviton emission vertex. Its key feature is that the soft momentum limit (of the emitted graviton) from the Lipatov vertex is the same as Weinberg's expression for gravitational radiation in the high energy limit. For complementary discussions, we refer the reader to \cite{Amati:1990xe, Ciafaloni:2015vsa, Ciafaloni:2015xsr}, where the matching of the Lipatov and the Weinberg regimes were discussed previously.

We first analyze the Lipatov vertex given by Eq.~\eqref{gravitational-lipatov-vertex},
\begin{align}
    C^{\mu\nu}(q_1, q_2) = \frac12 C^\mu(q_1, q_2)C^\nu(q_1, q_2) - \frac12 N^\mu(q_1, q_2)N^\mu(q_1, q_2)~.\no
\end{align}
Expanding out the first term, and keeping only the singular terms in the soft momenta $k$, we find 
\begin{align}
\label{expandingCmCn}
    & C^\mu(q_1, q_2)C^\nu(q_1, q_2) \sim \(-\bsq_1^\mu-\bsq_2^\mu +\frac{p_1^\mu q_1^2}{p_1\cdot k}-\frac{p_2^\mu q_2^2}{p_2\cdot k} \)\(-\bsq_1^\nu-\bsq_2^\nu +\frac{p_1^\nu q_1^2}{p_1\cdot k}-\frac{p_2^\nu q_2^2}{p_2\cdot k} \) \no\\[5pt]
    &=\frac{p_1^\mu p_1^\nu q_1^4}{(p_1\cdot k)^2} +\frac{p_2^\mu p_2^\nu q_2^4}{(p_2\cdot k)^2}- \frac{(p_1^\mu p_2^\nu+p_1^\nu p_2^\mu)q_1^2q_2^2}{(p_1\cdot k)(p_2\cdot k)}- \[ \(\bsq_1^\mu+\bsq_2^\mu\)\(\frac{p_1^\nu q_1^2}{p_1\cdot k}-\frac{p_2^\nu q_2^2}{p_2\cdot k}\) +(\mu \leftrightarrow \nu)\]~.
\end{align}
The third term in the r.h.s on the second line  is canceled by the contribution from $N^\mu(q_1, q_2)N^\mu(q_1, q_2)$, whose expansion gives

\begin{align}
\label{expandingNmNn}
N^\mu(q_1, q_2)N^\nu(q_1, q_2) & =  \bsq_1^2\bsq_2^2 \(\frac{p_1^\mu p_1^\nu }{(p_1\cdot k)^2} +\frac{p_2^\mu p_2^\nu }{(p_2\cdot k)^2}- \frac{(p_1^\mu p_2^\nu+p_1^\nu p_2^\mu)}{(p_1\cdot k)(p_2\cdot k)}\)~.
\end{align}
Note that when considering the $k\to 0$ limit, the momenta $q_1$ and $q_2$ are also constrained since $k = q_1-q_2$. This leads to a partial cancellation of the first two terms in Eq.~\eqref{expandingCmCn}. As a result, we obtain
\begin{align}
    \frac12(C^\mu C^\nu - N^\mu N^\nu) &\sim \bsq_1^2 \(\frac{p_1^\mu p_1^\nu }{(p_1\cdot k)^2} -\frac{p_2^\mu p_2^\nu }{(p_2\cdot k)^2}\)(\bsk\cdot \bsq_1) -\bsq_1^2 \[\bsq_1^\mu \(\frac{p_1^\nu }{p_1\cdot k}-\frac{p_2^\nu }{p_2\cdot k}\)+(\mu \leftrightarrow \nu)\]\,.
\end{align}
We define now $\gamma^{\m\n}$ as
\begin{align}
    C^{\mu\nu}(q_1, q_1-k) \equiv \bsq_1^2\gamma^{\mu\nu}(q_1, q_1-k)~,\qquad |\bsk^2| \ll q_1^2~.
\end{align}
where
\begin{align}
\label{lipatov-soft-limit}
\gamma^{\mu\nu}(q_1, q_1-k) = \(\frac{p_1^\mu p_1^\nu }{(p_1\cdot k)^2} -\frac{p_2^\mu p_2^\nu }{(p_2\cdot k)^2}\)(\bsk\cdot \bsq_1) - \[\bsq_1^\mu \(\frac{p_1^\nu }{p_1\cdot k}-\frac{p_2^\nu }{p_2\cdot k}\)+(\mu \leftrightarrow \nu)\]~.
\end{align}
Note that the emission vertex has a quadratic pole in $(p_i\cdot k)$. Such poles are not present in the Weinberg formula as discussed below and our goal is to reconcile the presence of such poles in the two formulas. 

Since the emitted graviton is on-shell, it is convenient to contract the emission vertex with the graviton polarization tensor $\ep^{\m\n}$. Noting the relation between the various momenta
\be
\label{momentum-constraints}
q_1 = p_1-p_1'~,\qquad\qquad q_2 = p_2'-p_2~,\qquad k=p_1+p_2-p_1'-p_2'~,
\ee
and fixing the polarization tensor to be in light cone gauge
\be
\label{polarization-conditions}
\ep_{\m\n}(k)p_1^\m = 0~,\qquad\qquad \ep_{\m\n}(k)k^\m=0\,,
\ee
the contraction of $\gamma^{\m\n}$ with it gives 
\begin{align}
\label{soft-lipatov-final}
    \gamma_{\m\n}\ep^{\m\n} &= -\frac{k^ik^j}{(p_2^-k^+)^2}\(\frac{p_2^-}{k^-}\)^2 \bsk\cdot\bsq_1\ep_{ij} - 2\frac{\bsq_1^\mu k^i}{p_2^-k^+}\(\frac{p_2^-}{k^-}\)\ep_{i\mu}= -4\[\frac{\bsk\cdot\bsq_1}{\bsk^4} \bsk^i\bsk^j + \frac{\bsk^i\bsq_1^j}{\bsk^2}\]\ep_{ij}~.
\end{align}
In deriving this expression, we made use of the identities
\begin{align}
\label{polarization-identity}
    \ep_{\m\n} p_2^\n = -\frac{p_2^-}{k^-}k^i\ep_{i\mu}~,\qquad \ep_{\m\n} p_2^\m p_2^\n = \(\frac{p_2^-}{k^-}\)^2 k^ik^j\ep_{ij}~,\qquad 2k^+k^- = \bsk^2~.
\end{align}

Now let us analyze the high energy limit of the Weinberg vertex. For every insertion of a soft graviton line (carrying momentum $k$) to an external leg of a non-radiative amplitude carrying momentum $p_i$, one attaches a factor  
\begin{align}
\label{WeinbergVertex}
    L^{\m\n}_{W} =  \sum_{i}\frac{\eta_i p^\mu_i p^\nu_i}{p_i\cdot k}~,
\end{align}
to reconstruct the radiative amplitude. Here $\eta_i=\pm 1$, depending on whether the momenta are incoming to the vertex or outgoing from it. For $2\to 2$ scattering with momenta $p_1+p_2\to p_1'+p_2'$, we get 
\begin{align}
\label{weinberg-expanded}
    L^{\m\n}_{W} =  \frac{ p^\mu_1 p^\nu_1}{p_1\cdot k}+\frac{ p^\mu_2 p^\nu_2}{p_2\cdot k}-\frac{ p'^{\mu}_1 p'^{\nu}_1}{p_1'\cdot k}-\frac{ p'^\mu_2 p'^\nu_2}{p'_2\cdot k} ~.
\end{align}
Contracting this expression with the polarization tensor then gives 
\begin{align}
\label{weinb-1}
    L^{\m\n}_{W}\ep_{\m\n} =  0+\(\frac{p_2^-}{k^-}\)^2\frac{k^ik^j}{p_2^-k^+}\ep_{ij}-\frac{q_1^\mu q_1^\nu}{k^-p_1^+-\bsk\cdot \bsq_1}\ep_{\m\n}+X
\end{align}
The zero in the first term follows from the gauge condition Eq.~\eqref{polarization-conditions}. The second term follows from the identity in Eq.~\eqref{polarization-identity}. For the third term, in the numerator, we used Eq.~\eqref{momentum-constraints}, whereas for the denominator we used the approximation
\be
p_1'\cdot k = p_1\cdot k - q_1\cdot k \approx p_1^+ k^- - \bsq_1\cdot \bsk~.
\ee
We replaced $q_1\cdot k\rightarrow \bsq_1\cdot \bsk$ because $q_{1,2}$ can be written as
\be
q_1^\mu=\boldsymbol{q}_1^\mu+\frac{k \cdot p_2}{p_1 \cdot p_2} p_1^\mu, \quad q_2^\mu=\boldsymbol{q}_2^\mu-\frac{k \cdot p_1}{p_1 \cdot p_2} p_2^\mu~,
\ee
which ensures $k=q_1-q_2$ and $k^+=q_1^+,~k^-=q_2^-$. Therefore, in the soft $k$ limit,
\be
q_1^\mu \approx \boldsymbol{q}_1^\mu~,\qquad q_2^\mu \approx \boldsymbol{q}_2^\mu~.
\ee
In the high energy limit, $p_1^+$ and $p_2^-$ are much larger than any other momenta. Doing a Taylor expansion, the third term in Eq.~\eqref{weinb-1} therefore becomes
\be
-\frac{q_1^\mu q_1^\nu}{k^-p_1^+-\bsk\cdot \bsq_1}\ep_{\m\n} \approx -\frac{\bsq_1^i \bsq_1^j}{k^-p_1^+}\ep_{ij}+O\(\(\frac{1}{p_1^+}\)^2\)~.
\ee
Finally, we come to the last term in Eq.~\eqref{weinb-1} denoted by $X$ and corresponding to the last term in Eq.~\eqref{weinberg-expanded}. To evaluate it, first note that
\begin{align}
    p'^\mu_2 p'^\nu_2 &= p^\mu_2 p^\nu_2+q^\mu_2 p^\nu_2+p^\mu_2 q^\nu_2+q^\mu_2 q^\nu_2\no\\[5pt]
    &\approx p^\mu_2 p^\nu_2+\bsq^\mu_2 p^\nu_2+p^\mu_2 \bsq^\nu_2+\bsq^\mu_2 \bsq^\nu_2~.
\end{align}
The second line follows from the soft $k$ limit, as discussed above. When contracted with the polarization tensor, it gives
\begin{align}
\label{weinb-2}
    p'^\mu_2 p'^\nu_2 \ep_{\m\n} \approx \[\(\frac{p_2^-}{k^-}\)^2 k^ik^j -\frac{2p_2^-}{k^-}\bsk^i \bsq_2^j + \bsq_2^i\bsq_2^j\]\ep_{ij}~.
\end{align}
Next, we analyze the denominator of the last term in Eq.~\eqref{weinberg-expanded}, which can be reexpressed as  
\begin{align}
\begin{split}
    \frac{1}{p'_2\cdot k} &= \frac{1}{p_2\cdot k+q_2\cdot k} \approx \frac{1}{p_2^- k^+-\bsq_2\cdot \bsk}\approx \frac{1}{p_2^- k^+}\(1-\frac{\bsq_2\cdot \bsk}{p_2^- k^+}+O\(\(\frac{1}{p_2^-}\)^2\)\)~.\\
\end{split}
\end{align}
In the above, we expanded to first nontrivial order in $p_2^-$ because there are terms upto $(p_2^-)^2$ present in the numerator--see Eq.~\eqref{weinb-2}. Putting everything together, we find  
\begin{align}
\begin{split}
    X = -\frac{p'^\mu_2 p'^\nu_2}{p'_2\cdot k} \ep_{\m\n} &= -\[\(\frac{p_2^-}{k^-}\)^2\frac{ k^ik^j}{p_2^- k^+}-\(\frac{p_2^-}{k^-}\)^2\frac{\bsq_2\cdot \bsk}{\(p_2^- k^+\)^2} k^ik^j -\frac{2p_2^-}{k^-}\frac{1}{p_2^- k^+}\bsk^i \bsq_2^j + \frac{\bsq_2^i\bsq_2^j}{p_2^- k^+}\]\ep_{ij}
\end{split}
\end{align}
Note that the first term on the r.h.s cancels against the second term in Eq.~\eqref{weinb-1}. We therefore get
\begin{align}
\label{weinb-3}
    L^{\m\n}_{W}\ep_{\m\n} =  \[4\frac{\bsq_2\cdot \bsk}{\bsk^4} k^ik^j + 4\frac{1}{\bsk^2}\bsk^i \bsq_2^j-\frac{\bsq_2^i\bsq_2^j}{p_2^- k^+}-\frac{\bsq_1^i \bsq_1^j}{k^-p_1^+}\]\ep_{ij}~.
\end{align}
The last two terms are suppressed w.r.t. the first two terms in the high-energy limit. Also, since $\bsk$ is small, we can replace $\bsq_2$ with $\bsq_1$ in the second term above. After these manipulations, our final expression for the Weinberg current in the high-energy limit is
\be
 L^{\m\n}_{W}\ep_{\m\n} = - \[4\frac{\bsq_1\cdot \bsk}{\bsk^4} \bsk^i\bsk^j + 4\frac{\bsk^i \bsq_1^j}{\bsk^2}\]\ep_{ij}~.
\ee
This matches precisely with the soft limit of the Lipatov vertex we derived in Eq.~\eqref{soft-lipatov-final}. Thus  the soft limit of Lipatov vertex is equivalent to the high-energy limit of the Weinberg soft graviton emission vertex. 


\subsection{A classical double-copy relation for the Lipatov vertex}
\label{sec:classical-double-copy}

We will discuss here a double-copy relation between the Lipatov vertices in gauge theory and gravity first observed in \cite{Raj:2023iqn}. In order to establish this relation, we will begin with an expression derived for the classical Yang-Mills radiation field produced in the collision of two massive colored charges $c_\alpha^a(\tau_\alpha)$, where $a$ is the color index, with masses $m_\alpha$, and velocities $v^\alpha$ ($\alpha=1,2$). The trajectories of the  particles are parameterized by their respective worldline parameters $\tau_\alpha$. Solving the Wong equations of motion of the classical charged particles in a slowly varying Yang-Mills background field, the result for the Yang-Mills radiation field is \cite{Goldberger:2016iau}
\begin{equation}
\label{GRMain}
\begin{gathered}
\cA^{\mu,a}(k) =-\frac{g^3}{k^2} \sum_{\substack{\alpha, \beta = 1,2 \\
\alpha \neq \beta}} \int \mu_{\alpha, \beta}(k)\left[\frac{c_\alpha \cdot c_\beta}{m_\alpha} \frac{q_\alpha^2}{k \cdot v_\alpha} c_\alpha^a\left\{-v_\alpha \cdot v_\beta\left(q_\beta^\mu-\frac{k \cdot q_\beta}{k \cdot v_\alpha} v_\alpha^\mu\right)+k \cdot v_\alpha v_\beta^\mu-k \cdot v_\beta v_\alpha^\mu\right\}\right. \\
\left.+i f^{a b c} c_\alpha^b c_\beta^c\left\{2\left(k \cdot v_\beta\right) v_\alpha^\mu-\left(v_\alpha \cdot v_\beta\right) q_\alpha^\mu+\left(v_\alpha \cdot v_\beta\right) \frac{q_\alpha^2}{k \cdot v_\alpha} v_\alpha^\mu\right\}\right]~.
\end{gathered}
\end{equation}
where $g$ denotes the Yang-Mills gauge coupling. The four-momentum of the emitted on-shell gluon is $k$ and $\m_{\alpha,\beta}(k)$ is the integration measure
\begin{equation}
\mu_{\alpha, \beta}(k)=\frac{d^4q_\alpha}{\(2\pi\)^4}\frac{d^4q_\beta}{\(2\pi\)^4}\left[(2 \pi) \delta\left(v_\alpha \cdot q_\alpha\right) \frac{e^{i q_\alpha \cdot b_\alpha}}{q_\alpha^2}\right]\left[(2 \pi) \delta\left(v_\beta \cdot q_\beta\right) \frac{e^{i q_\beta \cdot b_\beta}}{q_\beta^2}\right](2 \pi)^4 \delta^4\left(k-q_\alpha-q_\beta\right)~.
\label{measure}
\end{equation}
The timelike velocities of the massive particles are normalized to $v_\alpha^2=1$. The spacelike vectors $b_\alpha$ are for simplicity set to be purely transverse (perpendicular to the collision axis). 
 
We will now show that Eq.~\eqref{GRMain} encodes the QCD Lipatov vertex plus certain sub-eikonal corrections. To see this~\cite{Raj:2023iqn}, one needs to first take the ultrarelativistic limit of Eq.~\eqref{GRMain}, achieved by parameterizing the four-velocity $v$ of the incoming particles with the boost parameter $\gamma$. For finite $\gamma$ the particles are on a time-like trajectory; the ultrarelativistic limit corresponds to $\gamma\to\infty$. However, in order to keep the energies finite, we need to simultaneously take $m\to0$ such that $\gamma m$ is finite. We next perform the longitudinal integrals in $q^+$ and $q^-$, with the result~\cite{Raj:2023iqn}
\begin{align}
\label{QCDLVLCSubEik}
    \cA^{\mu,a}(k) = -\frac{g^3}{k^2}\int & \frac{d^2\bsq_{2}}{\(2\pi\)^2} \frac{e^{-i \bsq_{1} \cdot \bsb_{1}}}{\bsq_{1}^2} \frac{e^{i \bsq_{2} \cdot \bsb_{2}}}{\bsq_{2}^2} \no\\
    & \bigg[i f^{a b c} c_1^b c_2^c \(2\frac{k\cdot p_2}{p_1\cdot p_2} p_1^\m-2\frac{k\cdot p_1}{p_1\cdot p_2}p_2^\m +\frac{q_1^2}{k\cdot p_1}p_1^\m-\frac{q_2^2}{k\cdot p_2}p_2^\m-q_1^\m - q_2^\m\) \\[5pt]
    & +c_1 \cdot c_2 \bigg\{ \frac{q_1^2 c_1^a}{p_1\cdot k}\(q_2^\mu -\frac{k\cdot q_2}{k\cdot p_1} p_1^\m +\frac{k\cdot p_1}{p_1\cdot p_2}p_2^\m - \frac{k\cdot p_2}{p_1\cdot p_2} p_1^\m\)\no \\ 
    &\qquad \qquad + \frac{q_2^2 c_2^a}{p_2\cdot k}\(-q_1^\mu +\frac{k\cdot q_1}{k\cdot p_2} p_2^\m +\frac{k\cdot p_2}{p_1\cdot p_2}p_1^\m - \frac{k\cdot p_1}{p_1\cdot p_2} p_2^\m\)\bigg\}\bigg]~.\no
\end{align}
Here $p_1$ and $p_2$ are the (light-like) momenta of the colliding charges. Within the square brackets, the terms in parenthesis in the first line constitute the QCD Lipatov vertex obtained in Eq.~\eqref{QCD-Lipatov-vertex-1} through  dispersive techniques\footnote{To be consistent with our notations, we flipped the sign of $q_2$ here w.r.t. to the corresponding expressions in \cite{Raj:2023iqn}, such that $k=q_1-q_2$.}. The second line contains non-universal\footnote{These terms are non-universal because they are sensitive to the nature of the external particles, for instance their spin.} sub-eikonal $1/p_1^+$ and $1/p_2^-$ corrections. 

Next we show that there is a ``classical" color-kinematic replacement prescription \cite{Goldberger:2016iau} that reproduces the gravitational Lipatov vertex starting from  Eq.~\eqref{QCDLVLCSubEik}. The replacement rules are as follows\footnote{These rules have been generalized to higher orders in perturbation theory~\cite{Shen:2018ebu}.}
\begin{align}
\begin{split}
\label{DCprescription}
    &c^a_\a \to p^\m_\a  ~, \\
    &i f^{a_1 a_2 a_3} \rightarrow \Gamma^{\nu_1 \nu_2 \nu_3}\left(q_1, q_2, q_3\right)=-\frac{1}{2}\left(\eta^{\nu_1 \nu_3}\left(q_1-q_3\right)^{\nu_2}-\eta^{\nu_1 \nu_2}\left(q_2+q_1\right)^{\nu_3}+\eta^{\nu_2 \nu_3}\left(q_3+q_2\right)^{\nu_1}\right)~,  \\
    &g \to \kappa ~.
\end{split}
\end{align}
The motivation behind these replacements is the structural similarity between the classical equations of motion in QCD and gravity. Evidence for these replacement rules also stems from analyzing the double copy structure of classical fields due to boosted sources \cite{Akhoury:2013yua}, and from radiation in the soft regime that reproduces Weinberg's soft graviton theorem~\cite{PV:2019uuv}. We note there that the replacements in Eq.~\eqref{DCprescription} are not strictly of the BCJ-type\footnote{In the usual BCJ color to kinematic replacement \cite{Bern:2008qj, Bern:2010ue} one first writes a scattering amplitude (generically $m$-points and $L$-loop order) as 
\begin{align}
\label{BCJ-gauge-amplitude}
    \mathcal{A}_m^{(L)}=i^L g^{m-2+2 L} \sum_j \int \prod_{l=1}^L \frac{d^D p_l}{(2 \pi)^D} \frac{1}{S_j} \frac{n_j c_j}{\prod_{\alpha_j} p_{\alpha_j}^2},
\end{align}
Here the sum is over all distinct $m$-point $L$-loop graphs labeled by $j$ with only cubic vertices and $S_j$ is the symmetry factor associated to the graph. The denominator is a product over all Feynman propagators in the graph and the $c_i$ are the color factors that come from the structure constants appearing in the three-point vertices. The $n_j$ are the kinematic numerators that depend on the momenta and polarizations. In Eq.~\eqref{BCJ-gauge-amplitude}, the color factor triplet $(c_i, c_j, c_k)$ satisfy $c_i+c_j=c_k$, stemming from the Jacobi identity. 

Eq.~\eqref{BCJ-gauge-amplitude} satisfies the {\it BCJ duality} if there exist three associated kinematic numerators ${n_i,n_j,n_k}$, also related by a Jacobi identity: $n_i+n_j=n_k$. Furthermore, the $n_i$ factors must satisfy an anti-symmetry property that the color factors satisfy upon an interchange of two external legs attached to the same three-point vertex: $c_i\to-c_i\implies n_i\to-n_i$. 

A consequence of the BCJ duality is that one can combine two gauge theory amplitudes of the form Eq.~\eqref{BCJ-gauge-amplitude} $\mathcal{A}_m^{(L)}$ and $\tilde{\mathcal{A}}_m^{(L)}$, and obtain a gravity amplitude $\mathcal{M}_m^{(L)}$ 
\begin{align}
    \mathcal{M}_m^{(L)}=i^{L+1}\left(\frac{\kappa}{2}\right)^{m-2+2 L} \sum_j \int \prod_{l=1}^L \frac{d^D p_l}{(2 \pi)^D} \frac{1}{S_j} \frac{n_j \tilde{n}_j}{\prod_{\alpha_j} p_{\alpha_j}^2},
\end{align}
where one replaces $c_i\to \tilde{n}_i$ (the kinematic numerator in $\tilde{\mathcal{A}}_m^{(L)}$) and $g\to \kappa/2$ in the formula of $\mathcal{A}_m^{(L)}$. We point the reader to the reviews \cite{Carrasco:2011hw, Sondergaard:2011iv, Bern:2019prr, Adamo:2022dcm} for further details. 
\label{BCJ-footnote}
}
investigated in \cite{SabioVera:2011wy, SabioVera:2012zky, Johansson:2013nsa}. In these papers, the usual BCJ color-kinematic replacement for obtaining the gravitational Lipatov vertex from the gauge theory Lipatov vertex does not work fully. One only obtains the $C^\mu C^\nu$ part the gravitational vertex but not the $N^\mu N^\nu$ bit in Eq.~\eqref{gravitational-lipatov-vertex}. It was realized later in \cite{Raj:2023iqn} that the appropriate setup where there exists a manifest color-kinematic relation for the full Lipatov vertex is that of the Wong+Yang-Mills equations. 

Applying Eq.~\eqref{DCprescription} to Eq.~\eqref{QCDLVLCSubEik} gives $A^{\mu,a}\rightarrow A^{\mu\nu}$, the gravitational wave amplitude 
\begin{align}
\label{QCDLVLCSubEikGrav}
    &\cM^{\mu\nu}(k) = -\frac{\kappa^3}{k^2}\int  \frac{d^2\bsq_{2}}{\(2\pi\)^2} \frac{e^{-i \bsq_{1} \cdot \bsb_{1}}}{\bsq_{1}^2} \frac{e^{i \bsq_{2} \cdot \bsb_{2}}}{\bsq_{2}^2} \bigg[-\frac12 \(-2 p_2^\n (k \cdot p_1)+2 p_1^\n (k \cdot p_2)-\frac s2(q_2+q_1)^\n\)\no \\[5pt]
    &\qquad\qquad\qquad\(-\bsq_1^\m - \bsq_2^\m + \frac{k\cdot p_2}{p_1\cdot p_2} p_1^\m-\frac{k\cdot p_1}{p_1\cdot p_2}p_2^\m -\frac{\bsq_1^2}{k\cdot p_1}p_1^\m+\frac{\bsq_2^2}{k\cdot p_2}p_2^\m\) \\[5pt]
    & +\frac{s}{2} \bigg\{ \frac{q_1^2 p_1^\n}{p_1\cdot k}\(\bsq_2^\mu +\frac{k\cdot q_2}{k\cdot p_1} p_1^\m - \frac{k\cdot p_2}{p_1\cdot p_2} p_1^\m\) + \frac{q_2^2 p_2^\n}{p_2\cdot k}\(-\bsq_1^\mu +\frac{k\cdot q_1}{k\cdot p_2} p_2^\m - \frac{k\cdot p_1}{p_1\cdot p_2} p_2^\m\)\bigg\}\bigg]~.\no
\end{align}
The first line in the square brackets comes from the kinematic replacement of the structure constant $f^{abc}$ where the momenta $q_1~(q_2)$ are incoming (outgoing) and $k=q_1-q_2$ is outgoing. In writing this expression, we used the constraint $p_\alpha\cdot q_\alpha=0$ that follows from the delta function in Eq.~\eqref{measure} (after taking the ultrarelativistic limit) and replaced $2p_1\cdot p_2$ with the squared center-of-mass energy  $s$. The second line is simply the parenthesis in the first line of Eq.~\eqref{QCDLVLCSubEik} where we additionally used the identities $q_1^\mu=\boldsymbol{q}_1^\mu+\frac{k \cdot p_2}{p_1 \cdot p_2} p_1^\mu, \quad q_2^\mu=\boldsymbol{q}_2^\mu+\frac{k \cdot p_1}{p_1 \cdot p_2} p_2^\mu,$ to express $q_\alpha^\m$ in terms of the transverse momenta alone. When written in this way, the second line is identical to the expression for the QCD Lipatov vertex in Eq.~\eqref{QCD-Lipatov-vertex-1}. The last line in Eq.~\eqref{QCDLVLCSubEikGrav} follows from the last line in Eq.~\eqref{QCDLVLCSubEik} but with the replacement $c^a \to p^\m$. 

A straightforward algebraic simplification of the expression given in Eq.~\eqref{QCDLVLCSubEikGrav} gives 
\begin{align}
\label{QCDLVLCSubEikGrav1}
    \cM^{\mu\nu}(k) = \frac{\kappa^3s}{2\,k^2}\int & \frac{d^2\bsq_{2}}{\(2\pi\)^2} \frac{e^{-i \bsq_{1} \cdot \bsb_{1}}}{\bsq_{1}^2} \frac{e^{-i \bsq_{2} \cdot \bsb_{2}}}{\bsq_{2}^2} \frac{1}{2}\,\[C^\m C^\n - N^\m N^\n + k^\m\(\frac{p_1^\n}{p_1\cdot k}\bsq_1^2+\frac{p_2^\n}{p_2\cdot k}\bsq_2^2\)\]~.
\end{align}
The combination of the first two terms is precisely the gravitational Lipatov vertex. The term in the bracket that is proportional to $k^\m$ is unphysical and can be removed by a gauge transformation. This completes our demonstration of how one recovers the gravitational Lipatov vertex from a classical color-kinematic duality. 

An important observation is that the terms in the second line of Eq.~\eqref{QCDLVLCSubEik} that were subleading for large $p_{1}^+$, $p_{2}^-$ contribute at leading order after the color-kinematic replacement $c^a\to p^\m$. This also demonstrates that the eikonal approximation does not commute with the double copy procedure. One needs to keep the appropriate sub-eikonal terms in the gauge theory result before performing the double copy replacements to arrive at the gravity answer. 
\subsection{Fractionation and classicalization in \texorpdfstring{$2\to 2+n$}{} scattering amplitudes}
\label{sec:classicalization}

As we have discussed thus far, scattering processes in the Regge asymptotics of both QCD and gravity are characterized by the dominance of multi-particle exchanges and large logarithmic enhancements resulting in structures comprised of  Lipatov vertices and reggeized graviton propagators. Lipatov developed a 2-D reggeon effective field theory (EFT) for both QCD and gravity with these building blocks as the relevant degrees of freedom in the MRK regime~\cite{Lipatov:1991nf}. ACV significantly developed this 2-D EFT \cite{Amati:1993tb,Amati:2007ak} to calculate semi-classical contributions to the (generalized) eikonal phase in the $2\to 2$ S-matrix which, as previously alluded to, can be organized as an expansion in $R_S^2/b^2$. This expansion is understood to  be purely classical,
weighted by a factor of $1/\hbar$. While the Lipatov vertex  can be understood as a tree level classical quantity,  this is not apparent for the reggeized graviton. Indeed we saw earlier that reggeized gravitons are constructed from all-loop contributions at leading logarithmic acccuracy.  In this subsection, we will address  $\hbar$ counting in scattering phase shifts, and the phenomena of fractionation and classicalization in multi-particle production. 

We  first revisit the discussion of the $2\to 2$ S-matrix from the introduction in this section. 
Recall that in the eikonal framework of ACV all the classical contributions to the $2\to 2$ S-matrix are exponentiated to give the generalized eikonal phase $e^{2i\delta(b, E)}$, with $\delta(b, E)$ admitting a nontrivial expansion in powers of $(R_S^2/b^2)^n/\hbar$. The results for $n=0,1$ were computed in \cite{Amati:1987uf, Amati:1990xe}, as noted earlier in Eq.~\eqref{delta-1} and Eq.~\eqref{delta-2}. These results were obtain by ACV within the S-matrix framework in string theory~\cite{Gross:1987kza,Gross:1987ar,Muzinich:1987in,Amati:1987uf}, and taking the field theory limit. 

Their results are also obtained\footnote{Generically, loop-level amplitudes in QFT encode both classical ($O(\hbar^{-1})$) and quantum ($O(\hbar^{0})$ and higher) terms. See for example discussions in the context of QED and gravity in  \cite{Holstein:2004dn,Bjerrum-Bohr:2017dxw, Bjerrum-Bohr:2016hpa}. In the latter, the KMOC formalism connects classical observables to on-shell quantum amplitudes~\cite{Kosower:2018adc}. For an application of this formalism to the classicaly double copy, see \cite{delaCruz:2020bbn}.} from resumming the leading $\hbar$ terms in the loop expansion of the $\mathcal{M}_{2\to 2}$ amplitude in general relativity~\cite{Bjerrum-Bohr:2018xdl, Bjerrum-Bohr:2021vuf, Bjerrum-Bohr:2021din, DiVecchia:2023frv, Damgaard:2023ttc}.
For example, in Einstein gravity the one-loop correction to the $2\to 2$ amplitude ($\mathcal{M}_{1}$) receives contributions from not only box and the cross-box diagrams but also from the triangle, inverted triangle and bubble diagrams--see for instance \cite{Bjerrum-Bohr:2021din, DiVecchia:2023frv}. In the eikonal framework, these diagrams are computed with the propagators of the external particles between any two virtual gravitons replaced by eikonal propagators. The classical $\hbar$ expansion of $\mathcal{M}_{1}$ is then obtained by isolating the non-analytical terms in the small $q^2$ expansion, where $q$ is the total momentum transfer. The leading term in the expansion of $\mathcal{M}_{1}$ is $1/(2!\hbar^2) (2i\delta_0)^2$ which contributes towards the exponentiation of $\delta_0$. The subleading term proportional to $1/\hbar$ vanishes for case of massless external particles. Finally the order $\hbar^0$ term in the expansion of $\mathcal{M}_{1}$ contributes to the exponentiation of $\delta_1$ in Eq.~\eqref{delta-1} which is a quantum correction -- such corrections do not necessarily exponentiate. Similarly, higher post-Minkowskian (PM) contributions to the eikonal phase comes can be extracted from the resummation of the relevant terms appearing in the classical expansion of higher-loop diagrams. As noted previously, the first sub-leading classical contribution, at 3 PM, to the eikonal phase contributes to $\delta_2$ in Eq.~\eqref{delta-2}.  Much of the interest in quantum $\rightarrow$ classical dynamics discussed above is in the context of computing the potential in black hole inspirals to high PM order~\cite{Damour:2017zjx,Damour:2019lcq}, while our interest is in the Regge asymptotics of trans-Planckian scattering. 

Returning to our discussion of $2\rightarrow 2+n$ graviton scattering, how can we understand reggeization and multi-particle production as semi-classical features of an {\it ab initio} ``quantum first" QFT picture in perturbative gravity~\cite{Giddings:2018koz}.  We will argue here that the key to this is $t$-channel ``fractionation"~\cite{Giddings:2010pp} followed by $s$-channel ``classicalization"~\cite{Dvali:2010jz,Dvali:2011aa}. We will outline these ideas here and flesh out the arguments later in Sec.~\ref{sec:CGC} for the QCD case and in Sec.~\ref{sec:GR-shockwave-formalism} for gravity. 

For the latter case, while there are many subtleties involved as articulated in \cite{Giddings:2018koz}, a reasonable ``bottom up" approach is via Weinberg's soft graviton theorem, aspects of which  we discussed previously in Sec.~\ref{sec:Weinberg}. As pointed out there, the concept of infrared divergences in virtual graphs and their cancellation, {\it a la} Block-Nordsiek, is an intrinsically quantum mechanical concept that involves $\hbar$. However,  as Weinberg noted in his paper as ``Another remark", the power spectrum itself is classical; an example of such a semi-classical phenomenon is the classification of showers of partons as jets in QCD~\cite{Sterman:1977wj}. Weinberg's remark leads us to anticipate that reggeization in MRK kinematics can be understood similarly. We saw explicitly in Sec.~\ref{sec:Weinberg} that the Weinberg radiative factor in ultrarelativistic kinematics is the soft limit of the Lipatov vertex; therefore, due to the required cancellation of IR divergences the virtual contributions must therefore behave similarly in the Lipatov and Weinberg regimes. Indeed,  Addazi, Bianchi and Veneziano (ABV)~\cite{Addazi:2016ksu} showed that Weinberg's soft factor or ``cusp anomalous dimension"\footnote{\label{ABV-cusp}As we will see next in Sec.~\ref{sec:CGC}, in modern language the cusp factor in QCD can be interpreted as a Wilson line~\cite{Korchemskaya:1994qp}; the reggeized graviton, similarly, can be understood to be the logarithm of the gravitational Wilson line~\cite{Melville:2013qca}.} in the massless limit has precisely the Sudakov double log structure of Eq.~\eqref{one-loop-graviton}. Specifically, the factor $\sigma(\bsk^2)$ in Eq.~\eqref{rho-factor-5} along with Eq.~\eqref{alpha_grav} which is associated to attaching a virtual line obtained in MRK kinematics\footnote{This requires identifying the hard scale $\Lambda$ in Weinberg (and in ABV Eq.~(4.5) along with Eq.~(4.7)) with the momentum transfer $t$ in the hard $2\rightarrow 2$ amplitude separating the Weinberg $2\rightarrow 2+n$ regime from the Lipatov $2\rightarrow 2+n$ regime.}.

Further, as Weinberg showed, the infrared dynamics in QED and gravity behave nearly identically, leading, respectively, to soft photon and soft graviton exponentiation. In QED, it was shown by Faddeev and Kulish (FK)~\cite{Kulish:1970ut} that the IR divergent ``Coulomb tail" contributions to the Hamiltonian can be factored out and represent soft photon coherent state operators that act on the asymptotic vacuum. The S-matrix corresponding to these incoming and outgoing coherent states is infrared finite to all orders in perturbation theory\footnote{See \cite{Feal:2022iyn,Feal:2022ufw} and references therein to some of the extensive literature on this topic.}. A significant recent development is the interpretation of these  QED coherent states as the eigenstates of charges corresponding to the broken asymptotic symmetry of large gauge transformations~\cite{Kapec:2017tkm}. There exists an analogous  FK construction in  gravity~\cite{Choi:2017ylo} due to the broken asymptotic BMS symmetry of supertranslations and superrotations~\cite{Bondi:1962px,Sachs:1962wk,Ashtekar:2018lor}; these FK states are coherent states in gravity. The FK S-matrix was shown previously~\cite{Cachazo:2014fwa,Strominger:2017zoo} to be equivalent to Weinberg's soft graviton theorem. 

Unlike QED and gravity, QCD is a confining theory and the coherent state description of even perturbative phenomena is not fully robust~\cite{Catani:1985xt,Dixon:2008gr}. Nevertheless, there is an emergent semi-classical picture in Regge asymptotics. For the MRK kinematics discussed in Sec.~\ref{sec:2} where $\alpha_S \ln (s/|t|)\sim 1$, BFKL evolution leads to rapid growth in the phase space occupancy of produced gluons. In MRK kinematics,  where 
the $n$-final state particles have the same semi-hard transverse momentum but are ``angle ordered" in rapidity, the phase space occupancy peaks at 90 degrees orthogonal to the collision axis. In this gluon saturation regime we discussed in Sec.~\ref{sec:NLL-gluon-sat}, one observes emergent classicalization with phase space occupancies $N\sim 1/\alpha_S$;  the dynamics is nonperturbative, albeit weakly coupled. 

In particular, the power counting for multi-particle production in the CGC EFT is significantly modified compared to power counting in perturbative QCD,  as we will discuss at length in the next section. In brief, the dominant contribution to multi-particle production arises from cutting reggeons exchanged in horizontal ladders, which one may call $s$-channel fractionation, and leads to the formation of shockwaves\footnote{In the CGC EFT, the transition from the gluon saturation region to the BFKL regime  is smooth for inclusive distributions}. Further, ultrarelativistic hadron and nuclear collisions in this regime can be treated as collisions of shockwaves, with multi-particle production similarly obeying $s$-channel fractionation. This process creates strongly correlated overoccupied matter called the glasma~\cite{Lappi:2006fp,Gelis:2006dv,Dumitru:2008wn} that we will discuss further in Sec.~\ref{sec:glasma}.  Overoccupancy in the CGC EFT is characterized by the saturation scale $Q_S$, which screens color correlations of modes with momenta below $Q_S$.

The degrees of freedom in the EFT are strong classical color sources of order $O(1/g)$ and dynamical gauge and fermion fields (that begin to contribute at NLO in the power counting). A very important feature is that the color sources are stochastic on coherence times of interest, with the stochasticity represented by nonperturbative gauge invariant weight functionals that are many-body density matrices encoding the ultrasoft $\Lambda_{\rm QCD}$-scale dynamics (to be distinguished from the $Q_S$ scale dynamics) of the theory at high energies. Remarkably, there is a concrete map of the CGC EFT to Lipatov's 2-D reggeon EFT, with the color sources identified as the reggeon fields. 
As we will discuss at length in Sec.~\ref{sec:CGC}, loop corrections to LLx and NLLx accuracy can be absorbed in the renormalization of the sources, and of the weight functionals, retaining its structure as a classical-statistical EFT. 
The color sources/reggeized fields are semi-classical objects in this description. In the Schwinger-Keldysh formulation of QFT, appropriate for describing strong field QFT, defined in terms of sum and difference fields, genuine quantum effects only come in at cubic order in the difference fields; it is at this order that the classical-statistical framework breaks down. In the CGC EFT, this occurs starting at NNLLx accuracy and likely leads also to the breakdown of reggeization. 

In variance with the language of cut and uncut reggeons, the reinterpretation in the CGC EFT of multi-particle production as occuring in the presence of coherent sources has the advantage of being a straightforward extension of QFT to the strong field regime with introductory discussions available in classic QFT textbooks~\cite{Itzykson:1980rh,Weinberg:1995mt,Peskin:1995ev}. For instance, as we will discuss at length in Sec.~\ref{sec:glasma}, so-called AGK rules~\cite{Abramovsky:1973fm} for cut and uncut reggeons have a simple interpretation in terms of Cutkosky's rules in strong time-dependent fields.  The AGK rules were developed for QCD but were also employed extensively by ACV in gravity. In the strong-field gluon saturation regime, the dominant contributions in the power counting for multi-particle production are not from cuts to the $t$-channel ``vertical" gluon ladder, but instead from $s$-channel cuts to horizontal ladders. The transition from the one regime to the other was outlined in Sec.~\ref{sec:NLL-gluon-sat}, reflecting the breakdown of the OPE due to large power corrections. Its these $n$-gluon emissions that leads to  phase space overoccupancy ($N\sim O(1/\alpha_S)$): In other words, it leads to classicalization.

Our QCD discussion provides useful insight into fractionation and classicalization in gravity. However first, as  discussed above, the asymptotic in-out states in the S-matrix picture of gravity can already be thought of as FK coherent states absorbing the IR divergences manifest in the perturbative Weinberg picture\footnote{See \cite{Elkhidir:2024izo,Cristofoli:2022phh}, where the same conclusion is reached. For a further  interesting approach to the problem, we refer the reader to \cite{Neill:2013wsa}, which employs the BCFW recursion relations~\cite{Britto:2004ap,Britto:2005fq} and generalized unitary methods~\cite{Bern:1994cg,Bern:1994zx}. These approaches are complementary to Schwinger-Keldysh framework we will follow in this review.}; as noted, this formalism has powerful support when one takes into account the BMS extension of the Poincar\'{e} group. Further fractionation 
can occur via gravitational BFKL as discussed in the previous sub-sections in kinematics where $\lambda_{\rm GR}\ln(s/|t|)\sim 1$, and will follow the same patterns as outlined for QCD. However for most practical applications\footnote{Nevertheless, our discussion of FK coherent states and gravitational BFKL addresses the interesting {\it gedankan} problem of black hole formation in $2\rightarrow 2+n$ scattering in the QFT formulation of general relativity. }, the initial states  of relevance to black hole formation and dynamics are already classical corresponding to a huge overoccupancy of gravitons. 

Thus just as in the CGC EFT, the dominant contribution to multi-graviton production when $b\rightarrow R_S$ is from the $s$-channel ``multi-reggeon" cuts\footnote{Very similar arguments were advanced by ABV to reconcile their work with that of Dvali and collaborators~\cite{Dvali:2014ila}.} discussed by ACV~\cite{Amati:1993tb,Amati:2007ak}. As we will discuss in Sec.~\ref{sec:GR-shockwave-formalism}, the S-matrix problem in this context turns into the problem of multi-particle production in shockwave collisions. The new element is the same as in QCD, the nonperturbative weight functionals that naturally introduce stochastic multi-graviton initial conditions at $R_S$. Unlike the former though, the direction in gravity is IR$\rightarrow$ UV. In \cite{Dvali:2021ooc}, it is speculated that the physics of these overoccupied semi-classical states is universal, with $1/R_S$ playing the role of $Q_S$ in gravity. Indeed, as we will discuss further in Sec.~\ref{sec:GR-shockwave-formalism} the parallels are quite striking. One can however be agnostic about whether there is an imprint of UV physics on this IR horizon scale; in Sec.~\ref{sec:GR-shockwave-formalism}, we will briefly outline how our formalism can provide tests of different scenarios at future gravitational wave detectors.

\section{Color Glass Condensate approach to high energy scattering in QCD}
\label{sec:CGC}

As extensively discussed in Section~\ref{sec:2}, the outstanding achievement of Lipatov and collaborators was to show that $2 \to n$ scattering,  which gives the dominant contribution to the imaginary part of the $2\to 2$ gluon amplitude in MRK kinematics, can be described as iterations of one rung of the ladder containing reggeized gluons and the Lipatov vertex, to NLLx accuracy. The color singlet projection of the exchange of two reggeized gluons is the perturbative pomeron, the weak coupling counterpart of the soft pomeron often invoked to describe the systematics of total cross-sections. However as discussed in Section~\ref{sec:NLL-gluon-sat}, BFKL dynamics cannot be the complete story since its solution shows that the unintegrated gluon distribution $\mathcal{F}(x,\bsk)$ diffuses to infrared and ultraviolet momenta with increasing rapidity. The former is clearly troublesome since that’s the nonperturbative regime $\bsk \sim \Lambda_{\rm{QCD}}$ where weak coupling computations are invalid. Further, the rapid growth of the inclusive cross-section for a fixed impact parameter violates unitarity at large rapidites. Not least, the increasing phase space occupancy due to the rapid proliferation of gluons at small $x$ suggests that many-body (higher twist) correlations are important and not captured by the BFKL evolution. All of these issues persist at NLLx accuracy.

A solution to the aforementioned problems is the phenomenon of gluon saturation~\cite{Gribov:1983ivg,Mueller:1985wy} that we also discussed at some length in the DIS context in Section~\ref{sec:NLL-gluon-sat}. Its principal feature is an emergent close packing scale $Q_S(x)\gg \Lambda_{\rm QCD}$ at maximal occupancy that unitarizes the inclusive cross-section at fixed impact pararameter. In other words, for a weakly interacting probe of given fixed $Q^2$ with $\alpha(Q^2)\ll 1$, there is a corresponding value of $x$ for which the probe scatters of the hadron target with unit probability at occupancy $N\sim 1/\alpha_S\gg 1$. We will now discuss gluon saturation within the framework of the Color Glass Condensate effective field theory, and the resulting shockwave picture of multi-particle production at high energies\footnote{Another almost exactly contemporaneous early discussion can be found in \cite{Verlinde:1993te}, and was  applied to the gravitational shockwaves we will discuss in \cite{Verlinde:1993mi}.}. 

\subsection{Shockwave picture of DIS}
\label{sec:shockwave-QCD}

\begin{figure}[ht]
\centering
\includegraphics[scale=1]{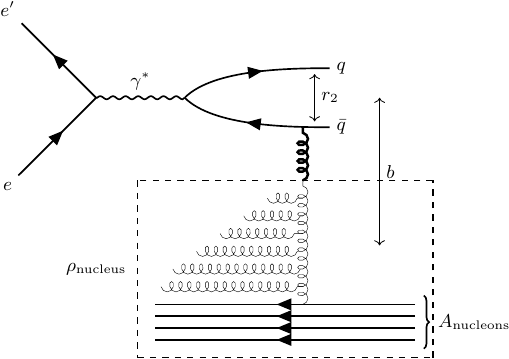} 
\caption{The scattering of a quark-antiquark dipole in DIS off a boosted heavy nucleus. The length of the emitted gluon lines indicate their distance in rapidity from ``valence" partons whose rapidities are close to the light cone of the heavy nucleus. At high energies, partons inside the dashed box can be represented by a coherent color charge density $\rho_{\rm nucleus}$ that is static on the times scales of the interaction. }
\label{dipole-interaction}
\end{figure}

We noted earlier that small $x$ physics at high occupancy is described by the Color Glass Condensate (CGC) EFT~\cite{Gelis:2010nm,Kovchegov:2012mbw}. Fig.~\ref{dipole-interaction} illustrates the emergent shockwave picture of the DIS process in Fig.~\ref{fig:DIS-fig}. In that figure, in the rest frame of the dipole target, the fast moving nucleus with momentum $P^+\rightarrow \infty$ emits a large number of gluons which in the Regge limit are ordered in rapidity. The fastest gluons (represented by the longest gluon lines) co-move with the valence degrees of freedom on the light cone and the slowest (small $x$) gluons scatter off the dipole projectile, with $x\sim x_{\rm Bj}$. In the figure shown here, 
the dipole of size $r_\perp \sim 1/Q$, at impact parameter 
$b \ll 1/\Lambda_{\rm QCD}$, is shown exchanging a colored gluon with a lump (represented by the 
dotted rectangle) of maximal size $1/Q_S$ consisting of static color sources on the relevant time scale of the scattering. 

On the light front, the dynamics illustrated can be understood to have a 2+1-D Galilean structure~\cite{Weinberg:1966jm,Susskind:1967rg,Bardakci:1968zqb,Majumdar:2024rxg}; the fast (large $x$) degrees of freedom are isomorphic to heavy masses, and conversely, the slow (small $x$) are light dynamical degrees of freedom. This naturally suggests that a Born-Oppenheimer EFT can be employed to describe small $x$ physics, which provides the essential kinematic motivation for the CGC EFT. A further key element is that the quasi-static lump represents a quasi-classical state of high occupancy\footnote{The energy separations of the high occupancy screened gluons occupying a momentum mode $Q_S$ are $\sim Q_S/N$, the characteristic decay time of the shock is $\sim \frac{1}{\alpha_S Q_S}$  which is considerably longer than the typical resolution scale $1/Q$ of the probe but much shorter than the eikonal time scale $\sim P^+/Q^2$~\cite{Dvali:2021ooc}. }; further, since the lump contains a large number of color charges, the most likely color charge representation is a high dimensional classical representation of the $SU(N_c)$ algebra~\cite{McLerran:1993ka,McLerran:1993ni,Jeon:2004rk}. 

Since the dipole perceives the lump as being Lorentz contracted in its rest frame, this lumpy ``shockwave" can be represented by a classical color charge density distribution $\rho_{\rm Y_0}(\bsx)$ which is a $\delta$-function in $x^-$ with support in spacetime rapidity $\eta_0= \ln(x^-/x_0)\approx Y_0$, where $Y_0$ is the corresponding momentum space rapidity. The expectation values of operators in the hadron in the CGC EFT corresponds to performing two sequential path integral averages~\cite{Iancu:2003xm}: 
\begin{itemize}
    \item The full path integral over dynamical gauge field modes (with $Y > Y_0$) in the presence of background gauge fields generated by a given configuration of stochastic sources with $Y\leq Y_0$, where 
    $Y_0 = \ln(k^+/\Lambda^+)$. Here $\Lambda^+$ is a scale in rapidity which corresponds to the large number of color sources generated either by BFKL-like evolution and/or an initial condition (such as a large nucleus) containing these. The action describing this dynamics is the full QCD action plus a gauge invariant term coupling the large $x$ current to the gauge fields in the path integral. 
    \item A classical path integral average over a gauge invariant stochastic functional of the distribution of the static color sources at the scale $Y_0$.
\end{itemize}
The expectation value of an operator ${\cal O}$ in the CGC EFT (an example being the time ordered product of electromagnetic currents in Eq.~\eqref{eq:DIS-hadron-tensor}) can be expressed as~\cite{McLerran:1993ka,Jalilian-Marian:2000pwi} 
\begin{eqnarray}
\label{eq:CGC-path-integral}
    \langle {\cal O}\rangle = \int [D\rho] W_{\Lambda^+}[\rho]\int_{\Lambda^+}\!\!\!\! [DA] {\cal O}[A]\exp\bigg[-\frac{i}{4}\int d^4 x\, F_{\mu\nu}^a F^{\mu\nu,a}- \frac{1}{N_c}\int d^2 x_\perp {\rm Tr}\(\rho(x_\perp) \ln K(x_\perp)\) \bigg]\,,
\end{eqnarray}
where $F_{\mu\nu}$ is the Yang-Mills field strength tensor and 
\begin{eqnarray}
\label{Wilson-Line-QCD}
    K(x_\perp) = P_{x^-}\exp\left( i\int dx^+ A_a^-(x_\perp,0,x^+) \,T_a\right)\,.
\end{eqnarray}
Note that dynamical quark-antiquark degrees of freedom (so-called ``sea quarks") are implicit in this small $x$ effective action; they are radiatively generated from the gauge fields. Since the hadron has momentum $P^+\rightarrow \infty$, the current represented by the large $x$ degrees of freedom is given by $J^\mu = \delta^{\mu+}\rho(x_\perp)$, with the contributions of the other components being sub-leading in powers of $1/P^+$. The matrix $V$ then represents the color rotation of the sources in the background gauge fields. The structure of the $\rho \ln V$ coupling term above is a gauge invariant generalization of the $J^+\cdot A^-$ eikonal coupling one obtains in QED - for an elegant discussion, see \cite{Bjorken:1970ah}. We note further that the sources and fields are Lorentz contracted in $x^-$ when $P^+$ is large. In a careful treatment, one needs to smear out the source distribution in $x^-$, which for consistency, requires path ordering in $V$~\cite{Jalilian-Marian:1996mkd}. 

For a given distribution of classical color sources, the path integral in Eq.~\eqref{eq:CGC-path-integral} has a nontrivial saddle point solution in Lorenz gauge $\partial_\mu A^\mu=0$: 
\begin{eqnarray}
\label{eq:classical-YM-Lorentz}
    A^+(x_\perp,x^-) = \int \frac{d^2 y_\perp}{4\pi}\, \ln\(\frac{1}{(x_\perp-y_\perp)^2\Lambda^2}\)\,\rho(y_\perp,x^-)\,\,;\,\, A^-= A_\perp = 0\,.
\end{eqnarray}
Here $\Lambda$ is the IR cutoff in the solution of the two-dimension Poisson equation $\nabla_\perp^2 A^+= -\rho$. Note that this result can also be understood as the saddle point solution in the ``wrong" light cone gauge $A^-=0$. 

In the ``right" light cone gauge $A^+=0$ corresponding to $P^+\rightarrow \infty$, the parton model interpretation of DIS is manifest. In this gauge, the saddle point solution takes the form 
\begin{eqnarray}
\label{eq:classical-light-cone}
    A^-=0\,\,;\,\,A_\perp(x_\perp) = \frac{-1}{ig} U\nabla_\perp U^\dagger\,\,\,{\rm where}\,\,\, U = P_{x^-}
    \exp\(ig \int_{-\infty}^{x^-} dz^-\, \frac{\rho(x_\perp,z^-)}{\nabla_\perp^2} \)\,.
\end{eqnarray}
Here the residual gauge freedom is fixed by imposing retarded boundary conditions $A_\perp\rightarrow 0$, as $x^-\rightarrow -\infty$ -- for a careful discussion, see \cite{Gelis:2008rw}.

Before we discuss this solution further, let us discuss the stochastic weight $W_{\Lambda^+}[\rho]$ in Eq.~\eqref{eq:CGC-path-integral}, which is the other ingredient we need to compute expectation values of operators in the gluon saturation regime. This gauge invariant functional clearly needs nonperturbative input on many-body distributions and correlations of strong classical color sources $\rho\sim 1/g$, and in general can be a very complicated quantity containing information on n-body correlations inside the hadron. 

It turns out however that for a large nucleus with atomic number $A\gg 1$, the problem is greatly simplified\footnote{This may seem counterintuitive since the nuclear physics may be thought of as ``complicated" or ``messy". However here, as with many apparently complex problems, careful consideration of time and length scales of interest, leads to unanticipated simplicity.}. At small $x$, the coherence length, or ``Ioffe time"~\cite{Ioffe:1969kf} of the probe $l_c= 1/(2m_N x) \gg 2 R_A$, where $R_A\sim A^{1/3}$ is the nuclear radius in Fermi. 
Hence, parametrically for $x\ll A^{-1/3}$, the small sized quark-antiquark dipole in DIS scattering of the high energy nucleus will interact nearly simultaneously ($l_c \gg l_{\rm mfp}$, where $l_{\rm mfp}$ is the mean free path separating nucleons) with partons from the different nucleons it scatters off at a given impact parameter. Because the partons in individual nucleons are confined, their correlations with partons in the other nucleons that the probe scatters off are strongly suppressed on the time scales of the scattering. Thus the probe coherently scatters of a large color charge configuration generated by {\it random color charges} in the $A\rightarrow \infty$ limit. 

The problem of the distribution of a large number of random $SU(N_c)$ color sources was discussed at length in \cite{Jeon:2004rk}. In the SU(3) case, for instance, labeling representations by $(m,n)$ where $m$ and $n$ are the integers corresponding to the number of upper and lower tensor indices, and using Young tableaux to represent these, one obtains recursion relations that can be solved to determine the distribution of representations. One finds that the typical representation is characterized by $m\sim n\sim \sqrt{k}$, where $k\sim A^{1/3}$ the number of SU(3) charges; for a large nucleus, this corresponds to a classical color charge representation. As a consequence, one can express the distribution of representations as the classical path integral in Eq.~\eqref{eq:CGC-path-integral}, with 
\begin{eqnarray}
\label{eq:Gaussian-functional}
  W_{\Lambda^+}[\rho] = \exp\left(- \int d^2 x_\perp \frac{\rho^a \rho^a}{2\mu_A^2}\right)  \,,
\end{eqnarray}
where $a=1,\cdots,8$ and 
\begin{eqnarray}
\label{eq:squared-color-charge}
    \mu_A^2 = \frac{g^2 A}{2\pi R^2}\,.
\end{eqnarray}
Here $\mu_A^2\sim A^{1/3}\,{\rm fm}^{-2}$ is the color charge squared per unit area, representing the typical color charge in the nucleus seen by a nuclear probe. In the large $A$ asymptotics that we introduced, $\mu_A^2$ is a large dimensionful scale. Note that in writing this expression we have assumed a strict $\delta(x^-)$ distribution of sources. In general, $\rho^a\equiv \rho^a(x_\perp,x^-)$ and $\mu_A^2= \int dy^- \lambda(y^-)$, with $\lambda$ here denoting the mean squared color charge density per unit volume. 

For $SU(3)$ color charges, there is in addition a term proportional to $d_{abc} \rho^a\rho^b\rho^c$, but it is parametrically suppressed at large $A$ relative to Gaussian term~\cite{Jeon:2005cf}. In the CGC framework, this provides the mechanism for the $t$-channel exchange of a C-odd Odderon, just as the Gaussian distribution provides the mechanism for C-even Pomeron exchange. We will discuss this point further later. 

For a large nucleus, the dipole scattering amplitude\footnote{Recall the discussion in Sec.~\ref{sec:NLL-gluon-sat} arriving at Eq.~\eqref{eq:DIS-cross-section}, and the subsequent discussion of the dipole cross-section.} at leading order in this EFT picture is then captured by the formula 
\be
\langle d\sigma_{\rm dipole}^{\rm LO}\rangle = 2 \int d^2 b \int [D\rho_{\rm A}]\, \exp\left(- \int d^2 x_\perp \frac{\rho^a \rho^a}{2\mu_A^2}\right) 
\,T_{\rm LO}\left(b+\frac{r_\perp}{2}, b-\frac{r_\perp}{2}\right)[\rho_{\rm A}]\,,
\label{eq:CGC-inclusive}
\ee
where
\begin{eqnarray}
\label{eq:T-S-matrix}
    &T_{\rm LO}(x_\perp,y_\perp) \equiv 1- S_{\rm LO}(x_\perp,y_\perp) = 1- D(x_\perp,y_\perp),\,\, 
    {\rm with}\,\, D(x_\perp,y_\perp) = \frac{1}{N_c}{\rm Tr}\left(V(x_\perp)V^\dagger(y_\perp)\right)\,.\no\\
\end{eqnarray}
Here $V$ is the counterpart in the fundamental representation of path ordered exponential defined in Eq.~\eqref{eq:classical-light-cone} (replacing $T^a\rightarrow \tau^a$), $r_\perp=x_\perp-y_\perp$, and $T_{\rm LO}$ ($S_{\rm LO}$) denotes (the imaginary part) of the tree level scattering amplitude (real part of the S-matrix) of the quark-antiquark dipole scattering off the classical ``Coulomb" background field generated by the source distribution $\rho_{\rm A}(\bsx)$. At LO, as we will discuss further later in Sec.~\ref{sec:JIMWLK}, the $S$-matrix is equivalent to the ``dipole correlator" $D(x_\perp,y_\perp)$ of Wilson lines. The Gaussian represents the stochastic distribution of color sources at some rapidity scale $Y_0= \ln(P^+/\Lambda^+)$ (satisfying $Y_0\gg \ln(A^{1/3})$) separating these sources from the classical fields that interact with the probe. 

For this Gaussian distribution, one can  compute Eq.~\eqref{eq:CGC-inclusive} explicitly, with the result (assuming for simplicity a homogeneous impact parameter profile)
\begin{eqnarray}
\label{eq:GBW+MV}
    \langle d\sigma_{\rm dipole}^{\rm LO}\rangle = 2\pi R_A^2 \left[1-\exp\left(-\frac{r_\perp^2 Q_S^2}{4}\right)\right]\,,
\end{eqnarray}
defining $Q_S^2 = \alpha_S\,C_F\,\mu_A^2 \ln(1/r_\perp^2\Lambda^2)$, where $C_F = (N_c^2-1)/2 N_c$ and $\Lambda$ is an infrared scale. In the large $A$ limit of QCD, we see thus already at LO in the CGC EFT an explicit realization of the saturation phenomenon we discussed earlier. Firstly, we note the emergence of the saturation scale $Q_S\sim \mu_A^2\propto A^{1/3}$ fm$^{-2}$. For $r_\perp \ll 1/Q_S$, we see that $T_{\rm LO}= r_\perp^2\, Q_S^2$; this LO perturbative QCD result  represents the weakly interacting ``color transparency" of very small sized probes. Conversely, for $r_\perp \gg 1/Q_S$, $S_{\rm LO}\rightarrow 0$, indicating that the probe interacts very strongly, and is ``color opaque". In fact, for $A\rightarrow \infty$, the LO cross-section saturates the quantum ``black-disc" limit allowed by unitarity: $\sigma_{\rm dipole}^{\rm LO} = 2\pi R_A^2$. 

As we outlined earlier, the reason one has strong interactions even when the coupling is weak is because the phase space occupancy is large, of $O(1/\alpha_S)$. One can check explicitly that this is the case by computing 
$\frac{1}{\pi R_A^2}\frac{dN}{d^2 k_\perp}\propto \int d^2 x_\perp e^{ik_\perp\cdot x_\perp}\langle A_\perp (x_\perp) A_\perp (0)\rangle$ in $A^+=0$ gauge. In addition to the multiplicity, one can analytically compute arbitrary many-body correlators in the CGC EFT, which display nontrivial intrinsic n-body correlations after subtracting off the trivial one-body distribution raised to the appropriate n$^{\rm th}$ - power. 

Going beyond LO in the CGC EFT, the bremsstrahlung of gluons as well as virtual self-energy and vertex corrections in the shockwave background have to be taken into account. As the only dimensionful scale in the problem, one anticipates that the QCD coupling must run as a function of this scale, with $\alpha_S(Q_S^2)\ll 1$ for $Q_S^2\gg \Lambda_{\rm QCD}^2$.  A systematic computation of the higher order contributions requires knowledge of small fluctuation propagators in the shockwave background. We will discuss these propagators in Sec.~\ref{sec:CGC-propagators}, and employ them in Sec.~\ref{sec:JIMWLK} to compute the RG equations for n-point Wilson line correlators. As we will discuss, the saturation scale in this case is a dynamical scale $Q_S\equiv Q_S(x)$ (that grows with decreasing $x$), characterizing the nontrivial fixed point of the RG evolution.

\subsubsection{The CGC and color memory}
\label{sec:color-memory}

A remarkable phenomenon in gravity is the gravitational memory effect that corresponds to the physical displacement of gravitational wave interferometers due to the relative ``kick" imprinted on them by the passage of a gravitation wave~\cite{Zeldovich:1974gvh,Braginsky:1987kwo,Christodoulou:1991cr}. This gravitational memory effect can be understood~\cite{Strominger:2017zoo} to be a consequence of a transition (a supertranslation) between different degenerate classical vacua of the Bondi-Metzner-Van der Burg-Sachs (BMS) extension of the Poincar\'{e} group in general relativity~\cite{Bondi:1962px,Sachs:1962wk,Ashtekar:2018lor}. Since the different vacua differ by zero energy gravitons, the memory effect is also directly related to the Weinberg soft graviton theorem~\cite{Strominger:2014pwa} we discussed previously. 

An exactly analogous color memory effect exists in Yang-Mills theory~\cite{Pate:2017vwa}. This color memory effect was first derived employing retarded coordinates ($u,r,z,\zbar$) where $u = t-r$ is the retarded time, $r$ is the radial coordinate, and $z$ and $\zbar$ are stereographic coordinates on the celestial sphere at null infinity, corresponding to $r=\infty$ at fixed $u$. The Minkowski metric can then be reexpressed in these ``Bondi coordinates" as 
\begin{eqnarray}
    ds^2 = -du^2 - 2dudr +2r^2 \gamma_{z\zbar}dzd\zbar \,,
\end{eqnarray}
with the ``round metric" on the celestial two-sphere $\gamma_{z\zbar} = 2/(1+z\zbar)^2$. The color memory effect in these coordinates is the relative color rotation experienced by a ``test" quark-antiquark pair on the celestial 2-sphere at null infinity  after experiencing the passage of a ``wave" of color flux in some finite retarded time interval $u_i < u < u_f$. Analogously to the gravitational memory case, this can be interpreted as a transition between two degenerate classical vacua.  The color rotation in Yang-Mills theory can be expressed as Wong's equations~\cite{Wong:1970fu} describing the color precession of colored quarks experiencing the color flux; a gauge covariant formulation of the net color rotation is given by the closed loop around the color flux traced by the colored quarks on the null 2-sphere. 

On the surface, color memory appears to be a mathematical curiosity since, unlike gravity, QCD is a confining theory and therefore the presence of colored degrees for freedom (let alone classical flux) at null infinity is impossible. However, as we just argued, there is a classical weak coupling regime of the theory that emerges in the Regge limit. Indeed, the discussion of color memory to the CGC EFT has a one-to-one map which becomes apparent by translating Bondi coordinates to light cone coordinates,
\begin{eqnarray}
\label{eq:lightcone-Bondi}
    x^+ = \frac{1}{\sqrt{2}}\left(u + \frac{2r}{1+z\zbar} \right)\,\,;\,\, x^- =\frac{1}{\sqrt{2}}\left(u + \frac{2rz \zbar}{1+z\zbar} \right)\,\,;\,\, x_1+ix_2 = \frac{2rz}{1+z\zbar}\,,
\end{eqnarray}
and taking the infinite momentum frame limit. This is achieved by scaling 
$$ (r,u,z,\zbar) \rightarrow (\zeta r,\zeta^{-1}u,\zeta^{-1}z,\zeta^{-1}\zbar),$$ 
where $\zeta$ parametrizes the boost operator, and taking $\zeta\rightarrow \infty$. In this limit, $\gamma_{z\zbar}=2$, and 
\begin{eqnarray}
    x^+= \sqrt{2}r\rightarrow \infty\,\,;\,\, x^- = \frac{1}{\sqrt{2}}\left(u +2rz\zbar\right)\rightarrow 0\,\,;\,\, x_1+ix_2 = 2 rz \,.
\end{eqnarray}
Thus in the IMF, the celestial sphere at null infinity maps on to the CGC shockwave kinematics, where the gauge fields are independent of $x^+$, are strongly localized around $x^-=0$, and have nontrivial dynamics in the transverse plane. With this map, we understand the trace over Wilson lines in Eq.~\eqref{eq:T-S-matrix} as the net color rotation acquired by the $q\bar q$ dipole as it crosses the shockwave. The solution in Eq.~\eqref{eq:classical-light-cone} represents distinct vacua under large gauge transformations at every point on the transverse plane that are separated by the shockwave. Further, one has a gauge invariant analog to the ``kick" experienced by the gravitational wave detectors, in the form of the saturation scale $Q_S$; an unambiguous extraction of  $Q_S$ would therefore be equivalent to measuring the color memory effect~\cite{Ball:2018prg}. 

The map we have established between the CGC and color memory is potentially powerful since it allows us to formulate scattering in the CGC in the language of soft theorems and asymptotic symmetries~\cite{He:2015zea}. The latter may be particularly useful when one moves away from  $A\rightarrow \infty$ asymptotics, where the Gaussian approximation breaks down and strongly correlated gluon dynamics on the celestial sphere is not analytically tractable any more. In particular, this strongly correlated dynamics should be represented by Goldstone modes reflecting the broken global Poincar\'{e} symmetry and those of large gauge transformations representing the CGC shockwave~\cite{Dvali:2021ooc}.

\subsection{Propagators in gluon shockwave backgrounds}
\label{sec:CGC-propagators}
We have argued thus far that perturbation theory breaks down in the gluon saturation regime of QCD with the strong color fields nevertheless described by a weakly coupled classical EFT. It is important to establish whether this EFT description is robust when quantum fluctuations are taken into account, and further, to establish where such a description breaks down. As first step, we will discuss here shockwave propagators in the classical CGC background that are necessary to compute NLO corrections to operators in the CGC EFT. 

The computation of shockwave propagators in the light cone gauge  $A^+ = 0$ , where  $P^+$ denotes the large longitudinal momentum component, is technically involved, as was demonstrated in \cite{Ayala:1995kg}. 
This is because of the presence of $P^+=0$ modes that require careful treatment of boundary conditions at spatial $x^-=\pm \infty$. By contrast, the formulation in the other light cone gauge  $A^- = 0$ is more tractable \cite{McLerran:1994vd}. This gauge admits the same classical background field configuration as the Lorenz gauge solution given in Eq.~\eqref{eq:classical-YM-Lorentz}, and moreover, this background exhibits a manifest double-copy correspondence with gravitational shockwave solutions \cite{Akhoury:2013yua}. We will therefore adopt the  $A^- = 0$  gauge for our analysis here.

Before computing the gluon propagator, it will be instructive to first compute the (retarded) propagator of an adjoint scalar field coupled minimally to this background\footnote{A more detailed analysis is provided in \cite{Raj:2024xsi}, which for completeness is summarized here.}. The derivation of quark and gluon propagators follow similarly. Our starting point is to obtain the solution to the small fluctuation equations of the scalar field, given by
\begin{align}
    \p_\mu \p^\mu \phi - 2ig[A_-,\p_+\phi]=0~,
\end{align}
Noting that $[A, B]_a=i f^{a b c} A_b B_c=-\left(T^b\right)_{c a} A_b B_c=[(A \cdot T) B]_a$, one can write the above equation as
\begin{align}
\label{scalar-gauge-small-fluct}
    \square \phi(x)-2ig(A_-(x)\cdot T) \partial_{+}\phi(x)  = 0~.
\end{align}
The equation for the retarded Green's function $G_R(x,y)$ is then 
\begin{align}
\label{retarded-prop-conv}
    \square_x G_R(x,y) - 2ig(A_-(x)\cdot T) \partial_{x^+} G_R(x,y) = \delta(x-y)~,
\end{align}
supplemented by boundary data specified at a fixed $x^-$ slice in the region $x^-<0$. 

Expressing the solution for the field in terms of the retarded Green's function, one can show that the latter generically satisfies the recursion relation~\cite{Blaizot:2004wu,Raj:2024xsi}
\begin{align}
\label{ret-gauge-prop-rec}
    G_R(x, y)=\int d^4 z \,G_R(x, z)\, \delta (z^- -z^-_0)\,2\,\partial_z^{-} G_R(z, y)~,
\end{align}
where the surface $z^- = z^-_0$ is the spacetime slice where the initial data is defined.
The shockwave background is pure gauge everywhere except at $x^-=0$. Therefore the propagators $G_R(x, y)$ in this background are free propagators everywhere, except for the paths corresponding to $x^->0, y^-<0$ or $x^-<0, y^->0$. The propagator crosses the shockwave background for these configurations and is therefore nontrivial. We can compute $G_R(x, y)$ for $x^->0, y^-<0$ using the identity in Eq.~\eqref{ret-gauge-prop-rec}, and writing it as  
\begin{align}
\label{G-conv-0}
G_R(x,y) &=\int d^4z ~d^4w \,G_R^0(x,z)\, \delta(z^{-}-z^-_0)\,2\,\p^-_z\, G_R(z,w)\, \delta(w^{-}-w^-_0)\,2\,\p^-_w\,G_R^0(w,y)~.
\end{align}
We set here $z_0^-=\delta$, $w^-_0=0$, where $\delta$ specifies the shockwave width in the $x^-$ direction -- we will set it to zero at the end of the calculation. These choices of $z^-_0$ and $w^-_0$ allows us to  replace  $G_R(x,z)$ and $G_R(w,y)$ by $G_R^0(x,z)$, $G_R^0(w,y)$ in Eq.~\eqref{G-conv-0}, where 
\be
\label{free-retarded-scalar-propagator}
G_R^0(x,y) = -\int \frac{d^4k}{\(2\pi\)^4} \frac{e^{-ik\cdot(x-y)}}{k^2+i\epsilon k^-} = \frac{1}{2\pi}\Theta(x^--y^-)\Theta(x^+-y^+)\delta\((x-y)^2\)~.
\ee
An explicit derivation of the last equality is given in \cite{Blaizot:2004wu,Raj:2024xsi}. 

To determine $G_R(z,w)$ in Eq.~\eqref{G-conv-0}, we first express it as 
\be
\label{retarded-gf-0}
G_R(x, y) = -\int \frac{d^4k}{\(2\pi\)^4} \frac{1}{k^2+i\epsilon k^-} \phi_k(x)\phi_k^*(y)~,
\ee
where $\phi_k(x)$ denotes eigenfunctions of the small fluctuation equation in Eq.~\eqref{scalar-gauge-small-fluct}; in the absence of the shockwave background, these are simply plane waves and one recovers Eq.~\eqref{free-retarded-scalar-propagator}. The solution of Eq.~\eqref{scalar-gauge-small-fluct} to determine $\phi_k(x)$ simplifies greatly if we employ the eikonal approximation, where transverse gradients of order $P_\perp$ are negligible compared to those of order $P^+$ in the longitudinal direction when crossing the shockwave. Eq.~\eqref{scalar-gauge-small-fluct} then simplifies to 
\be
\label{scalar-gauge-small-fluct-1}
2\p_+\p_-\phi(x) -2ig(A_-(x) \cdot T) \partial_{+}\phi(x) = 0\,.
\ee
With the input from the scalar field solution in  the region $x^-<0$, the solution for $x^- > 0$ is 
\be
\phi^a(x^->0) = U(\bsx)^a_{~b} ~\phi^b(x^-<0)~,\qquad {\rm with}\qquad \phi^a(x^-<0) =  e^{-ikx} c^a~.
\ee
Here $c^a$ is a pointlike color charge and $U(\bsx)$ is a light-like Wilson line, path ordered in $x^-$: 
\be
\label{eq:QCD-Wilsonline}
U(\bsx) = P \exp \(ig\int dz^- A^+(z^-, \bsx)\cdot T\)~.
\ee
Thus the result of the scattering of the colored scalar with the shockwave is to rotate it in color space. The full solution is  
\be
\phi_k^a(x) = \Theta(-x^-) \,e^{-ikx} c^a + \Theta(x^-)\, e^{-ikx} U^a_{~b}(\bsx) c^b~.
\ee
This gives  the free propagator in Eq.~\eqref{retarded-gf-0} for  $\theta(-x^-)\theta(-y^-)$ or the color rotated free propagator for $\theta(x^-)\theta(y^-)$. For the  nontrivial component of the Green function in the r.h.s of Eq.~\eqref{G-conv-0}, we obtain  
\begin{align}
\label{scalar-nontrivial}
    G_R^{ab}(z^->0, ~w^-<0) = -\int \frac{d^4k}{\(2\pi\)^4} \frac{e^{-i k\cdot(z-w)}}{k^2+i\epsilon k^-} c^a \[U(\bsz)\]^b_{~d} c^d~, 
\end{align}
where we need to take the limit $\delta\to 0$. This gives 
\begin{align}
    \lim_{\delta\to0^+} G_R^{ab}(z^-=\delta, ~w^-=0) = \frac12 \Theta(z^+-w^+)\delta^{(2)}(\bsz-\bsw) c^a \[U(\bsz)\]^b_{~d} c^d~.
    \label{sw-prop}
\end{align}
We obtained the r.h.s by first performing the $k^+$ integral closing the contour from below, taking $\delta\to 0^+$, and employing the identity 
\be
\int \frac{dk^-}{2\pi} \frac{e^{-ik^-(z^+-w^+)}}{-ik^-} = \Theta(z^+-w^+)~.
\ee
Substituting this result back into Eq.~\eqref{G-conv-0}, we obtain 
\begin{align}
\label{scalar-gauge-prop}
G_R(x,y) &=\int d^4z ~G_R^0(x,z)\, \delta(z^{-})\, U(\bsz)\, 2\,\p^-_z G_R^0(z,y)~,
\end{align}
which agrees with Eq. (9) of  \cite{Hebecker:1998kv}. 

\begin{figure}[ht]
\centering
\raisebox{-12pt}{\includegraphics[scale=1.0]{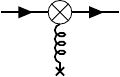}}
=
\raisebox{-0pt}{\includegraphics[scale=1.0]{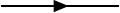}}
+ $\sum_{n=1}^\infty$
\raisebox{-12pt}{\includegraphics[scale=1.0]{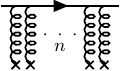}}
\caption{Dressed propagator of a scalar field in the shockwave background of overoccupied gluons. The first term on the r.h.s is the free scalar propagator corresponding to the no scattering ``1" term in the expansion of the Wilson line in Eq.~\eqref{T-matrix-g}. The other terms in the expansion of the Wilson line are represented by the coherent multiple scattering of the scalar field off color charges within the shockwave.}
\label{gluon-shockwavepropagator}
\end{figure}

The Fourier transform 
\begin{align}
\label{fourier-transform-def}
    \tilde{G}_R(p,p') = \int d^4x~d^4y~ G_R(x,y) e^{ipx}e^{-ip'y}\,,
\end{align}
of the propagator then takes the compact form
\begin{align}
\label{scalar-gauge-prop-mom}
    \tilde{G}_R(p,p') = \tilde{G}_R^0(p) (2\pi)^4\delta^{(4)}(p-p') + \tilde{G}_R^0(p) \mathcal{T}(p,p')\tilde{G}_R^0(p')\,,
\end{align}
where $\tilde{G}_R^0(p)$ is the free propagator in momentum space,
\be
\label{free-G0-prop}
\tilde{G}_R^0(p) = \int d^4x ~e^{ipx} ~G^0_R(x) = -\frac{1}{p^2+i\epsilon p^-}~,
\ee
and the effective vertex is 
\be
\label{T-matrix-g}
\mathcal{T}(p,p') \equiv - 4\pi i(p')^-\delta(p^- -(p')^-)  \(\int d^2\bsz ~e^{i(\bsp-\bsp')\cdot \bsz}\(U(\bsz)-1\) \)~.
\ee
The dressed propagator can be represented pictorially as shown in Fig.~\ref{gluon-shockwavepropagator}.
Recall that from the classical field solution in Eq.~\eqref{eq:classical-light-cone},  
$U(x_\perp) = P_{x^-} \exp\big(ig \int_{-\infty}^{x^-} dz^-\, \frac{\rho(x_\perp,z^-)}{\nabla_\perp^2} \big)$, where the argument of the exponent, as shown in Eq.~\eqref{eq:classical-YM-Lorentz} represents the Coulomb propagator in Lorentz gauge. The corresponding gauge field $A^+$ in this gauge is sometimes called a Glauber gluon in the literature~\cite{Boussarie:2023izj}; specifically, the momentum exchange corresponding to the shockwave therefore corresponds to the coherent (path ordered) product of Glauber gluon operators. Since this $A^+$ field is independent of $x^+$ and has delta-function support in $x^-$, its Fourier modes have $k^-=0$, are independent of $k^+$, and satisfy $k^-k^+ \ll k_\perp^2 \ll Q^2$, where $Q^2$ is the resolution scale of the external probe. In perturbative QCD, the Glauber regime is power suppressed for inclusive observables.  However, as we have argued, in the saturation regime, all order Glauber contributions are contained in the path ordered exponentials that describe the classical EFT. 

The results for  the colored scalar propagator in the gluon shockwave background can be straightforwardly extended to fermion and gluon propagators~\cite{McLerran:1994vd,Balitsky:1995ub,McLerran:1998nk,Balitsky:2001mr}. 
We obtain for the retarded gluon shockwave propagator in $A^-=0$ gauge the result, 
\begin{align}
\label{gluonPropPosFinal2}
    \tilde{G}_{R,\mu \nu}\left(p, p^{\prime}\right)=(2 \pi)^4 \delta^{(4)}\left(p-p^{\prime}\right)\tilde{G}^0_{R,\mu \nu}(p)+\tilde{G}^0_{R,\mu \rho}(p) \mathcal{T}_R^{\rho \sigma}\left(p, p^{\prime}\right)\tilde{G}^0_{R,\sigma \nu}\left(p^{\prime}\right)~,
\end{align}
with the gluon-shockwave effective vertex 
\cite{Balitsky:2001mr, Roy:2018jxq, Roy:2019hwr, Blaizot:2004wv}
\begin{align}
\label{T-matrix-gluon}
    \mathcal{T}_R^{\mu \nu}\left(p, p^{\prime}\right)=-4 \pi i \Lambda^{\mu \nu} p^- \delta\left(p^{-}-(p')^{-}\right) \int d^2 \boldsymbol{z}~ \mathrm{e}^{i \boldsymbol{z} \cdot\left(\boldsymbol{p}-\boldsymbol{p}^{\prime}\right)}\left( U\left(\boldsymbol{z}\right)-1\right)~.
\end{align}
Here $\tilde{G}^0_{R,\mu \nu}(p) = \Lambda_{\m\n}\tilde{G}^0_{R}(p)$ and $\Lambda_{\m\n} = \eta_{\m\n}-\frac{p^\m n^\n+p^\n n^\m}{p\cdot n}$. 
As noted previously, computations in $A^-=0$ gauge should be distinguished from the light cone gauge $A^+=0$, where the parton interpretation of DIS structure functions in the IMF with large hadron $P^+$ is more transparent.

Likewise, for the momentum space quark propagator in the shockwave background, we obtain 
\begin{eqnarray}
\label{eq:quark-shockwave-propagator}
    S_{ij}(k,l)=S_{im}^0(k)\, {\cal T}_{mr}^q(k,l) \, S_{rj}^0(l)\,,
\end{eqnarray}
where $S_{ij}^0$ is the free quark propagator with color indices $i$ ($j$) for the outgoing (incoming) quark, and the effective quark-shockwave vertex is 
\begin{eqnarray}
\label{eq:T-matrix-quark}
    {\cal T}_{mr}^q(k,l)= 2\pi \delta(k^--l^-)\gamma^- {\rm sgn}(l^-)\int d^2 x_\perp e^{-i (k-l)\cdot x_\perp} V_{mr}^{\rm sgn (l^-)}(x_\perp)\,.
\end{eqnarray}
Here $V_{mr}(x_\perp)$ is the counterpart in the fundamental representation to the adjoint Wilson line in 
Eq.~\eqref{eq:QCD-Wilsonline}, as noted previously in Eq.~\eqref{eq:T-S-matrix}. 

In Sec.~\ref{sec:2}, we introduced the phenomenon of reggeization that becomes manifest in MRK kinematics, and the concept of the reggeized gluon. As we saw further, the pomeron can be understood as the color singlet projection of the exchange of two reggeized gluons in the forward scattering amplitude. In \cite{Lipatov:1995pn,Lipatov:1996ts}, Lipatov discussed a reggeon EFT that captures the BFKL dynamics of high energy QCD and beyond; for a discussion of the Feynman rules in this EFT, see for instance \cite{Antonov:2004hh}. What is the relation of this framework with the language of Wilson lines and the CGC EFT that we outlined? Because the two frameworks address the small $x$ problem from different perspectives, a map between the two formalisms is not straightforward. Specifically, the reggeon EFT is formulated in the limit where perturbation theory is applicable, while the CGC EFT has as its starting point the gluon saturation regime, and is most robust for large nuclei for the reasons we discussed previously. 

Nevertheless, there are regimes where the two frameworks overlap and there is much value in exploring this more deeply as we discussed in the GR context in Sec.~\ref{sec:classicalization}. An interesting suggestion~\cite{Caron-Huot:2013fea} is that the reggeized gluon field $R^a(x_\perp)$ can be thought of as the logarithm of the adjoint Wilson line\footnote{It was also shown previously in \cite{Jalilian-Marian:2000pwi} in the context of a worldline EFT that the logarithm of Wilson line is the relevant quantity in the EFT derivation of the BKFL equation.}:
\begin{eqnarray}
\label{eq:reggeon-field}
 R^a(x_\perp)=   f^{abc} \ln \left(U^{bc}\right) = \int dx^- A^{+,a}(x^-,x_\perp) 
 \equiv \int \frac{dx^- d^2 y_\perp}{4\pi}\, \ln\(\frac{1}{(x_\perp-y_\perp)^2\Lambda^2}\)\,\rho^a(y_\perp,x^-)\,.\no\\
\end{eqnarray}
If we assume $\rho^a(x^-,x_\perp) = {\tilde \rho}^a(x_\perp)\delta(x^-)$, we see then the reggeized gluon field is the convolution of the 2-D Coulomb propagator with the classical color charge density, which is the key element in the CGC EFT. As we will discuss shortly, one can interpret reggeization as the modification of $\rho^a$ induced by quantum fluctuations at small $x$. 
A further important connection was pointed out in \cite{Hentschinski:2018rrf} - see also \cite{Bondarenko:2017ern,Bondarenko:2018eid} -- where it was shown that the reggeized gluon-gluon vertex and the reggeized gluon-quark-quark vertex are identical to corresponding effective vertices in the CGC EFT given in Eqs.~\eqref{T-matrix-gluon} and \eqref{eq:T-matrix-quark}, respectively. 

\subsection{Renormalization group evolution in the Color Glass Condensate}
\label{sec:JIMWLK}
As we have discussed thus far, the physics of gluon saturation is the physics of high phase space occupancy of partons, for which a classical EFT description is a natural starting point. We have argued that the rapid growth of the $2\rightarrow n$ cross-section creates classical closely packed lumps of size $1/Q_S$ corresponding to maximal occupancy. It is these strongly correlated lumps that make up the shockwave that the DIS probe scatters off. It is important to ascertain how robust this picture is under quantum evolution. As a first step, in Sec.~\ref{sec:CGC-propagators}, we discussed the computation of computation of adjoint color scalars, gluons and fermions in the shockwave background. We will now discuss NLO computations in the CGC, and outline how they lead to a renormalization group treatment that resums leading logarithmic contributions at small $x$ to all loop order in the shockwave background.  

This RG framework was developed in a series of papers resulting in the Balitsky-JIMWLK hierarchy for n-point Wilson line corrrelators and corresponding RG equation~\cite{Balitsky:1995ub,Jalilian-Marian:1996mkd,Jalilian-Marian:1997qno,Jalilian-Marian:1997ubg,Jalilian-Marian:1998tzv,Iancu:2000hn,Iancu:2001ad,Ferreiro:2001qy}, as well as the Balitsky-Kovchegov equation describing specifically the RG evolution of the dipole Wilson line corelator we introduced in Eq.~\eqref{eq:T-S-matrix}~\cite{Balitsky:1995ub,Kovchegov:1999yj,Kovchegov:1999ua} that describes the inclusive DIS cross-section. The less inclusive the process, the more sensitive it is to higher-point Wilson line correlators. For instance, employing Eq.~\eqref{eq:quark-shockwave-propagator}, an explicit computation of the scattering amplitude for the LO DIS inclusive cross-section for a quark-antiquark pair of jets in the shockwave background reveals~\cite{Caucal:2021ent},
\begin{eqnarray}
\label{eq:dijet-LO}
    d\sigma_{\rm LO}^{\gamma^*A\rightarrow q\bar q X}\propto \langle Q(x_\perp,y_\perp;y_\perp^\prime,x_\perp^\prime) - D(x_\perp,y_\perp)-D(y_\perp^\prime,x_\perp^\prime)+1\rangle_{Y_0}\,.
\end{eqnarray}
In addition to the dipole correlator $D$ that appeared in the DIS inclusive cross-section in Eq.~\eqref{eq:CGC-inclusive}, the cross-section is sensitive to the ``quadrupole" 4-point Wilson line correlator~\cite{Jalilian-Marian:2004vhw,Blaizot:2004wu}
\begin{eqnarray}
    Q(x_\perp,y_\perp;y_\perp^\prime,x_\perp^\prime)= \frac{1}{N_c}{\rm Tr}\,\left(V(x_\perp)V^\dagger(y_\perp)V(y_\perp^\prime)V^\dagger(x_\perp)^\prime\right)\,. 
\end{eqnarray}

\begin{figure}[ht]
    \centering
    \subfigure[]{\includegraphics[scale=1]{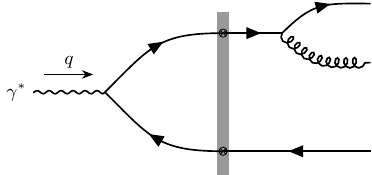}}
    \qquad\qquad
    \subfigure[]{\includegraphics[scale=1]{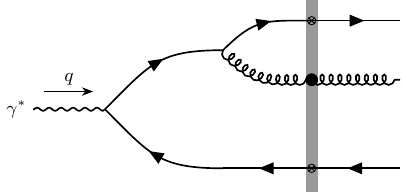}}
    \vspace{1em}
    \subfigure[]{\includegraphics[scale=1]{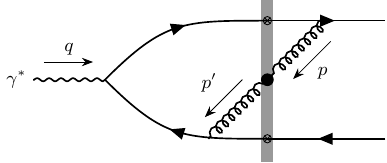}}
    \qquad\qquad
    \subfigure[]{\includegraphics[scale=1]{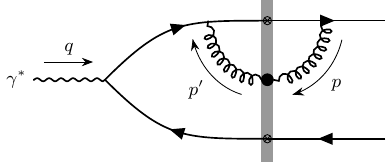}}
    \caption{Real and virtual diagrams for a dipole interacting with a shockwave. (a) Real gluon emission from the quark after scattering off the shockwave (b) Real gluon emission from quark before scattering from shockwave (c) Vertex correction from the gluon crossing the shockwave (d) Gluon self energy with gluon crossing the shockwave}
    \label{fig:dijet}
\end{figure}
Its easy to see from the identity $V(x_\perp)t^aV^\dagger(x_\perp) = t^b U^{ab}(x_\perp)$ that further gluon emission from the quark dipole (as would be the case for tri-jet production at LO, or di-jet production at NLO) leads to more nontrivial combinations of Wilson line correlators. The latter is illustrated in 
Fig.~\ref{fig:dijet}. Remarkably, the RG evolution of all such correlators with rapidity is described in the CGC EFT by the Balitsky-JIMWLK hierarchy, or more compactly by the JIMWLK Hamiltonian, which we will now discuss. 

The RG evolution in the CGC EFT follows from Eq.~\eqref{eq:CGC-path-integral}, where taking the classical saddle point $A_{\rm cl}\equiv A_{\rm cl}(\rho)$ as we discussed previously, gives for a generic observable, the leading order expression 
\begin{eqnarray}
\label{eq:LO-operator}
    \langle {\cal O}\rangle_{Y_0} = \int [d\rho] W_{\rm Y_0}[\rho] {\cal O}_{\rm LO}[\rho]\,,
\end{eqnarray}
where $Y_0 = \ln(P^+/\Lambda^+$), is the rapidity separating the shockwave sources $\rho$ from the classical field. Eq.~\eqref{eq:CGC-inclusive} is a specific example of Eq.~\eqref{eq:LO-operator} for inclusive DIS scattering, with $W_{Y_0}$ corresponding to the Gaussian distribution of sources. At next-to-leading order in the CGC EFT, including only the leading $O(\alpha_S \delta Y$), imposing $\alpha_S \delta Y\sim O(1)$ real and virtual corrections to the shockwave generated with the change of rapidity $\delta Y= Y_1-Y_0$, can be expressed formally as 
\begin{eqnarray}
\label{eq:NLO-operator}
    \langle \delta {\cal O}\rangle_{Y_0+\Delta Y} = \alpha_S \delta Y\,\int [D\rho] W_{\rm Y_0}[\rho] \left({\cal H}_{\rm LO} \,{\cal O}_{\rm LO}\right)\,. 
\end{eqnarray}
Here $H_{\rm LO}$ is the JIMWLK Hamiltonian which can be expressed as the functional operator
\begin{eqnarray}
\label{eq:Hamiltonian}
    {\cal H}_{\rm LO} = \frac{1}{2}\int_{u_\perp,v_\perp}\frac{\delta}{\delta A^{+,a}(u_\perp)} \chi^{ab}(u_\perp,v_\perp) \frac{\delta}{\delta A^{+,a}(v_\perp)}\,,
\end{eqnarray}
with the (cut) propagator  
\begin{eqnarray}
\label{eq:JIMWLK-propagator}
    \chi^{ab}(u_\perp,v_\perp)= \frac{1}{\pi}\int_{z_\perp}\!\! \frac{1}{(2\pi)^2}\frac{(u_\perp-z_\perp)\cdot (v_\perp-z_\perp)}{(u_\perp-z_\perp)^2(v_\perp-z_\perp)^2}\left[1+ U_{u_\perp}^\dagger U_{v_\perp}-U_{u_\perp}^\dagger U_{z_\perp}-U_{z_\perp}^\dagger U_{v_\perp}\right]^{ab}\,.
\end{eqnarray}
The $U_{x}$'s here are the path ordered exponentials of the classical field $A^+$ in Lorenz gauge that we defined in Eq.~\eqref{eq:QCD-Wilsonline}, where we have, for brevity, written their arguments as subscripts, as we have for the integral measures ($\int_x = \int d^2 x$). Further, $A^{+}(x_\perp)= \int_{-\infty}^{\infty} dz^- A^{+}(z^-,x_\perp)$.

It was shown in the original JIMWLK papers that the functional expression for the Hamiltonian in Eq.~\eqref{eq:Hamiltonian} is a combination of two contributions, one arising from the leading one-loop correction\footnote{The sub-leading ``pure" $\alpha_S$ contribution is nothing but the QCD polarization tensor in the background field, which in the UV generates the one-loop beta function~\cite{Ayala:1995hx}. Thus running coupling corrections to JIMWLK only appear at next-to-leading-logarithmic accuracy.} to the classical field in the interval $\Delta Y \sim 1/\alpha_S$, and the other from the small fluctuation cut propagator in the classical background field. However to leading logarithmic accuracy, the former, represented as $\nu^a(x_\perp)$, satisfies the identity~\cite{Weigert:2000gi,Gelis:2008rw}
\begin{eqnarray}
    \nu^a(x_\perp) = \frac{1}{2}\int_{y_\perp}\!\frac{\delta}{\delta A^{+,b}(y_\perp)}\chi^{ab}(x_\perp,y_\perp)\,,
\end{eqnarray}
allowing one to combine the two contributions\footnote{Since only $\chi^{ab}(x_\perp,y_\perp)$ appears in Eq.~\eqref{eq:Hamiltonian}, this has on occasion caused the misleading impression that the evolution of the shockwave does not include contributions from the polarization tensor in the background field.} in the form shown in Eq.~\eqref{eq:Hamiltonian}. 

Combining Eqs.~\eqref{eq:LO-operator} and \eqref{eq:NLO-operator}, and performing a (functional) integration by parts, one obtains
\begin{eqnarray}
\label{eq:classical-statistical-JIMWLK}
    \langle {\cal O}+\delta{\cal O}\rangle_{\rm Y_1}= \int [d\rho]\left(\left[1+ \alpha_S\delta Y {\cal H}_{\rm LO}\right]W_{\rm Y_0}[\rho]\right) {\cal O}_{\rm LO}\,.
\end{eqnarray}
Defining 
\begin{eqnarray}
    W_{Y_1}[\rho] = \left[1+ \alpha_S\delta Y {\cal H}_{\rm LO}\right]W_{\rm Y_0}[\rho]\,,
\end{eqnarray}
one obtains
\begin{eqnarray}
\label{eq:JIMWLK-RG}
    \frac{\delta}{\delta Y}W_{Y_1}[\rho]= {\cal H}_{\rm LO} W_{Y_1}[\rho]\,.
\end{eqnarray}
Thus to LLx accuracy, the small $x$ evolution of arbitrary operators, to all-loop order is captured by a functional RG - that describes the universal evolution of the weight functional $W_Y[\rho]$ convoluted with the corresponding leading order operator. This requires an appropriate nonuniversal initial condition that incorporates the nonperturbative many-body correlations in hadron wavefunctions. We argued earlier that for a large nucleus with atomic number $A$, this initial condition is given by the Gaussian weight functional in Eq.~\eqref{eq:Gaussian-functional}, where the typical scale of the color correlations is given by its variance $\mu_A^2\propto A^{1/3}$. For protons and light nuclei, it is unclear what the appropriate choice of $W$ is, and indeed, even if the mean field picture of large color charges is applicable. We will return to this point later.

A key question in the above discussion is the validity of Eq.~\eqref{eq:NLO-operator}. For the simplest dipole operator in Eq.~\eqref{eq:CGC-inclusive}, this was confirmed in multiple approaches~\cite{Balitsky:1995ub,Kovchegov:1999ua,Mueller:2001uk,Kovner:2001vi,Gelis:2008rw,Kovchegov:2012mbw}, in addition to the JIMWLK papers. In particular, one finds that the dipole correlator in Eq.~\eqref{eq:T-S-matrix} satisfies the evolution equation~\cite{Balitsky:1995ub}
\begin{eqnarray}
\label{eq:JIMWLK-dipole}
    \frac{\langle D(x_\perp,y_\perp)\rangle}{\partial Y} = -\frac{\alpha_S}{2\pi^2}\int_{z_\perp} \frac{(x_\perp-y_\perp)^2}{(x_\perp-z_\perp)^2(y_\perp-z_\perp)^2}\langle D(x_\perp-y_\perp) - D(x_\perp,z_\perp) D(z_\perp,y_\perp)\rangle_{Y}\,,
\end{eqnarray}
where 
\begin{eqnarray}
\label{eq:BFKL-coordinate-kernel}
    {\tilde K}_{\rm BKFL}=\frac{(x_\perp-y_\perp)^2}{(x_\perp-z_\perp)^2(y_\perp-z_\perp)^2}\,,
\end{eqnarray}
is the coordinate space counterpart of the momentum space BFKL kernel we derived in Eq.~\eqref{BFKL-eq-q0-1}. Alternately, one can recover Eq.~\eqref{eq:JIMWLK-dipole} by applying ${\cal H}_{\rm LO}$ in Eq.~\eqref{eq:Hamiltonian} to the dipole operator, and using the identities $\frac{\delta U(x_\perp)}{\delta A^{+,a}(x_\perp)}= -ig \delta(x_\perp-z_\perp)U(x_\perp) T^a$ and $\frac{\delta U^\dagger(x_\perp)}{\delta A^{+,a}(x_\perp)}= ig \delta(x_\perp-z_\perp)T^a U^\dagger (x_\perp)$, along with $\tau^b U_{ba} = V(x_\perp)\tau^a V^\dagger(x_\perp)$. Eq.~\eqref{eq:JIMWLK-dipole}  reduces to a closed form expression in terms of the S-matrix (=$\langle D\rangle$) in the $A\rightarrow \infty$ and $N_c\rightarrow \infty$ because the second term on the r.h.s can be replaced by the factorized form $\langle DD\rangle\rightarrow \langle D\rangle  \langle D\rangle$ in these limits. This closed form, albeit nonlinear, equation is known as the Balitsky-Kovchegov (BK) equation~\cite{Balitsky:1995ub,Kovchegov:1999ua,Kovchegov:1999yj}. 

Now recall that in Eq.~\eqref{eq:T-S-matrix}, we defined the dipole scattering amplitude for a fixed configuration of color charges as $T= 1-S \equiv \frac{1}{N_c}{\rm Tr}(V(x_\perp)V^\dagger(y_\perp)$. If we further expand out $V= P\exp\left(ig \frac{\rho}{\nabla_\perp^2}\right)$ to lowest order (assuming $\rho/\nabla_\perp^2 \ll 1$) and keep only terms up to quadratic order in $\rho$, one can further simplify Eq.~\eqref{eq:JIMWLK-dipole} and rewrite it in terms of the scattering amplitude as 
\begin{eqnarray}
\label{eq:BFKL-coordinate}
    \frac{\partial \langle T(x_\perp,y_\perp)\rangle}{\partial Y} = {\bar \alpha}_S\int_{z_\perp} {\tilde K}_{\rm BFKL}\,
    \left[\langle T(x_\perp,z_\perp)+ T(z_\perp,y_\perp)-T(x_\perp,y_\perp)\rangle\right]\,,
\end{eqnarray}
which is the coordinate space BFKL equation for the dipole scattering amplitude. Here ${\bar \alpha}_S=\frac{\alpha_S N_c}{\pi}$. Indeed, identifying the $\langle T\rangle $'s in terms of quadratic correlators of $\rho$'s, and defining ${\bar \phi}(k_\perp) = {\tilde \mu}_Y^2(k_\perp)/k_\perp^2 $ (where ${\tilde \mu}_Y^2(k_\perp)$ is the Fourier transform of $\mu_Y^2(r_\perp)=\int d^2 X\langle \rho^a(X_\perp+r_\perp/2)\rho^a(X_\perp-r_\perp/2)\rangle/(\pi R_A^2)/(N_c^2-1)$, assuming a homogeneous color charge distribution in the nucleus), the Fourier transform of Eq.~\eqref{eq:BFKL-coordinate} gives\footnote{This unintegrated distribution is often called the ``dipole" gluon distribution to distinguish it from the Weizs\"{a}cker-Williams gluon distribution of the nucleus-see also footnote \ref{footnote:two-dists} in Sec.~\ref{sec:2}.} 
\begin{eqnarray}
\label{eq:BK-momentum}
    \frac{\partial {\bar \phi}(k_\perp)}{\partial Y} = \frac{{\bar \alpha}_S}{\pi}\int_{p_\perp} \frac{1}{(k_\perp-p_\perp)^2}\left[{\bar \phi}(p_\perp)- \frac{k_\perp^2}{2p_\perp^2}{\bar \phi}(k_\perp)\right]\,,
\end{eqnarray}
which, replacing the r.h.s of Eq.~\eqref{fraction-substitution} by its l.h.s, is equivalent to Eq.~\eqref{BFKL-eq-q0-1}. It is striking to see how simply the BFKL equation for the inclusive cross-section is obtained in the CGC EFT. 

The full BK equation (containing contributions to all orders in $\rho/\nabla_\perp^2$) includes an additional nonlinear term $-T(x_\perp,z_\perp) T(z_\perp,y_\perp)$ on the r.h.s of Eq.~\eqref{eq:BFKL-coordinate}. We noted previously the momentum space expression in Eq.~\eqref{eq:JIMWLK-dipole-momentum-space}. This term is responsible for generating interactions of the fan diagram illustrated in Fig.~\ref{fig:fan-diagram}. The BK equation also confirms the conjecture of gluon saturation in \cite{Gribov:1983ivg,Mueller:1985wy}, based on the analysis of the leading power suppressed Feynman diagrams. Specifically, we see that Eq.~\eqref{eq:BFKL-coordinate} has a nontrivial IR fixed point for $T=1$, which causes the growth of the scattering amplitude with rapidity to saturate (at fixed impact parameter). Generically, the BK equation follows a reaction-diffusion process, and indeed, to a good approximation, can be mapped on to the so-called Fisher-Kolmogorov-Piscounov-Petrovsky (FKPP) equation describing the motion of traveling wave fronts in statistical mechanics - for a comprehensive recent review, see \cite{Angelopoulou:2023qdm}. 

As a related feature, the BK equation exhibits the phenomenon of geometric scaling, whereby the dipole scattering amplitude scales as $r_\perp^2 Q_S^2(Y)$, generalizing Eq.~\eqref{eq:GBW+MV} to include the rapidity evolution of $Q_S$. Remarkably, the idea of geometric scaling came from the observation~\cite{Stasto:2000er} that the HERA inclusive cross-section data\footnote{However it is premature to claim discovery of gluon saturation on this basis since the extracted saturation scales are not robustly in the weak coupling regime for which the framework is applicable~\cite{Aschenauer:2017jsk}. Because of the access to large nuclei, DIS at the EIC is more promising in this regard.} for $x\leq 0.01$ (see Eq.~\eqref{eq:DIS-cross-section}) scaled as $\sigma_{\gamma^* A}(x_{\rm Bj},Q^2)\propto \sigma_{\gamma^* A}(Q^2/Q_S^2(x_{\rm Bj}))$. The BK equation in full generality can only be solved numerically but one can exploit the FKPP analogy to extract analytic expressions for the saturation scale to good accuracy~\cite{Munier:2003vc,Munier:2003sj}. Specifically, Eq.~\eqref{eq:JIMWLK-dipole-momentum-space} can be approximated at LLx by~\cite{Beuf:2010aw}
 \begin{eqnarray}
 \label{eq:traveling-wave-QS}
 &\phi(Y,k_\perp^2)=\phi\left(\ln\left(\frac{k_\perp^2}{Q_S^2(Y)}\right)\right)\,\, {\rm with}\no\\
     &\ln \left(\frac{Q_S^2(Y)}{Q_0^2}\right) = 2 {\bar \alpha}_s Y -\frac{3}{2\gamma_c}\ln\left({\bar \alpha}_S Y\right) +{\rm Constant}-\frac{3}{\gamma_c^2}\sqrt{\frac{2\pi}{\chi''(\gamma_c){\bar \alpha}_S Y}} + O\left(\frac{1}{{\bar \alpha}_S Y}\right)\,,
 \end{eqnarray}
 where $\gamma_c\approx 0.63$ and $\chi''(\gamma)$ is the second derivative of the BFKL eigenvalue expressed as 
 $\chi(\gamma) = 2\Psi(1)-\Psi(\gamma)-\Psi(1-\gamma)$, with $\Psi$ here the digamma function. 
 This expression is valid in a wide ``geometrical scaling" rapidity window $k_\perp^2 \ll Q_S^2(Y)\exp\left(\sqrt{2\chi''(\gamma_c){\bar \alpha}_S Y} \right)$ that extends parametrically well beyond $k_\perp^2 \sim Q_S^2(Y)$. 

All of our discussion thus far of JIMWLK and BK LLx RG evolution was for the DIS inclusive dipole cross-section, or equivalently, the dipole correlator. One can extend this to more nontrivial quantities, such as the inclusive di-jet cross-section illustrated in Fig.~\ref{fig:dijet}. As noted, the LO expression in Eq.~\eqref{eq:dijet-LO} is sensitive to a 4-point quadrupole correlator in addition to the 2-point dipole correlator. At NLO, the real and virtual Feynman graphs contributing to the di-jet cross-section were computed explicitly in \cite{Caucal:2021ent}, employing the quark and gluon shockwave propagators given in Sec.~\ref{sec:CGC-propagators}. In the slow gluon (small $x$) limit of the expressions, one observes a large number of terms with dipole $D$, quadrupole $Q$, and 
bilinear operator combinations 
$DD$, $DQ$. However remarkably one sees that this complex structure is reproduced exactly\footnote{One can similarly compute the single jet cross-section and arrive at the same conclusion~\cite{Caucal:2024cdq}; however here one has only terms linear and bilinear in $D$. It can further be confirmed that integrating over the single jet phase space in the slow gluon limit recovers Eq.~\eqref{eq:JIMWLK-dipole}.} by applying the JIMWLK Hamiltonian in Eq.~\eqref{eq:Hamiltonian} to the leading order di-jet cross-section in Eq.~\eqref{eq:dijet-LO}, thereby providing a highly nontrivial confirmation of Eq.~\eqref{eq:NLO-operator}. 

In Sec.~\ref{sec:NLL-gluon-sat}, we discussed briefly the extension of the BFKL framework to NLLx and the related small $x$ resummation of large double transverse logarithms that improve the accuracy of the framework. Likewise, the BK and JIMWLK RG equations have been extended to NLLx accuracy~\cite{Balitsky:2007feb,Balitsky:2013fea,Kovner:2013ona,Lublinsky:2016meo}, with the most recent progress in this direction being key pieces of the NNLLx BK equation~\cite{Caron-Huot:2016tzz}. Similarly to the BFKL case, the small $x$ resummation of large double transverse logarithms, so key to the stability of NLLx BFKL, have been applied to the BK framework~\cite{Ducloue:2019ezk} and state-of-the-art NLLx numerical simulations compared successfully to HERA data~\cite{Beuf:2020dxl}. An interesting result is that the traveling wave structure of the BK equation persists when NLLx running coupling effects are included;  one can extract an expression for the saturation scale analogous to the fixed coupling case in Eq.~\eqref{eq:traveling-wave-QS}. The leading term on the r.h.s grows more slowly now as $Y^{1/2}$, with further sub-leading contributions, as shown in \cite{Beuf:2010aw}. 

We focused thus far on DIS observables in the CGC EFT. A powerful feature of this shockwave framework, as we will discuss further in Secs.~\ref{sec:Gluon-shockwaves} and \ref{sec:glasma}, is that it can  be applied to compute multi-particle production in the glasma~\cite{Lappi:2006fp,Gelis:2006dv,Dumitru:2008wn} formed in hadron-hadron/nucleus and nucleus-nucleus collisions at high energies. A natural formalism implicit in our shockwave discussion, but especially key to  understanding shockwave collisions, is the Schwinger-Keldysh (SK) closed time path formalism. In this formalism, the path integral is expressed on an upper contour from an initial time to a fixed time of interest, and then back to the initial time on a lower contour, with ``+" and ``-" fields defined on the upper and lower contour respectively. 
For a generic interacting field theory, one can denote the sum and difference fields as $\varphi_r$ (retarded) and $\varphi_a$ (advanced) fields, respectively. Reexpressing the path integral on the contour in terms of these fields, one find that one obtains an initial density matrix entirely in terms of the retarded field and its time derivative (for the example of an interacting scalar theory). The classical equations of motion {\it and} the small fluctuation equations of motion both result from integrating out terms linear in $\varphi_a$; the next order $\varphi_a^3$ contain the genuine O($\hbar$) corrections which are suppressed when the leading classical $\varphi_r$ fields are large. Keeping only the leading linear contributions in $\varphi_a$ is called the classical-statistical approximation. 

The LLx JIMWLK equations describing the RG evolution of a single shockwave can be derived entirely in this formalism~\cite{Gelis:2008rw,Jeon:2013zga}, providing a natural segue to more complex dynamics of shockwave collisions. Indeed, Eq.~\eqref{eq:classical-statistical-JIMWLK} provides a specific illustration. The leading LLx quantum corrections can be absorbed into evolution of the stochastic weight functional with rapidity, which is then convoluted by the LO operator computed using the classical Yang-Mills evolution. Genuine (sub-leading in $\hbar$) quantum effects appear only at cubic order in the difference fields~\cite{Jeon:2013zga}. Many features of reggeon field theory  can be understood in terms of the combinatorics\footnote{An example are the so-called Abramovksy-Gribov-Kancheli (AGK) rules~\cite{Abramovsky:1973fm} we alluded to earlier in Sec.~\ref{sec:classicalization}, that are employed extensively in the ACV framework for shockwave collisions in gravity. We will discuss these connections further in Sec.~\ref{sec:glasma}.}

As a simple example, as we noted in the discussion following Eq.~\eqref{eq:Gaussian-functional}, from the interpretation in Eq.~\eqref{eq:reggeon-field} of the color charge density $\rho$ as the reggeon field, the pomeron and odderon can be understood, respectively, as bilinear and trilinear gauge invariant products of the $\rho$'s. 
Given this semi-classical interpretation, it seems very plausible that one can understand the phenomenon of reggeization that we discussed extensively, in the QCD context in Sec.~\ref{sec:2} and for gravity in Sec.~\ref{sec:3}, as a semi-classical phenomenon despite the apparent multi-loop resummation. As has been noted previously, identifying $\hbar$ with loops can be fallacious~\cite{Holstein:2004dn}. This argument is simply realized in the SK formalism. 

A corollary to this question is whether the breakdown of reggeization, shown to occur beyond NLLx accuracy due to cuts\footnote{See \cite{Fadin:2016wso,Caron-Huot:2017fxr,Falcioni:2021dgr,Fadin:2024eyf} for discussions of such Regge cut contributions at NNLLx accuracy.} from multiple reggeized gluon exchanges~\cite{DelDuca:2001gu} can be understood as the breakdown of the classical-statistical approximation, at NNLx accuracy. In SK language, this would occur when the back-reaction of the $\varphi_a^3$ fields in the action become important; such contributions can be computed in simpler field theories using 2-particle irreducible functional techniques~\cite{Berges:2004yj}. In other words, can this phenomenon be related to $1/n$ corrections that control the decay of the shockwave~\cite{Dvali:2021ooc}? We speculate that, if so, such corrections could be interpreted as arising from the rescattering of the decay products of the shockwave with the NLO emissions from the DIS probe shown in Fig.~\ref{fig:dijet}. 

\subsection{Gluon shockwave collisions: derivation of Lipatov vertex}
\label{sec:Gluon-shockwaves}

For strong sources comprising the large $x$ modes of the scattering nuclei 
(which are order $\rho_{\rm nucleus}\sim 1/g$), the leading term in the power counting is the produced classical field $A_{\rm cl}^\mu$, which too is of order $O(1/g)$; the single-inclusive distribution in the produced glasma is therefore of $O(1/\alpha_S)$. At NLO in the SK formalism, just as for the DIS case, there are two sorts of contributions: a) the one loop correction to the classical field $a_{\rm quant}^\mu$, and 
b) the small fluctuation propagator $\langle a_{\rm quant}^\mu a_{\rm quant}^\nu\rangle$. 
As we discussed previously,  the logarithmic enhancements $\alpha_S \ln(1/x) \sim O(1)$ to these contributions are what contribute to the JIMWLK Hamiltonian, and are therefore absorbed in the evolution of the single-inclusive gluon distribution. Thus at each step in the rapidity evolution of the individual nuclei before the collision, the problem of n-particle inclusive gluon production at a given rapidity is simply the solution of the QCD Yang-Mills equations in the presence of the static source distributions of each of the nuclei evolved up to that scale. Note that this assumes that the wee partons of each of the nuclei don't talk to each other before the collision; in other words, the 
weight functionals $W[\rho_{\rm nucleus}]$ of each of the nuclei factorize in the collision. This factorization holds to LLx when $\rho_{\rm nucleus}\sim 1/g$~\cite{Gelis:2008rw,Gelis:2008ad}. We will provide further detailed context, and outline the steps that justify the above statements, in Sec.~\ref{sec:glasma}. 

With this understood, the leading order contribution to the problem of shockwave collisions is given by the solution of the Yang-Mills (YM) equations 
\begin{align}
\label{YM:dense-dense}
D_\m F^{\m\n}=J_{\rm HI}^\n~,
\end{align}
where $F_{\m\n}=\p_\m A_\n-\p_\n A_\m+ig[A_\m,A_\n]$ is the field strength tensor and $J_{\rm HI}^\m$ is the covariantly conserved current, $ D_\m J_{HI}^\m = 0$. For shockwave scattering of nuclei (heavy-ion collisions at ultrarelativistic energies), the shockwave currents can be represented as 
\be
J_{\rm HI}^{\n,a} = \delta^{\n+}\rho^a_A(x_\perp)\delta(x^-) + \delta^{\n-}\rho^a_B(x_\perp) \delta(x^+)
\,.
\ee
Here $\rho^a_A(x_\perp)$ and $\rho^a_B(x_\perp)$ are the quasi-classical color charge distributions of each of the nuclei corresponding to a higher dimensional representation of the color charges depicted in Fig.~\ref{dipole-interaction}, and distributed in the transverse plane of the scattering. For $A\gg 1$, the weight functional\footnote{Unless required, we will not specify the rapidity $Y$ of the inclusive radiation spectrum.} $W[\rho_{A,B}]$ is Gaussian distributed such that $\langle \rho_{A,B}^a(x_\perp)\rho_{A,B}^b(y_\perp)\rangle = Q_S^2\, \delta^{ab}\, \delta^{(2)}(x_\perp-y_\perp)$, with $Q_S^2 \propto A^{1/3}\, \Lambda_{\rm QCD}^2$. For simplicity, we will assume that the nuclei are identical. The $\delta(x^\mp)$ terms represent eikonal currents, for which classical sub-eikonal corrections are $O(1/P^\pm)$ respectively. The currents, to leading order, are independent of the light cone times $x^\pm$, respectively; this reflects the fact that the interaction of the 
two pure gauges corresponding to the light cone sources is not a pure gauge in QCD, and is sufficient to generate finite field strength in the forward light cone. 

The nucleus-nucleus scattering problem thus formulated~\cite{Kovner:1995ja,Kovner:1995ts}, can in full generality only be solved numerically~\cite{Krasnitz:1998ns,Krasnitz:2000gz,Berges:2020fwq}. However one can identify the expansion parameters $\rho_A/\nabla_\perp^2, \rho_B/\nabla_\perp^2$ in the YM equations that one can expand in to obtain analytic solutions. These are the dilute-dilute YM asympotics of $\rho_A/\nabla_\perp^2, \rho_B/\nabla_\perp^2 \ll 1$ (corresponding to the regime of large transverse momenta $k_\perp \gg Q_S$)~\cite{Kovner:1995ja,Kovner:1995ts,Kovchegov:1997ke,Gyulassy:1997vt} and dilute-dense asymptotics  $\rho_A/\nabla_\perp^2\ll 1, \rho_B/\nabla_\perp^2 \sim 1$~\cite{Dumitru:2001ux,Blaizot:2004wu,Gelis:2005pt}, or $Q_{S,A}\ll k_\perp \ll Q_{S,B}$. 
The dense-dense regime of $\rho_A/\nabla_\perp^2, \rho_B/\nabla_\perp^2 \sim 1$, as noted previously, is not analytically tractable and corresponds to fully nonlinear solutions of the YM equations. 

\begin{figure}[ht]
\centering
\includegraphics[scale=1]{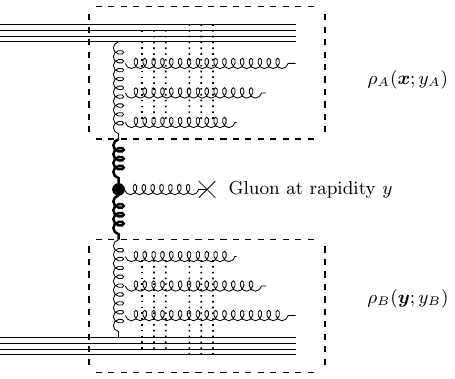} 
\caption{Dilute-dilute regime of shockwave scattering in QCD. The inclusive gluon radiative 
distribution (depicted by the emission of a gluon line at rapidity y)  is insensitive to  eikonal exchanges (depicted with dashed lines) within the color charge densities $\rho_A$ and $\rho_B$. The emission of reggeized gluons (in bold font) from these sources interact via the effective Lipatov vertex.}
\label{dilute-dilute}
\end{figure}

\begin{figure}[ht]
\centering
\includegraphics[scale=1]{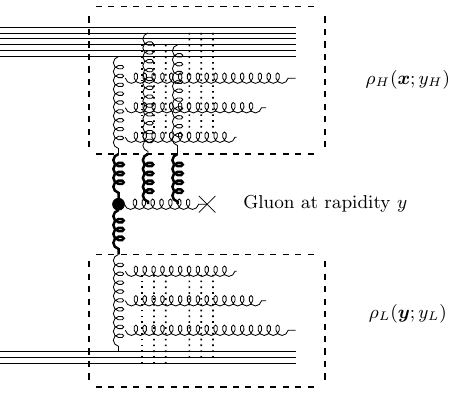}
\caption{Dilute-dense scattering ($\rho_L/\nabla_\perp^2 \ll 1$ and $\rho_H/\nabla_\perp^2\sim 1$). For $\rho_H/\nabla_\perp^2\sim 1$, coherent multiple scatterings from the nucleus can be resummed into a lightlike Wilson line. }
\label{dilute-dense}
\end{figure}

The dilute-dense scattering case is illustrated in Fig.~\ref{dilute-dilute} and the dilute-dense case in Fig.~\ref{dilute-dense}. In the former case, since $\rho_A/\nabla_\perp^2, \rho_B/\nabla_\perp^2 \ll 1$, coherent multiple scattering is suppressed in both of the colliding nuclei. This is the ``BFKL regime" of high energy scattering since in this limit the rapidity evolution of  both ``classical lumps" is described by the BFKL equation. In the dilute-dense case, switching henceforth $\rho_{A,B}\rightarrow\rho_{L,H}$, with  $\rho_L/\nabla_\perp^2\ll 1$ and $\rho_H/\nabla_\perp^2\sim 1$, multiple scattering insertions on the emitted gluon, as we will elaborate, can be absorbed into a Wilson line. In the ``dense-dense" case of $\rho_{L,H}/\nabla_\perp^2\sim 1$ it is not feasible to factorize coherent multiple scatterings from both nuclei into separate Wilson lines, and as noted, this configuration of can only be evaluated numerically. 

The gluon shockwave with transverse source distribution $\rho_H(\bsx)$ moving in the positive $z$ direction is generated by the static current
\begin{equation}
J_{\mu}=g \delta_{\mu-} \delta\left(x^{-}\right) \rho_H(\bsx) ~.
\end{equation}
is given by
\be
\label{gShockBgnd}
\bar{A}_\m(x^-,\bsx) = -g \delta_{\mu-} \delta\left(x^{-}\right) \frac{\rho_{H}\left(\boldsymbol{x}\right)}{\nabla_\perp^2}~,
\ee
with nonvanishing field strength only at $x^-=0$. 

In the shockwave collision problem, one introduces the current of the incoming shockwave with the transverse color charge distribution $\rho_L(\bsx)$ moving in the negative $z$ direction:
\be
J_{\mu} = g\delta_{\mu+} \delta(x^+)\rho_L\(\bsx\)~.
\ee
\begin{figure}[ht]
\centering
\includegraphics[scale=1]{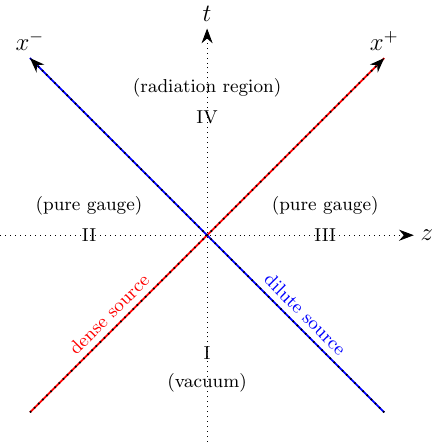}
\caption{Spacetime diagram of collision of two gluon shockwaves. Red and blue lines represent the lightlike trajectories of the incoming shockwaves. The future light cone of the collision point $t=z=0$ is where gluon radiation occurs. Gauge fields are pure gauges in regions I, II and III.}
\label{spacetimeg}
\end{figure}
For $t>0$, in the dilute-dilute approximation, one simply linearizes the YM equations to linear order in the sources $\rho_H$ and $\rho_L$ and solves for the radiation field $a_\mu$. In light cone gauge $a_+=0$, the physical components of the gauge field can be expressed as  
\begin{align}
    &\square a_{i,c} = -g^3\(\Theta(x^+)\Theta(x^-) \p_i\(\frac{\rho_H}{\nabla_\perp^2}\rho_L\)-2\delta(x^+)\delta(x^-) \frac{\rho_H}{\nabla_\perp^2} \frac{\p_i\rho_L}{\nabla_\perp^2}\)T^a T^b f_{abc}~.
\end{align}
The Fourier transform of this equation (after putting the momenta of the emitted gluons  on-shell $k^2 = 2k_+k_--\bsk^2=0$) gives 
\begin{align}
    \label{gaugeLV}
    a_{i,c}(k) &= -\frac{2ig^3}{k^2+i\epsilon k^-}\int \frac{d^2\bsq_2}{(2\pi)^2}\(q_{2i}-k_i\frac{\bsq_2^2}{\bsk^2}\)\frac{\rho_H}{\bsq_1^2}\frac{\rho_L}{\bsq_2^2}T^a T^b f_{abc}~.
\end{align}
Here,  $1/k^2$ corresponds to the emitted gluon propagator,  $1/\bsq_1^2$ and $1/\bsq_2^2$ are the exchanged reggeized gluon propagators and the Lipatov vertex is the term in the parenthesis -- see Fig.~\ref{dilute-dilute}. 

To see the latter explicitly, we will recast the covariant expression for the Lipatov vertex in Eq.~\eqref{QCD-Lipatov-vertex-1} in light cone gauge. We first partially gauge fix the gluon polarization vector $\varepsilon_+=0$,  which implies $\ep_-k^-=\ep_ik_i~,\ep_\m p_1^\m =0~,\ep_\m p_2^\m=\ep_-p_2^-$ since $p_1 = (p_1^+,0,0,0)$ and $p_2 = (0,p_2^-,0,0)$. It is then straightforward to deduce the form of $C^\mu$ in light cone gauge \cite{Ioffe:2010zz}
\begin{align}
\label{QCDLVLC1}
&\ep^*_{\m}(k) C^\m(q_1,q_2) = -2\ep^*_i \(q_{2i}-k_i\frac{\bsq_{2 }^{2}}{\bsk^{2}}\) \equiv \ep^*_i C_i(\bsq_1,\bsq_2)~.
\end{align}
The light cone gauge expression makes transparent the fact that the dependence of this vertex is only on $\bsq_1$ and $\bsq_2$ and not the external momenta $p_1,p_2$.

In \cite{Blaizot:2004wu}, and later in \cite{Gelis:2005pt}, it was shown that the Lipatov vertex is contained in the classical YM solutions in both dilute-dilute and dilute-dense scattering regimes. For the dilute-dense case, one obtains
\begin{align}
\label{dilutedenseQCD}
a_i(k)  =  -\frac{2ig}{k^2+i\epsilon k^-}\int \frac{d^2\bsq_{2}}{(2\pi)^2} \(q_{2i}-k_i\frac{\bsq_2^2}{\bsk^2}\) \frac{\rho_L(\bsq_{2})}{\bsq_{2}^2}\bigg(U(\bsk+\bsq_{2})-(2\pi)^2 \delta^2(\bsk+\bsq_{2})\bigg)\,,
\end{align}
where $U(\bsk)$  is the Fourier transform of 
\begin{align}
    U(x^-, \bsx)  \delta\left(x^{+}\right)  = \exp\(ig \int_{-\infty}^{x^-} dz^- \bar{A}_-(z^-, \bsx) \cdot T \)~,
\end{align}
with $\bar{A}_-(z^-, \bsx) $ defined in Eq.~\eqref{gShockBgnd}. The Wilson line encodes the coherent multiple scattering of the emitted gluon off the dense source $\rho_H$ in Fig. \ref{dilute-dense}. Expanding the above result to lowest order in $\rho_H$ allows one to recover the dilute-dilute result in Eq.~\eqref{gaugeLV}.
In summary, we see that the Lipatov vertex we first encountered in Eq.~\eqref{QCD-Lipatov-vertex-1} in our discussion of the  $2\rightarrow 3$ scattering amplitude in Sec.~\ref{sec:2.1} can be obtained from solutions of the classical YM equations in the presence of nontrivial classical color sources that evolve with rapidity via the BFKL/BK/JIMWLK equations. 

\subsection{Multi-particle production in shockwave collisions}
\label{sec:glasma}
\begin{figure}[ht]
    \centering    
    \includegraphics[]{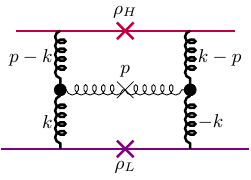}
    \caption{Illustration of Eq.~\eqref{gaugeLV} for dilute-dilute single inclusive gluon production in the CGC EFT with classical fields/reggeized gluons (dark curly lines) and the Lipatov vertex (black blobs).  This contribution is the imaginary part of a two-loop Feynman diagram, with the crosses representing the on-shell final states.}
    \label{glittering-glasma-1}
\end{figure}

We established in the previous subsection that the gauge field produced in dilute-dilute shockwave collisions is given by Eq.~\eqref{gaugeLV}, where the r.h.s includes a convolution of the fields 
${\cal A}^+(\bsk) = \frac{\rho_L(\bsk^2)}{\bsk^2}$,   
${\cal A}^-(\bsp-\bsk) = \frac{\rho_H((\bsp-\bsk)^2)}{(\bsp-\bsk)^2}$, 
and the Lipatov vertex in light cone gauge, $C^i(k,p-k)=(k_i-p_i\frac{\bsk^2}{\bsp^2})$. 
Here ${\cal A}^+$ and ${\cal A}^-$, as noted previously, can be interpreted as reggeon fields in Lipatov's EFT; they interact via the Lipatov vertex (the reggeon-regggeon-gluon vertex in this language) to produce the on-shell gluon. 
This is  illustrated in Fig.~\ref{glittering-glasma-1}. 

The average number of gluons produced in a dilute-dilute shockwave collision is~\cite{Blaizot:2004wu,Gelis:2006yv}  
\begin{eqnarray}
\label{eq:gluon-single-inclusive}
    {\bar n}_g = \int \frac{d^3 p}{(2\pi)^3 2 E_p}\langle | {\cal M}^c (p)|^2\rangle_{\rm \rho_L,\rho_H}\,,
\end{eqnarray}
where 
\begin{eqnarray}
    {\cal M}^c(p) = p^2 a_i^c(p)\,,
\end{eqnarray}
with the expression for $a_i^c$ given in Eq.~\eqref{gaugeLV}.  Here 
\begin{eqnarray}
\label{eq:dilute-dilute-average}
    \langle {\cal O} \rangle_{\rho_L,\rho_H} = \int [d\rho_L][d\rho_H]\, W_{\rm Y_L + Y}[\rho_L]\, W_{\rm Y_H-Y}[\rho_H]\,{\cal O}[\rho_L,\rho_H]\,,
\end{eqnarray}
where the l.h.s is computed at rapidity $Y$. Eq.~\eqref{eq:dilute-dilute-average} is a generalization of the DIS expression in Eq.~\eqref{eq:LO-operator}  and is highly nontrivial. 
This is because it is not clear that the leading logarithmic ($\alpha Y\sim O(1)$) quantum corrections to this classical-statistical leading order result can be similarly factorized, as in Eq.~\eqref{eq:NLO-operator} for the DIS case. In \cite{Gelis:2008rw}, it was shown that Eq.~\eqref{eq:dilute-dilute-average} is robust to leading logarithmic accuracy in rapidity. In other words, quantum corrections that are accompanied by large logarithms $\ln(P^\pm/\Lambda^\pm)$ (where $P^\pm$ denotes the momenta of the ultrarelativistic nuclei and $\Lambda^\pm$, the momentum modes specifying the rapidity of interest), can be factorized to all loop orders, and absorbed in the weight functionals $W$. As in the DIS case, the rapidity evolution of each of these weight functionals is governed by the JIMWLK RG equation. The computation of an operator ${\cal O}[\rho_L,\rho_H]$ is then obtained from solutions of the classical Yang-Mills equations. 

The power counting in the CGC EFT underlying Eq.~\eqref{eq:dilute-dilute-average} is as follows. The color sources 
$\rho_{L,H}$ are $O(1/g)$. The amplitude shown in Fig.~\ref{glittering-glasma-1} is therefore also of $O(1/g)$ and the inclusive multiplicity is then $O(1/g^2)$. Further, as noted previously, an expansion parameter is 
$\rho_{L,H}/\nabla_\perp^2$, arising from expanding Wilson lines, leading to our dilute-dilute, dilute-dense and dense-dense classification, depending on the kinematic regions of interest. Finally, as outlined in our prior RG discussion, the leading logs in all orders, preserve the structure of this power counting. All further tree and higher order diagrams are sub-leading either by powers of the coupling, in next-to-leading log contributions, or sub-eikonal contributions in powers of the energy. Note that Fig.~\ref{glittering-glasma-1} corresponds to a particular leading in $\hbar$ contribution to a two-loop Feynman diagram but is manifestly a tree level contribution in the CGC EFT. 

\begin{figure}[ht]
    \centering    
    \includegraphics[]{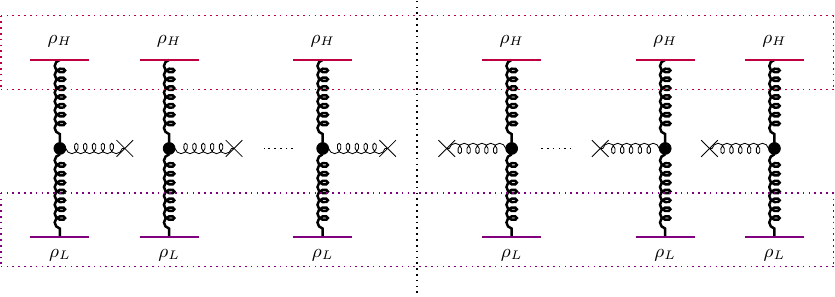}
    \caption{Illustration of Eq.~\eqref{eq:kth-moment-gluons} for the n-gluon inclusive multiplicity in the glasma, computed in the dilute-dilute limit of the CGC EFT, generalizing the structure shown in  Fig.~\ref{glittering-glasma-1}. The dashed rectangles indicate stochastic averaging over the sources. }
    \label{glittering-glasma-2}
\end{figure}

Remarkably, Eq.~\eqref{eq:dilute-dilute-average} also holds for multiple-gluon emission in the glasma\footnote{One can alternatively  straightforwardly instead discuss energy distributions and energy-energy correlators. 
There is a significant revival of interest in such correlators in the context of jet physics~\cite{Moult:2025nhu}. Nonperturbative contributions to energy-energy correlators in the CGC EFT were first computed in \cite{Krasnitz:1998ns}, and can in the dilute-dilute approximation, be mapped to recent computations in the DGLAP/BFKL overlap regime \cite{Chang:2025zib,Budhraja:2024tev}.}, and one can write the $k$'th factorial moment of the multiplicity distribution as \cite{Gelis:2008ad}
\begin{eqnarray}
\label{eq:kth-moment-gluons}
    \langle n (n-1)\cdots (n-k-1)\rangle = \int \frac{d^3 p_1}{(2\pi)^3 2 E_{p_1}}\cdots \frac{d^3 p_k}{(2\pi)^3 2 E_{p_k}} \left \langle \frac{d^k N}{d^2 p_{\perp,1}dy_1\cdots d^2 p_{\perp,k}dy_k}\right \rangle_{\rho_L,\rho_H}\,,
\end{eqnarray}
where $\langle\cdots\rangle$ corresponds to Eq.~\eqref{eq:dilute-dilute-average}. This contribution to multi-particle production is illustrated in Fig.~\ref{glittering-glasma-2}. It is valid for all the kinematic regimes in the power counting. This is inclusive of the dense-dense regime (at early times $t\leq 1/Q_S$) even though the factorized objects in Eqs.~\eqref{gaugeLV} and \eqref{dilutedenseQCD} are not clearly identifiable in this case. 
Important caveats in this derivation are a), it is only valid when all the particles are produced in  a window of rapidity $\Delta Y\leq 1/\alpha_S$, and b), it is only valid to LLx. The extension of this framework to $\Delta Y > 1/\alpha_S$, and to NLLx, will require significant conceptual and technical developments. 

This mechanism for multi-particle production, while highly sub-leading at low gluon occupancies, becomes the dominant mechanism when occupancies approach $1/\alpha_S$. Nevertheless, it  is important to keep in mind that the approach to classicalization is driven by BFKL evolution, and similarly, is made up of Lipatov vertices and reggeized propagators. 
For the Gaussian weight functional\footnote{This combinatorics also works for the RG evolved weight functional, which is a nonlocal Gaussian, whose momentum-dependent variance satisfies the BK equation. See the discussion following  Eq.~\eqref{eq:BFKL-coordinate}. } in Eq.~\eqref{eq:Gaussian-functional}, the combinatorics of the color charge densities in Eq.~\eqref{eq:dilute-dilute-average} can be worked out explicitly~\cite{Gelis:2009wh};  the n-particle probability distribution is the negative binomial distribution (NBD)~\cite{DeWolf:1995nyp,Dremin:2000ep}
\begin{eqnarray}
\label{eq:NBD-QCD}
    P_n = \frac{\Gamma(n+q)}{\Gamma(q)\Gamma(n+1)}\frac{{\bar n}^n q^q}{({\bar n}+q)^{{ n}+q}}\,,
\end{eqnarray}
where $\bar{n}$ is the mean of the distribution and $q$ is defined as
\begin{eqnarray}
\label{eq:NBD-parameter}
    q = \zeta \frac{(N_c^2-1) S_\perp}{2\pi}\,.
\end{eqnarray}
Here $\zeta$ is a nonperturbative $O(1)$ constant and $S_\perp$ is the transverse overlap area of the shockwaves at a fixed impact parameter. The  NBD distribution~\cite{Dremin:2000ep} was employed previously to fit multiplicity distributions with $q$ as a phenomenological parameter interpolating between a Bose-Einstein distribution for $q=1$ and a Poisson distribution for $q\rightarrow \infty$. 

In the Gaussian approximation of the CGC, we see that $q$ can be computed {\it ab initio}.  The structure of $q$ in Eq.~\eqref{eq:NBD-parameter}, in particular its dependence on the saturation scale $Q_S$, is a result of the IR divergence of the single-inclusive distribution, which is an artifact of the dilute-dilute approximation. Such divergences are absent in the full dense-dense case (all orders in $\rho_{L,H}/\nabla_\perp^2$), which is  a more appropriate framework when transverse momenta $\leq Q_S$ are probed. However in this fully nonlinear framework, shockwave collisions have to be treated numerically employing classical-statistical real-time lattice simulations~\cite{Krasnitz:1998ns}. Such simulations do not qualitatively alter the NBD structure obtained in the analytical computations~\cite{Lappi:2009xa}. Multiplicity distributions in the CGC EFT have been compared to the data in hadron-hadron collisions at colliders~\cite{Tribedy:2011aa,Schenke:2013dpa}. When such distributions are plotted in terms of the  Koba-Nielsen-Olesen (KNO) scaling variable $n/{\bar n}$~\cite{Dremin:2000ep}, good agreement is found at colliders~\cite{Tribedy:2011aa,Schenke:2013dpa} for moderate values of this ratio, but deviate significantly for larger values corresponding to rare events\footnote{The multi-particle production computation in the CGC EFT is valid parametrically for $\Delta Y \leq 1/\alpha_S$. Some of the data is over wide ranges in rapidity where corrections to these expressions may be of quantitative importance.}. This can signal the breakdown of the classical statistical framework, requiring modifications to the framework to treat such events. An example of such  configurations are the pomeron loops we mentioned previously that go beyond our ``mean field" treatment.

We will conclude this section by discussing the connections of this formalism with that of the AGK cutting rules in reggeon field theory we alluded to previously in this section, and in the gravity context, in Sec.~\ref{sec:classicalization}. Instead of formulating our discussion in terms of multiplicity moments, we can directly formulate it in the language of $2\rightarrow n$ probabilities, where unitarity constraints, for instance, are more transparent. In the SK closed time path formalism, the $n$-particle probability for strong time-dependent sources can be expressed as~\cite{Gelis:2006yv} 
\begin{eqnarray}
\label{eq:master-Pn}
    P_n = \exp(-a/g^2)\sum_{p=0}^n \frac{1}{p!}\sum_{\alpha_1+\cdots\alpha_p=n} 
    \frac{b_{\alpha_1}\cdots b_{\alpha_p}}{g^{2p}}\,.
\end{eqnarray}
In this compact expression, $a$ is the imaginary part of the sum over all connected vacuum-vacuum graphs in the presence of external sources. (In the absence of such sources, it would be zero.) The sum in $p$ is over the 
number of disconnected graphs producing the $n$ particles. Lastly, $b_r$ is the probability corresponding to the sum of all $r$-particle cuts that contribute to a given multiplicity. With this definition, $a= \sum_r^\infty b_r$. As a simple example illustrating this formula, $b_1$ is the sum over all 1-particle cuts that contribute to $P_1$ but $b_1^n/n!$ also contributes to $P_n$. Indeed $P_n = e^{-b_1/g^2}\frac{1}{n!}\big(\frac{b_1}{g^2}\big)^n$ is just the Poisson distribution if $b_r=0$ for $r\geq 2$. This is a good approximation to $P_n$  due to the power counting in $g$ for strong sources $O(1/g)$; however, it leads to a significant error in the moments of the multiplicity. For instance,  ${\bar n} = g^{-2}\sum_r r b_r$, can receive a significant quantitative correction depending on the theory specific relative size of the multi-particle cuts $b_r$ ($r>1$) to $b_1$. Likewise,  for the connected $p$th moments 
$\langle n^p\rangle_{\rm conn.} = g^{-2}\sum_r r^p b_r$.

The master formula in Eq.~\eqref{eq:master-Pn} contains a wealth of useful information. For instance, one can expand out $a$ (the sum of the imaginary part of all connected vacuum-vacuum graphs) in the exponential and write the formula as 
\begin{eqnarray}
\label{eq:AGK1}
    P_n = \sum_{m=0}^{\infty} P_{n,m} \,\,\,\,{\rm with}\,\,\,\, P_{n,m} = \frac{1}{g^{2m}}\sum_{p+l=m}
    \frac{(-a)^l}{l!}\frac{1}{p!}\sum_{\alpha_1+\cdots+\alpha_p=n} 
    b_{\alpha_1}\cdots b_{\alpha_p}\,.
\end{eqnarray}
In the expression on the r.h.s., $m$ represents the number of disconnected vacuum-to-vacuum diagrams contributing to $P_n$ (with each disconnected diagram starting at $O(1/g^2)$), $l$ is the number of such ``uncut" diagrams  from expanding out the exponential to $l$'th order, and $p$ denotes the number of cut subdiagrams that contribute to the multiplicity. In contrast, the $l$ diagrams do not contribute to the multiplicity but are ``shadowing" or absorptive contributions. In the language of AGK~\cite{Abramovsky:1973fm}, $m$ represents the number of reggeons, $l$ the number of uncut reggeons and $p$ the number of cut reggeons. From Eq.~\eqref{eq:AGK1}, 
\begin{eqnarray}
    \label{eq:AGK2}
    \sum_{n=1}^{\infty}n P_{n,m} =0\,, \,\forall \,\,m\geq 2,\,\,\,{\rm and}\,\,\, \sum_{n=1}^{\infty}n (n-1) P_{n,m} =0\,,\, \forall\, \,m\geq 3,\,\,\, {\rm etc.}
\end{eqnarray}
In arriving at this relation, we implemented the unitarity condition $P_n=1$, which gives  $a=\sum_r b_r$ from Eq.~\eqref{eq:master-Pn}. 
This result has a simple theory-independent interpretation. Two or more disconnected sub-graphs do not affect the mean multiplicity, only the 1-particle irreducible graphs; likewise, three or more disconnected graphs do not impact the variance, and similarly,  for the higher moments. 

An equivalent statement one can extract from Eq.~\eqref{eq:master-Pn}~\cite{Gelis:2006yv} is that diagrams with two or more cut disconnected sub-graphs cancel in the computation of the multiplicity, and so on. In particular, ${\bar n} = \langle n_{\rm cut}\rangle \langle n\rangle_1$: the average multiplicity is the average number of cut sub-diagrams ($= a$) times the average multiplicity of an individual cut sub-diagram. This is unsurprising since the multiplicity of cut sub-diagrams (unlike the particle multiplicity) is Poissonian; the disconnected graphs are by definition independent of each other. The reason that one obtains these model-independent relations is that one has summed over the multiplicity in obtaining them. A corollary is that they contain limited information on the dynamics of the theory, which instead is contained in the $b_r$ in Eq.~\eqref{eq:master-Pn} specifying the magnitude of the intrinsic $r$th connected sub-graph squared amplitudes as an expansion in $g$. 

For a more microscopic understanding of Eq.~\eqref{eq:gluon-single-inclusive}, recall the LSZ relation\footnote{For this discussion, to avoid clutter with indices we will work with a self-interacting scalar theory. The arguments generalize straightforwardly to gauge theories.}
\begin{eqnarray}
    \langle p_1\cdots p_n\, {\rm out}|0_{\rm in}\> = \frac{1}{Z^{n/2}}\left[\prod_i^n d^4 x_i \,e^{ip_i\cdot x_i}(\Box_{x_i}+m^2)\frac{\delta}{i\delta J(x_i)}\right]\exp\left(i{\cal Z}(J)\right)\,.
\end{eqnarray}
Here $Z$ is the wavefunction renormalization constant and ${\cal Z}(J)$ is the sum over all connected vacuum-vacuum graphs. The probability to produce $n$ particles is 
\begin{eqnarray}
    P_n = \frac{1}{n!}\int \prod_i^n \frac{d^3 p_i}{(2\pi)^3 2 E_i}| \langle p_1\cdots p_n\, {\rm out}|0_{\rm in}\rangle|^2\,,
\end{eqnarray}
which can be reexpressed as~\cite{Gelis:2006yv}
\begin{eqnarray}
    P_n=\frac{1}{n!} {\cal D}^n[J_+,J_-]\exp\left(i {\cal Z}[J_+]-i{\cal Z}^*[J_-]\right)|_{J_-=J_+= J}\,.
\end{eqnarray}
Here $+$ and $-$ denote the sources on the upper  and lower parts of the SK contour, which are set to the physical value $J$ at the end of the computation, and 
\begin{eqnarray}
    {\cal D}[J_+,J_-]=\int d^4 x \int d^4 y \,Z\,G_{+-}^0(x,y)\, \frac{(\Box_{x}+m^2)}{Z} \frac{(\Box_{y}+m^2)}{Z}\frac{\delta}{i\delta J_+(x)}\frac{\delta}{i\delta J_-(y)}\,,
\end{eqnarray}
with the SK cut propagator (or Wightman function) defined to be 
\begin{eqnarray}
  G_{+-}^0(x,y)=\int   \frac{d^3 p}{(2\pi)^3 2 E_p}\, e^{ip\cdot (x-y)}\,.
\end{eqnarray}

Defining the generating function, 
\begin{eqnarray}
    {\cal F}(z) = \sum_n z^n P_n = e^{z {\cal D}[J_+,J_-]}e^{i {\cal Z}[J_+]}e^{-i{\cal Z}^*[J_-]} = e^{{\cal Z}_{\rm SK}[z,J_+,J_-]}\,,
\end{eqnarray}
we obtain,
\begin{align}
\label{eq:SK-mult}
    {\bar n} =& \int d^4 x \int d^4 y \,Z\,G_{+-}^0(x,y)\, \frac{(\Box_{x}+m^2)}{Z} \frac{(\Box_{y}+m^2)}{Z}\nonumber\\
    &\times\left[\frac{\delta i{\cal Z}_{\rm SK}[z=1,J_+,J_-]}{\delta J_+(x) }\frac{\delta i{\cal Z}_{\rm SK}[z=1,J_+,J_-]}{\delta J_-(y)}+\frac{\delta^2 i{\cal Z}_{\rm SK}[z=1,J_+,J_-]}{\delta J_+(x) J_-(y)}\right]_{J_+=J_-=J}\,.
\end{align}
We note that all moments of the multiplicity can similarly be generated by further functional differentiation of ${\cal Z}_{\rm SK}$ with respect to $J_+/J_-$. These generate in-in connected correlators in the strong field vacuum as opposed to the time-ordered in-out connected correlators that appear in the LSZ computation of $P_n$. 

The LO computation of ${\bar n}$ is of order $g^{-2} (gJ)^n$ and only involves the first term in the brackets on the r.h.s of Eq.~\eqref{eq:SK-mult}. This contribution can be understood as follows. Taking the functional derivative of ${\cal Z}_{\rm SK}$ with respect to $J_+$ and, likewise with respect to $J_-$, amputates a source, replacing it with the fields $\phi_+$ and $\phi_-$, respectively, which are sewn together with the cut propagator $G_{+-}^0$. Each of these fields connects to arbitrary numbers of sources summed over all possible $\pm$ signs for the intermediate vertices. Because of the identities $G_{++}^0- G_{+-}^0 = G_R^0=G_{-+}^0 - G_{--}^0$, where $G_{++}^0 (G_{--}^0)$ are the time ordered (anti-time ordered) Feynman propagators, the successive sum over $++$ and $+-$ (or $--$ and $-+$) vertices converts all the intermediate propagators into the retarded propagators $G_R^0$. One then obtains 
\begin{align}
    {\bar n}_{\rm LO} = \int d^4 x \int d^4 y \,G_{+-}^0(x,y)\, (\Box_{x}+m^2) (\Box_{y}+m^2)\phi_{\rm cl.}(x)\phi_{\rm cl}(y)\,.\nonumber\\
\end{align}
For initial conditions at $t=-\infty$, where the field and its first derivative vanish, using integration by parts, this equation can be reexpressed as
\begin{align}
\label{eq:classical-SK}
{\bar n}_{\rm LO}=\int \frac{d^3 p}{(2\pi)^3 2 E_p} \bigg|\lim_{t\rightarrow \infty} \int d^3 x\, e^{ip\cdot x}(\partial_t - iE_p)\phi_{\rm cl}(x)\bigg|^2 \,.
\end{align}
The expression $|\cdots|^2$ is nothing but $\langle 0_{\rm in}|a^\dagger(p)a(p)|0_{\rm in}\rangle$, as one might anticipate.

For classical Yang-Mills, the equivalent expression is Eq.~\eqref{eq:gluon-single-inclusive}, with the gauge field in the forward light cone given by Eqs.~\eqref{gaugeLV} and \eqref{dilutedenseQCD}, for the dilute-dilute and dilute-dense solutions of the classical Yang-Mills equations. In the dense-dense case, as noted, the Yang-Mills equations can only be solved numerically. Diagrammatically, this would correspond to having arbitrary numbers of classical fields/reggeized gluons from both upper and lower sources rescattering off the produced gluons.  Their solution leads to the overoccupied non-equilibrium glasma, which flows to a nonthermal fixed point~\cite{Berges:2013eia,Berges:2013fga,Berges:2014yta} before thermalizing to form a quark-gluon plasma (QGP).
Remarkably, one sees rapid thermalization at weak coupling $\alpha_S(Q_S)\ll 1$ in the Regge limit, as long as $\alpha_S N\sim O(1)$, where $N$ here is the phase space occupancy. A review of this nonequilibrium dynamics and  interdisciplinary connections is provided in \cite{Berges:2020fwq}.

Going beyond LO, both terms in the bracket in Eq.~\eqref{eq:SK-mult} contribute. The first term has the structure $\phi_{\rm quant}\phi_{\rm cl}$ and the second term is the cut propagator $G_{+-}$ in the background field. We discussed the latter previously in Sec.~\ref{sec:CGC-propagators} for the background of a single shockwave. The computation for the case of two shockwaves is similar and has been discussed previously in \cite{Gelis:2006yv,Gelis:2008ad,Gelis:2008rw}. One still has an initial value problem where instead of solving the classical equations of motion for the classical field at LO, one has to solve the small fluctuation equations of motion for the small fluctuation field, with plane wave initial conditions. A similar but more involved set of steps can be employed to solve for $\phi_{\rm quant}$; in this case, one has UV divergences from the closed loop that have to be regulated. However, to leading logarithmic accuracy, one simply obtains the result in Eq.~\eqref{eq:kth-moment-gluons}. 

In summary, to one-loop accuracy (and leading logarithmic contributions to all orders in a limited kinematic region in rapidity) one can compute moments of the multiplicity as an initial value problem, solving classical equations of motion and small fluctuation equations. This is equivalent to a statement of the Feynman tree theorem~\cite{Feynman:1963ax}. As emphasized in \cite{Caron-Huot:2010fvq}, the extension of this framework to two loops is nontrivial. This is because, as we noted previously in Sec.~\ref{sec:JIMWLK},  genuine $\hbar$ corrections of order $\phi_a^3$ in the difference fields contribute; this requires extensions of the Schwinger-Keldysh framework beyond the classical-statistical approximation.

\section{Trans-Planckian gravitational scattering of shockwaves}
\label{sec:GR-shockwave-formalism}

In Secs.~\ref{sec:2} and ~\ref{sec:3}, we discussed  $2\rightarrow 2+n$ scattering in QCD and in Einstein gravity, respectively, in the multi-Regge asymptotics within the framework of perturbation theory around Minkowski spacetime. While this was crucial for computing the $n$ particle contribution to the $2\to 2$ scattering in the leading log approximation, these methods encounter significant limitations when applied to very high-energy regimes. The inherent nature of these high-energy interactions, characterized by copious parton production, drives the evolution of produced particles towards over-occupied dense states. This extreme density (or over-occupancy) fundamentally challenges the applicability of the perturbative framework, as its underlying assumptions, valid at weak coupling and small occupancies, break down under such conditions. 

The inability of conventional perturbation theory to accurately describe these dense environments necessitates the exploration of nonperturbative techniques. In Sec.~\ref{sec:CGC}, we discussed how this problem was tackled in the case of gauge theory using semi-classical techniques that led to a shockwave picture of high energy scattering. In this semi-classical shockwave framework, we showed that  the Balitsky-Fadin-Kuraev-Lipatov (BFKL) equation, which drives the evolution of parton distributions towards high densities at high energies, can be  recovered by taking a low-density limit. Conversely, this Color Glass Condensate EFT facilitated the derivation of nonlinear completions of the BFKL equation, providing a robust solution to the unitarity problem that plagued earlier perturbative descriptions. 

Further, we showed that the Lipatov vertex describing the single inclusive gluon radiation spectrum is recovered in so-called dilute-dilute and dilute-dense occupancy semi-classical limits in the shockwave collisions. The dense-dense single inclusive spectrum is fundamentally nonperturbative but can be computed numerically. We  discussed how one understands $2\rightarrow n$ scattering in the shockwave framework, and provided a complementary QFT understanding and generalization of the AGK rules employed in the reggeon field theory literature. We showed that multi-particle production in strong fields can be computed, to all orders in leading logarithmic accuracy, in a classical-statistical formalism; the spacetime evolution of operators, for each configuration of sources, are described by solutions of classical equations of motion with retarded boundary conditions. Because of the similarities between QCD and gravity developed in the previous sections, it appears likely that insights from this EFT description of QCD in Regge asymptotics can be applied to gravity. 

Specifically, given the common dispersive techniques in the Regge limit, and the emergent double-copy relations,  a compelling objective is to establish an analogous {\it ab initio} EFT framework within Einstein gravity. The purpose of this section is to take steps in this direction. A specific aim is to show that the gravitational Lipatov vertex, a crucial element in describing particle emission in multi-Regge kinematics, can be recovered through the computation of radiation produced in gravitational shockwave collisions, as was the case for gluon shockwave collisions.  The subsequent subsections will detail shockwave derivations, and discuss the implications of this semi-classical CGC-inspired approach to high-energy scattering in gravity. We will also discuss further its connections to the ACV framework for trans-Planckian gravitational scattering. 

\subsection{Gravitational shockwave properties}

Before addressing the problem of recovering the gravitational Lipatov vertex in gravitational shockwave collisions, we will briefly review the properties of the shockwave. Gravitational shockwaves are exact solutions of Einstein's equations in the presence of delta-function sources that represent ultrarelativistic particles propagating at the speed of light. They are described by the Aichelburg and Sexl (AS) metric \cite{Aichelburg:1970dh} given by
\begin{align}
\label{ASmetric}
ds^2 = &~ 2\,dx^+dx^- -\delta_{ij}dx^i dx^j + 8\,\mu_H \,G \,\delta(x^-)\log(\Lambda |\bsx|) \(dx^-\)^2~.
\end{align}
Here $\Lambda$ is an IR cutoff scale. This spacetime is a solution to Einstein's equation with a nonvanishing energy-momentum (EM) tensor given by 
$$
T_{\m\n} = \mu_H\delta_{\mu-}\delta_{\nu-} \delta(x^-) \delta^{(2)}(\bsx)~.
$$ 
This is the EM tensor of a massless point particle moving in the positive $z$ direction carrying energy $\mu_H$ and located at the origin $\bsx=0$ in transverse space. 

The point particle approximation is accurate only in classical mechanics where there is no particle production. This is problematic for shockwave collisions since the classical spacetime in the post-collision region has large curvature along the transverse collision plane; one therefore  cannot ignore  quantum gravitational effects, in particular higher derivative corrections to Einstein's gravity \cite{Rychkov:2004sf, Rychkov:2004bm}. This argument nevertheless does not invalidate the use of semi-classical frameworks of black hole formation in the scattering of highly boosted particles. If one smears the $\delta(x^-)$ functions, giving it some width (along the $z$ direction), the curvatures remain small, allowing  the classical theory to remain applicable  post-collision even at small impact parameters \cite{Giddings:2004xy}. 

While the longitudinal spread in the EM current is a desired property, it is not crucial if impact parameters are large compared to $R_S$ where curvatures are small. There is however another feature of the EM current that one would like to incorporate, which is the spread in the transverse direction. Since the incoming particles are highly boosted they form a large occupancy cloud of gravitons resulting from bremsstrahlung across a wide range of rapidities. This requires us to treat their dynamics as a static extended mass distribution in the transverse directions. This point is analogous to the discussion in the CGC where such nontrivial transverse source profiles are automatically generated via the rapidity renormalization group equations. 

Therefore, motivated by this discussion, we will generalize the above form of the EM current to a source with a generic transverse spatial density $\rho_H(\bsx)$ with the shockwave profile
\begin{align}
\label{EMtensor1}
T_{\m\n} = \delta_{\mu-}\delta_{\nu-} \mu_H \delta(x^-) \rho_H(\bsx)~,
\end{align}
with a more general shockwave spacetime
\be
\label{denseBgnd1}
ds^2 = 2dx^+dx^- -\delta_{ij}dx^i dx^j + f(x^-,\bsx)\(dx^-\)^2 ~.
\ee
where $f(x^-,\bsx)$ is given by
\begin{align}
\label{backgroundg}
f(x^-,\bsx) &= 2\kappa^2\mu_H \delta(x^-)\frac{\rho_H(\bsx)}{\nabla_\perp^2} = \frac{\kappa^2}{\pi}\mu_H \delta(x^-) \int d^2\bsy ~\ln\Lambda |\bsx-\bsy| \rho_H(\bsy)~,
\end{align}
We used the Green function of the two-dimensional Laplacian $\nabla_\perp^2\equiv\delta_{ij}\p_i\p_j$ in the second equality. This form of the metric is analogous to the classical shockwave field $A^+$ in Eq.~\eqref{eq:classical-YM-Lorentz}.

As is apparent from Eq.~\eqref{denseBgnd1}, the spacetime is flat everywhere except at $x^-=0$ where the curvature (field strength) is infinite. Further, the vacua in the regions $x^->0$ and $x^-<0$ are not identical but are related by a coordinate transformation of the Minkowski vacuum. This is to be expected intuitively since the passing shock should affect spacetime measurements differently in these regions \cite{Dray:1984ha,tHooft:1996rdg}. Let us define a $y$-coordinate system, related to the $x$-coordinate in which the metric Eq.~\eqref{denseBgnd1} is written, by the discontinuous transformation 
\begin{align}
\label{xytransformation}
\begin{split}
x^- =& ~ y^-,\qquad x^i = ~ y^i - \kappa^2\m_H y^- \Theta(y^-)\frac{\p_i}{\nabla_\perp^2} \rho_H(\bsy) ~,\\[5pt]
x^+ =& ~ y^+ - \kappa^2\m_H \Theta(y^-) \frac{\rho_H(\bsy)}{\nabla_\perp^2} + \frac12  \kappa^4 \m_H^2~y^-\Theta(y^-) \(\frac{\p_i}{\nabla_\perp^2} \rho_H(\bsy)\)^2~.
\end{split}
\end{align}
This transformation follows from the solution of the null geodesics equation for the spacetime Eq.~\eqref{denseBgnd1}. The metric in the $y$-coordinate system then takes the form
\begin{align}
\label{denseBgnd2}
    ds^2 = 2dy^+dy^- -g_{ij}dy^idy^j~,
\end{align}
where $g_{ij}$ now depends nonlinearly on the source density $\rho_H$
\begin{align}
    g_{ij} = &~ \delta_{ij}-y^-\Theta(y^-)\bigg[2\kappa^2\m_H~\frac{\p_i\p_j}{\nabla_\perp^2}\rho_H(\bsy)- \kappa^4 \m_H^2~y^-\(\frac{\p_i\p_k}{\nabla_\perp^2}\rho_H(\bsy)\) \(\frac{\p_j\p_k }{\nabla_\perp^2}\rho_H(\bsy)\)  \bigg]~.
\end{align}

The spacetime Eq.~\eqref{denseBgnd2} is again an exact solution of Einstein's equations with the EM tensor in Eq.~\eqref{EMtensor1}. This form of the shockwave metric makes manifest that the region in front of the shock ($y^-<0$) is the Minkowski vacuum while the region after the shock ($y^->0$) is a pure gauge transformation of the Minkowski vacuum. To confirm the latter, one can explicitly compute the Riemann tensor $R_{\m\n\rho\sigma}$  of the metric in the $y^->0$ region, which vanishes even though its connection coefficients do not vanish. This discussion is analogous to the shockwave solution in the gauge theory case for which the field strength tensor vanishes before and after the shockwave even though the gauge fields do not vanish. They are distinct pure gauge solutions separated by the gluon shockwave~\cite{McLerran:1993ka,McLerran:1993ni}.

AS shockwaves can also be obtained from boosting a Schwarzschild black hole of mass $m_H$. To see this, one starts with the Schwarzschild metric in the isotropic coordinate system
\be
\label{SBH-metric}
ds^2 \equiv -\left(\frac{1-Gm_H / 2 r}{1+Gm_H / 2 r}\right)^2 d t^2 +\left(1+\frac{Gm_H}{2 r}\right)^4\left(dr^2+r^2\left(d \theta^2+\sin ^2 \theta d \phi^2\right)\right)\,.
\ee
We then make an appropriate coordinate transformation \cite{Aichelburg:1970dh} such that the black hole is boosted in the positive $z$ direction with the boost parameter $\beta = (1-\gamma^2)^{-1/2}$ ($\gamma$ being the boost factor). Keeping the energy of the black hole $(\mu_H = \gamma m_H)$ fixed, and taking $\beta\to 1$ limit, results in the shockwave metric in Eq.~\eqref{ASmetric}.

The AS shockwave serves as a background metric for the eikonal approximation in high energy scattering \cite{Akhoury:2013yua}. Further, the propagation of a probe particle through the shockwave background induces a phase shift (well-known in the GR literature as the Shapiro time delay), directly yielding the eikonal amplitude \cite{tHooft:1987vrq}. This connection highlights the importance of shockwaves in bridging classical gravitational physics to scattering amplitudes in the ultrarelativistic limit. In the ensuing subsection, we will address gravitational shockwave collisions and recover the gravitational Lipatov vertex in the radiation spectrum.

\subsection{Gravitational Lipatov vertex from shockwave collisions}
The problem of gravitational shockwave collisions has been studied extensively in the literature. In particular, D'Eath and Payne \cite{DEath:1992plq, DEath:1992mef, DEath:1992nmz} developed a perturbative framework for analytically computing the spacetime metric in the future light cone following the collision of two AS shockwaves. Their approach employs a perturbative ansatz for the metric, expressed as $g_{\mu \nu}=\eta_{\mu \nu}+\sum_{k=1}^{\infty} h_{\mu \nu}^{(k)}$, where $h_{\mu \nu}^{(k)}$ is the $k$-th order contribution to the metric perturbation. The initial data for this expansion are specified on a null hypersurface of the background spacetime incorporating the exact solution of Einstein's equations before the collision. The vacuum Einstein field equations are then solved iteratively, order by order, as a tower of wave equations. The source term for each order is generated by the lower-order perturbations, allowing for a systematic solution using the Green function method. 

A primary observable in these collisions is the inelasticity ($\ep$), defined as the percentage of the initial center-of-mass energy that is radiated away as gravitational waves. D'Eath and Payne's first-order perturbation theory yielded $\ep = 20\%$. Their more precise second-order result gave $\ep_{\text{2nd order}} = 16.4\%$, showing good agreement with subsequent numerical relativity results \cite{Sperhake:2008ga}. These results are often compared against theoretical upper bounds on radiated energy, such as those derived from apparent horizons\footnote{An apparent horizon is a surface in spacetime where outgoing null (light) rays instantaneously stop expanding and remain parallel. The presence of an apparent horizon signals the inevitable formation of an event horizon.} ($\ep \sim 29.3\%$) \cite{Penrose, Eardley:2002re}. Further, the hoop conjecture for black hole formation, requiring that the energy in such collisions be compressed within a hoop of circumference $2\pi R$, where $R\leq R_S$, provides a critical 
$\gamma$ above which black hole formation must take place~\cite{Klauder:1972je}, Numerical GR simulations of head-on collisions of boosted bosonic stars show that this occurs already at boosts about 1/3 the critical value~\cite{Choptuik:2009ww}; see also \cite{Rezzolla:2012nr,East:2012mb,Page:2022bem}.

A crucial aspect of black hole formation through gravitational shockwave collisions is the appearance of large curvatures and large quantum fluctuation on this apparent horizon~\cite{Eardley:2002re,Rychkov:2004sf, Rychkov:2004bm}. As discussed previously, one may  question the validity of the classical approximation in this process. In \cite{Giddings:2004xy}, the curvature invariant $R_{\m\n\rho\sigma}R^{\m\n\rho\sigma}$ in the collision region was studied to determine the regime of validity of the semi-classical computation. It was established that the classical approximation is valid when $b\gg R_S$. In the $b\sim R_S$ region, relevant for black hole formation, its important to treat particles as wave packets/extended matter distributions; in this case, the formation of trapped surfaces is accompanied by low curvatures and small quantum fluctuations~\cite{Giddings:2004xy}. However, since in this section our focus will be in the regime $b>R_S$, we can continue to work with the AS metric in Eq.~\eqref{ASmetric}; one is then not sensitive to the black hole formation in the final state. This simplification allows us to analytically compute gravitational radiation produced in gravitational shockwave collisions thereby recovering the gravitational Lipatov vertex.  We will revisit this issue in subsequent work. 

For our purpose here, since the covariant form of the gravitational Lipatov vertex in Eq.~\eqref{gravitational-lipatov-vertex} is quite complicated in form, it will be useful to recast it in light cone gauge where its form simplifies considerably. Let $\eps_\m(k)$ be the gluon polarization vector and $\eps_{\m\n}(k)$ be the graviton polarization tensor such that (suppressing helicity labels) $\eps_{\m\n}(k) \sim \eps_\m(k) \eps_\n(k)~,~\eps_\m(k)k^\m=0$. The precise relation between the two was given previously in Eq.~\eqref{graviton-pol}.  In  light cone gauge, the polarization tensor is $\eps_{+\mu}=0$. 

Recall the form of the gravitational Lipatov vertex
$$
C^{\m\n} = \frac12(C^\m C^\n - N^\m N^\n)~,
$$
with $C^\m$ given in Eq.~\eqref{QCD-LV-cov} and $N^\m$ in Eq.~\eqref{QED-factor}. Contracting $N^\mu(k_1, k_2)$ with the gluon polarization vector in light cone gauge gives
\begin{align}
\label{QCDLVLC2}
\ep^*_{\m}(k) N^\m(k_1,k_2) = -2\sqrt{\bsk_{1}^2\bsk_{2}^2} \frac{\ep^*_i \ell_i}{\bel^2} \equiv \ep^*_i N_i(\bsk_1,\bsk_2)~,
\end{align}
with $\ell=k_1+k_2$\footnote{The sign convention of $k_2$ in this section is opposite to that in Sec. \ref{sec:3}.}.
Using this result, the expression in Eq.~\eqref{QCDLVLC1} for the QCD Lipatov vertex in light cone gauge, and the gauge invariance condition $\Gamma_{\m\n}\ell^\n=0$, one can obtain all nonvanishing components of the gravitational Lipatov vertex in light cone gauge. These are as follows:
\begin{align}
\label{LCLV1}
C_{ij}(\bsk_1,\bsk_2)&=2\(k_{2i}-\ell_i\frac{\bsk_{2 }^{2}}{\bel^{2}}\) \(k_{2j}-\ell_j\frac{\bsk_{2 }^{2}}{\bel^{2}}\) - 2\ell_i\ell_j\frac{\bsk_{1}^2\bsk_{2}^2}{\bel^4}\,,\\[5pt]
\label{LCLV2}
C_{-i}(\bsk_1,\bsk_2)&=\frac{4\ell_-}{\bel_\perp^2}\bigg[ (\boldsymbol{k}_1\cdot \boldsymbol{k}_2) \(k_{2i}-\ell_i\frac{\bsk_{2 }^{2}}{\bel^{2}}\) -\ell_i\frac{\bsk_{1}^2\bsk_{2}^2}{\bel^2}\bigg]\,,\\[5pt]
\label{LCLV3}
C_{--}(\bsk_1,\bsk_2)&=\frac{8\ell_-^2}{\bel_\perp^4}\bigg[(\boldsymbol{k}_1\cdot \boldsymbol{k}_2)^2 -\bsk_{1}^2\bsk_{2}^2\bigg]~.
\end{align}

\begin{figure}[ht]
\centering
\includegraphics[scale=1]{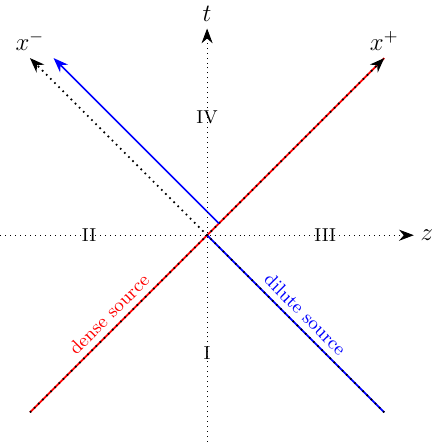}
\caption{Trajectories of colliding gravitational shockwaves. The trajectory in red (blue) is the shock with the dense (dilute) source $\rho_H$ ($\rho_L$). In the dilute-dense approximation we treat $\rho_L$ as a probe in the background created by $\rho_H$. The shift in the blue trajectory is the Shapiro time delay to the light shockwave.}
    \label{spacetime1}
\end{figure}

We will  now outline the derivation~\cite{Raj:2023irr} of these expressions in a semi-classical framework of colliding gravitational shockwaves in Einstein-Hilbert gravity in a manner exactly analogous to the QCD discussion in Sec.~\ref{sec:CGC}. We consider two incoming shocks from opposite directions along the $z$ axis. We take one of these shocks to be ``heavy" (or dense in the QCD parlance we introduced in Sec.~\ref{sec:CGC}) and the other ``light" (or dilute). The heavy shockwave acts like a background for the `light' shockwave which is a probe. These shockwaves are separated by the impact parameter $b$ in the transverse plane. The collision point is at $z=t=0$. The spacetime diagram for this process is depicted in Fig.~\ref{spacetime1}.

The EM tensor sourcing these shockwaves is given by,
\begin{align}
\begin{split}
\label{EMtensor2}
T_{\m\n} = &~ \delta_{\mu-}\delta_{\nu-} \m_H \,\delta(x^-) \rho_H(\bsx) +\delta_{\mu+}\delta_{\nu+} \m_L \delta(x^+) \rho_L(\bsx)~.
\end{split}
\end{align}
This expression is valid in the pre-collision region $t<0$ where the waves have not yet collided. In the region $t>0$, the EM tensor will get corrections due to the backreaction of the changes in the spacetime. This is depicted in Fig.~\ref{spacetime1} where the region I is the common Minkowski vacuum shared by both the shockwaves and regions II and III are spacelike regions of respectively dense and dilute shockwaves before the collision that correspond to the respective coordinate transformation of Minkowski vacuum (as discussed in the previous section). Finally, region IV corresponds to the future of the collision point in which the EM tensor of each of the shocks will get modified and backreact to create a radiative spacetime with a nonvanishing curvature.

We begin by setting up the equations that govern the changes in the spacetime metric in the dilute-dense approximation. Ultimately, our goal will be to calculate the correction to the EM tensor and the modification to the metric in region IV in the dilute-dilute approximation where we keep terms to linear order in $\rho_H$ and $\rho_L$ in the solution. This approximation will be sufficient for recovering the gravitational Lipatov vertex. The derivation outlined below is discussed in more detail in \cite{Raj:2023irr}.


\subsubsection{Equations of motion}
\label{sec:EOM}

Treating the spacetime created by shockwave $H$ as background ($\bar{g}_{\m\n}$), we consider small perturbations $h_{\m\n}$ around it:
\begin{align}
\label{perturb}
g_{\m\n} = \bar{g}_{\m\n} + h_{\m\n}~.
\end{align}
Here $\bar{g}_{\m\n}$ is given by Eq.~\eqref{denseBgnd1}. The perturbation $h_{\m\n}$ is further decomposed into a term $h_{\mu\nu}^{(1)}$ linear in $\rho_L$ and a term $h_{\mu\nu}^{(2)}$ of order $O(\rho_H\rho_L)$: 
\begin{align}
\label{eq:dilute-dense-expansion}
    h_{\mu\nu} = h_{\mu\nu}^{(1)}+\kappa \,h_{\mu\nu}^{(2)}~.
\end{align}
The former is sourced by the $T_{++}$ part of the pre-collision EM tensor in Eq.~\eqref{EMtensor2}, whereas the latter comes from the backreaction of the corrected EM tensor at this order. In light cone gauge ($h_{\mu+}=0$), linearizing Einstein's field equations gives, for the traceless field $\tilde{h}_{ij}$ (defined as $\tilde{h}_{ij} \equiv h_{ij}-\frac12 \delta_{ij} h~, h=\delta_{ij}h_{ij}$),
\begin{align}
\label{metricEq}
&\bar{g}_{--}\p_+^2 \tilde h_{ij}- \square \tilde h_{ij}= \kappa^2\bigg[\(2\p_i\p_j-\nabla_\perp^2 \delta_{ij}\)\frac{1}{\p_+^2}T_{++}  +2T_{ij}-\delta_{ij}T-\frac{2}{\p_+} \(\p_iT_{+j}+\p_jT_{+i}-\delta_{ij}\p_k T_{+k}\) \bigg]~.
\end{align}
Here $\square$ is the d'Alembertian operator in Minkowski background and $T\equiv \delta_{ij}T_{ij}$. This equation is accompanied by the first order equations of the remaining components of the metric:
\begin{align}
\label{constraintrel}
\p_+h_{-i} &=\p_j\tilde h_{ij}+\kappa^2\[\frac{2}{\p_+}T_{+i} - \frac{\p_i}{\p_+^2} T_{++}\]~\,,\\[5pt]
\p_+^2 h_{--} &= \p_i\p_j\tilde h_{ij} -\kappa^2\[\frac{\nabla_\perp^2}{\p_+^2} T_{++}  - T  - 2T_{+-} +\bar{g}_{--} T_{++}\]~.
\end{align}
To solve these equations, we need to determine how the various components of the energy momentum tensor evolve into region IV. For this, we will need to consider the geodesic motion~\cite{Taliotis:2010pi, Constantinou:2013tia, Goldberger_2017}  of the ultrarelativistic distribution of particles $\rho_L$ as they cross the shockwave background created by $\rho_H$.


\subsubsection{Spacetime evolution of the energy-momentum tensor}
\label{sec:EM-tensor-spacetime}
We need to derive how the EM tensor of a shockwave $L$ (comprised of ultrarelativistic particles) changes as it passes through the gravitational field of shockwave $H$. For simplicity, we treat shockwave $L$ as a point particle, represented by a delta function: $\rho_L(\bsx)=\delta^{(2)}(\bsx-\bsb)$. Since this point particle will follow a geodesic trajectory when it crosses shockwave $H$, one can express the EM tensor as a function of this geodesic as
\be
\label{PPEMT}
T^{\m\n}(x) = \frac{\mu_L}{\sqrt{-\bar{g}}} \int_{-\infty}^\infty d\l~ \dot{X}^\m \dot{X}^\n~ \delta^{(4)}(x-X(\l))~.
\ee
Here $\bar{g}=-1$ is the determinant of the background metric in Eq.~\eqref{denseBgnd1} and the dot denotes differentiation with respect to  the worldline parameter $\lambda$. This expression of the EM tensor follows from the relativistic action of a massless point particle. 

The problem then is to solve the geodesic equation and determine the trajectory $X^\mu(\lambda)$. It is important to acknowledge that this point particle treatment is a simplified model. In our point particle approximation, the transverse source density $\rho_L(\bsx)$ remains unchanged, effectively ``freezing" the transverse dynamics. While this detail is not central to our discussion here, we'll revisit it later when we derive the Lipatov vertex from our shockwave model. A more complete picture would involve the boosted point particles being accompanied by their own ``graviton cloud", as realized in the Faddeev-Kulish construction we discussed in Sec.~\ref{sec:classicalization}. This entire configuration would trace a congruence of null geodesics, rather than a single one. We will return to this topic in Sec.~\ref{sec:Raychaudhuri}.

The energy momentum tensor in Eq.~\eqref{PPEMT} is covariantly conserved only when the worldline $X^\m(\l)$ satisfies the geodesic equation
\be
\label{geodesicEq}
\ddot{X}^\m+\Gamma^\mu_{\nu\rho} \dot{X}^\n \dot{X}^\rho=0~,\qquad g_{\nu\rho}\dot{X}^\n\dot{X}^\rho=0~.
\ee
The second relation ensures that the trajectory is a null geodesic. Solving Eq.~\eqref{geodesicEq} with appropriate boundary condition at negative times one finds \cite{Raj:2023irr}, 
\begin{align}
\begin{split}
\label{geodesicSol}
    X^- &= \lambda~,\\[5pt]
    X^+ &= -  \kappa^2\mu_H \Theta(X^-) \frac{\rho_H(\bsb)}{\nabla_\perp^2} + \frac{\kappa^4 \mu_H^2}{2}X^- \Theta(X^-) \(\frac{\p_i\rho_H(\bsb)}{\nabla_\perp^2}\)^2~,\\[5pt]
    X^i &= b^i -  \kappa^2\mu_H X^-\Theta(X^-) \frac{\p_i\rho_H(\bsb)}{\nabla_\perp^2}~.
\end{split}
\end{align}
It can be checked that the solution satisfies the null constraint in Eq.~\eqref{geodesicEq}. Notice that the null geodesic is continuous along the $x^-$ and transverse directions (as a function of $\lambda$ or $X^-$) but acquires a discontinuity along the $x^+$ direction after crossing the shockwave $H$ at $X^-=0$. This behavior of the null trajectory is shown in Fig. \ref{spacetime1} (where motion in the transverse direction is not shown) and is precisely the content of the coordinate transformation in Eq.~\eqref{xytransformation}.

The result in  Eq.~\eqref{geodesicSol} and Eq.~\eqref{PPEMT} allows us to obtain all the nonvanishing components of particle $L$'s EM tensor in the {\it dilute-dense} approximation. For our presentation here, we require the complete expression for the EM tensor within the dilute-dilute approximation. This requires summing the EM tensor for both particle $H$ and particle $L$. Interestingly, it can be demonstrated that, when working in light cone gauge, particle $H$'s EM tensor gets no corrections from particle $L$'s gravitational field, even under the dilute-dilute approximation. The nonvanishing components of the total EM tensor are 
\begin{align}
\label{EMlower}
\begin{split}
T_{++} &=\mu_L \delta\left(x^{+}\right) \rho_L+\kappa^2 \mu_H \mu_L \Theta\left(x^{-}\right)\left[\delta^{\prime}\left(x^{+}\right) \frac{\rho_H}{\square_{\perp}} \rho_L+x^{-} \delta\left(x^{+}\right) \frac{\partial_i \rho_H}{\square_{\perp}} \partial_i \rho_L\right]~,\\[10pt]
\qquad T_{--} &= \mu_H \delta(x^-)\rho_H~, \\[10pt]
T_{-+} &= \kappa^2\mu_H\mu_L \delta(x^+)\delta(x^-) \frac{\rho_H }{\nabla_\perp^2} \rho_L ~,\\[10pt]
\qquad T_{+i} &= \kappa^2\mu_H\mu_L \delta(x^+)\Theta(x^-) \frac{\p_i\rho_H}{\nabla_\perp^2} \rho_L~.
\end{split}
\end{align}
We have replaced in these results the transverse delta functions with the finite transverse source distribution $\rho_L$. This is a harmless change with the cautionary remark that the transverse dynamics of the source distribution are frozen, which a good approximation for small positive $X^-$.

When working in the point particle approximation at a finite impact parameter, the solution for the EM tensor contains a contact term ambiguity, that is proportional to $\rho_H\rho_L$. This term, expressed as $\delta^{(2)}(\vec{x})\delta^{(2)}(\vec{b}-\vec{x}) = \delta^{(2)}(\vec{b})\delta^{(2)}(\vec{b}-\vec{x})$, technically vanishes for $|\vec{b}|>R_S$, which lies the region where our point particle model is applicable. However, despite vanishing in position space, this ``smeared" provides a finite contribution in momentum space. The precise coefficient of this term will not be fixed by the point particle analysis and the ambiguity in the solution should ideally be resolved by appealing to other physical considerations such as the unitarity of multi-particle production. With this mind, we include the following term 
\be
\label{eq:smearing}
\kappa^2\mu_H\mu_L x^-\Theta(x^-)\delta(x^+) \rho_H\rho_L~,
\ee
in the solution 
which vanishes in the point particle limit for large impact parameters. As a consequence, 
\begin{align}
    T_{++} &= \mu_L\delta(x^+)\rho_L +\kappa^2\mu_H\mu_L \Theta(x^-)\bigg[ \delta'(x^+)\frac{\rho_H}{\nabla_\perp^2} \rho_L + x^-\delta(x^+) \p_i\(\frac{\p_i\rho_H}{\nabla_\perp^2} \rho_L\)\bigg]\,.
\end{align}
We will now show that this addition is essential for obtaining the correct expression for the Lipatov emission vertex.

\subsubsection{Gravitational Lipatov vertex from shockwave collisions }
We start by determining $h_{ij}^{(1)}$ which represents the fluctuation in the background induced by the light shockwave. In this approximation, Eq.~\eqref{metricEq} simplifies to
\be
-\square \tilde h_{ij}^{(1)}= \kappa^2\mu_L x^+\Theta(x^+)\left(2\p_i\p_j-\nabla_\perp^2 \delta_{ij}\right)\rho_L+O(\kappa^4)~.
\ee
We substituted here the $O(\rho_L)$ expression for $T_{++}$ from Eq.~\eqref{EMlower} into the r.h.s. Since it is independent of $x^-$, ${\tilde h}_{ij}^{(1)}$ too is $x^-$-independent. Hence $\square h_{ij}^{(1)}= -\nabla_\perp^2h_{ij}^{(1)}$, which gives
\be
\label{htij1}
\tilde h_{ij}^{(1)} = \kappa^2\mu_L x^+\Theta(x^+)\left(\frac{2\p_i\p_j}{\nabla_\perp^2}-\delta_{ij}\right)\rho_L+O(\kappa^4)~.
\ee

At order $O(\rho_L\rho_H)$, the solution for the field ${\tilde h}_{ij}^{(2)}$ is found by plugging the solution in  Eq.~\eqref{htij1}, the corrected EM tensor from Eq.~\eqref{EMlower}, and Eq.~\eqref{backgroundg} into Eq.~ \eqref{metricEq}, which yields
\begin{align}
\label{metricSol}
\square\tilde{h}_{ij}^{(2)} =& \kappa^3 \mu_H\mu_L\bigg(2\delta(x^+)\delta(x^-) \frac{\rho_H}{\nabla_\perp^2}P_{ij}\frac{\rho_L}{\nabla_\perp^2}-\Theta(x^+)\Theta(x^-)\bigg[P_{ij}\left( \frac{\rho_H}{\nabla_\perp^2} \rho_L + x^+x^- \p_k\left(\frac{\p_k\rho_H}{\nabla_\perp^2} \rho_L\right)\right)\no \\[5pt]
&-2\bigg\{\p_i \left( \frac{\p_j\rho_H}{\nabla_\perp^2}\rho_L \right)+\p_j \left( \frac{\p_i\rho_H}{\nabla_\perp^2}\rho_L \right)-\delta_{ij} \p_k \left( \frac{\p_k\rho_H}{\nabla_\perp^2}\rho_L \right)\bigg\}\bigg]\bigg)~,
\end{align}
where $P_{ij}=2\p_i\p_j-\delta_{ij}\nabla_\perp^2$. This equation can be readily integrated in Fourier space, with the result
\begin{align}
\label{metricSolFt}
k^2 \tilde{h}_{ij}^{(2)}(k) = &\kappa^3 \mu_H\mu_L\int \frac{d^2\bsq_2}{\left(2\pi\right)^2} \frac{\rho_H(\bsq_1)}{\bsq_1^2}\frac{\rho_L(\bsq_2)}{\bsq_2^2}\no\\[10pt]
&\times \bigg( 2P_{ij}(\bsq_2)-\frac{\bsq_2^2}{k_+k_-}\bigg\{P_{ij}(\bsk)\left(1+\frac{\bsk\cdot \bsq_1}{k_+k_-}\right)-2\left(k_i q_{1j}+k_j q_{1i}-\delta_{ij}\bsk\cdot \bsq_1\right)\bigg\}\bigg)\,,
\end{align}
with $P_{ij}(\bsp) \equiv 2\,p_ip_j-\delta_{ij}\bsp^2$. The transverse momenta are constrained by $\bsk=\bsq_1+\bsq_2$. To extract the Lipatov vertex, we need to put the graviton's momentum $k$ on-shell ($2k_+k_- - \bsk^2=0$), where the Lipatov vertex is the residue of the $1/k^2$ pole. After straightforward manipulations of the above expression, we find 
\be
\label{hijfinalresult}
\tilde{h}_{ij}^{(2)}(k) = \frac{2\kappa^3\mu_H\mu_L}{k^2+i\epsilon k^-} \int \frac{d^2\bsq_2}{\left(2\pi\right)^2}\, C_{ij}(\bsq_1,\bsq_2) \frac{\rho_H}{\bsq_1^2}\frac{\rho_L}{\bsq_2^2} ~,
\ee
where $C_{ij}(\bsq_1, \bsq_2)$ is the gravitational Lipatov vertex in the lightcone gauge as defined in Eq.~\eqref{LCLV1}.

Recall that this result relies heavily on the inclusion of the contact term in the $T_{++}$ solution, as discussed previously. Without this term, we would only correctly reproduce the strict Yang-Mills double copy $C^\m C^\n$ part of the Lipatov vertex. However, the $N^\m N^\n$ term, which we understood from Sec.~\ref{sec:3} to be essential for unitarity, necessitates introducing this contact term in $T_{++}$. This provides the coordinate space interpretation of the introduction of a similar term in momentum space we discussed in Sec. \ref{sec: Gravitational-Lipatov}, necessary to fix the ambiguity in the reconstruction of the gravitational Lipatov vertex from the poles of the $2\to 3$ amplitude. 

Finally, we can similarly work out the expressions for $h_{-i}$ and $h_{--}$: 
\begin{align}
\label{hfinalresult}
h_{-i}^{(2)}(k) &= \frac{\kappa^3 s}{k^2+i\epsilon k^-} \int \frac{d^2\bsq_2}{\left(2\pi\right)^2} C_{-i}(\bsq_1,\bsq_2) \frac{\rho_H}{\bsq_1^2}\frac{\rho_L}{\bsq_2^2} ~,\\[10pt] 
h_{--}^{(2)}(k) &= \frac{\kappa^3 s}{k^2+i\epsilon k^-} \int \frac{d^2\bsq_2}{\left(2\pi\right)^2} C_{--}(\bsq_1,\bsq_2) \frac{\rho_H}{\bsq_1^2}\frac{\rho_L}{\bsq_2^2} ~,
\end{align}
where $C_{-i}$ and $C_{--}$ were provided in Eq.~\eqref{LCLV2} and Eq.~\eqref{LCLV3}  and $s=2\m_H\m_L$ represents the center-of-mass energy squared. These results show that a purely semi-classical approach, directly analogous to the Yang-Mills computations in \cite{Blaizot:2004wu} and \cite{Gelis:2005pt} we reviewed in the previous section, successfully recovers the gravitational Lipatov vertex.

In the following subsections, we will discuss the construction of the propagators of various quantum fields in the background of a gravitational shockwave, which will be relevant for the rapidity renormalization group construction in gravity. 


\subsection{Linearized fluctuations and the gravitational Wilson line}
\label{sec:3b}

Before discussing propagators of various quantum fields, we shall first analyse how a spin-2 wavepacket propagates across a gravitational shockwave. We start by treating the spin-2 wavepacket as a small perturbation about the background metric,
\begin{align}
    g_{\m\n} = \bar{g}_{\m\n}+\kappa\, h_{\m\n}~.
\end{align}
Here the fluctuation field $h_{\m\n}$ is normalized by the coupling $\kappa$ so that its kinetic term is canonically normalized. Working in light cone gauge $h_{\mu+}=0~$, linearizing  Einstein's equations results in  Eq.~\eqref{metricEq}, along with the first order relations in Eq.~\eqref{constraintrel} with all the EM components set to zero. 

The $++$ component of Einstein's equations\footnote{See Appendix B of \cite{Raj:2023irr} for further details.} sets $h\equiv \delta_{ij}h_{ij}=0$. It then follows that the physical spin-2 degrees of freedom are specified by $h_{ij}$, with the other components related to $h_{ij}$ by first-order constraint relations. This corresponds to the fact that the graviton has only two physical polarizations. Eq.~\eqref{metricEq} (after dropping the EM tensor terms) can be solved in the vicinity of $x^-=0$ where the transverse derivatives acting on $h_{ij}$ can be neglected, which gives
\be
\p_-h_{ij}-\frac12\bar{g}_{--}\p_+ h_{ij} =0~,
\ee
whose solution in terms of the initial condition $h_{ij}(x^+,x^-=x^-_0,\boldsymbol{x})$ is simply 
\be
\label{solij}
h_{ij}(x^+,x^-,\boldsymbol{x}) = V(x^-,\boldsymbol{x}) h_{ij}(x^+,x^-=x^-_0,\boldsymbol{x})\,.
\ee 
Here $V$ is the gravitational Wilson line operator given by
\be
\label{VWilson}
V(x^-, \boldsymbol{x}) \equiv \exp \(\frac12 \int_{x^-_0}^{x^-} dz^- \bar{g}_{--}(z^-, \boldsymbol{x})\,\p_+\)\,.
\ee
Using the constraint relations, the solution for the other two components can be evaluated as
\begin{align}\label{solim}
h_{-i}(x^-) &= V(x^-)h_{-i}(x_0^-) +\(\p_j V\) \frac{1}{\partial_+} h_{ij}(x_0^-)~,\\[5pt]
\label{solmm}
h_{--}(x^-) &= V(x^-)h_{--}(x_0^-) + 2 \(\p_i V\) \frac{1}{\partial_+} h_{-i}(x_0^-)+\(\p_i\p_j V\) \frac{1}{\partial_+^2} h_{ij}(x_0^-)~.
\end{align}
The solutions in Eqs.~\eqref{solij}, \eqref{solim} and \eqref{solmm} glue together formulae that connect plane wave evolution from one side of the shockwave to plane wave evolution on the other side. The gravitational Wilson line operator $V$ appearing in these formulas are shift operators that act along the shockwave whose magnitude is a function of the energy of the shockwave and its transverse distribution. 

These equations are analogous to equations in the Yang-Mills case that were used to obtain the QCD Lipatov vertex in \cite{Gelis:2005pt}. Further, these results admit a ``classical" double copy to the gauge theory formulas. In particular, the gauge theory Wilson line and the gravitational Wilson line are related by the color-kinematic replacement rule. This can be seen by comparing Eq.~\eqref{Wilson-Line-QCD} with Eq.~\eqref{VWilson}, where we observe that the color factor $T^a\to \p_+$ and the classical gauge field (of the gluon shockwave) is replaced by the classical gravitational field (of the gravitational shockwave). The factor of $1/2$ is explained as follows. Recall from footnote \ref{BCJ-footnote} that the double copy specifies the gauge coupling $g$ is replaced by $\kappa/2$. This replacement is applicable when the fields are canonically normalized, following which, the respective couplings appear in the classical solutions linearly.

\subsection{Propagators in gravitational shockwave backgrounds}
\label{sec:GR-propagators}
In this subsection, we present results for the retarded propagators of various quantum fields within gravitational shockwave background. Furthermore, we draw a comparison between these results and those obtained for propagators in a gluon shockwave background, revealing a straightforward connection via a color-kinematic replacement. For further details we refer to \cite{Raj:2024xsi}. 

Recalling the definition of the retarded propagator given in Sec.~\ref{sec:CGC-propagators} (in Eq.~\eqref{retarded-gf-0}), we start by adapting the convolution formula Eq.~\eqref{G-conv-0} for the retarded scalar Green's function to the shockwave spacetime 
\begin{align}
\begin{split}
G_R(x,y) &= \int d^4z~ d^4w ~G_R(x,z) \,\delta(z^{-}-z_0^-)\,2\,\p^-_z G_R(z,w) \delta(w^{-}-w^{-}_0)\,2\,\p^-_w \,G_R(w,y)~.\no
\end{split}
\end{align}
With $\delta>0$, we set $z^-_0=\delta$ and $w^-_0=0$. $\delta$ is introduced to give an infinitesimal width to the gravitational shockwave in the $x^-$ direction and will eventually be set to zero. 

Our focus here is on the off-shell propagation of quantum fields through a shockwave. Therefore, we consider the regions where $x^->\delta$ and $y^-<0$. Since the regions $(y^-, 0)$ and $(\delta, x^-)$ have a trivial background, in these regions, we can effectively replace the full retarded propagator $G_R(x,z)$ with the free propagator $G^0_R(x,z)$ in the convolution formula
\begin{align}
\label{G-conv-1}
G_R(x,y) =\int d^4z ~d^4w ~G_R^0(x,z)\, \delta(z^{-}-\delta)\,2\,\p^-_z \,G_R(z,w) \delta(w^{-})2\p^-_w \,G_R^0(w,y)~.
\end{align}
We now need to just determine $G_R(z,w)$ using its definition in Equation $\eqref{retarded-gf-0}$. 

Following a method similar to that outlined in the previous section we start with the solution of the small scalar fluctuation equation around the gravitational shockwave background. This was previously addressed in \cite{Raj:2023irr}, with the solution being
\be
\label{scalar-full-solution}
\phi_k(x) = \Theta(- x^-)\, e^{-ikx} + \Theta(x^-)\, e^{-ikx} \,U_k(\bsx)~,
\ee
where the phase $U_k(\bsx)$ is defined as:

\be
\label{grav-WL}
U_k(\bsx) = \exp \bigg(i\kappa^2 \mu \frac{\rho(\bsx)}{\square_{\perp}}k^- \bigg)~.
\ee
Plugging this result into the definition of the Green's function, with the conditions $x^->\delta$ and $y^-<0$ gives us
\be
\label{nontrivial-prop}
G_R(x,y) = -\int \frac{d^4k}{\left(2\pi\right)^4} \frac{e^{-ik(x-y)}}{k^2 +ik^-\epsilon} U_k(\bsx)~.
\ee
The result of this expression in the limit $\delta\to 0$ is
\be
\label{temp1}
\lim_{\delta\to 0 } G_R(x^-=\delta,y^-=0) = \frac12 \,e^{f\p_{y^+}}\, \Theta(x^+-y^+) \,\delta^{(2)}(\bsx-\bsy)~,
\ee
where the function $f$ is defined as $f(\bsx) = \kappa^2 \mu \frac{\rho(\bsx)}{\square_{\perp}}~.$ Substituting the result from Eq. \eqref{temp1} into Eq. $\eqref{G-conv-1}$ and performing the $w$-integral, results in the following expression
\be
G_R(x,y) = G_R^0(x,y) + \int d^4z~ G_R^0(x,z) \left(e^{-f(\bsz)\p_{z^+}}-1\right)\delta(z^-) 2\p_{z^+} G_R^0(z,y)~.
\ee
The identity term is separated from $e^{-f(\bsz)\p_{z^+}}$ which yields the full result for the retarded propagator as a free propagator plus an interacting part which captures all order interactions with the shockwave. Finally, we can perform the Fourier transformation and obtain the propagator in momentum space
\be
\label{scalar-prop-mom}
\tilde{G}_R\left(p, p^{\prime}\right)=\tilde{G}_R^0(p)(2 \pi)^4 \delta^{(4)}\left(p-p^{\prime}\right)+\tilde{G}_R^0(p) \mathcal{T}\left(p, p^{\prime}\right) \tilde{G}_R^0\left(p^{\prime}\right)~.
\ee
where the effective vertex $\mathcal{T}$ is given by (see also \cite{deGioia:2022fcn})
\be
\label{T-matrix}
\mathcal{T}(p,p') = - 4\pi i (p')^- \delta(p^- -(p')^-) \int d^2\bsz ~e^{i(\bsp-\bsp')\cdot \bsz} \left(e^{i f(\bsz)p'_+}-1\right)~,
\ee
and $G_R^0(p)$ is specified in Eq.~\eqref{free-G0-prop}.

Just as in the gauge theory case, the form of the shockwave propagator in Eq.~\eqref{scalar-prop-mom} clearly illustrates the physical interpretation of multiple scatterings off the shockwave. The first term accounts for the case of no interaction, while the $\mathcal{T}(p,p')$ function, which resums all such interactions into an exponential, captures the multiple interactions with the shockwave (see the pictorial representation in Fig.~\ref{gluon-shockwavepropagator}). The form of the propagator in Eq.~\eqref{scalar-prop-mom} is also what appears in the case plane-wave backgrounds that have been discussed in the literature long ago in \cite{Gibbons:1975jb} and more recently in \cite{Adamo:2022qci, Adamo:2024oxy}.

We now move on to the propagator of a spin-2 particle in a gravitational shockwave background. The retarded Green function for a spin-2 particle is (the subscript $R$ is dropped below to avoid clutter in the formulas)
\begin{align}
\label{gravPropDef1}
    G_{\m\n\rho\sigma}(x, y)=  -\int \frac{d^4k}{\(2\pi\)^4} \frac{1}{k^2+i\epsilon k^-}\sum_{\lambda\lambda'} h_{\m\n, k}^{(\lambda\lambda')}(x)h_{\rho\sigma, k}^{*(\lambda\lambda')}(y)~.
\end{align}
In flat space, the spin-2 wave packet $h_{\m\n, k}^{(\lambda\lambda')}(x)$ is given by $h_{\m\n, k}^{(\lambda\lambda')}(x) = \epsilon^{(\lambda\lambda')}_{\m\n}(k) e^{-ikx}$ where $\epsilon^{(\lambda\lambda')}_{\m\n}$ is the graviton polarization tensor which satisfies the completeness relation
\be
\label{polSum1}
\sum_{\lambda\lambda'} \epsilon^{(\lambda\lambda')}_{\m\n}(k) \epsilon^{*(\lambda\lambda')}_{\rho\sigma}(k) =\frac12 \(\Lambda_{\m\rho}\Lambda_{\n\sigma}+\Lambda_{\m\sigma}\Lambda_{\n\rho}-\Lambda_{\m\n}\Lambda_{\rho\sigma} \)~, \qquad \Lambda_{\m\n} = \eta_{\m\n} -\frac{n_\m k_\n + n_\n k_\m}{n\cdot k}~.
\ee
Here $n^\m$ is an arbitrary null vector. This expression can be derived by using its relation to the spin-1 polarization vector ($\epsilon_\mu^{(\lambda)}$) which is given by
\be
\label{polarization-double-copy}
\epsilon_{\m\n}^{(\lambda\lambda')} = \frac{1}{2}\(\epsilon_\mu^{(\lambda)} \epsilon_\nu^{*(\lambda')}+\epsilon_\nu^{(\lambda)} \epsilon_\mu^{*(\lambda')} - \epsilon_\mu^{(\omega)} \epsilon_\nu^{*(\omega)}\delta^{\lambda\lambda'}\)~.
\ee
The spin-2 polarization tensor thus constructed satisfies transversality and tracelessness condition $k^\m\epsilon_{\m\n}^{(\lambda\lambda')}=\eta^{\m\n}\epsilon_{\m\n}^{(\lambda\lambda')} =0$ which follows by making use of $k^\mu \epsilon_\m^{(\lambda)}=0$ and $\eta^{\m\n}\epsilon_\mu^{(\lambda)} \epsilon_\nu^{*(\lambda')}=
\delta^{(\lambda\lambda')}$. Further, the graviton polarization is automatically in light cone gauge if the photon polarization vector satisfies the light cone gauge condition: $n^\mu\epsilon_\m^{(\lambda)}=0$ for a null vector $n^\mu$. The result in Eq.~\eqref{polSum1} can be straightforwardly derived using the completeness relation of the photon polarization vectors $\sum_\lambda \epsilon_\mu^{(\lambda)}(k)\left(\epsilon_\nu^{(\lambda)}(k)\right)^*=\eta_{\mu \nu}-\frac{k_\mu n_\nu+k_\nu n_\mu}{k \cdot n}$. Substituting the polarization sum in Eq.~\eqref{polSum1} into Eq.~\eqref{gravPropDef1}, we obtain the following result for the free graviton propagator
\begin{align}
\label{free-graviton-propagator}
    G^0_{\m\n\rho\sigma}(x, y) &= -\frac12 \int \frac{d^4k}{\(2\pi\)^4} \frac{\(\Lambda_{\m\rho}\Lambda_{\n\sigma}+\Lambda_{\m\sigma}\Lambda_{\n\rho}-\Lambda_{\m\n}\Lambda_{\rho\sigma} \)}{k^2+ik^-\epsilon} e^{-ik(x-y)} \no\\[5pt]
    &\equiv \int \frac{d^4k}{\(2\pi\)^4} \tilde{G}_{\m\n\alpha\beta}^0(k) e^{-ik(x-y)}~.
\end{align}
Similar to the scalar case, this result for the free graviton propagator can then be used to calculate the graviton propagator in a gravitational shockwave background. We present this result below in the eikonal approximation and refer to \cite{Raj:2024xsi} for details:
\begin{align}
\label{gravitonPropPosFinal2}
\tilde{G}_{\m\n\rho\sigma}(p,p') = \tilde{G}_{\m\n\rho\sigma}^0(p) (2\pi)^4\delta^{(4)}(p-p')+ \tilde{G}_{\m\n\alpha\beta}^0(p) \mathcal{T}^{\a\b\gamma\delta}(p,p')\tilde{G}_{\gamma\delta\rho\sigma}^0(p')~,
\end{align}
where the effective vertex $\mathcal{T}_{\m\n\rho\sigma}$ (or T-matrix in the language of scattering amplitudes) is
\begin{align}
\label{T-matrix-graviton}
    \mathcal{T}_{\m\n\rho\sigma}(p,p') =& -\frac12 \(\Lambda_{\m\rho}\Lambda_{\n\sigma}+\Lambda_{\m\sigma}\Lambda_{\n\rho}-\Lambda_{\m\n}\Lambda_{\rho\sigma} \)4\pi i (p')^- \delta(p^- -(p')^-) \no\\
    &\times\int d^2\bsz ~e^{i(\bsp-\bsp')\cdot \bsz}  \(e^{i f(\bsz)p'_+}-1\)~.
\end{align}
The function $f$ was defined earlier as $f(\bsx) = \kappa^2 \mu \frac{\rho(\bsx)}{\square_{\perp}}~.$

Having computed the eikonal gravitational shockwave propagator, we can now compare it against the corresponding results in the QCD case. Specifically, for the case of scalar propagators, we will compare Eq.~\eqref{scalar-gauge-prop-mom} against Eq.~\eqref{scalar-prop-mom}. The classical double-copy relationship discussed in Sec. \ref{sec:classical-double-copy} is evident between them, manifest in their respective $\mathcal{T}$ amplitudes. To see this, note that the phase factor in Eq. \eqref{grav-WL}, can be written as the gravitational Wilson line 
\begin{align}
\label{grav-WL-1}
    U_k(\bsx) = P\exp \left(\frac{i}{2} \int d z^{-} g_{--}\left(z^{-}, \bsx\right) k^-\right)~.
\end{align}
This expression can be contrasted with the Wilson line for the colored scalars in Eq. \eqref{eq:QCD-Wilsonline}. Recall from the discussion in Sec.~\ref{sec:classical-double-copy}, the  shockwave background itself exhibits the property $gA_-\to-\frac12 g_{--}$. When the color charge density $\rho(\bsx)$ in QCD case is replaced by the mass density $\mu\,\rho(\bsx)$ in GR, and  $g\to \kappa$ and $T^a\to -k^-$, the QCD Wilson line operator precisely maps onto the gravitational Wilson line operator \cite{Melville:2013qca}. Hence, we see that the double-copy relationship between the scalar propagators Eq.~\eqref{scalar-gauge-prop-mom} and Eq.~\eqref{scalar-prop-mom} is inherited from the classical double-copy of the underlying shockwave backgrounds. Similarly, one can compare the retarded gluon and graviton shockwave propagators in Eq. \eqref{gluonPropPosFinal2} and Eq. \eqref{gravitonPropPosFinal2}, respectively where the same relationship holds \cite{Raj:2024xsi}.

We now provide a brief remark on the sources of sub-eikonal corrections to the shockwave propagators. As demonstrated in Sec. \ref{sec:classical-double-copy}, sub-eikonal corrections to the gluon Lipatov vertex were important for successfully recovering the gravitational Lipatov vertex via the classical double copy prescription in Eq.~\eqref{DCprescription}. Further, we recall from the discussion in Sec.~\ref{sec:3} that the gravitational Lipatov vertex is crucial for subleading classical corrections to the eikonal phase which are suppressed in $R_S^2/b^2$. This suggests that in our above analysis of the shockwave propagators where we omitted terms involving transverse derivatives of the mass distribution $\rho(\bsx)$, one should keep terms upto the first nontrivial order in $R_S^2/b^2$ that would be required for a consistent construction of the rapidity RG. 

For scalar propagators, incorporating these sub-eikonal effects is straightforward. The derivation of the full solution for the scalar field was done in \cite{Raj:2024xsi} where it was found that while the solution takes the form as before in Eq.~\eqref{scalar-full-solution}, the associated gravitational Wilson line in Eq.~\eqref{grav-WL} gets generalized to the following 
\begin{align}
\label{general-wilson-line}
U_k(x^-,\bsx) = \exp i\bigg( k^- f_1(\bsx)+ k^- x^-f_2(\bsx) -x^- k_i f_{3,i}(\bsx) \bigg)~.
\end{align}
Here $f_1$, $f_2$, and $f_{3,i}$ are functions of the transverse coordinate $\bsx$ with the last two containing transverse derivatives of $\rho(\bsx)$ that are $O(1/b)$ suppressed with respect to $f_1$. These are given by 
\begin{align}
f_1(\boldsymbol{x})=\kappa^2 \mu \frac{1}{\square_{\perp}} \rho(\boldsymbol{x}), \quad f_2(\boldsymbol{x})=-\frac{1}{2} \kappa^4 \mu^2 x^{-}\left(\frac{\partial_i}{\square_{\perp}} \rho(\boldsymbol{x})\right)^2, \quad f_{3, i}(\boldsymbol{x})=\kappa^2 \mu x^{-} \frac{\partial_i}{\square_{\perp}} \rho(\boldsymbol{x}) .
\end{align}

The term $f_1$ corresponds to the previously computed eikonal contribution (that was defined as $f(\bsx)$ below Eq.~\eqref{temp1}), while $f_2$ and $f_{3,i}$ account for the sub-eikonal effects. This form for the Wilson line follows from the solution to the geodesic equations in Eq.~\eqref{geodesicSol}. A scalar wave packet propagating through the shockwave follows a geodesic path described by these solutions; the complete solution for small scalar fluctuations can be formulated with the Wilson line in Eq.~\eqref{general-wilson-line}. Following the methodology outlined above, this generalization then allows for the computation of the retarded scalar propagator that  incorporates sub-eikonal terms. The sub-eikonal contributions encoded in $f_2$ and $f_{3,i}$, are suppressed relative to the eikonal term ($f_1$) by powers of $R_S/b$ and $(R_S/b)^2$ respectively (considering $\rho(\bsx)/\square_\perp \sim \log(b)$ for large $b$). 

For graviton propagators, one needs to account for changes to the polarization tensor when considering sub-eikonal corrections. This effect, which is more complicated to derive, was neglected in our prior discussion. A spin-2 wave packet before interacting with the shockwave satisfies $h_{\m\n, k}^{(\lambda\lambda')}(x) = \epsilon^{(\lambda\lambda')}_{\m\n}(k) e^{-ikx}$. To determine how these basis elements evolve after passing through the shockwave, one must transform the spin-2 fluctuations along the null geodesics crossing the shockwave. This can be achieved using the transformation of a rank-two tensor. An examination of the transformed components \cite{Raj:2024xsi} reveals that, when terms with transverse derivatives acting on $\rho(\bsx)$ are neglected, the transverse components of the graviton  remain effectively unchanged. Thus at leading eikonal order the graviton polarization tensor is not modified by scattering from the shockwave. As a result, the calculation of the graviton propagator at leading eikonal order is analogous to that of the scalar propagator;  it will however differ at sub-leading orders.

\begin{figure}[ht]
\centering
\includegraphics[scale=0.75]{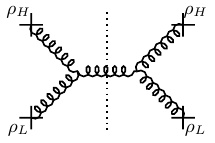}
\caption{Cut vacuum-to-vacuum H diagram contributing at leading order to the observable $\mathcal{O}^{ij}(x,y) \equiv \<A^i(x)A^j(y)\>$. The crosses depict the source densities $\rho_H$ and $\rho_L$.}
\label{classical-field-0}
\end{figure}

\begin{figure}[ht]
\centering
\raisebox{-32pt}{\includegraphics[scale=0.75]{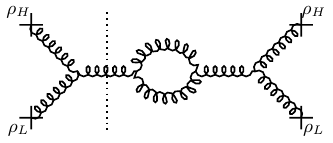}}
+
\raisebox{-32pt}{\includegraphics[scale=0.75]{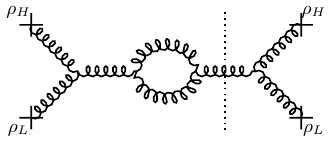}}
+
\raisebox{-32pt}{\includegraphics[scale=0.75]{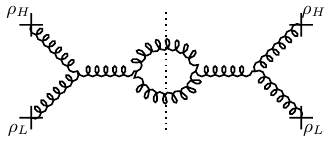}}
\caption{These cut vacuum-vacuum diagrams depict both higher order insertions of the classical field that contribute to a resummation of $(g\rho)^n$ terms as well as next-to-leading order $O(g^2)$ corrections to the leading order contribution shown in Fig.~\ref{classical-field-0}.}
\label{one-loop-classical-field}
\end{figure}

We conclude this section with remarks on corrections to the leading order gravitational wave radiation spectum in shockwave collision computed in \cite{Raj:2023irr}. One of the two NLO contributions  to the inclusive spectrum in the SK formalism is given by
\begin{align}
\langle N\rangle_{\mathrm{NLO}(1)}=\int \frac{d^3 p}{(2 \pi)^3 2 E_p} \int \frac{d^3 q}{(2 \pi)^3 2 E_q}|\mathcal{T}(-q, p)|^2~.
\end{align}
As discussed in \cite{Gelis:2008rw}, this specific contribution corresponds to the cut ``Wightman" propagator ($G_{+-}$). The other NLO contribution stems from the interference between the leading-order result and its one-loop correction. This latter term can be derived from the Feynman propagator ($G_{++}$). The leading-order cut H-diagram is  depicted in Fig.~\ref{classical-field-0}, while the NLO contributions are illustrated in Fig.~\ref{one-loop-classical-field}.

Interestingly, as demonstrated for gluon radiation in shockwave scattering within the CGC EFT, the NLO contributions to the radiation spectrum can be framed entirely as an initial value problem. The key quantity in this approach is the retarded dressed Green function in the shockwave background. The computation of the NLO contributions should follow analogously to the QCD case, as detailed in Secs.~\ref{sec:CGC-propagators} and \ref{sec:glasma}. Note that as suggested by the rightmost figure in Fig.~\ref{one-loop-classical-field}, this contribution corresponds to pair production in a time-dependent strong field background; this can be understood as Hawking radiation for $b> R_S$ in shockwave collisions~\cite{Hartle:1976tp}.

\subsection{Multi-particle radiation in shockwave collisions}
\label{sec:Raychaudhuri}

In Sec.~\ref{sec:CGC}, we discussed the shockwave formalism in the Regge asymptotics of QCD in the context of deeply inelastic collisions and in high energy hadron-hadron collisions. We discussed in particular the CGC EFT, where static color sources are separated by a scale $\Lambda^+$ (whose logarithm relative to  the beam momentum $P^+$ is the rapidity of interest) from dynamical gauge fields. Since physical quantities must be independent of this scale, this requirement lends itself to an RG in rapidity corresponding to the Balitsky-JIMWLK equations for n-point Wilson line correlators. As noted, one recovers the BFKL equation for the 2-point ``dipole" correlator in the low-density limit of these equations. These equations have a nontrivial IR fixed point, that correspond to the formation of an overoccupied classical lump characterized by an energy (or $x$) dependent emergent ``saturation scale" $Q_S(x)$. 

We explored $2\rightarrow n$ scattering in the emergent shockwave description; at high occupancies, the leading contribution is from  $t$-channel fractionation (coherent multiple scattering);  subsequent $s$-channel radiation (inelastic particle production) is parametrically suppressed at low occupancies but becomes large with increasing occupancy of the sources.
Specifically, gluon emissions appear formally at NLO in $\alpha_S$; however, there are large contributions from the phase space at small $x$, with $(\alpha_S\ln(1/x))^n$ contributions to all orders in perturbation theory. This is the BFKL regime of multi-particle production.  However as we saw in Sec.~\ref{sec:glasma}, when the sources emitting radiation have large occupancies $O(1/g)$, the power counting is significantly modified; multi-particle production occurs from $s$-channel cuts, and can be described in a semi-classical framework.

Given the strong similarities between this EFT formalism and $2\rightarrow n$ scattering in gravity, articulated in Sec.~\ref{sec:3}, and in this section, one may ask whether a similar RG can be formulated in gravity, whose nontrivial fixed point is a black hole with the emergent scale $R_S\sim 1/Q_S$; indeed, it is speculated that these overoccupied states in the two theories have universal properties~\cite{Dvali:2021ooc}. In this sub-section, we will outline some  lessons from this double copy that may be relevant for strong field gravity. We emphasize that this discussion is less grounded than those in previous sub-sections and should be understood as a collection of ideas resulting from this correspondence, and their connections to other approaches, that can and should be explored further.

In gravity, the leading contribution is also from coherent multiple scattering of gravitons. While gravitational BFKL contributes to multi-graviton emission, the kinematic window for such contributions is much smaller than in QCD.  Rather, as discussed in Sec.~\ref{sec:classicalization}, the initial states emitting the (relatively) hard radiation\footnote{The typical momentum is set by saturation scale $Q_S \sim \sqrt{s}/n\sim 1/R_S$, which also sets the temperature scale of Hawking radiation. The power spectrum of the radiation will however be very broad with a significant hard tail extending out to the energies of the colliding shockwaves.}
can be described as classical coherent states of softer gravitons. Just as in the CGC EFT, emissions from these classical sources are what dominate inelastic multi-graviton production as $b\rightarrow R_S$. Likewise, to compute expectation values of the full density matrix, one has to average over all possible configurations of such classical coherent sources\footnote{For an interesting related discussion in the ACV framework, see \cite{Ademollo:1989ag,Ademollo:1990sd,Ciafaloni:2009in,Ciafaloni:2018uwe}. }. This averaging entangles the hard emitted radiation with soft radiation (forming the coherent states)~\cite{Strominger:2017aeh,Carney:2017oxp}.  Thus similarly to the CGC EFT one can in the GR case begin with an initial condition which is a boosted overoccupied mass distribution, corresponding to a stochastic distribution\footnote{A similar distribution ${\bar W}_{\rm GR} [\rho_L]$ can be understood to correspond to the graviton cloud represented by $\rho_L$. } of configurations ${\bar W}_{\rm GR}[\rho_H]$. For reviews of earlier work on classical-statistical configurations in the GR context, see \cite{Hu:1999mm,Hu:2020luk}, and references therein; 
for a recent discussion, see \cite{Chawla:2021lop}.

For a fixed distribution of sources $\rho_L$ and $\rho_H$, multi-particle production can be determined from Cutkosky's rules in strong time-dependent fields, as discussed previously in Sec.~\ref{sec:glasma}. This corresponds to the combinatorics of cut and uncut graphs in the presence of the strong sources. The nontrivial upshot of this discussion is a very simple result~\cite{Gelis:2006yv,Gelis:2006ye,Dumitru:2008wn,Gelis:2008rw,Gelis:2008ad,Gelis:2009wh}, which at first blush, should also apply to gravity: for n-particle production, for a fixed distribution of sources, it is given by Eq.~\eqref{hijfinalresult} containing the gravitational Lipatov vertex, with the n-particle distribution represented as 
\begin{align}
\label{eq:n-part-GR-glasma}
    \left \langle \frac{d^n N}{d^2 k_{\perp 1} d y_1 \cdots d^2 k_{\perp n} dy_n}\right \rangle = \int [d\rho_L] [d\rho_H] {\bar W}_{\rm GR}[\rho_L]\,{\bar W}_{\rm GR}[\rho_H]\, \frac{dN}{d^2 k_{\perp 1} dy_1}[\rho_L,\rho_H]\cdots \frac{dN}{d^2 k_{\perp 1} dy_1}[\rho_L,\rho_H]\,,
\end{align}
where 
\begin{eqnarray}
\label{eq:1-part-GR-glasma}
    \frac{dN}{d^2 k_\perp dy}[\rho_L,\rho_H] = \frac{1}{(2\pi)^6}|k^2 \tilde{h}_{ij}^{(2)}(k)|^2\,.
\end{eqnarray}
From the structure of Eq.~\eqref{hijfinalresult}, we see immediately that this expression is proportional to the gauge-invariant scalar product of the gravitational Lipatov vertex. 

\begin{figure}[ht]
    \centering    
    \includegraphics[]{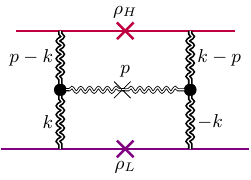}
    \caption{Illustration of dilute-dilute single inclusive graviton production with classical fields/reggeized gravitons (dark curly lines) and the Lipatov vertex (black blobs).  This contribution is the imaginary part of a two-loop Feynman diagram (the H-diagram), with the crosses representing the on-shell final states.}
    \label{ZL-figure84}
\end{figure}

\begin{figure}[ht]
    \centering    
    \includegraphics[]{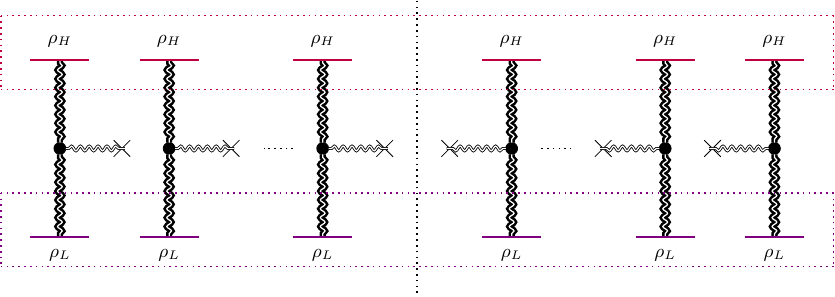}
    \caption{Illustration of dilute-dilute inclusive multi-graviton production representing Eq.~\eqref{eq:n-part-GR-glasma}. The dashed red lines represent the stochastic averaging over coherent sources.} 
    \label{ZL-figure85}
\end{figure}

Further, one can also understand Eq.~\eqref{eq:1-part-GR-glasma} as the inelastic piece of the H-diagram (in the language of ACV~\cite{Amati:1990xe}), and Eq.~\eqref{eq:n-part-GR-glasma}, as cuts of n-such ``horizontal" H-ladders. As noted in Sec.~\ref{sec:glasma}, these correspond to a negative binomial distribution, which interpolates between a Poisson distribution and a Bose-Einstein distribution~\cite{Gelis:2009wh}. Not least, in the strong field limit, Eq.~\eqref{eq:1-part-GR-glasma} is parametrically of order $1/\lambda_{\rm GR}$. This is the largest contribution at high occupancies; in this case, the multi-particle distribution, as in a laser, is closer to the Bose-Einstein distribution. 

We note in conclusion that the above construction can be simply mapped to the discussion in the introduction of Sec.~\ref{sec:3} on the expansion ({\it a la} ACV) of the $S$
-matrix in terms of the phase shifts $\delta= \delta_0+\delta_1+\delta_2+\cdots$. As observed there, $\delta_0$ is purely real arising from the (eikonal) exponentiation of multiple exchanges of classical fields (reggeized gravitons) between the coherent state (classical) sources. The next contribution $\delta_1$ in the power counting is analogous to the QCD Feynman graphs in  Fig.~\ref{leading-virtual}. Because of the closed loop, this contribution is sensitive to the Planck scale, which suppresses it by $l_{\rm Planck}^2/b^2$ at fixed impact parameter. For $\delta_2$, at two-loops, one obtains the first absorptive (inelastic) contribution from the H-diagram in Fig.~\ref{ZL-figure84}. This can be understood in reverse as resulting from the integration over the phase space of the emitted graviton; in our language, this would be the integral over the bremsstrahlung phase factor $\int \frac{d^2k_\perp}{k_\perp^2}\frac{dx}{x}$, whose exponentiation will correspond to the double log in Eq.~\eqref{delta-2}. Since it is manifest in our approach that this is a classical contribution, the impact parameter integral must be cut off at the scale $R_S$ of the stochastic sources. Eq.~\eqref{eq:n-part-GR-glasma} further tells us that there must be such classical absorptive contributions at post-Minkowsian order $O(G^p)$, where $p=3,5,\cdots$, each with an additional double log power, the net result of which results in their exponentiation. We will return to this discussion in future work\footnote{We thank Ira Rothstein for illuminating discussions on this topic.}.

\subsubsection{Self-forces and tidal deformations  in shockwave collisions}
\label{sec:tides}

Eqs.~\eqref{eq:n-part-GR-glasma} and \eqref{eq:1-part-GR-glasma}, though providing a useful guide to our thinking, are far from the full story. One  observes for instance that the expressions are  IR divergent in transverse momenta. In the framework outlined, these are regulated on the scale of the source distribution ${\bar W}_{\rm GR}[\rho_H]$; for a black hole, we anticipate this would be of order $k_\perp\sim (R_S)^{-1}$. In other words, they are absorbed into the ``Weinberg coherent states" as we noted previously. 

\begin{figure}[ht]
\centering
\includegraphics[scale=1]{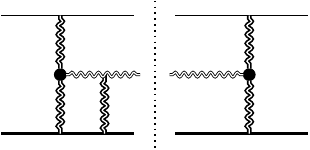}
\caption{Illustration of the rescattering of the graviton produced in shockwave collision with a reggeized graviton in the dilute-dense power counting.}
\label{grav-LPM-effect}
\end{figure}

More importantly, our earlier discussion was for dilute-dilute scattering, valid for large impact parameters. 
When one considers radiation closer to the source with the larger mass density ($\rho_H$), higher order ($R_S/b$) contributions in our expansion of ${\tilde h}_{ij}$ in Eq.~\eqref{eq:dilute-dense-expansion} become important. One of the diagrams illustrating such a correction is shown in Fig.~\ref{grav-LPM-effect}.
To begin to address this, we observe that the conditions under which Eq.~\eqref{hijfinalresult} was derived change significantly as occupancy increases with decreasing impact parameter. The point particle approximation to the energy momentum tensor post-collision in Eq.~~\eqref{PPEMT} to the geodesic equations is no longer robust.  This will modify Eqs.~\eqref{geodesicEq}-\eqref{EMlower}, and therefore the equations of motion in Sec.~\ref{sec:EOM}. 

An interesting question  is whether the structure of Eq.~\eqref{eq:n-part-GR-glasma}, with the Lipatov vertex and reggeized gravitons as building blocks, will be preserved. In QCD, this is still the case, with the modified expression for the gauge field given in Eq.~\eqref{dilutedenseQCD}, with the rescattering contributions from the metric  exponentiated to all orders into a Wilson line. The corresponding question in gravity is whether Eq.~\eqref{hijfinalresult} is similarly modified by the gravitational Wilson line in Eq.~\eqref{VWilson}. There are however differences in the QCD and gravity cases that make this simple double copy replacement unlikely. Firstly, as we saw even in our discussion of the dilute-dilute case in Sec.~\ref{sec:EOM}, the  r.h.s of Eq.~\eqref{metricEq} contains contributions from the stress-tensor that are formally sub-eikonal (in $t/s$) 
relative to the leading $T_{++}$ term; this terms are necessary to obtain a closed form for the equations of motion.
(This was prefigured  in the classical double copy discussion in Sec.~\ref{sec:classical-double-copy} where similarly one observes that sub-eikonal contributions have to be retained to recover the gravitational Lipatov vertex.)

A further key point is that in going from the dilute-dilute to dilute-dense approximation with decreasing impact parameter is that one has to simultaneously consider both so-called ``self-force" and ``tidal deformation" contributions\footnote{For a nice complementary discussion of these within a worldline EFT~\cite{Goldberger:2004jt}, with similar conclusions, see \cite{Barack:2018yvs,Barack:2023oqp}. Other EFT based approaches include \cite{Galley:2006gs, Galley:2008ih,Cheung:2024byb}. For discussions in the language of scattering amplitudes, see \cite{Kosmopoulos:2023bwc}. A key difference of these works from ours is that we are working not at finite boost but in the shockwave $\gamma\rightarrow \infty$ limit. In this limit, the computation of self-force and tidal effects may simplify considerably~\cite{Galley:2013eba}. }. As we noted earlier, the dilute-dense approximation corresponds to keeping $\rho_L/\nabla_\perp^2\ll 1$ and $\rho_H/\nabla_\perp^2\sim O(1)$. In most black hole merger scenarios, both $\rho_L/\nabla_\perp^2$ and $\rho_H/\nabla_\perp^2$ are becoming large simultaneously with decreasing impact parameter so one is going from dilute-dilute to dense-dense kinematics rapidly. As we discussed in Sec.~\ref{sec:Gluon-shockwaves},  the latter is only tractable numerically in QCD; the situation in this regard is even more severe in gravity. It is therefore convenient to work in a dilute-dense limit corresponding to the case where the gravitational radiation being emitted close to the heavy-mass compact object and far from the light one\footnote{Besides theoretical interest, this kinematics may be relevant for extreme inspiral ratios or for hyperbolic encounters of light black holes off heavy ones~\cite{Escriva:2022duf}.}. This is because the geodesic deviations of the light sources are small but that of the radiated graviton are large, providing potential analytical insight into the dynamics of the ``chirp", namely, the critical impact parameter where gravitational radiation ceases~\cite{Page:2022bem}. 

To obtain the closed form of equations for $h_{\mu\nu}^{(3)}$,  the tidal correction illustrated in Fig.~\ref{grav-LPM-effect} (at $O(\rho_L \rho_H^2)$) will induce a time-dependence to the static sources, which in turn needs to be accounted for in null geodesic equations. The multipoles of this deformation correspond to so-called dynamical Love numbers; an elegant discussion of these within the worldline EFT framework is given in \cite{Ivanov:2024sds}. An important constraint in the shockwave formalism would be to match these Love numbers to those obtained in the EFT framework at finite boost. We will shortly outline the modifications to the geodesic equations in our context.  

Before we do so, we briefly mention a potentially interesting feature of the coherent rescattering illustrated in Fig.~\ref{grav-LPM-effect} (which does appear to have been discussed previously in the GR literature) is the gravitational variant of the Landau-Pomeranchuk-Migdal (LPM) effect~\cite{Landau:1965ksp,Migdal:1956tc,Baier:1996vi}, originally  discovered in QED. In the usual Bethe-Heitler description of energy loss in QED media, the radiative amplitude is simply the sum over the radiative amplitude from all the medium charges. However if the energy of the emitted photon is small, there is a characteristic ``formation time" where the photon is collinear to the fast charge. If the mean free path between the scattering centers is shorter than this time scale the relative phase between the multiple scatterings matters, leading to a destructive interference contribution in the radiative cross-section relative to the Bethe-Heitler result. This is the LPM effect. 

In QCD, a similar analysis applies but one has to take the rescattering of the emitted gluons into account; one obtains a qualitative change in the power spectrum of the emitted radiation relative to the QED case~\cite{Baier:2000mf}. As shown in explicitly in \cite{Baier:1996kr}, the analysis corresponds to taking into account multiple scattering corrections similarly to 
Fig.~\ref{dilute-dense} in the QCD case, but further taking the relative phases in the sum of the individual radiative amplitudes into account. These provide the LPM destructive interference; they are sub-eikonal but relevant for finite boost and a large number of scattering centers. Given the strong similarities between the QCD and gravity analyses of radiation in the respective (weakly coupled) strong field regimes, one anticipates a similar LPM contribution to be relevant to gravitational radiation when tidal forces become important. As in QCD, the LPM effect will modify the shape of the power spectrum.   

\subsubsection{Geodesic congruence of wee gravitons and stretched horizons}
The study of geodesics, and families of geodesics (referred to as geodesic congruence) in general relativity, is important for understanding how spacetime curvature influences the motion of test particles and light rays. Two key equations that describe the behavior of geodesics are the geodesic deviation equation, and the Raychaudhuri equation,  which can both be related to the stated necessary modifications to the dilute-dilute framework underlying Eq.~\eqref{eq:n-part-GR-glasma}. These equations are well-known to provide insights into gravitational lensing, singularity theorems, and the structure of spacetime \cite{Wald:1984rg, Witten:2019qhl}.

For multi-particle $2\rightarrow n$ scattering in the regime of high occupancies, the Raychaudhuri equation will be especially relevant. It governs the evolution of the expansion along a congruence of geodesics
\begin{equation}
\label{raychaudhuri-eq}
\frac{d\theta}{d\lambda} = -\frac{1}{2} \theta^2 - \sigma_{\m\n} \sigma^{\m\n} + \omega_{\m\n} \omega^{\m\n} - R_{\m\n} \xi^\m \xi^\n.
\end{equation}
The l.h.s of this equation defines the change in $\theta$, the expansion scalar, with the variation in the affine parameter $\lambda$ introduced previously in the context of Eq.~\eqref{EMlower} and the solution of Eq.~\eqref{geodesicEq}. The expansion scalar $\theta$ is defined as the trace of the null extrinsic curvature, $\theta = \nabla_\mu \xi^\mu$, where $\xi^\mu$ is the tangent vector field to the null congruence. On the r.h.s, in addition to $\theta$,  $\sigma_{\mu\nu}$ is the shear tensor, $\omega_{\mu\nu}$ the vorticity, and $R_{\m\n}$ the Ricci curvature tensor of the spacetime. Detailed expressions for these will not interest us here - we refer the reader to \cite{Wald:1984rg, Poisson:2009pwt}. 

The physical interpretation of the l.h.s is  that of the logarithmic derivative of the area element of a congruence of geodesics with respect to the affine parameter \cite{Poisson:2009pwt}, while the r.h.s represents the forces whose action distorts this area. The focusing theorem states that when the vorticity of the congruence is zero, and the spacetime satisfies the null energy condition ($T_{\m\n} \xi^\m \xi^\n \geq 0$ for arbitrary, future-directed vectors $\xi^\mu$), the expansion scalar does not increase at all regular points of the geodesic congruence.

In the shockwave spacetime\footnote{Since this refers to a single shockwave, we will refer to the mass distribution generically as $\rho$, rather than $\rho_{L,H}$ in the collision. } in Eq.~\eqref{denseBgnd1}, the null geodesic congruence is generated by the null vector field  $\xi \equiv \xi^\m\p_\m$:
\begin{align}
\label{vector-field}
    \xi = \p_{-} + \[-\kappa^2\mu\delta(x^-) \frac{\rho(\bsx)}{\nabla_\perp^2} + \frac{\kappa^4 \mu^2}{2}\Theta(x^-) \(\frac{\p_i\rho(\bsx)}{\nabla_\perp^2}\)^2\]\p_{+} + \[-  \kappa^2\mu \Theta(x^-) \frac{\p_i\rho(\bsx)}{\nabla_\perp^2}\]\p_{i}~.
\end{align}
The components of this vector field satisfy the null condition $k_\mu k^\mu = 0$, and the expansion scalar $\theta = \nabla_\mu\xi^\mu$ can be explicitly evaluated to be
\begin{align}
    \theta = -\kappa^2\mu\Theta(x^-) \rho(\bsx)~.
\end{align}
We see that this quantity undergoes a discontinuous jump upon crossing the shockwave at $\lambda = x^- = 0$, seen previously in the solution of Eq.~\eqref{geodesicSol}. The magnitude of this discontinuity is proportional to the transverse distribution $\rho(\bsx)$ that measures the transverse area spanned by the congruence as discussed above.

On can check the validity of the Raychaudhuri equation  for a null congruence in shockwave spacetime. This is most easily done by first defining the tensor 
\be
B_{\m\n} = \nabla_\m \xi_\n~,
\ee
and noting that the first three terms on the r.h.s of Eq.~\eqref{raychaudhuri-eq} can be packaged together as~\cite{Poisson:2009pwt}:
\begin{align}
    B^{\m\n}B_{\m\n} = \frac{1}{2} \theta^2 + \sigma^{\m\n}\sigma_{\m\n} -\omega^{\m\n}\omega_{\m\n}~,
\end{align}
allowing us to express Eq.~\eqref{raychaudhuri-eq} as
\begin{align}
\label{RI-1}
    \frac{d\theta}{d\lambda} = -B^{\m\n}B_{\m\n} - R_{\m\n}\xi^\m\xi^\n~.
\end{align}
For the spacetime in Eq.~\eqref{denseBgnd1}, the only nonvanishing component of the Ricci tensor is $R_{--} = \kappa^2 \mu \delta(x^-) \rho(\bsx).$ The quantities on the right side of Eq.~\eqref{RI-1} are evaluated to be
\begin{align}
    B^{\m\n}B_{\m\n} &= \kappa^4\mu^2\Theta(x^-)\(\p_i\p_j\frac{\rho(\bsx)}{\nabla_\perp^2}\)\(\p_i\p_j\frac{\rho(\bsx)}{\nabla_\perp^2}\)~,\\[5pt]
    R_{\m\n}\xi^\m\xi^\n &= \kappa^2 \mu \delta(x^-) \rho(\bsx)~.
\end{align}
To verify the Raychaudhuri equation for the expansion scalar in Eq.~\eqref{RI-1}, we need to evaluate the total derivative on the left side of Eq.~\eqref{RI-1}:
\begin{align}
    \frac{d\theta}{d\lambda} &= \frac{\p x^-}{\p\lambda}\frac{\p\theta}{\p x^-} + \frac{\p x^i}{\p\lambda}\frac{\p\theta}{\p x^i}~,\no\\[5pt]
    &= -\kappa^2\mu\delta(x^-) \rho(\bsx) + \(-  \kappa^2\mu \Theta(x^-) \frac{\p_i\rho(\bsx)}{\nabla_\perp^2}\)\(\kappa^2\mu\Theta(x^-) \p_i\rho(\bsx)\)~,\no\\[5pt]
    &= -\kappa^2\mu\delta(x^-) \rho(\bsx) - \kappa^4\mu^2 \Theta(x^-) \frac{\p_i\rho(\bsx)}{\nabla_\perp^2} \p_i\rho(\bsx)
\end{align}
We see that the first term above cancels against the Ricci curvature term in the Raychaudhuri equation. For the remaining terms, we make use of the identity
\begin{align}
\label{temp-1}
    \(\p_i\p_j\frac{\rho(\bsx)}{\nabla_\perp^2}\)\(\p_i\p_j\frac{\rho(\bsx)}{\nabla_\perp^2}\) = \p_i\[\(\p_j\frac{\rho(\bsx)}{\nabla_\perp^2}\)\(\p_i\p_j\frac{\rho(\bsx)}{\nabla_\perp^2}\)\] - \(\p_j\frac{\rho(\bsx)}{\nabla_\perp^2}\)\p_j\rho(\bsx)~.
\end{align}
This implies that up to a total derivative (in the transverse direction), $B^{\m\n}B_{\m\n}$ can be written as
\begin{align}
    B^{\m\n}B_{\m\n} = -\kappa^4\mu^2\Theta(x^-)\(\p_i\frac{\rho(\bsx)}{\nabla_\perp^2}\)\p_i\rho(\bsx)~,
\end{align}
which cancels against the second term in $d\theta/d\lambda$. Hence the shockwave spacetime with the transverse mass distribution $\rho(\bsx)$ satisfies the Raychaudhuri equation, provided the total derivative term in Eq.~\eqref{temp-1} vanishes. 

For the shockwave collision case of interest, the rapid $s$-channel radiation we outlined generates a large phase space occupancy of wee gravitons that has been conjectured to form a black hole when the phase space occupancy $n\sim 1/\lambda_{\rm GR}(Q_S)$, where $Q_S=\sqrt{s}/n\sim 1/R_S$~\cite{Dvali:2010jz,Dvali:2011aa,Addazi:2016ksu}. As discussed in Sec.~\ref{sec:glasma}, this is completely analogous to the formation of an overoccupied glasma state in ultrarelativistic heavy-ion collisions, where $Q_S$ is the saturation scale. A promising approach to study this process quantitatively in gravity is within the dilute-dense formalism for radiation at $b\approx R_S$. The large phase space occupancy of emitted gravitons suggests that their evolution can be similarly described by the null Raychaudhuri equation. As observed in \cite{Susskind:1994vu}, the overoccupied wee gravitons form a 2-D area close to $R_S$. We now recall from Eq.~\eqref{eq:n-part-GR-glasma} that from the $2\rightarrow n$ perspective, this is a stochastic process with  ${\bar W}_{\rm GR}[\rho_H]$, as remarked in Sec.~\ref{sec:classicalization}, corresponding to the density matrix of the screened UV modes comprising the dynamics at $b\leq R_S$. This setup is reminiscent of the dynamics of wee partons in the CGC EFT, and it would be interesting to explore whether the focusing and absorption of  of wee gravitons\footnote{For a similar discussion of the classical and quantum gravitational phase space in the context of the null Raychaudhuri equation, see \cite{Ciambelli:2025flo}. } induced by the Raychaudhuri equation can similarly allow one to interpret the event horizon as the RG fixed point of their evolution. 

Another relevant comparison is the role of semi-classical quantum fluctuations in the shockwave background. We showed in Sec.~\ref{sec:GR-propagators} that the corresponding Green functions have a double copy structure to their QCD counterparts in Sec.~\ref{sec:CGC-propagators}. In the latter case, the leading ``BFKL logs" are absorbed in the JIMWLK RG evolution of the stochastic weight functionals which satisfy a Fokker-Planck equation in the functional space\footnote{A discussion of such fluctuations in shockwave backgrounds is given in \cite{Verlinde:2022hhs} and it would be interesting to explore whether our discussion can be formulated in this language.}; the double copy (following our discussion in Secs.~\ref{sec:2} and \ref{sec:3}) implies that this is also the case in gravity. Therefore both classical and quantum noise contribute to the single-inclusive and higher factorial moments of the gravitational wave spectrum. It has been shown that quantum noise induces an additional quantum contribution to the null Raychaudhuri equation~\cite{Bak:2023wwo,Cho:2023dmh}; this quantum contribution does not violate its focusing properties and is consistent with a quantum null energy condition~\cite{Bousso:2015mna}. 

Quantum noise is significantly suppressed relative to classical noise. However it has been argued that in scenarios where the stochastic distribution is that of squeezed states in contrast to coherent states, there is a significant enhancement of quantum noise that may be detected\footnote{Amusingly, the application of squeezed vacuum states in gravitational wave interferometers is an an important technique in {\it reducing} the quantum noise that limits these measurements~\cite{Barsotti:2018hvm}.} at future gravitational wave detectors~\cite{Parikh:2020nrd,Parikh:2020kfh,Kanno:2020usf}. Just as the distinction between quantum and classical optics is only manifest in higher point correlators, noise correlations in gravitational wave detectors may also provide a way to distinguish between their quantum versus classical origins~\cite{Parikh:2023zat}. With these improvements, we can be optimistic that future gravitational wave detectors will be able to do so-see for example~\cite{Berti:2015itd,Hild:2010id,Evans:2021gyd,Vermeulen:2024vgl}, and references therein.


\section{Bookends and loose threads}
\label{sec:Future}
These lectures  provide an introduction to $2\to n$ scattering in QCD and gravity in high energy Regge asymptotics. In the QCD case, discussed in Secs.~\ref{sec:2} and \ref{sec:CGC}, we outlined an explicit derivation of the BFKL equation, using dispersive techniques, to all orders in perturbation theory in the leading logarithmic approximation. The emergent building blocks of the $2\to n$ amplitude are nonlocal Lipatov vertices and reggeized propagators. The solution of the BFKL equation shows that the cross-section grows rapidly with energy, accompanied by a slow UV and IR diffusion of transverse momenta with rapidity. The former is cured  by running coupling effects appearing at next-to-leading log accuracy, while infrared diffusion is cured by many-body screening and recombination effects arising from the overpopulation of phase space. 

This gluon saturation phenomena signals the breakdown of the operator product expansion and conventional perturbation theory in the Regge asymptotics of QCD since all-order power corrections contribute~\cite{Mueller:1996hm}. Specifically, this occurs when the exchanged squared momenta  $Q^2\leq Q_S^2(x)$, where $Q_S^2(x)\gg \Lambda_{\rm QCD}^2$ is the dynamical emergent saturation scale. The presence of this large scale indicates that weak coupling methods are applicable, and is quantified in the semi-classical CGC EFT, where the weak coupling expansion is around a strong field vacuum corresponding to large  $\rho\sim O(1/g)$ classical color charge densities. In the shift from the BFKL paradigm to the CGC one, we switched as well from the ``in-out" amplitude formalism for the former to the ``in-in" Schwinger-Keldysh (SK) formalism that more efficiently captures the strongly-correlated  many-body dynamics of the latter. 

In high energy kinematics, the strong classical ``shockwave" field comprised of overoccupied wee gluons is the non-Abelian analog of the Weizs\"{a}cker-Williams equivalent photon field in electrodynamics. However, unlike electrodynamics, the wee gluons are strongly correlated, with their many-body (2-D) dynamics on the celestial sphere described by correlators of  Wilson line ``vertex operators" sourced by color sources that are static on the dynamical time scales of interest. This CGC construction is robust for a large nucleus which acts as a coherent source of large color charge, with the saturation scale 
$Q_S^2\propto A^{1/3}$ in units of $\Lambda_{\rm QCD}^2$. In an RG picture, this scale depends dynamically on the rapidity, since what we call sources or fields depends on the separation between large $x$ and small $x$ degrees of freedom within the EFT. 

The resulting Balitsky-JIMWLK equations describe the RG evolution of the aforementioned rapidity scale-dependent vertex operators that appear in DIS off nuclei at collider energies. In the ``dilute" limit $\rho/k_\perp^2\ll 1$, where $k_\perp$ is transverse momentum of a parton in the dense gluon cloud, the two-point ``dipole" correlator satisfies the BFKL rapidity evolution equation, thereby recovering a key feature of $2\to n$ scattering. Precision computations of DIS processes to next-to-leading order and next-to-leading logarithmic accuracy are now available, which will allow one to systematically test and refine the CGC EFT at the Electron-Ion Collider. 

The CGC classical fields have a one-to-one map to the reggeized fields, and  likewise the quark (gluon) shockwave propagators to quark-quark-reggeon (gluon-gluon-reggeon) propagators, in Lipatov's reggeon field theory. The former is however more versatile because it can be straightforwardly applied to treat multi-particle production in the hadron-hadron, hadron-nucleus and nucleus-nucleus collisions in the language of shockwave scattering. An  important feature of the BFKL $\to$ CGC shift at high occupancies is from $s$-channel to $t$-channel fractionation.  In other words, it represents the change from the fractionation of energy through radiation in a single ``vertical" $t$-channel ladder to fractionation of the overall $t$-channel momentum exchange via multiple $t$-channel exchanges in the ``horizontal" ladder generated in shockwave scatting. 

Multi-particle production from the time-dependent strong fields in shockwave scattering can be computed in the SK formalism by systematic application of Cutkosky's rules for a fixed distribution of sources, followed by averaging over weight functional representing the distribution of sources. In analytical dilute-dilute and dilute-dense limits, the multi-particle spectrum is a negative binomial distribution generalizing the $2\to 3$ emission amplitude controlled by the Lipatov vertex. The so-called AGK cutting  rules  of reggeon field theory are straightforwardly recovered in this framework. In the dense-dense limit of shockwave scattering (appropriate for describing collisions of dense sources of color charge such as heavy-ions), an analytical treatment is no longer feasible. Detailed numerical simulations describe the generation of an overpopulated glasma in these collisions,  and its evolution through turbulent and hydrodynamic attractors to a thermal quark-gluon plasma. The QGP was discovered at RHIC and confirmed in higher energy collisions at the LHC. Quantitative comparisons of theory to experiment implement the shockwave scattering framework outlined here. 

Remarkably, $2\to n$ scattering in gravity has strong mathematical and conceptual correspondence to the QCD case. In Sec.~\ref{sec:3}, we showed that an identical dispersive approach to that developed in Sec.~\ref{sec:2} applies in the multi-Regge asymptotics of gravity, which leads to the gravitational analog of the BFKL equation with the principal elements being the gravitational Lipatov vertices and reggeized $t$-channel propagators. Most strikingly, the gravitational Lipatov vertex can be expressed as the difference of the bilinear double copy of the  QCD Lipatov vertex and that of the photon bremsstrahlung vertex. However, unlike QCD, the regime of applicability of gravitational BFKL is  limited because the dimensionless gravitational coupling is very small, with $\lambda_{\rm GR} \ln(s/|t|)\ll 1$ even at trans-Planckian energies. 

Nevertheless, the elements of the BFKL construction are important because they contain exactly the same structure of real graviton emission and virtual graviton exchanges that comprise the Weinberg soft theorem. Specifically, in the ultrarelativistic regime, the Weinberg emission vertex is the soft limit of the Lipatov vertex, and the likewise, the Sudakov double logs (whose exponentiation is responsible for reggeization) have an identical structure to the Weinberg's result for soft virtual exchanges. These contributions dress the graviton cloud accompanying the incoming asymptotic graviton states in $2\to n$ scattering, replacing them with Faddeev-Kulish coherent states. At trans-Planckian energies, the shockwave comprised of this soft graviton cloud breaks the asymptotic BMS symmetry of large gauge transformations in gravity, a  quantitative consequence of which  is the gravitational memory effect. (In Sec.~\ref{sec:color-memory}, we discussed the analogous color memory effect  manifest in the CGC.) In summary, though BFKL dynamics does not drive inelastic multi-particle production, its effects contribute to the shockwave description of $2\to n$ scattering, in an exactly analogous manner to the CGC shockwave framework in QCD. 

Multi-particle production in gravitational shockwave collisions was discussed in 
Sec.~\ref{sec:GR-shockwave-formalism} in the gravitational dilute-dilute framework developed for QCD. Solutions of the Einstein equations in this set up demonstrate that the gravitational single inclusive spectrum is a classical double copy of its gluon counterpart, with the QCD Lipatov vertex replaced by the gravitational Lipatov vertex. The same result  is arrived by computing gluon radiation emitted by classical color charges in a Wong+classical Yang-Mills set up, and performing a classical double copy replacement; interestingly, to arrive at this result, it is important to keep leading sub-Eikonal terms on the QCD side of the double copy. 

In the ``quantum first" SK construction of the coherent state of wee gravitons, there are a distribution of quantum paths in configuration space that comprise the initial density matrix. Since their energy level 
separations are $1/n$, they are not distinguishable, except on the  asymptotically long time scales that are sensitive to these low frequencies. They are however of relevance in constructing the S-matrix for multi-particle production, as is the case for the weight functional over classical color charges in QCD. They are especially important for a systematic treatment of Hawking radiation, and entangle its spectrum with the many-body wee graviton spectrum making up the initial density matrix. More simply, classical multi-particle production in the dilute-dilute framework follows the same pattern as established in QCD, with a similar combinatorics applicable to the $n$-particle  final state. Unlike QCD, the extension to the dilute-dense scenario involving the rescattering of radiation with reggeized gravitons from the compact dense source is not as simple and requires taking the physics of tidal forces and geodesic congruences into account. However, as in QCD one anticipates that the latter will display collective hydrodynamics behavior resulting in black hole formation. 

There are several loose threads in the bookends to this narrative. For $2\to n$ scattering, the extension of the BFKL (and CGC) frameworks beyond NLLx accuracy is challenging, and will involve taking into account the breakdown of reggeization. In shockwave language, this requires going beyond the classical-statistical framework for multi-particle production, and taking into account the decay of the shockwave. An interesting possibility is to reformulate this problem in terms of the Goldstone modes corresponding to the broken global symmetries in the shockwave construction. More generally, the extension of this multi-particle production framework beyond weak coupling remains challenging despite the many developments in strong coupling holographic approaches. This is especially important since the coupling in collider experiments is not particularly weak in the relevant phase space. 

While the discovery and characterization of an emergent classical regime of QCD is a goal of current and future colliders, the opposite is true in gravity, with a major goal being the discovery of quantum effects in gravity. Despite considerable developments in gravitational wave astronomy, empirical constraints on 
such effects is elusive. As we noted, the statistics of multi-graviton production is non-Poissonian, and likely close to squeezed states comprising a Bose condensate. A better understanding of the interplay between the multi-graviton mechanism we have outlined, the null  Raychaudhuri equation, and rescattering effects as $b\rightarrow R_S$ is a promising approach to arrive at quantitative predictions towards detecting  quantum features of gravity. 

An outstanding loose thread is to connect our  discussion to entanglement and quantum information. This is clearly a vast subject but there are some aspects that are fundamental to the topic of these lectures. A starting point is `t Hooft's postulate of the $S$-matrix formulation of gravity as a rigorous way to approach the apparent puzzle posed by information loss  in black holes. The tools developed here are useful in this regard. The double copy is an especially valuable guide since one can pose similar questions in QCD, where we have a well-defined quantum theory. We have discussed here  how classical states of high occupancy emerge in $2\to n$ scattering, and further, how these wee parton states unitarize the cross-section a fixed impact parameter. It has long been argued that wee partons satisfy the holographic principle, with their information context forming a two-dimensional surface that is accessible in Regge asymptotics. The important development since is the understanding that wee parton correlations are screened on a semi-hard dynamical emergent scale $R_S= 1/Q_S(x)$. 

One naively expects the formation of such ``macroscopic" classical states to be exponentially suppressed in $2\to n$ scattering. The fact they form with unit probability means they are comprised of a large number of microstates which compensate for this exponential suppression~\cite{Dvali:2021ooc}. In other words, they must have a very large entropy. What are these microstates? We understand that they are gapped with gap sizes $\sim Q_S/n$, and are classical only in the sense of $n\rightarrow \infty$. For finite $n\sim 1/\alpha_S$, they must decay on a  characteristic time scale $t_{\rm G}\sim n/Q_S = 1/(\alpha_S Q_S)$. In Regge asymptotics, where $\sqrt{s}\rightarrow \infty$,  $t_{\rm G}\ll 1/\Lambda_{\rm QCD}$; the shockwave decays on time scales much shorter than hadronization time scales. Broken global Poincar\'{e} and color symmetries (of large gauge transformations) are restored on this time scale, with a careful search of correlations in the asymptotic final state necessary to detect the imprints of the metastable classical state. 

From Bekenstein, we know that the entropy of macroscopic configurations of microstates $S \leq 2\pi E R$, where $E$ is the energy contained in a region of radius $R$~\cite{Bekenstein:1973ur}; in our case, $E=n\,Q_S$ and $R=R_S$, which gives 
$S=O(1/\alpha_S)$, since $n=1/\alpha_S$. From this perspective, the number of microstates $e^S= e^{1/\alpha_S}$ compensates for the $e^{-1/\alpha_S}$ suppression one would obtain from the combinatorics of perturbative Feynman diagrams at large $n$. This is often misunderstood to be a fine tuning argument but it is not the case in QCD: We have independent computations of the dynamics of the theory that demonstrate that macroscopic classical states can form with unit probability. A useful way to think about this problem is to extend the CGC EFT and treat the nonperturbative dynamics of the weight functionals $W[\rho]$ in terms of the Goldstone dynamics of the excitations of the condensate, where the Goldstone scale (corresponding to the global symmetries broken by the shockwave) can be estimated to be $f_G=\sqrt{n}\, Q_S$~\cite{Dvali:2021ooc}. The Bekenstein-Hawking area law~\cite{Bekenstein:1973ur,Hawking:1976de} for the entropy in units of this scale is then $S=4\pi R_S^2\,f_G^2 \sim 1/\alpha_S$, recovering our estimate above. 

The double copy suggests a similar argument may hold in gravity, and this has indeed been argued previously to be the case~\cite{Dvali:2011aa}. However, as we have seen, the double copy logic is fraught sooner in gravity than in QCD and the strong field corrections appear sooner as $b\to R_S$. An important challenge would be to match the Goldstone framework of broken global symmetries with RG and null energy conditions informing a microscopic derivation of the Raychaudhuri equation that systematically treats the $1/n$ quantum corrections we alluded to. For promising interesting work in this direction, see \cite{Ciambelli:2024swv}; a discussion of its embedding in the language of complexity theory and quantum information can be found in \cite{Brown:2022kum}.

\section*{Acknowledgments}
\addcontentsline{toc}{section}{Acknowledgments}
The two of us have benefited greatly from useful comments and discussions with  ``large n" helpful colleagues  that have informed these lectures. We would like to thank in particular Tim Adamo, Abhay Ashtekar, P.V. Athira, Juergen Berges, Fabian Bautista, Michal Heller, Simon Caron-Huot, JJ Carrasco, Lance Dixon, Vittorio Del Duca, Leonardo de la Cruz, John Donoghue, Sergey Dubovsky, Gia Dvali, Einan Gardi, Nava Gadddam, Victor Gorbenko, Grisha Korchemsky, Florian Kuhnel, Alok Laddha, Larry McLerran, Tuomas Lappi, Matt Lippert, Hong Liu, A. Manu, Ian Moult, V. P. Nair, Jorge Noronha, Donal O'Connell, Maulik Parikh, Julio Parra-Martinez, Monica Pate, Jan Pawlowski, Siddharth Prabhu, Ana Raclariu, Radu Roiban, Ira Rothstein, Ananda Roy, Bj\"{o}rn Schenke, Andy Strominger and Gabriele Veneziano.  Special thanks are due to  Isabelle Fite and Anna Stasto for their collaboration and insights on aspects of this work. 

This work developed from  lectures delivered by R.V. at the 2024 Crakow School of Physics in Zakopane. He is grateful to Michal Praszalowicz for his organization of this outstanding school, his encouragement in writing up this extended version of the lectures and his patience with the inevitable delay. R.V would also like to thank Martin Beneke and Robert Szafron for inviting him to the exciting program on ``EFT and Multi-Loop Methods for Advancing Precision in Collider and Gravitational Wave Physics" at the Munich Institute for Astro-, Particle and BioPhysics (MIAPbP),  funded by the Deutsche Forschungsgemeinschaft (DFG, German Research Foundation) under Germany´s Excellence Strategy – EXC-2094 – 390783311.

R.V. is supported by the U.S. Department of Energy, Office of Science under contract DE- SC0012704 and within the framework of the SURGE Topical Theory Collaboration. He acknowledges partial support from an LDRD C grant (25-043) from BNL. R.V was also supported at Stony Brook by the Simons Foundation as a co-PI under Award number 994318 (Simons Collaboration on Confinement and QCD Strings). H.R is a Simons Foundation postdoctoral fellow at Stony Brook supported under Award number 994318.


\bibliographystyle{JHEP.bst}
\addcontentsline{toc}{section}{References}
\bibliography{reference}

\end{document}